\documentclass[namedreferences]{solarphysics}

\usepackage[hyperref,optionalrh]{spr-sola-addons} 
\usepackage{graphicx}        
\usepackage{color}           
\usepackage{breakurl}        
\usepackage{rotating}
\usepackage{booktabs}
\usepackage{makecell}
\usepackage{longtable}
\usepackage{geometry}
\usepackage{multirow}
 \usepackage{amssymb}
 \usepackage{bm}
\usepackage{subfig}
\usepackage{lscape}

\newcommand{\speed}[1]{#1 km~s${}^{-1}$}
\newcommand{\accel}[1]{#1 km~s${}^{-2}$}

\newcommand{\fig}[1]{Figure~\ref{#1}}
\newcommand{\rsun}[1]{${#1}\,R_\odot$}
\newcommand{\tbl}[1]{Table~\ref{#1}}

\newcommand\arcsec{\mbox{$^{\prime\prime}$}}

\newcommand{\aap}{    {\it Astron. Astrophys.}}
\newcommand{\aaps}{   {\it Astron. Astrophys. Suppl.}}
\newcommand{\aapr}{   {\it Astron. Astrophys. Rev.}}

\newcommand{\aj}{     {\it Astron. J.}} 
\newcommand{\apj}{    {\it Astrophys. J.}}
\newcommand{\apjl}{   {\it Astrophys. J. Lett.}}

\newcommand{\grl}{    {\it Geophys. Res. Lett.}}

\newcommand{\jgr}{    {\it J. Geophys. Res.}}
\newcommand{\mnras}{  {\it Mon. Not. Roy. Astron. Soc.}}
\newcommand{\nat}{    {\it Nature}}
\newcommand{\pasp}{   {\it Pub. Astron. Soc. Pac.}}
\newcommand{\pasj}{   {\it Pub. Astron. Soc. Japan}}

\newcommand{\solphys}{{\it Solar Phys.}}
 
\newcommand{\ssr}{    {\it Space Sci. Rev.}}
\newcommand{\zap}{{\it Z. Astrophys.}}
 \newcommand{\araa}{    {\it Annu. Rev. Astron. Astrophys}}
\chardef\us=`\_

\begin{document}

\begin{article}
\begin{opening}

\title{Coronal Quasi-periodic Fast-mode Propagating Wave Trains}

\author[addressref={aff1,aff2,aff4},email={ydshen@ynao.ac.cn}]{\inits{Y. D.}\fnm{Yuandeng}~\lnm{Shen}}
\author[addressref={aff1,aff4},email={ }]{\inits{X. P.}\fnm{Xinping}~\lnm{Zhou}}
\author[addressref={aff1,aff4},email={ }]{\inits{Y. D.}\fnm{Yadan}~\lnm{Duan}}
\author[addressref={aff1,aff4},email={ }]{\inits{Z. H.}\fnm{Zehao}~\lnm{Tang}}
\author[addressref={aff1,aff4},email={ }]{\inits{C. R.}\fnm{Chengrui}~\lnm{Zhou}}
\author[addressref={aff1,aff4},email={ }]{\inits{S.}\fnm{Song}~\lnm{Tan}}
\address[id=aff1]{Yunnan Observatories, Chinese Academy of Sciences,  Kunming, 650216, China}
\address[id=aff2]{State Key Laboratory of Space Weather, Chinese Academy of Sciences, Beijing 100190, China}
\address[id=aff4]{University of Chinese Academy of Sciences, Beijing 100049, China}
\runningauthor{Shen et al.}
\runningtitle{Coronal Quasi-periodic Fast-mode Wave Trains}

\begin{abstract}
Quasi-periodic fast-mode propagating (QFP) wave trains in the corona have been studied intensively in the past decade, thanks to the full-disk, high spatiotemporal resolution, and wide-temperature coverage observations taken by the Atmospheric Imaging Assembly (AIA) onboard the {\em Solar Dynamics Observatory} ({\em SDO}). In AIA observations, QFP wave trains are seen to consist of multiple coherent and concentric wavefronts emanating successively near the epicenter of the accompanying flares; they propagate outwardly either along or across coronal loops at fast-mode magnetosonic speeds from several hundred to more than \speed{2000},  and their periods are in the range of tens of seconds to several minutes. Based on the distinct different properties of QFP wave trains, they might be divided into two distinct categories including narrow and broad ones. For most QFP wave trains, some of their periods are similar to those of quasi-periodic pulsations (QPPs) in the accompanying flares, indicating that they are probably different manifestations of the same physical process. Currently, candidate generation mechanisms for QFP wave trains include two main categories: pulsed energy excitation mechanism in association with magnetic reconnection and dispersion evolution mechanism related to the dispersive evolution of impulsively generated broadband perturbations. In addition, the generation of some QFP wave trains might be driven by the leakage of three and five minute oscillations from the lower atmosphere. As one of the new discoveries of {\em SDO}, QFP wave trains provide a new tool for coronal seismology to probe the corona parameters, and they are also useful for diagnosing the generation of QPPs, flare processes including energy release and particle accelerations. This review aims to summarize the main observational and theoretical results of the spatially-resolved QFP wave trains in extreme ultraviolet observations, and states briefly a number of questions that deserve further investigations.
\end{abstract}
\end{opening}

\section{Introduction}
The solar atmosphere is divided into the photosphere, chromosphere, transition region and corona based on the distinct different physical properties. The outermost atmosphere layer of the Sun, the corona, is made of high-temperature magnetized plasma, which extends at a height of about 5 Mm above the photosphere into the heliosphere. In the low corona ($\leq$ \rsun{1.3}), the magnetic field strength ranges from 0.1--0.5 gauss in the quiet Sun and in coronal holes to 10--50 gauss in active region resolved elements, with typical temperature (electron densities) of 1--2 MK ($10^{9} ~\rm cm^{-3}$ ) in the quiet Sun and 2--6 MK ($10^{11} ~\rm cm^{-3}$) in active regions. These physical parameters determine that the coronal plasma, consisting of electrons and ions, is magnetically confined where charged particles are guided by magnetic field lines in a helical gyromotion along the magnetic field lines \citep[e.g.,][]{2005psci.book.....A}. 

The tenuous and hot corona stores a large amount of energy mainly in the highly non-potential magnetic field of active regions. Generally, the stored energy can be released impulsively through magnetic reconnection and thus cause large-scale solar eruptions such as flares \citep{2011LRSP....8....6S}, filament/jet eruptions \citep{2010SSRv..151..333M,2021RSPSA.47700217S}, coronal mass ejections \citep[CMEs,][]{2011LRSP....8....1C}. These energetic solar eruptions will inevitably excite various types of magnetohydrodynamic (MHD) waves in the corona \citep[e.g.,][]{2005LRSP....2....3N,2020SSRv..216..136L,2020SSRv..216..140V,2020ARA&A..58..441N,2021SSRv..217...73N,2021SoPh..296...47T,2021SSRv..217...34W}. In addition, the leakage of photospheric and chromospheric oscillations into the corona can also led to the generation of coronal waves \citep[e.g.,][]{1969SoPh....7..351B,2002A&A...387L..13D,2009A&A...505..791S,2012ApJ...753...53S}, and the mode conversion should occur in the chromosphere where the plasma pressure is approximately equal to the magnetic pressure \citep[e.g.,][]{2003ApJ...599..626B}. Generally, there are three MHD wave modes, including Alfv\'{e}n wave, slow and fast magnetosonic waves. Alfv\'{e}n waves are incompressible in the linear regime and can only cause doppler shifts in observed line measurements, while slow and fast magnetosonic waves are compressional and can cause compression and rarefaction of the plasma density. Hence that compressional magnetosonic waves can be directly imaged through detecting intensity variations, since the optically thin emission measure in extreme ultraviolet (EUV) and soft X-rays is directly proportional to the square of electron density, and thus to the observed flux \citep{2005psci.book.....A}. However, one should be cautions with this, as the column. depth perturbations should also be taken into account \citep[e.g.,][]{2003A&A...409..325C, 2012A&A...543A..12G}. MHD waves not only carry energy away from their excitation sources and dissipate it into the medium where they propagate, but also reflect the physical properties of the waveguides and the background corona. Therefore, the investigation of MHD waves is very important for understanding the heating of the upper solar atmosphere, the acceleration of solar wind, and the physical parameters of the solar atmosphere with the method of coronal seismology \citep[e.g.,][]{2005LRSP....2....3N,2012RSPTA.370.3193D,2016SSRv..200...75N,2021SSRv..217...34W}. In addition, since MHD waves are accompanying phenomena of solar eruptions, they are also important for diagnosing the driving mechanism and energy release process of solar eruptions.

Rapidly propagating large-scale disturbances in the solar atmosphere were firstly observed in the chromosphere with group-based H$\alpha$ telescope; they show as arc-shaped bright fronts and were dubbed as Moreton waves \citep[e.g.,][]{1960AJ.....65U.494M,1960PASP...72..357M}. Moreton waves propagate rapidly at a  speed of \speed{500-2000} so that they can reach a long distance of the order of $10^5$ km and cause the oscillation of remote filaments \citep[e.g.,][]{2002PASJ...54..481E,2014ApJ...786..151S,2014ApJ...795..130S}. Since it is hard to understand the long distance propagation of Moreton waves in the dense chromosphere \citep[see,][]{2016GMS...216..381C}, \cite{1968SoPh....4...30U} interpreted them as chromospheric response of coronal fast-mode magnetosonic waves or shocks. Uchida's model not only naturally explained the observed features of Moreton waves, but also predicted the existence of large-scale fast-propagating magnetosonic waves or shocks in the low corona. The high temperature of the corona makes the coronal plasma mainly radiates in the EUV and X-ray wavebands. However, due to the strong absorption of these radiations by the Earth's atmosphere, the routine observation of the low corona can only be made in the outer space. Therefore, the large-scale fast-propagating disturbances in the corona were discovered until 1998 by the {\em Extreme-ultraviolet Imaging Telescope} \citep[EIT:][]{1995SoPh..162..291D} onboard the {\em Solar and Heliospheric Observatory} ({\em SOHO}), delayed the discovery of chromospheric Moreton waves about 40 years \citep{1997SoPh..175..571M,1998GeoRL..25.2465T}. The observational characteristics of large-scale corona disturbances are similar to those of chromospheric Moreton waves, such as the arc-shaped or circular diffuse wavefronts centered around the epicenter of the associated flares. Therefore, they were quickly thought to be the long-awaited coronal counterparts of chromospheric Moreton waves, i.e., fast-mode MHD waves or shocks excited by flare-ignited pressure pulses \citep[e.g.,][]{1999ApJ...517L.151T,2000ApJ...543L..89W,2001JGR...10625089W}. However, this interpretation was challenged by many follow-up studies, due to some abnormal characteristics such as much lower speeds compared to Moreton waves \citep{2000A&AS..141..357K} and stationary wavefronts \citep{1999SoPh..190..107D}. During the past two decades, observational and theoretical studies were intensively performed to study the driving mechanism and physical nature of these large-scale coronal disturbances. Thanks to the high spatiotemporal resolution observations taken by the {\em Atmospheric Imaging Assembly} \citep[AIA:][]{2012SoPh..275...17L} onboard the {\em Solar Dynamic Observatory} ({\em SDO}), now we have recognized that a large-scale propagating coronal disturbance is typically composed of a fast-mode magnetosonic wave or shock followed by a slower wavelike feature, in which the former is often driven by a CME, corresponding to the coronal counterpart of a chromospheric Moreton wave \citep[e.g.,][]{2012ApJ...752L..23S,2011ApJ...738..160M,2012ApJ...745L...5C}, while the origin and physical nature of the latter is still unclear \citep{2014SoPh..289.3233L,2015LRSP...12....3W,2016GMS...216..381C,Shen2020}. It should be pointed out that a bewildering multitude of names were used in history for large-scale fast-propagating coronal disturbances, such as ``EIT waves'', ``(large-scale) coronal waves'', ``(large-scale) coronal propagating fronts'', and ``EUV waves''. In this paper, we tend to use the term ``EUV waves'' based on their main observation waveband.

The launch of the {\em SDO} started a new booming era in the research of coronal MHD waves, mainly due to its unprecedented observation capability. The AIA onboard the {\em SDO} observes the Sun uninterruptedly with a full-Sun (1.3 solar diameters) field of view, which has seven EUV channels covering a wide temperature range from $6\times10^{4}$ to $2\times10^{7}$ K and a high signal-to-noise (sensitivity) for two- to three-seconds exposures. The temporal cadence and spatial resolution of the images taken by the AIA are respectively of 12 seconds and 1\arcsec.2 \citep{2012SoPh..275...17L}. The combination of these high observation capabilities makes AIA the best ever instrument for detecting coronal MHD waves with small intensity amplitudes. Since the launch of the {\em SDO} in 2010, besides the great achievements in the study of single pulsed global EUV waves, quasi-periodic fast-mode propagating (QFP) wave trains have also been directly imaged \citep{2010ApJ...723L..53L,2011ApJ...736L..13L}. In history, directly imaging observations of QFP wave trains were very scarce, although they had long been theoretically predicted \citep{1984ApJ...279..857R} and confirmed by numerical studies with similar characteristics \citep[e.e.,][]{1993SoPh..144..101M,1994SoPh..151..305M,1998SoPh..179..313M}. This was mainly attributed to the low observation capability of previous solar telescopes, such as their low spatiotemporal resolution, low sensitivity, narrow temperature coverage and small fields of view. As one of the new discoveries of the {\em SDO}, QFP wave trains have attracted a lot of attentions since the discovery; they were identified as fast-mode magnetosonic waves by using a three-dimensional MHD simulation \citep{2011ApJ...740L..33O}. So far, there are dozens of QFP wave trains that have been analyzed in detail with multi-wavelength observations, and remarkable theoretical attention has been given to their excitation, propagation, and damping mechanisms. The investigation of QFP wave trains is very important at least in the following aspects. Firstly, as a new accompanying phenomenon of solar eruptions, it is worthy to study their basic physical properties and physical connections to solar eruptions. Secondly, since QFP wave trains often show some common periods with those of the quasi-periodic pulsations (QPPs) in the accompanying flares, the study of QFP wave trains can shed light on our understanding of the unresolved generation mechanisms of flare QPPs that show as quasi-periodic intensity variation patterns with characteristic periods typically ranging from a few seconds to several minutes and can be seen in a wide range of wavelength bands from radio to gamma-ray light curves \citep[e.g.,][]{1961ApJ...133..243Y,1969ApJ...155L.117P, 1983ApJ...271..376K, 2010ApJ...708L..47N, 2010SoPh..267..329K, 2011ApJ...740...90V, 2014SoPh..289.1239N, 2016ApJ...832...65Z, 2017ApJ...848L...8M, 2019ApJ...878...78C, 2019ApJ...886L..25Y, 2020ApJ...895...50H, 2020A&A...642A.195K, 2020ApJ...893L..17L, 2020A&A...639L...5L, 2020ApJ...893....7L, 2020ApJ...888...53L, 2021ApJ...921..179L, 2021ApJ...910..123C,2021SoPh..296..130L,2021RAA....21...66L,Li:2021vc}. Thirdly, QFP wave trains provide a new seismological tool for diagnosing the physical parameters of the solar corona that are currently difficult, or even impossible to measure. In addition, since the damping of fast-mode magnetosonic waves is fast, they are thought to be important for balancing the typical radiative loss rates of active regions \citep[e.g.,][]{1994ApJ...435..482P,2011ApJ...736L..13L}. 

The aim of this review is intended to summarize the main theoretical and observational results of the spatially-resolved QFP wave trains in the EUV wavelength band, focusing on recent advances and the seismological applications. \cite{2014SoPh..289.3233L} had published a preliminary review on QFP wave trains 7 years ago based on only 6 published events at that time. The present review mainly focuses on new observational and theoretical advances in recent years, but also includes previous theoretical and observational studies for without sacrificing completeness. Other types of coronal MHD waves are not covered in this review, interested readers can refer to many excellent review papers published in recent years \citep[e.g.,][]{2005LRSP....2....3N,2020ARA&A..58..441N,2021SSRv..217...73N,2015LRSP...12....3W,2018SSRv..214...45M,2020SSRv..216..136L,2020SSRv..216..140V,Shen2020,2021SSRv..217...34W,2021SSRv..217...66Z}.

\section{Observational Signature}
\subsection{Pre-{\em SDO} Observation}
Space-borne solar telescopes before the {\em SDO} are not good for detecting QFP wave trains, mainly because of their lower observing capabilities such as  spatiotemporal resolution and sensitivity. Although the {\em TRACE} has a superior spatial resolution, but its lower time resolution, poorer sensitivity, and smaller field of view are all not conducive for the detection of QFP wave trains. Therefore, sporadic imaging detections of possibly coronal QFP wave trains were mainly reported during solar total eclipse observations and coronagraph observations, through detecting the intensity, velocity and line width fluctuations \citep[e.g.,][]{1984SoPh...90..325P,1999SoPh..188...89C,2002SoPh..207..241P,2003A&A...406..709K}. In addition, indirect signals of QFP wave trains were also studied by using radio observations. During the total solar eclipse on 1995 October 24,  \cite{1997SoPh..170..235S} detected intensity variations with periods of 5--56 seconds. Possible evidence of periodic MHD waves was also reported in other studies using coronagraph observations \citep[e.g.,][]{1983A&A...120..185K,1997ApJ...491L.111O,2002SoPh..209..265S}. Some estimations showed that these observed oscillations and periodic MHD waves carry carry enough energy for heating the active region corona and could contribute significantly to solar wind acceleration in open magnetic field structures, if they are Alfv\'{e}n or fast-mode magnetosonic waves.

Probably, more reliable imaging detection of QFP wave trains was detected during the total solar eclipse on 1999 August 11, by using the {\em Solar Eclipse Corona Imaging System} \citep[SECIS:][]{2001MNRAS.326..428W}. Detailed analysis results showed that the detected oscillations could be a QFP wave train that travels along active region loops \citep{2002MNRAS.336..747W}, whose period, speed, wavelength, and intensity amplitude were about 6 seconds, \speed{2100}, 12 Mm, and 5.5\%, respectively. In a subsequent paper, the authors further detected more periods of the wave train in the range of 4--7 seconds, indicating the wave train's periodicity nature \citep{2003A&A...406..709K}. After the launch of the {\em Transition Region And Coronal Explorer} \citep[{\em TRACE}:][]{1999SoPh..187..229H}, \cite{2005A&A...430L..65V} probably observed a QFP wave train that propagated along an open magnetic field structure above a post-flare arcade, using the 195 \AA\ wavelength images. Measurements showed that the wave train had a period of 90--220 seconds and propagated at a speed of \speed{200--700} at a height of 90 Mm above the solar surface. In addition, a quasi-periodic large-scale global EUV wave train was also reported in \cite{2010A&A...522A.100P}, by using the 171 \AA\ imaging observations taken by the Extreme Ultraviolet Imager \citep[EUVI;][]{2004SPIE.5171..111W} onboard the {\em Solar Terrestrial Relations Observatory} \citep[{\em STEREO}][]{2008SSRv..136....5K}. In their observation, multiple large-scale coherent EUV wavefronts propagating over the disk limb were ahead of the CME bubble, and the authors proposed that the quasi-periodic EUV wave train was driven by the fine expanding pulse-like lateral structures in the CME bubble, because the wavefronts appeared as the lateral expansion of the CME bubble slows down and terminates.

\subsection{General Properties}
Unambiguous signatures of QFP wave trains were directly imaged in EUV images taken by the AIA instrument onboard the {\em SDO} \citep{2010ApJ...723L..53L,2011ApJ...736L..13L,2012ApJ...753...53S}, and they were identified as fast-mode magnetosonic waves by \cite{2011ApJ...740L..33O} using a three-dimensional MHD model of a bipolar active region structure. Since the initial discovery, QFP wave train has attracted a lot of attentions, and a mass of observational and numerical studies have been performed to investigate their excitation mechanisms and physical properties. The occurrence of QFP wave trains is rather common and is frequently associated with single pulsed global EUV waves, flares and CMEs. According to the first 4.5 years observation of the {\em SDO}, \cite{2016AIPC.1720d0010L} performed a simple statistical study of QFP wave trains based on the database of global EUV waves cataloged at LMSAL \citep[][\url{https://www.lmsal.com/nitta/movies/AIA_Waves}]{2013ApJ...776...58N}, and the authors found that about one third of global EUV waves in association with flares and CMEs are accompanied by QFP wave trains. This occurrence rate is clearly undervalued for all flare activities, because in fact the occurrence of many QFP wave trains do not accompanied by global EUV waves and CMEs. Until now, more than thirty QFP wave trains have been analyzed in detail in the literature. The physical parameters and the main associated solar activities of published QFP wave trains are listed in \tbl{tbl1}. In these events, QFP wave trains exhibit recurrence characteristics in some active regions along specific trajectories \citep[e.g.,][]{2013AA...554A.144Y,2015AA...581A..78Z,2020ApJ...889..139M,2021arXiv210909285Z} and refraction and reflection effects during their interaction with coronal  structures or at the remote footpoints of closed loop systems \citep[e.g.,][]{2011ApJ...736L..13L,2018MNRAS.480L..63S,2018ApJ...853....1S,2019ApJ...873...22S}. In particularly,  turbulent cascade caused by the counter-propagating of two QFP wave trains along the same closed loop system was also observed \citep{2018ApJ...860...54O}.

Based on \tbl{tbl1}, we can make a simple statistical study to QFP wave trains. It can be seen that QFP wave trains propagate at fast speeds of about \speed{305--2394} and with strong decelerations of \accel{0.1--4.1}; they can propagate a long distance over 500 Mm ($\textgreater$ \rsun{0.7}) before their disappearance. It should be noted that the values of these parameters could be higher, since they are typically measured on the plane of the sky.Their occurrence are typically accompanied by flares and commonly appear firstly at a distance greater than 100 Mm away from the flare epicenter. Such a distance is consistent with the theoretical prediction of the initial periodic phase of an impulsively generated fast magnetosonic wave, during which the intensity amplitude needs time to be amplified for detection \citep{1983Natur.305..688R,1984ApJ...279..857R}. In addition, the observability of fast-mode magnetosonic waves is also significantly affected by observation angle \citep{2003A&A...409..325C}. In observation, the amplitude of QFP wave trains does show a first increase and then decrease trend as they propagate outward along funnel-like loops, and this might be due to the combined result of the amplification caused by density stratification and attenuation result from geometric expansion of the waveguide \citep{2013AA...554A.144Y}. According to  \tbl{tbl1}, QFP wave trains are typically associated with large-scale solar activities including flares, CMEs and global EUV waves. One can see that the associated flares can either be energetic GOES soft X-ray M class \citep[e.g.,][]{2014AA...569A..12N,2017ApJ...844..149K}, or low-energy events such as small brightening patches \citep{2018MNRAS.480L..63S,2020ApJ...889..139M} and features of possible reconnection events that can not even cause small GOES flares \citep{2017ApJ...851...41Q, 2018ApJ...868L..33L}. This result might indicate that the occurrence of QFP wave trains does not need too much energy. Alternatively, the presence of special physical condition might be an important factor instead, because in some active regions recurrent flares at the same location are often associated with recurrent QFP wave trains along the same trajectory. 

We checked the correlation between QFP wave trains and CMEs based on the CACTUS (\url{https://wwwbis.sidc.be/cactus/}) and CDAW (\url{https://cdaw.gsfc.nasa.gov/CME_list/}) databases. For the 32 published QFP wave trains, 26 of them are associated with CMEs, which means that the association rate of QFP wave trains with CMEs is about $26/32~\approx~80\%$. The average speeds of those CMEs that are  accompanied by QFP wave trains are in the range of \speed{174--1466}, which suggests that QFP wave trains are associated with both slow and fast CMEs, and no clear correlation preference with the two types of CMEs can be found. For the QFP wave trains propagating along coronal loops, we also checked their correlation with global EUV waves. It is found that 18 of the 27 QFP wave trains were associated with global EUV waves, which translates to an association rate of about $18/27~\approx~70\%$. Moreover, for the global EUV waves that were accompanied by QFP wave trains, 5 of them were not associated with CMEs \citep{2013AA...553A.109K,2018MNRAS.480L..63S,2018ApJ...860L...8S,2020ApJ...889..139M}. In other word, these QFP wave trains were associated with failed solar eruptions without association to CMEs, and the ratio of this kind of QFP wave train is about $5/18~\approx~30\%$. For those QFP wave trains that are not associated global EUV waves, almost all of them were associated with CMEs. This is probably the reason why the association rate between QFP wave trains and CMEs (80\%) is higher than that between global EUV waves (70\%). We noted that in \cite{2016AIPC.1720d0010L} the authors found that all the QFP wave trains associated with global EUV waves are also associated with flares and CMEs. In addition, based on the simple survey of two flare productive active regions of AR12129 and AR12205, the authors also found an interesting trend of preferential association of QFP wave trains with successful solar eruptions accompanied by CMEs. Here, based on the survey of the published events, we would like to point out that not all QFP wave trains are simultaneously accompanied by both global EUV waves and CMEs, and the association rate with successful solar eruptions does higher than that with failed ones (say, 80\% vs 20\%). 

Because the occurrence of QFP wave trains is tightly associated with flares, we further checked the temporal relationship between the start time of QFP wave trains and the start and peak times of the associated flares (see \tbl{tbl1}). One can see that the start of QFP wave trains can either be before or after the peak times of the accompanying flares. For those QFP wave trains that appeared before the flare peak times, their start times are usually about 1--57  minutes later

\begin{landscape}
\begin{table}
\caption{Physical parameters of the published QFP wave trains}
\centering
\resizebox{1.0\columnwidth}{!}{
\scalebox{1.}{
\begin{tabular}{ccccccccccccccc}
\Xhline{1.pt}
\rule{0pt}{15pt}{\bf Event} &\multicolumn{4}{c}{\bf {Associated Phenomena}}   &\multicolumn{8}{c}{\bf{Physical Parameters}}                                               &\\
\cmidrule(r){1-1} \cmidrule(r){2-5} \cmidrule(r){6-14}
Date&\multicolumn{2}{c}{Flare} &CME &EUV Wave & Start Time & Duration&Angular Width & Speed  & Deceleration &Period     &Wavelength   & Intensity & Energy Flux&Reference\\
yyyy-mm-dd &Start/Peak Time [UT] &GOES Class & [km s$^{-1}$] &[Y/N] &[UT]& [Min.]&[Degree] & [km s$^{-1}]$ & [km s$^{-2}]$ & [Sec.] &[Mm]  &Amplitude &[$\times 10^5$ erg cm$^{-2}$ s$^{-1}$] \\
\Xhline{1.pt}
\rule{0pt}{15pt}&\multicolumn{13}{c}{\bf Narrow QFP wave trains\tabnote{Wave trains propagate along coronal loops.}}&\\
\Xhline{1.pt}
2010-04-08  &02:32/03:25     &B3.8                  &174    &Y   &03:15      &65        &40       &450--1200   &0.2--5.8      &40--240   &---             &---              &---           &\cite{2010ApJ...723L..53L}\\
2010-08-01  &07:25/08:57    &C3.2                  &260     &N   &07:45      &60       &60        &2200           &---               &40--181   &133           &1\%--5\%   &0.1-2.6   &\cite{2011ApJ...736L..13L}\\
2010-08-01  &07:25/08:57    &C3.2                  &260     &N   &08:06      &25       &---        &1000--2000 &---               &---            &---             &---              &---           &\cite{2011ApJ...736L..13L}\\
2010-09-08  &23:00/23:33    &C3.3                  &818     &Y   &23:11      &18       &25        &1020--1220 &3--4            &30--240    &---             &1\%--5\%   &---           &\cite{2012ApJ...753...52L}\\
2011-02-14  &05:35/05:50    &BP\tabnote{Brightening Patches}                     &---         &Y   &05:53      &26       &40       &322             &0.138          &390          &100           &2\%--4\%   &---           &\cite{2018MNRAS.480L..63S}\\
2011-02-15  &04:29/04:49    &C8.3                  &---         &Y   &04:38      &32       &25       &388             &0.38            &200          &110          &1\%--4\%  &---           &\cite{2018ApJ...860L...8S}\\
2011-03-09  &23:47/23:51    &BP                    &---         &Y   &23:48      &22       &20       &718             &---               &40             &40            &---              &---           &\cite{2020ApJ...889..139M}\\
2011-03-10  &04:05/04:09    &BP       &---         &Y   &04:05      &12       &10       &876             &---               &50            &30              &---              &---           &\cite{2020ApJ...889..139M}\\
2011-03-10  &06:39/06:48    &C4.0                  &408      &Y   &06:40      &6         &25        &682--837    &---               &45           &40              &---              &---           &\cite{2019ApJ...871L...2M}\\
2011-03-25  &23:08/23:22     &M1.0                 &---         &Y   &23:12      &6         &---       &1011--1296 &---               &180         &20--30        &---             &---           &\cite{2013AA...553A.109K}\\
2011-05-30  &10:48/10:57     &C2.8                 &248      &Y   &10:50      &12       &25        &834             &---              &25--400   &23.8           &2\%--8\%    &---           &\cite{2012ApJ...753...53S}\\
2011-05-30  &10:48/10:57     &C2.8                 &248      &Y   &10:50      &12       &25        &740-850      &1.3--2.3     &38--58     &24--34        &2\%--8\%   &---           &\cite{2013AA...554A.144Y}\\
2011-06-02  &06:30/06:36     &C1.4                 &253      &Y   &06:30      &13       &20        &776             &---              &120         &110             &1\%--4\%    &---         &\cite{2015AA...581A..78Z}\\
2011-06-02  &07:22/07:46     &C3.7                 &976      &Y   &07:35      &20       &20        &978             &---              &120         &110             &1\%--6\%    &---          &\cite{2015AA...581A..78Z}\\
2011-09-23  &23:48/23:56     &M1.9                 &617      &N   &00:06      &12       &15        &320            &---              &130          &40              &--                &--           &\cite{2016ApJ...822....7K}\\
2011-11-09  &11:40/11:45      &RE\tabnote{Reconnection Events}                    &627      &N  &11:45       &33       &35        &305            &0.715         &74-390     &30              &1\%--5\%    &---          &\cite{2017ApJ...851...41Q}\\
2011-11-09  &11:40/11:45       &RE                   &627      &N  &11:45       &35       &20        &343            &1.17           &54-458     &30              &1\%--8\%    &---          &\cite{2017ApJ...851...41Q}\\
2012-04-23 &17:37/17:51      &C2.0                 &355      &Y  &17:40       &23       &18        &689             &1.0             &80           &33               &1\%-5\%     &1.2-4.0   &\cite{2013SoPh..288..585S} \\
2012-07-14 &09:07/09:12      &C1.4                 &561      &N  &09:14       &6       &10          &538--719    &---               &180         &120             &---               &---          &\cite{duan2021} \\
2013-04-23 &18:10/18:33      &C3.0                 &403      &Y  &18:20       &60       &10         &474            &---               &110         &40               &1\%--4\%    &0.43        &\cite{2021SoPh..296..169Z}\\
2013-05-22 &12:35/13:32      &M5.0                &1466     &Y &13:32        &120     &---        &1860           &---               &120--180 &320            &2\%-4\%     &1.8         &\cite{2018ApJ...860...54O}\\
2013-05-22 &13:00/13:57      &C5.0                 &---         &N &13:05        &120     &---        &1670           &---              &120--180  &---              &2\%-4\%     &1.8         &\cite{2018ApJ...860...54O}\\
2013-12-07 &07:17/07:29      &M1.2                &909       &Y &07:25        &15       &10        &538--2394   &---              &50--180    &46--429     &---               &---          &\cite{2014AA...569A..12N}\\
2013-12-07 &07:17/07:29      &M1.2                &909       &Y &07:26        &32       &55        &941--1851   &---              &50--130    &62--181     &---               &---          &\cite{2014AA...569A..12N}\\
2014-03-23 &03:05/03:48       &C5.0                &820      &N &03:08        &60       &80        &884--1485    &---             &25--550    &6-20           &2\%-4\%     &---          &\cite{2018ApJ...853....1S}\\
2015-07-12 &17:34/17:44      &B4.0                 &416      &N  &17:37       &6         &35        &1100             &2.2            &43--79     &47--87        &---               &---          &\cite{2018MNRAS.477L...6S}\\
2019-03-08 &03:07/03:18       &C1.3                &239      &Y   &03:33      &7         &35        &1083--1366  &---              &62--66     &69--87        &---               &---          &\cite{2021ApJ...908L..37M}\\
2019-03-08 &03:07/03:18       &C1.3                &239      &Y   &03:35      &7         &25        &536--656      &---              &65-66      &35--43        &---               &---          &\cite{2021ApJ...908L..37M}\\
\Xhline{1.pt}
\rule{0pt}{15pt}{\bf Range}  &   &BP/RE--M    &174--1466 &19/9 & &6--120 &10--80  &305--2394 &0.1--5.8    &25--550   &24--429    &1\%--8\% &0.1--4.0\\
\Xhline{1.0pt}
\rule{0pt}{15pt}&\multicolumn{13}{c}{\bf Broad QFP wave trains \tabnote{Wave trains propagate on the solar surface.}} &\\
\Xhline{1.0pt}
2010-09-08  &23:00/23:33      &C3.3                &818      &    &23:11      &12       &360\tabnote{By assuming that the wave train has a dome-shaped structure  propagating in all directions.}      &370--650     &0.1--0.4        &36--212    &80--140               &10\%--20\%      &---       &\cite{2012ApJ...753...52L}\\
2012-04-24  &07:38/07:45      &C3.7                &443      &    &07:41      &9         &200      &747              &---                &163          &84---100             &10\%--35\%      &---        &\cite{2019ApJ...873...22S}\\    
2012-05-07  &14:03/14:31       &M1.9                &665     &     &14:06     &20       &90         &664--1416   &---                &120--240  &150                    &---                     &---        &\cite{2017ApJ...844..149K}\\
2011-02-24  &07:23/07:35       &M3.5                &1186    &    &07:30      &30      &270       &668              &---                &90            &58                      &25\%--35\%      &10-19  &\cite{zhou2021b} \\      
2013-04-23  &18:10/18:33       &C3.0                &403      &   &18:16       &7        &140       &1100            &4.1               &120          &170                      &15\%                &12        &\cite{2021SoPh..296..169Z}\\
\Xhline{1.pt}
\rule{0pt}{15pt}{\bf Range}  &   &C--M    &403--1186 & & &7--30 &90--360       &370--1416 &0.1--4.1    &36--240       &58--170            &10\%--35\% &10--19\\
\Xhline{1.0pt}
\end{tabular}}}
\label{tbl1}
\end{table}
\end{landscape}

\noindent  than the beginning of the accompanying flares, but about 3--51 minutes earlier than the flare peak times. For QFP wave trains that occurred after the peak times of the accompanying flares, their beginning times are about 0--17 minutes later than the flares' peak times. In the 32 published QFP wave trains, there are 24 (8) cases that occurred before (after) the peak times of the accompanying flares. Therefore, we can reach a preliminary conclusion that most of QFP wave trains occur during the impulsive phase of flares (say, 24/32 = 75\%). It seems that the energy level of flares are not the key physical condition for determining the start time of a QFP wave train, because for energetic GOES soft X-ray M class flares, the associated QFP wave trains can occur either in the impulsive \citep[e.g.,][]{2013AA...553A.109K, 2014AA...569A..12N, 2017ApJ...844..149K, zhou2021b} or the decay \citep[e.g.,][]{2016ApJ...822....7K,2018ApJ...860...54O} phases. Whereas, the start times of QFP wave trains are probably associated with the durations of flares. Taken the cases accompanied by M class flares as examples, one can find that the impulsive phase of those flares with short durations tends to launch QFP wave trains during their impulsive phase \citep[e.g.,][]{2013AA...553A.109K, 2014AA...569A..12N, 2017ApJ...844..149K, zhou2021b}, while those with long durations are likely to excite QFP wave trains during their decay phase \citep[e.g.,][]{2016ApJ...822....7K, 2018ApJ...860...54O}. The lifetimes of the published QFP wave trains are typically in the range of 6--65 minutes, which is comparable to that of the impulsive phases of the accompanying flares (3--68 minutes). The longest duration among all published QFP wave trains was reported by \cite{2018ApJ...860...54O}, which reached up to about 2 hours. In this case, the flare had a long impulsive phase of about 57 minutes, and the QFP wave train started at the beginning of the decay phase of the accompanying flare. 

\subsection{Classification}
\begin{figure}[!t] 
\centerline{\includegraphics[width=0.9\textwidth]{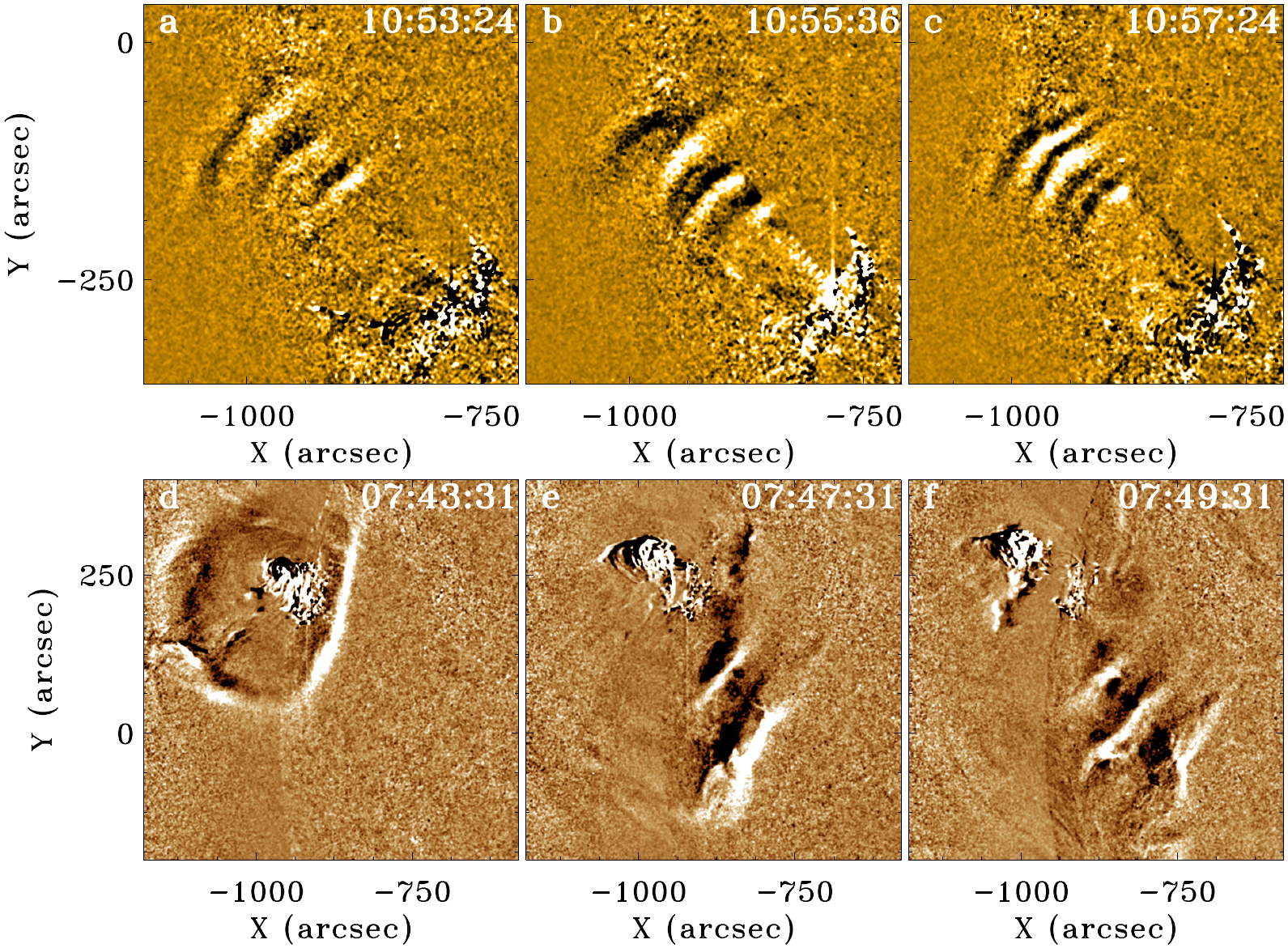}}
\caption{Examples for the two types of QFP wave trains. The top row shows the narrow QFP wave train on 2011 May 30 using the AIA 171 \AA\ running-difference images, which occurred close to the east limb of the solar disk and was analyzed in details in \cite{2012ApJ...753...53S} and \cite{2013AA...554A.144Y}. The bottom row shows the broad QFP wave train on 2012 April 24 using the AIA 193 \AA\ running-ratio images, which occurred on the east limb of the solar disk and propagated along the solar surface \citep[see][for details]{2019ApJ...873...22S}. The wave trains manifest as a chain of arc-shaped bright fronts propagating outwardly from the accompanying flare epicenter.}
\label{fig1}
\end{figure}

A typical QFP wave train is composed of multiple coherent and concentric arc-shaped wavefronts emanating successively near the epicenter of the accompanying flare and propagate outwardly either along or across coronal loops \citep[e.g.,][]{2011ApJ...736L..13L,2012ApJ...753...53S,2012ApJ...753...52L,2019ApJ...873...22S}. Imaging observational results based on high spatiotemporal resolution AIA data indicate that QFP wave trains might be broken down into two main categories based on their significant different physical characteristics: namely, narrow and broad QFP wave trains. The main distinct difference between the two types of QFP wave trains mainly include physical parameters of their observation waveband, propagation direction, angular width, intensity amplitude and energy flux (see \tbl{tbl1}). Narrow QFP wave trains are typically observed at the AIA 171 \AA\ channel \citep[occasionally appeared at the AIA 193 \AA\ and 211 \AA\ channels, see][]{2010ApJ...723L..53L,2013SoPh..288..585S}; they propagate along the apparent direction of the magnetic field within a relatively small angular extent of about 10--80 degree and typically result in intensity fluctuations with a small amplitude of about 1\%--8\% relative to the background corona (see the top row of \fig{fig1}). The energy flux carried by narrow QFP wave trains is basically in the range of $0.1-4.0 \times 10^{5}~ {\rm erg~cm}^{-2}~{\rm s}^{-1}$ \citep[e.g.,][]{2011ApJ...736L..13L,2012ApJ...753...53S,2018MNRAS.477L...6S}. Broad QFP wave trains are frequently observed at the AIA's all EUV channels and can cause intensity fluctuations with a large amplitude of about 10\%--35\% relative to the background corona  (see the bottom row of \fig{fig1}). They propagate across magnetic field lines in the quiet-Sun region with a large angular extent of about 90--360 degree and carry an energy flux of about $10-19 \times 10^{5}~ {\rm erg~cm}^{-2}~{\rm s}^{-1}$ \citep{2012ApJ...753...52L,2019ApJ...873...22S,2021SoPh..296..169Z,zhou2021b}. In comparison, the two types for QFP wave trains have distinct different propagation preferences with respect to the magnetic field orientation, and the temperature coverage range of broad QFP wave trains is significantly wider than narrow QFP wave trains. In addition, all physical parameters including angular width, intensity amplitude and energy flux of broad QFP wave trains are evidently greater than those of narrow QFP wave trains.

Besides the above differences, the two types of QFP wave trains also show some similarities such as their physical parameters of propagation speed, deceleration, period and wavelength (see \tbl{tbl1}). Specifically, for narrow (broad) QFP wave trains, the physical parameters of propagation speed, deceleration, period and wavelength are in the ranges of \speed{305--2394 (370--1416)}, \accel{0.1--5.8 (0.1--4.1)}, 25--500 (36--240) seconds and 24--429 (58--170) Mm, respectively. In some events, broad QFP wave trains can be captured by coronal loops and become narrow QFP wave trains, which might mean the transformation of the former into the latter. For example, \cite{2019ApJ...873...22S} reported a broad QFP wave train propagating across the solar surface, whose east portion was trapped in a closed loop system and propagated at a speed relatively faster than the on-disk component. In other two events reported by \cite{2018ApJ...860L...8S} and \cite{2019ApJ...871L...2M}, the authors evidenced the transformation of single pulsed global EUV waves into narrow QFP wave trains along coronal loops. The successful capture of global EUV waves by coronal loops was also reported in \cite{2021SoPh..296..169Z}, where the trapped EUV wave showed an interesting firstly slowed down but then speeded up process owning to the variations of the physical parameters along the loop structure. In such a case, the global EUV waves are captured by coronal loops during their interaction, and the formation of the narrow QFP wave trains are probably due to the dispersive evolution of the initial disturbances caused by global EUV waves. In a one-dimensional numerical simulation performed by \cite{2015ApJ...799..221Y}, the authors showed that weak, fast wave trains can be formed by dispersion due to a series of partial reflections and transmissions of single pulsed EUV wavefronts during their interaction with loop-like coronal structure \citep{2016ApJ...828...17Y}. As what had been pointed out by \cite{2015ApJ...799..221Y}, the successful capture of an EUV wave may require the width of the coronal loop system approximately equal to half of the initial width of the EUV wavefront. We note that fast-mode global EUV waves were evidenced to convert into slow-mode magnetosonic waves during their interaction with coronal loops \citep[][]{2016ApJ...822..106C,2017ApJ...834L..15Z,2018ApJ...863..101C}. \cite{2016SoPh..291.3195C} numerically studied this phenomenon and found that the conversion occurs near the plasma $\beta\approx~1$ layer in front of the magnetic quasi-separatrix layer; the authors argued that such a mode-conversion process can account for the so-called stationary wavefronts formed when global EUV waves passing through quasi-separatrix layers \citep{1999SoPh..190..107D}.

\subsection{Kinematics}
Kinematics is the most fundamental property of any propagating disturbance, which is generally characterized by physical parameters of speed and acceleration. If the propagating disturbance is an MHD wave, it should exhibit wave phenomena such as reflection, refraction and diffraction effects during its interaction with coronal structures with a steep speed gradient \citep[e.g.,][]{2012ApJ...754....7S,2013ApJ...773L..33S}. Therefore, one can simply base on the speed and propagation behavior to determine the physical nature of a propagating disturbance in the solar atmosphere. For example, if a propagating intensity disturbance in the solar atmosphere that exhibits wave phenomena and propagates at slow (fast) magnetosonic wave speed, one can simply say that the disturbance is probably a slow (fast) magnetosonic wave \citep[e.g.,][]{2012ApJ...752L..23S}. 

\begin{figure} 
\centerline{\includegraphics[width=0.9\textwidth]{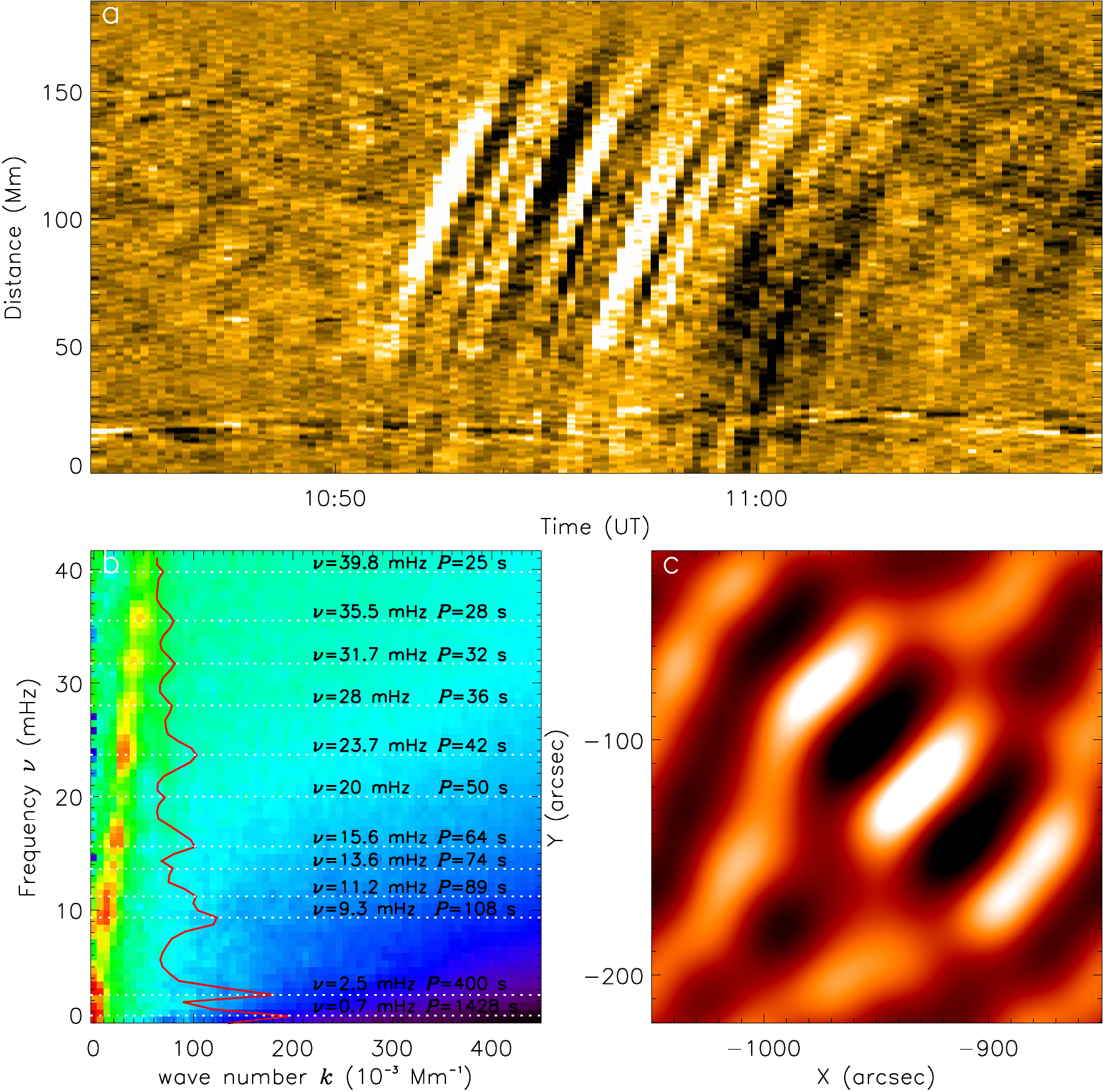}}
\caption{Kinematics analysis of the narrow QFP wave train on 2011 May 30. Panel (a) is the time-distance diagram along the propagation direction of the wave train, in which each bright intensity ridge represents a wavefront. Panel (b) is the $k-\omega$ map, in which the red curve shows the power peaks along the straight ridge. Panel (c) is a Fourier-filtered image around the dominating frequency of 15.6 Hz.}
\label{fig2}
\end{figure}

For QFP wave trains, there are two frequently-used methods to measure their speeds. The most popular method is the generation of time-distance diagrams along straight paths or sectors across the wavefronts by composing the one-dimensional intensity profiles at different times along a specific path using running- or base-difference time sequence images (see \fig{fig2}(a)). In a time-distance diagram, the wavefronts display as enhanced bright ridges, and the average speed can be obtained by fitting these ridges with a linear function, while the acceleration can be estimated by fitting the ridges with a quadratic function. The other method is the generation of a $k-\omega$ map using the method Fourier analysis of a three-dimensional data tube in (x, y, t) coordinates, where the field-of-view should cover the propagation region of the QFP wave train (see \fig{fig2}(b)). The details of this method can be found in many papers \citep[e.g.,][]{2004ApJ...617L..89D,2011ApJ...736L..13L,2012ApJ...753...53S}. In the $k-\omega$ map, the wave signature is represented by a steep  narrow ridge that describes the dispersion relation of the QFP wave train, and the slope of the ridge gives the average phase ($v_{ph}=\nu/k$) and group ($v_{gr}=d\nu/dk$) velocities \citep[e.g.,][]{2011ApJ...736L..13L,2012ApJ...753...53S}. The ridge in the $k-\omega$ map also reveals the frequency distribution in the QFP wave train, which exhibits as some discrete power peaks representing the dominating frequencies of the wave train \citep[e.g.,][]{2018MNRAS.477L...6S,2018ApJ...853....1S}. For a specific dominating frequency, one can obtain the Fourier-filtered images with a narrow Gaussian function centered at the dominating frequency (see \fig{fig2} (c)).

As shown in \tbl{tbl1} for the published events, the projection speeds of narrow and broad QFP wave trains are in the range of \speed{305--2394} and  \speed{370--1416}, while their decelerations are in the the range of \accel{0.1--5.8} and \accel{0.1--4.1}, respectively. These results indicate that the deceleration of QFP wave trains is significantly strong, and it seems that the faster waves are accompanied by stronger decelerations, consistent with the statistical result of global EUV waves \citep{2017SoPh..292..185L}. The speed of QFP wave trains do not show any preferential correlation with neither successful nor failed solar eruptions. Specifically, the speeds of the six QFP wave trains that were not associated with CMEs (i.e., failed eruptions) are in the range of \speed{322--1670}, while that of the other events accompanied by CMEs (successful eruptions) are in a similar range of \speed{305--2394}. Even for events that are associated with fast CMEs whose average speeds are greater than \speed{1000}, the speeds of the accompanying QFP wave trains can either be slow \citep[\speed{668};][]{zhou2021b} or fast \citep[\speed{1860};][]{2018ApJ...860...54O}. The speed of QFP wave trains does not show any preferential correlation with the energy class of the accompanying flares. For both low- and high-energy flares, the speeds of the accompanying QFP wave trains are all in the same range from several hundred to over \speed{2000}. These results might imply that the speed of QFP wave trains is mainly determined by the physical property of the medium in which they propagate, such as plasma density and magnetic strength as defined by the dispersion relation of fast magnetosonic waves. In addition, these results also suggest that QFP wave trains should be freely-propagating linear, or a slightly nonlinear fast magnetosonic waves, as suggested by the small Mach number (1.01) of a narrow QFP wave train \citep{2021SoPh..296..169Z}.

\subsection{Periodicity and Origin} \label{sp}
The periodicity in QFP wave trains carries important physical information in the eruption source regions and the medium in which they propagate. Investigating the generation and characteristics of the periodicity in QFP wave trains can help us to probe the eruption mechanism of solar eruptions and the physical property of the supporting medium. Generally, the periods of a QFP wave train can be isolated by using the methods of Fourier analysis and wavelet analysis \citep[][\url{https://atoc.colorado.edu/research/wavelets}]{1998BAMS...79...61T}. Sometimes, one can also directly measure the periods from time-distance diagrams. Based on the published events (see \tbl{tbl1}), the periods of narrow QFP wave trains are in a wide range of 25--550 seconds, while those of broad QFP wave trains are in the range of 36--240 seconds. Because the time cadence of the EUV channels of the AIA is 12 seconds, we are not able to detect periods lower than 24 seconds \citep{2014SoPh..289.3233L}. However, this insufficiency can be compensated by high temporal resolution radio observations. For example, some spatially unresolved events observed in radio similar to QFP wave trains demonstrated short periods of seconds \citep[e.g.,][]{2013A&A...550A...1K,2018ApJ...861...33K} and even subsecond \citep[e.g.,][]{2011SoPh..273..393M,2019ApJ...872...71Y}. In addition, high temporal resolution data taken during solar eclipses is also important for detecting short periods of QFP wave trains \citep[e.g.,][]{2002MNRAS.336..747W,2003A&A...406..709K,2016SoPh..291..155S}.

Observational studies showed that a QFP wave train often contains multiple periods. It has been confirmed in many events that some prominent periods of QFP wave trains are temporally correlated to QPPs in the accompanying flares, but the others are not \citep[e.g.,][]{2011ApJ...736L..13L, 2012ApJ...753...53S}. In particular cases, the periods of a QFP wave train are all associated with the QPPs in the accompanying flare \citep[e.g.,][]{2013SoPh..288..585S,2018ApJ...853....1S,zhou2021b}. However, there are still many cases whose periods are completely not associated with the accompanying flares \citep[e.g.,][]{2018MNRAS.477L...6S,2019ApJ...873...22S}. These results suggest that the periodicity origin of QFP wave trains should be diversified, and some of them are probably associated with flare QPPs. 

Generally, flare QPP is loosely defined as the periodic intensity variations in flare light curves seen in a wide wavelength range from radio to gamma-rays, with characteristic periods ranging from a fraction of a second to several tens of minutes \citep{2019PPCF...61a4024N}. In addition, since a light curve is obtained by observing the Sun as a star, i.e., without spatial resolution, it is hard to say what kind of physical process is responsible for the appearance of QPPs in the light curve. Because of these reasons, so far scientists have proposed a handful of possible mechanisms to account for the generation of flare QPPs \citep[see][and references therein]{2009SSRv..149..119N, 2016SoPh..291.3143V, 2018SSRv..214...45M, 2019PPCF...61a4024N, 2020STP.....6a...3K, 2021SSRv..217...66Z}. As pointed out by \cite{2009SSRv..149..119N}, the possible mechanisms for QPPs can be divided into two categories including pulsed energy release and MHD oscillations, and both can be relevant to the generation of QFP wave trains \citep{2011ApJ...736L..13L, 2012ApJ...753...53S, 2013SoPh..288..585S, 2018ApJ...853....1S}. Pulsed energy release can take place in different situations and forms, but it is commonly associated with various nonlinear processes in magnetic reconnections, such as the dynamic evolution of plasmoids  \citep[e.g.,][]{2000A&A...360..715K, 2015ApJ...799...79N, 2013ApJ...767..168L, 2018ApJ...868L..33L, 2018ApJ...866...64C, 2021ApJ...908L..37M}, oscillatory reconnection \citep[e.g.,][]{1991ApJ...371L..41C, 2009A&A...493..227M, 2012A&A...548A..98M, 2012ApJ...749...30M, 2019ApJ...874..146H, 2019ApJ...874L..27X, 2017ApJ...844....2T, 2019A&A...621A.106T}, and modulation resulting from external quasi-periodic disturbances \citep[e.g.,][]{2006A&A...452..343N, 2006SoPh..238..313C, 2009A&A...505..791S, 2012ApJ...753...53S, 2012ApJ...757..160J, 2019A&A...625A...3J}. MHD oscillations are relevant to the inherent physical properties of the wave hosts and their surrounding background medium, which can modulate flare energy release (or plasma emission) and therefore result in QPPs and QFP wave trains whose periodicities are prescribed either by certain resonances or by dispersive narrowing of initially broad spectra \citep{1983Natur.305..688R,2005A&A...440L..59F,2009SSRv..149..119N}.

Observationally, the periods of QFP wave trains are comparable to the typical period of flare QPPs, both are in the range of a few seconds to several minutes.   In addition, while QFP wave trains are mainly observed in flare impulsive and decay phases, QPPs can appear in all flare stages from pre-flare to decay phases. In some cases, the two phenomena can occur simultaneously and with the similar periods, indicating their intimate physical connection. However, the detailed physical relationship between the two phenomena is yet to be resolved. In our view, QFP wave trains and the simultaneous flare QPPs might represent the different aspects of a common physical process such as pulsed energy release or MHD oscillations in flares. In terms of their origins, QFP wave trains could be viewed as a subclass of QPPs in general, since some proposed physical processes for the generation of QPPs might not cause simultaneous QFP wave trains (for example, the oscillation of coronal loops). In addition, in some studies \citep[e.g,][]{2009ApJ...697L.108M, 2018ApJ...861...33K}, QPPs observed in radio were thought to be produced by the modulation of the local plasma density by QFP wave trains. In this case, QPPs are actually the result or the indirect signal of QFP wave trains. Because of these correlations, currently the proposed generation mechanisms for QFP wave trains are mainly analogy of those for flare QPPs (see Section \ref{theo} for details), since the latter have been investigated for more than half a century after the discovery \citep[see][and references therein]{2009SSRv..149..119N, 2016SoPh..291.3143V, 2018SSRv..214...45M, 2019PPCF...61a4024N, 2020STP.....6a...3K, 2021SSRv..217...66Z}.

\subsection{Amplitude and Intensity Profile}
The physical nature of QFP wave trains is also characterized by the special variation pattern of the wavefront intensity profiles. For example, the intensity profile of global EUV and Moreton waves often show a simultaneous increasing width and decreasing amplitude during the initial propagation stage, consistent with the nature of nonlinear fast-mode or shock waves. For freely-propagating linear or weakly nonlinear fast-mode magnetosonic waves, the integral over the entire wave pulse should be constant, as what has been reported in several studies of global EUV waves \citep[see][and references therein]{2015LRSP...12....3W}. Commonly, an intensity profile is defined as the intensity distribution along a specific path perpendicular to the wavefronts, which is a function of distance at a particular time. The intensity profile is often expressed as relative intensity change (i.e., $I/I_{0}$) or percentage change (i.e., $(I-I_{0})/I_{0}$) over the pre-event background. Here, $I$ and $I_{0}$ are the emission intensities at a certain time and the pre-event background emission intensity, respectively.

\begin{figure}[!t]
\begin{minipage}{0.52\linewidth}
\centering
\vspace{-0.3cm}
\subfloat{}\includegraphics[width=0.9\linewidth]{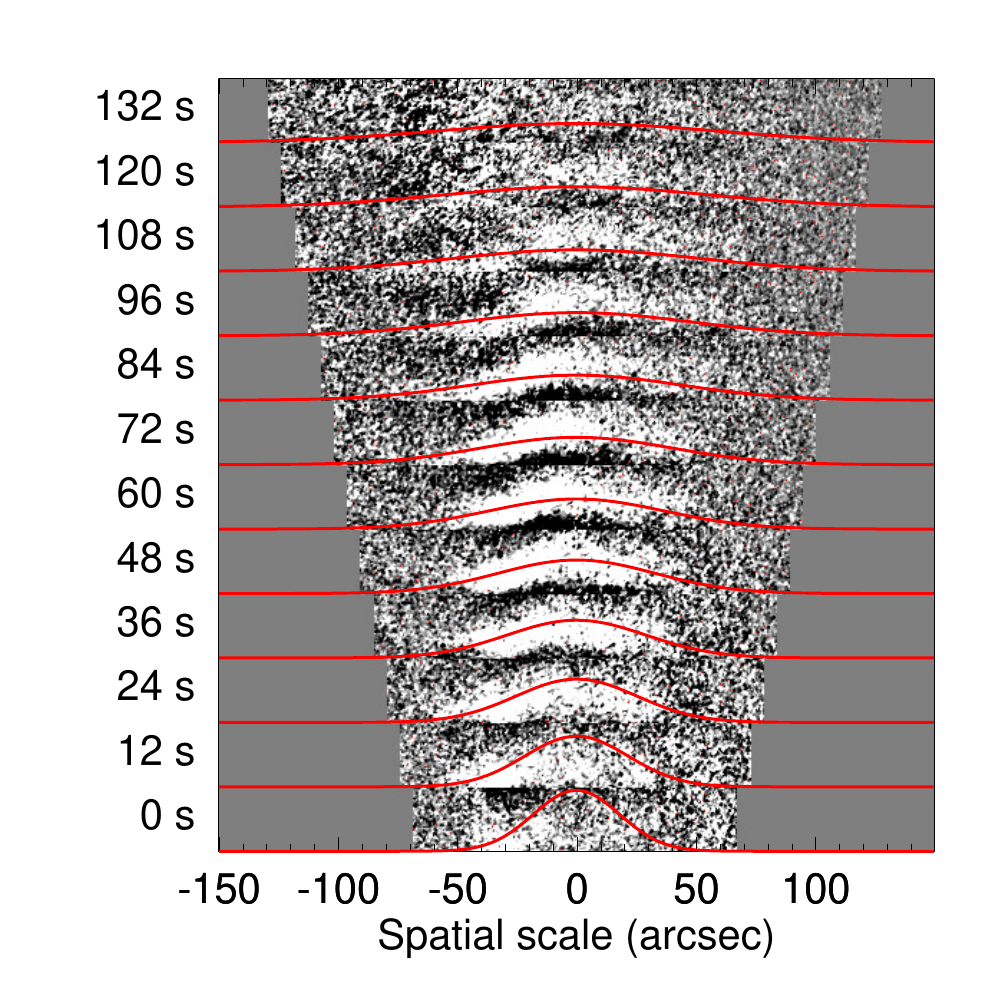}\vspace{-0.6cm}
\subfloat{}\includegraphics[height=3.95cm, width=0.9\linewidth]{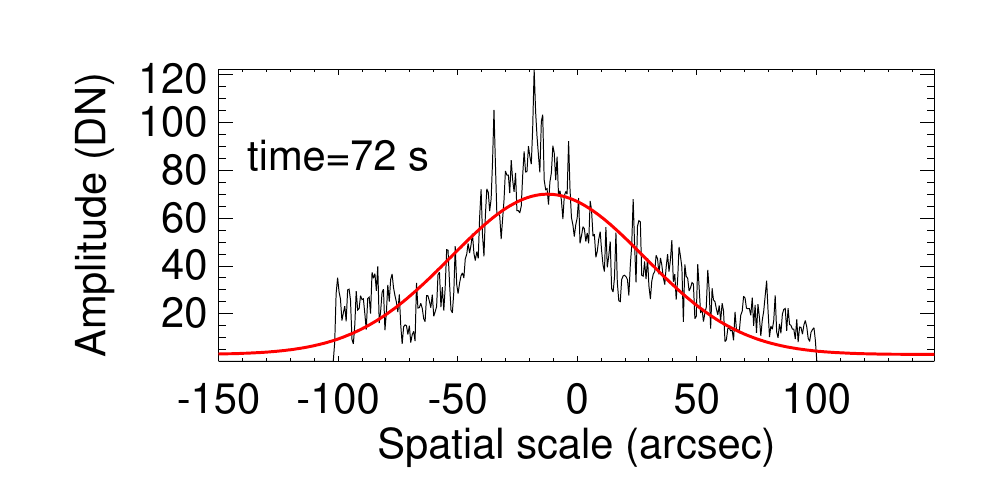}
\end{minipage}
\medskip
\hspace{-0.8cm}
\begin{minipage}{0.48\linewidth}
\centering
\subfloat{}\includegraphics[width=0.9\linewidth]{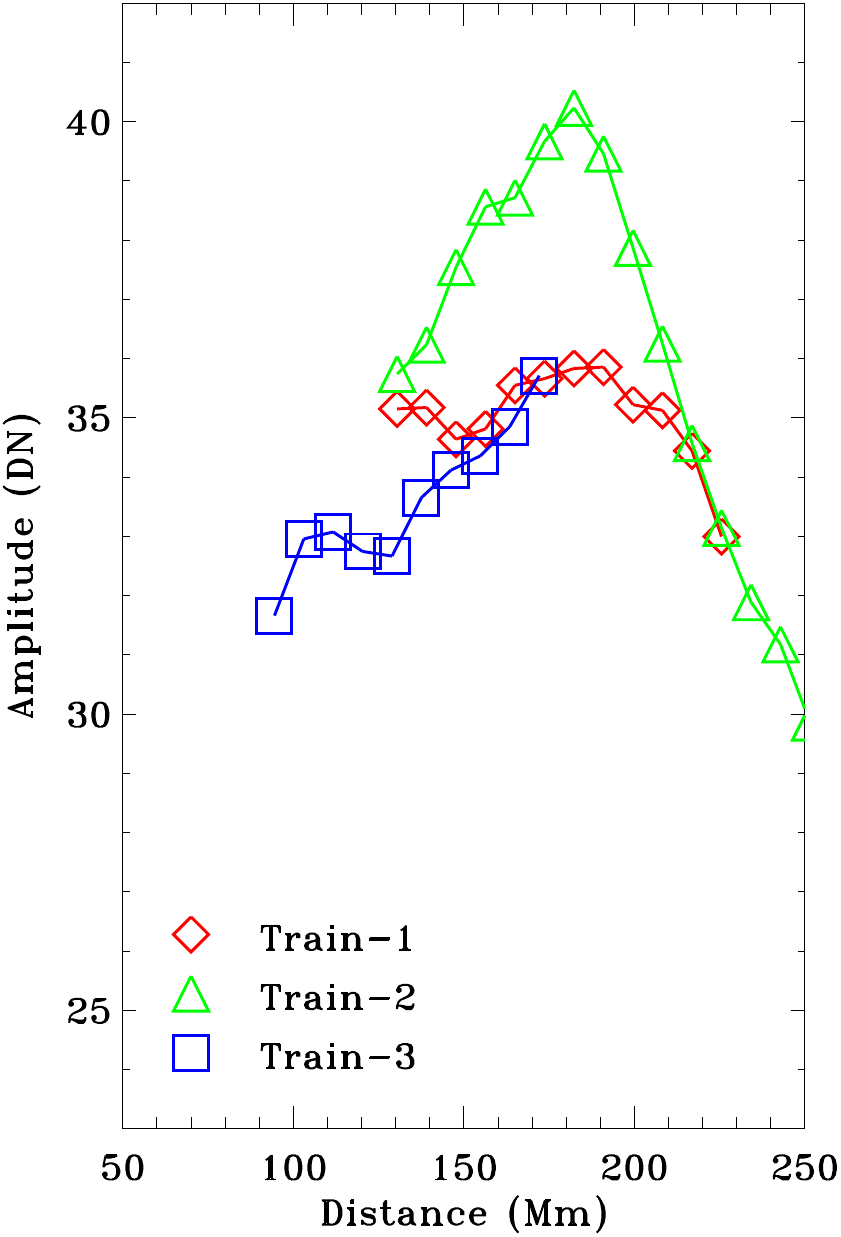}
\end{minipage}
\caption{Intensity profile and amplitude of the narrow QFP wave train on 2011 May 30  \citep{2013AA...554A.144Y}. The upper left panel is the temporal evolution of a specific wavefront at different times from the bottom up, while the left lower panel is the intensity profile of the wavefront at the time of 72 s. The red curves in the left panels are the corresponding Gaussian fitting curves of the intensity profiles. Right panel is the wave amplitudes of the three sub-QFP wave trains plotting as a function of the distance from the flare epicenter. The red diamonds, green triangles, and blue squares denote the parameters of Train-1, -2, and -3, respectively.}
\label{fig3}
\end{figure}

In practice, one often firstly generates a time-distance diagram and then obtains an intensity profile at a specific distance from the excitation source of a QFP wave train. In this case, the intensity profile is as a function of time as each wavefront traveling to the same distance. Observational results indicate that the peak intensity amplitudes of narrow and broad QFP wave trains are very different. Taking the published events as an example (\tbl{tbl1}), the values of the peak intensity amplitudes for narrow and broad QFP wave trains are in the ranges of 1\%--8\% and 10\%--35\%, respectively. It is noted that both narrow and broad QFP wave trains retain their variation ranges of the peak intensity amplitudes on stable levels for different events, and they do not show any notable physical connection with other parameters and the accompanying activities such as flares and CMEs. This may suggest that the intensity amplitudes of QFP wave trains are basically determined by the physical parameters of the supporting medium. Since narrow QFP wave trains propagate along corona loops in which the magnetic field strength and plasma density are typically higher than the quiet-Sun region where broad QFP wave trains propagate, we propose that the peak intensity amplitudes of QFP wave trains are probably affected by physical parameters such as magnetic field strength and plasma density of the medium, and the propagation direction of QFP wave trains with respect to the magnetic field direction. The very different intensity amplitudes of the two types of QFP wave trains are probably mainly caused by their different propagation mediums. As found in \cite{2017ApJ...847L..21P}, the geometrical waveguide dispersion suppresses the nonlinear steepening of trapped narrow QFP wave trains, while broad QFP wave trains propagate in the quiet-Sun region does not experience dispersion and can steepen significantly into shocks.

\begin{figure}[!t]
\centerline{\includegraphics[width=0.9\textwidth]{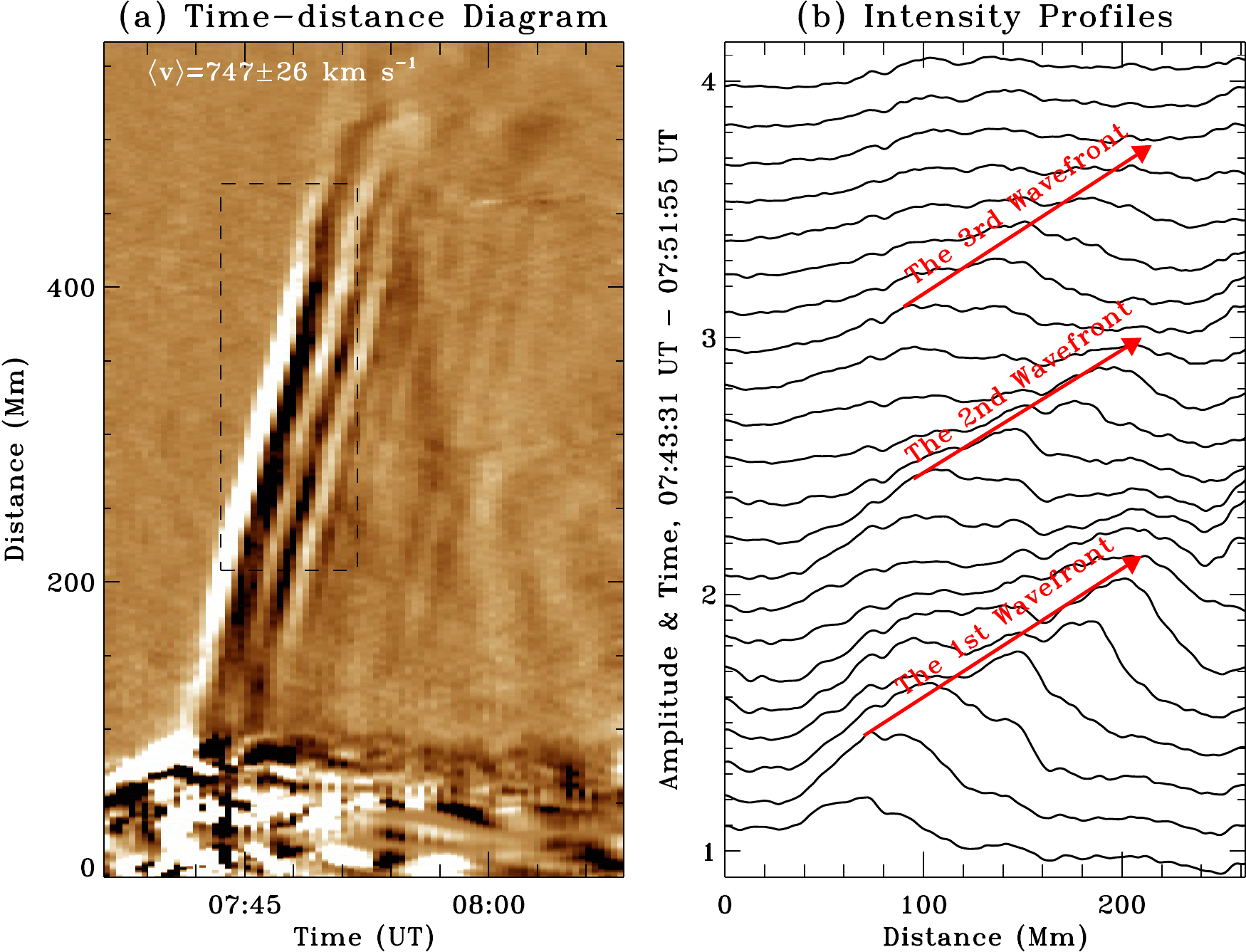}}
\caption{Intensity profile of the broad QFP wave train along the solar surface on 2012 April 24 \citep{2019ApJ...873...22S}. The left panel is a time-distance diagram made from AIA 193 \AA\ running-ratio images, in which the black dashed box show the region where the intensity profiles are checked. The right panel shows the percentage intensity profiles of the wave train at different times based on the AIA 193 \AA\ images, in which the red arrows indicate the first three wavefronts of the wave train.}
\label{fig4}
\end{figure}

For narrow QFP wave trains, \cite{2011ApJ...736L..13L} and \cite{2012ApJ...753...53S} checked the intensity profiles in the propagation direction at several consecutive times and found that the spatial profiles can be fitted with a sinusoidal function, from which physical information of phase speed, period, wavelength and amplitude can be obtained. Moreover, a variation trend of weak broadening width and decreasing amplitude of the wavefronts can be identified during the propagation. In addition, the authors also checked the temporal variation of intensity profiles, which are then used to periodic analysis with the aid of the wavelet analysis technique. \cite{2018MNRAS.480L..63S} reported the successive interactions of a narrow QFP wave train with two strong magnetic regions, they found that although the propagation direction was changed significantly after the interactions, the peak intensity amplitudes of the wave train remain at the same level. \cite{2013AA...554A.144Y} traced the detailed temporal evolution of the intensity amplitude of the narrow QFP wave train on 2011 May 30, they found that the intensity amplitude underwent a first increasing and then decreasing process (see also \cite{2018ApJ...853....1S} and the right panel of \fig{fig3}). The authors further check the evolution of a specific wavefront, it was found that the wavefront extended gradually along the waveguide, and the transverse distribution of the intensity profile perpendicular to the wave vector exhibited as a Gaussian profile (see the left panels of \fig{fig3}). For broad QFP wave trains, investigation of the variations of the intensity profiles are scarce. \cite{2019ApJ...873...22S} found the obvious broadening width and decreasing amplitude of the intensity profiles during the propagation of the broad QFP wave train on 2012 April 24 (see \fig{fig4}), and the initial steep intensity profiles weakened quickly with increasing time \citep[][]{2017ApJ...844..149K,zhou2021b}. In addition, the Alfv\'{e}n Mach number of the broad QFP wave train was estimated to be 1.39 in \cite{2019ApJ...873...22S}, indicating that the wave train was shocked significantly. These characteristics suggest that broad QFP wave trains are more similar to global EUV waves that are strong shocks during the initial stage but then quickly decay into linear or weak non-linear fast-mode magnetosonic waves \citep[e.g.,][]{2012ApJ...752L..23S}.

\subsection{Thermal Characteristics}
The AIA takes EUV images in seven channels covering a wide temperature range from 0.05 MK in the transition region to 20 MK in the flaring corona \citep{2012SoPh..275...17L}. The EUV observing channels of AIA and their peak response temperatures are 
304 \AA\ (He \uppercase\expandafter{\romannumeral 2}; ${\rm T} \approx 0.05 ~{\rm MK}$), 171 \AA\ (Fe \uppercase\expandafter{\romannumeral 9}; ${\rm T} \approx 0.6 ~{\rm MK}$), 193 \AA\ (Fe \uppercase\expandafter{\romannumeral 12}; ${\rm T} \approx 1.6 ~{\rm MK}$; Fe \uppercase\expandafter{\romannumeral 24}; ${\rm T} \approx 20 ~{\rm MK}$), 211 \AA\ (Fe \uppercase\expandafter{\romannumeral 14}; ${\rm T} \approx 2 ~{\rm MK}$), 335 \AA\ (Fe \uppercase\expandafter{\romannumeral 16}; ${\rm T} \approx 2.5 ~{\rm MK}$), 94 \AA\ (Fe \uppercase\expandafter{\romannumeral 18}; ${\rm T} \approx 6.3 ~{\rm MK}$), 131 \AA\ (Fe \uppercase\expandafter{\romannumeral 8}; ${\rm T} \approx 0.4 ~{\rm MK}$; Fe \uppercase\expandafter{\romannumeral 21}; ${\rm T} \approx 10 ~{\rm MK}$). Such a wide temperature coverage provides an unprecedented opportunity for diagnosing the thermal properties of QFP wave trains. Observations showed that narrow QFP wave trains are best seen in the AIA 171 \AA\ channel (occasionally in the AIA 193 \AA\ and 211 \AA\ channels), indicating the narrow temperature dependence. On the contrary, broad QFP wave trains cover a wider temperature range, which can be observed in all AIA's EUV channels (best seen in 193 \AA\ and 211 \AA\ channels) as global EUV waves.

According to the explanation given by \cite{2016AIPC.1720d0010L}, the narrow temperature dependence of narrow QFP wave trains are possibly due to two reasons. The first reason is owning to the physical property in the waveguide structures and the small intensity amplitude of narrow QFP wave trains. It is probably that the temperature of the wave-hosting plasma is close to the AIA 171 \AA\ channel's peak response temperature. In addition, due to the small intensity amplitude of narrow QFP wave trains, they are hard to cause large temperature departures, unlike the large intensity amplitude cased by broad QFP wave trains. These possible conditions might account for the absence of narrow QFP wave trains in other AIA's EUV channels. The other reason is possibly due to the detectability of the detectors used for different AIA channels. Since the AIA 171 \AA\ channel has a much higher photon response efficiency than any other channels by at least one order of magnitude, it is particularly sensitive to small intensity variations. The two reasons might work either separately or together. However, so far the exact reasons for the narrow temperature dependence of narrow QFP wave trains are still remain unclear.

\begin{figure} [!t]
\centerline{\includegraphics[width=0.9\textwidth]{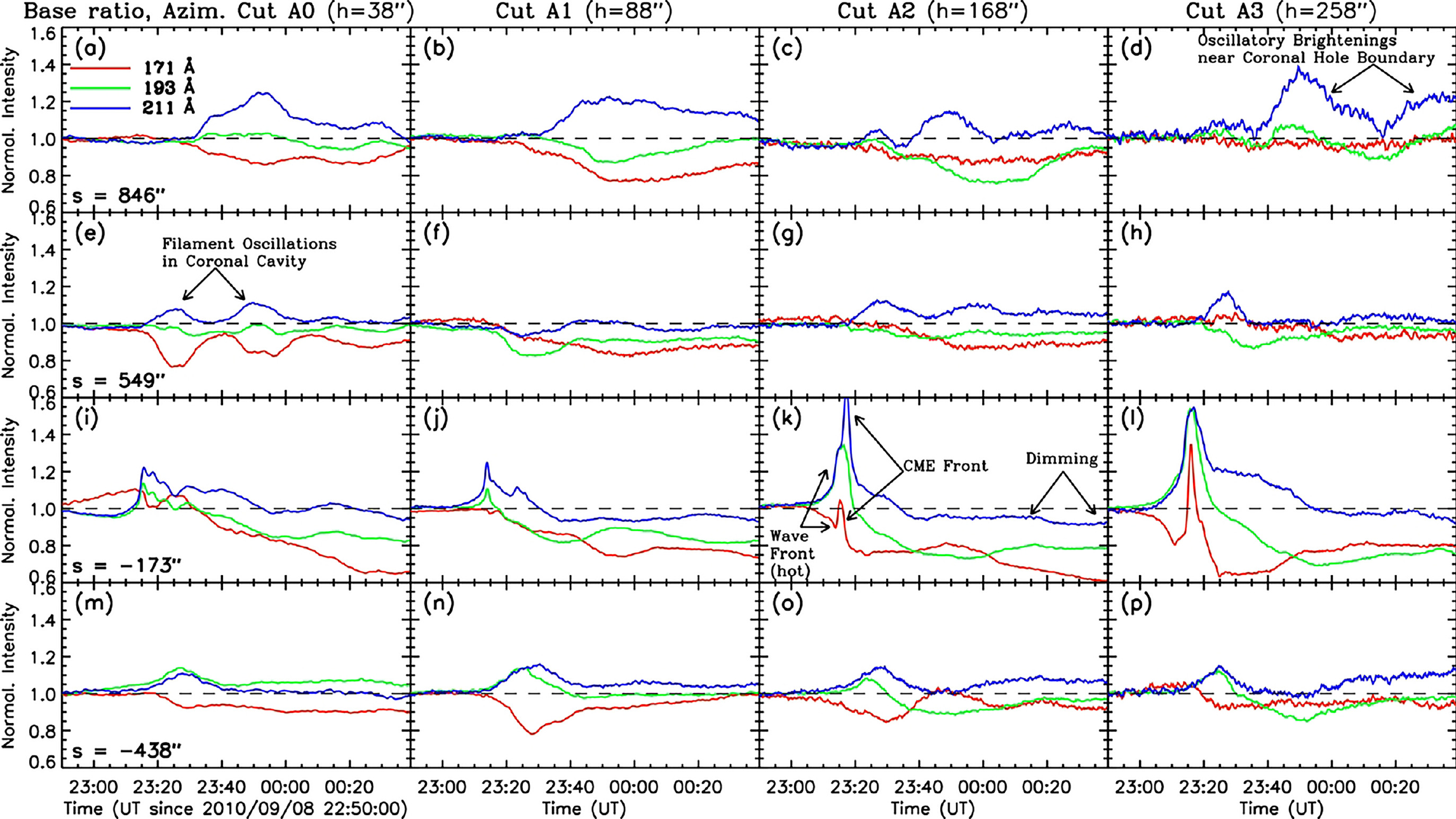}}
\caption{Base-ratio temporal profiles of emission intensity from azimuthal cuts at selected positions shown by the plus signs in Figure 4 in \cite{2012ApJ...753...52L}. The general trend of darkening at 171 \AA\ and brightening at 193 and 211 \AA\ indicates heating in the EUV wave pulse ahead of the CME.}
\label{fig5}
\end{figure}

In the broad QFP wave train on 2010 September 08, \cite{2012ApJ...753...52L} observed the darkening at 171 \AA\ and brightening at 193 \AA\ and 211 \AA\ of the wavefronts, which followed by a recovery in the opposite direction (see \fig{fig5}). This process indicates the initial heating and then subsequent cooling of coronal plasma, and it can be interpreted by adiabatic heating due to compression followed by cooling with subsequent expansion/rarefaction driven by a restoring pressure gradient force. A similar signature was previously reported in global EUV waves \citep[see][and references therein]{2014SoPh..289.3233L}. Such an adiabatic compression caused by EUV waves can cause a considerable heating to the coronal plasma. For example, \cite{2011ApJ...738..167S} estimated that a mild adiabatic compression can result in a maximum density increase of about $10\%$ and a temperature increase of about $7\%$.

\subsection{Energy Flux and Coronal Heating}
QFP wave trains carry energy away from their excitation sources, and the energy will be dissipated into the corona in which the waves propagate. Therefore, QFP wave trains can inevitably result in the heating of corona. Earlier observations have suggested that short period oscillations might make a significant contribution to the energy input of the coronal loops \citep[e.g.,][]{2001MNRAS.326..428W}. The {\em SDO}/AIA observational results show that the energy flux carried by narrow and broad QFP wave trains are in the range of about $(0.1-4.0) \times 10^5 {\rm ~erg~ cm}^{-2} {\rm ~s}^{-1}$ and $(1-2) \times 10^6 {\rm ~erg~ cm}^{-2} {\rm ~s}^{-1}$, respectively. Obviously, such an energy flux level is sufficient for sustaining the temperature of active region coronal loops, because the typical energy flux density requirement for heating coronal loops is estimated to be about  $10^5 {\rm ~erg~ cm}^{-2} {\rm ~s}^{-1}$ \citep{1977ARA&A..15..363W,2005psci.book.....A}. It is noted that the energy flux carried by broad QFP wave trains is at least one order of magnitude higher than that of narrow QFP wave trains. While narrow QFP wave trains are mainly attributed to the plasma heating of active region coronal loops, broad QFP wave trains are more efficient for the plasma heating in the quiet-Sun regions.

The energy flux carried by a QFP wave train can be estimated from the kinetic energy of the perturbed plasma that propagate with phase speed through a volume element. The energy of the perturbed plasma is 

\begin{equation}
\label{ef}
E = (\frac{1}{2} \rho v_{\rm 1}^2) v_{\rm gr},
\end{equation}

\noindent where $v_{\rm 1}$ is the disturbance speed of the locally perturbed plasma \citep{2004ESASP.575...97A}, and $v_{\rm gr}$ is the group speed of the wave. Generally, for a rough estimation, one can use the measurable phase speed ($v_{\rm ph}$) of a dispersive wave train to replace the group speed ($v_{\rm gr}$) in equation (\ref{ef}). For non-dispersion wave trains, their phase speeds are equal to the values of the group speeds. In addition, in optically thin corona, the emission intensity $I$ is directly proportional to the square of the plasma density $\rho$, i.e., $I \propto \rho ^{2}$. Therefore, the density modulation of the background density $\frac{\rm d \rho}{\rho}$ can be written as $\frac{\rm dI}{2I}$. So, the energy flux of the perturbed plasma could be written as 

\begin{equation}
E \geq \frac{1}{8} \rho v_{\rm ph}^{3} (\frac{\rm dI}{I})^{2},
\end{equation}

\noindent if we assume that $\frac{v_{\rm 1}}{v_{\rm ph}}$ is equal to or greater than $\frac{\rm d \rho}{\rho}$. Obviously, the energy flux estimated by this equation is determined by the coronal electron density $\rho$, perturbation amplitude of the emission intensity $\rm dI$, and the phase speed $v_{\rm ph}$ of QFP wave trains. Since the intensity amplitude of narrow QFP wave trains are all in the range of 1\%--8\%, their corresponding energy fluxes estimated based on this equation are all on the order of $\approx 10^5 {\rm ~erg~ cm}^{-2} {\rm ~s}^{-1}$. On the contrary, the energy fluxes of broad QFP wave trains are about one order of magnitude higher than narrow QFP wave trains, which mainly result from their higher perturbation amplitude of the emission intensity (10\%--35\%). Here, we would like to point out that the estimated energy fluxes of QFP wave trains are underestimated, since the energy flux decreases quickly by orders of magnitude with height due to the spreading of the waves over a large area as a result of magnetic field divergence  \citep{2011ApJ...740L..33O}. However, in practice, many estimations are based on the measurement of the intensity variation far away from their origin. 

Observations showed that the occurrence of QFP wave trains are quite common in the corona, although many of them can still not be detected based on our current telescopes \citep{2016AIPC.1720d0010L}. Besides the association with relatively stronger flares (GOES soft X-ray C- and M classes), they can also be excited by many low-energy small flares \citep[GOES soft X-ray B-class,][]{2010ApJ...723L..53L,2018MNRAS.477L...6S}, small coronal brightenings \citep[][]{2018MNRAS.480L..63S,2020ApJ...889..139M}, and some signatures of possible magnetic reconnection events that can not even be recognized as flares in the GOES soft X-ray light curves \citep[e.g.,][]{2017ApJ...851...41Q,2018ApJ...868L..33L}. In addition, due to the large-scale propagation nature of QFP wave trains, they are expected to further trigger a plenty of subsequent nano-flares \citep{1988ApJ...330..474P} or magnetic reconnection events in the corona with the complicated magnetic field, and these small flaring activities can probably further cause mini QFP wave trains. The energy dissipation of these undetected small-scale energetic events can further contribute more heating to the coronal plasma. Therefore, the contribution of QFP wave trains to the heating of coronal plasma might be more significant than our current perception \citep{2020SSRv..216..140V}.

\begin{figure}[!t]
\centerline{\includegraphics[width=0.9\textwidth]{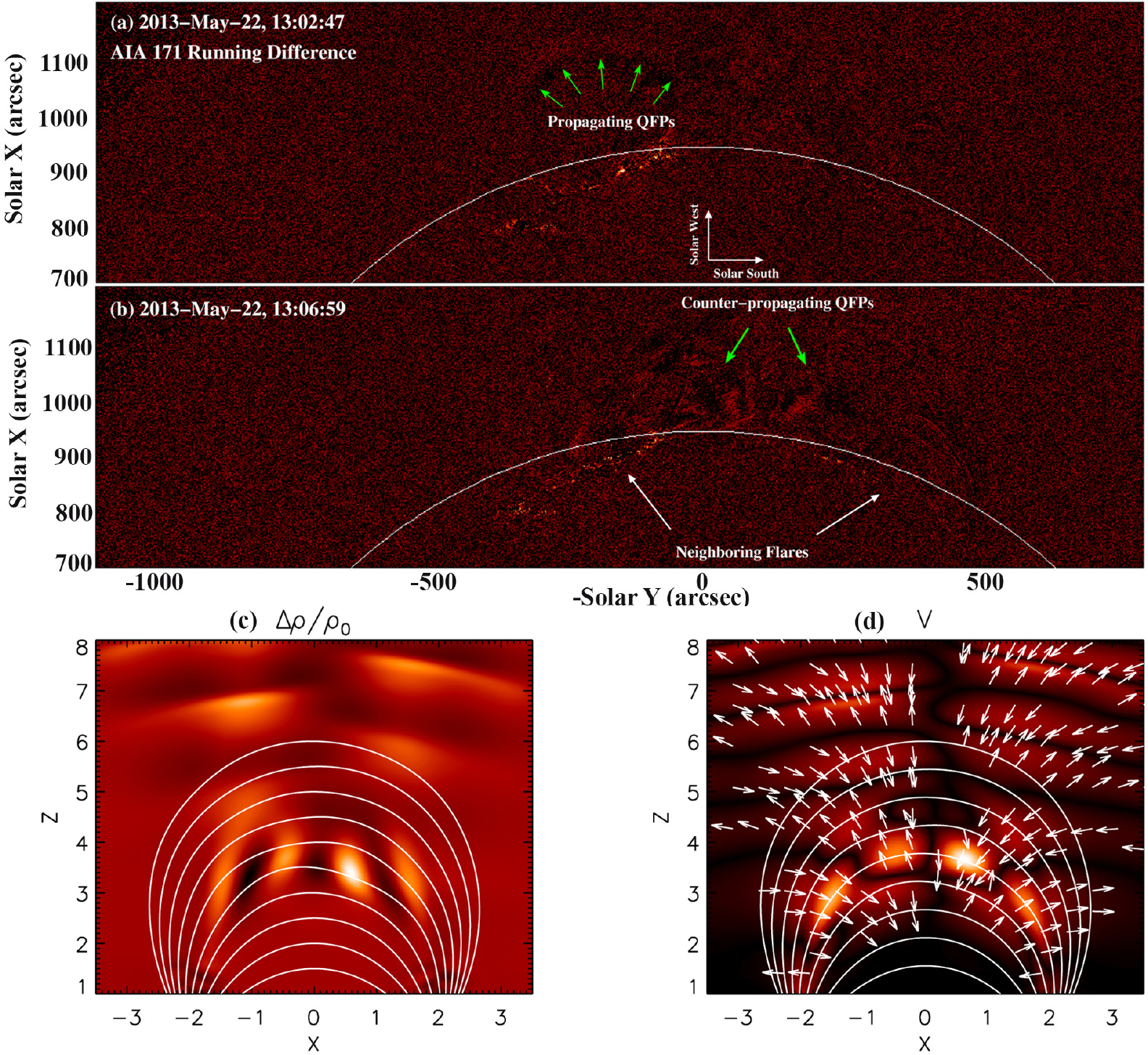}}
\caption{Interaction between counter-propagating narrow QFP wave trains in the same trans-equatorial coronal loop system on 2013 May 22 \citep{2018ApJ...860...54O}. Panel (a) shows the outward-propagating QFP wave train from the primary flare, while panel (b) shows the interaction between the two counter-propagating QFP wave trains from the primary flare on the left and the second flare on the right. The green arrows in panel (a) indicate the outward-propagating wavefronts, the two white arrows in panel (b) indicate the locations of the two flares, and the two green ones indicate the interaction sites. The bottom row show the corresponding numerical simulation results of the event, in which the left and the right panels are the density and velocity perturbations in the $x-z$ plane at $y=0$, respectively. The magnetic field lines and velocity direction (right panel only) are overlaid as white curves and arrows, respectively.}
\label{fig6}
\end{figure}

\subsection{Interaction with Coronal Structure}
The highly structured corona is an inhomogeneous and anisotropic medium full of hot magnetized plasma, which strews with strong magnetic structures such as active regions, coronal holes and filaments. The Alfv\'{e}n and fast-mode magnetosonic speeds at the boundary of these structures exist a strong speed gradient owning to the sudden changes of the magnetic field strength and plasma density. In addition, due to the gravitational stratification of the solar atmosphere, the plasma density falls off faster than the magnetic field strength in the low corona. Therefore, the Alfv\'{e}n and fast-mode magnetosonic speeds in the low corona increase with height on the quiet-Sun regions \citep{1999A&A...348..614M}. The large-scale propagation of QFP wave trains will inevitably interact with regions of strong gradients of Alfv\'{e}n and fast-mode magnetosonic speeds, and exhibit wave phenomena such as reflection, refraction and transmission effects. In addition, QFP wave trains can also excite oscillations of filaments and coronal loops during their propagation. The evidence of reflection, refraction and transmission effects of single pulsed global EUV waves have been reported in many studies; interested readers can refer to several recent reviews \citep{2014SoPh..289.3233L,2015LRSP...12....3W,2017SoPh..292....7L,Shen2020}.

For narrow QFP wave trains propagating along open funnel-like coronal loops, they do no interact with coronal structures. However, their propagation speed will be alerted by the increase of characteristic fast-mode speed with height. In some cases, QFP wave trains propagate along closed coronal loops, which will be reflected at the remote end of the loop system. \cite{2011ApJ...736L..13L} observed the bidirectional propagation of QFP wave trains in a closed loop system that connects the conjugate flare ribbons, but the authors were unclear whether the bidirectional wave trains were generated independently or the same wave train reflected repeatedly between the conjugated loop footpoints. \cite{2018ApJ...860...54O} firstly reported the detection of counter-propagating QFP wave trains along the same closed trans-equatorial coronal loop system, which were associated with two flares successively occurred in two neighboring active regions on 2013 May 22. The counter-propagating QFP wave trains propagated at large speeds of the order of $\textgreater$\speed{1000} and interacted at the middle section of the loop system, which further excited trapped kink-mode and slow-mode MHD waves in the coronal loops (see the top and the middle rows of \fig{fig6}). The authors further performed a three-dimensional MHD simulation for this event, and the results are well in agreement with the observations (see the bottom row of \fig{fig6}). The unambiguous reflection of a QFP wave train at the far end of the closed guiding coronal loop was observed by \cite{2019ApJ...873...22S}, in their case the incoming and reflected waves propagate at a similar speed of about \speed{900}, and the guiding closed loop system exhibited obvious kink oscillations. In addition, single pulse global EUV waves trapped in closed loops are also observed in some events, which can also trigger the transverse kink oscillation of the guiding loops \citep{2015ApJ...803L..23K,2021SoPh..296..169Z}.

\begin{figure} 
\centerline{\includegraphics[width=0.9\textwidth]{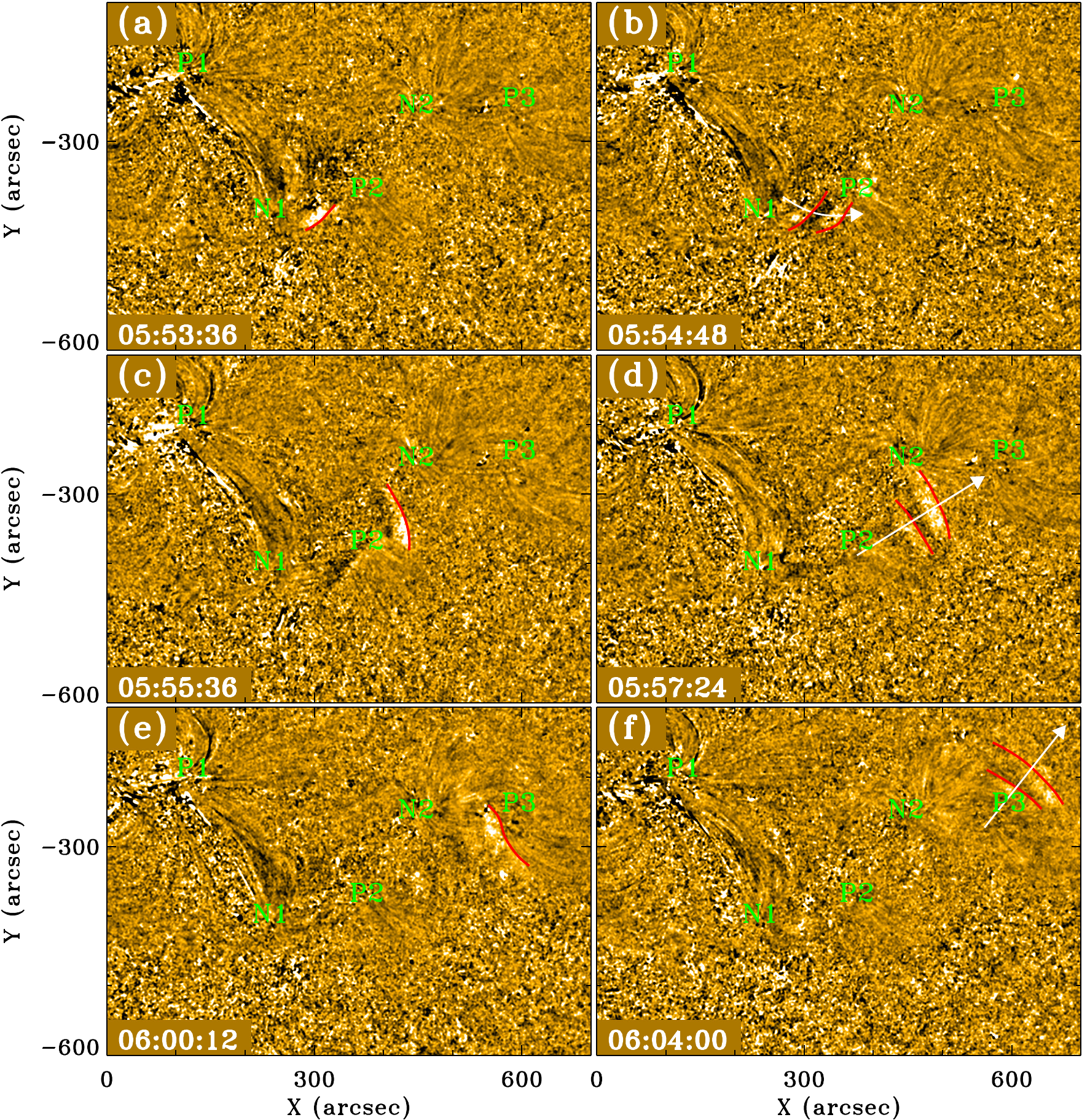}}
\caption{AIA 171 \AA\ running-ratio images show the interaction of the narrow QFP wave train on 2011 February 14 to remote strong magnetic polarities \citep{2018MNRAS.480L..63S}. The red curves marks the forefront of the wavefronts, and the white arrows indicate the propagation direction. The green symbols of P1, N1, P2, N2, and P3 mark the regions of strong magnetic fields, where letters P and N represent positive and negative magnetic polarities, respectively.}
\label{fig7}
\end{figure}

When multiple active regions exist simultaneously on the Sun, they are often connected by interconnecting coronal loops. \cite{2018MNRAS.480L..63S} reported a special narrow QFP wave train propagates along such closed interconnecting coronal loops, which passed through two different magnetic polarities and its propagation direction also changed significantly after each interaction with the magnetic polarities (see \fig{fig7}). It was noted that the propagation speeds before and after each of interactions showed little difference. This interesting phenomenon was interpreted as the refraction effect of the QFP wave train due to the strong speed gradients around the strong magnetic regions on the path. The refraction of narrow QFP wave trains was also evidenced in \cite{2018ApJ...853....1S}, in which the north part of the wavefronts became broader and more bent during their passing through a strong magnetic field region. This also result in the different propagation speeds of the north (\speed{1485}) and south (\speed{884}) parts of the wave train. 

For large-scale broad QFP wave trains propagating across the solar surface, they are more liable to interact with remote coronal structures. In the event studied by \cite{2019ApJ...873...22S}, the on-disk propagating wavefronts interacted with a remote active region and showed a significant deformation around  the middle section of the wavefronts, similar to what had been observed in global EUV wave \citep{2012ApJ...746...13L, 2013ApJ...773L..33S, 2013ApJ...775...39Y}. This phenomenon was interpreted as the transmission of a fast-mode magnetosonic wave through an active region in which the central characteristic fast-mode magnetosonic wave speed is faster than the rim. It was noted that the QFP wave train also result in the transverse oscillation of a remote filament and a closed coronal loop. \cite{2012ApJ...753...52L} studied a limb event in which broad QFP wave trains were observed in both south and north directions over the limb. The propagating wavefronts caused an uninterrupted chain sequence of deflections and/or transverse oscillations of remote coronal structures including a flux-rope coronal cavity and its embedded filament with delayed onsets consistent with the wave travel time at an elevated speed (by $\approx$50\%) within it, which indicates that the wavefronts penetrated through a topological separatrix surface into the cavity. The sequential response of remote coronal structures to the arrival of large-scale broad QFP wave trains reminds us that  global EUV waves can also cause the chain of oscillations of separate filaments \citep{2014ApJ...786..151S} and even simultaneous transverse and longitudinal oscillations of different filaments \citep{2014ApJ...795..130S,2016SoPh..291.3303P}. Recently, \cite{zhou2021b} observed the interaction of an on-disk broad QFP wave train with a remote low latitude coronal hole. During the successive transmission of the wavefronts through the coronal hole, intriguing refraction and reflection effects of the wave were identified around the coronal hole's west boundary. Since the coronal hole had a C-shape, the north and south arms of refracted wavefronts propagated towards each other and finally merged into one on the east side of the coronal hole. This phenomenon was interpreted as the interference effect of broad QFP wave trains, where the coronal hole acts as a concave lens. As mentioned above, observations and wave effects provide compelling evidence for supporting the interpretation of QFP wave trains as fast-mode magnetosonic waves.

\subsection{Possible Manifestations of QFP Wave Trains in Radio}
\begin{figure} [!t]
\centerline{\includegraphics[width=0.9\textwidth]{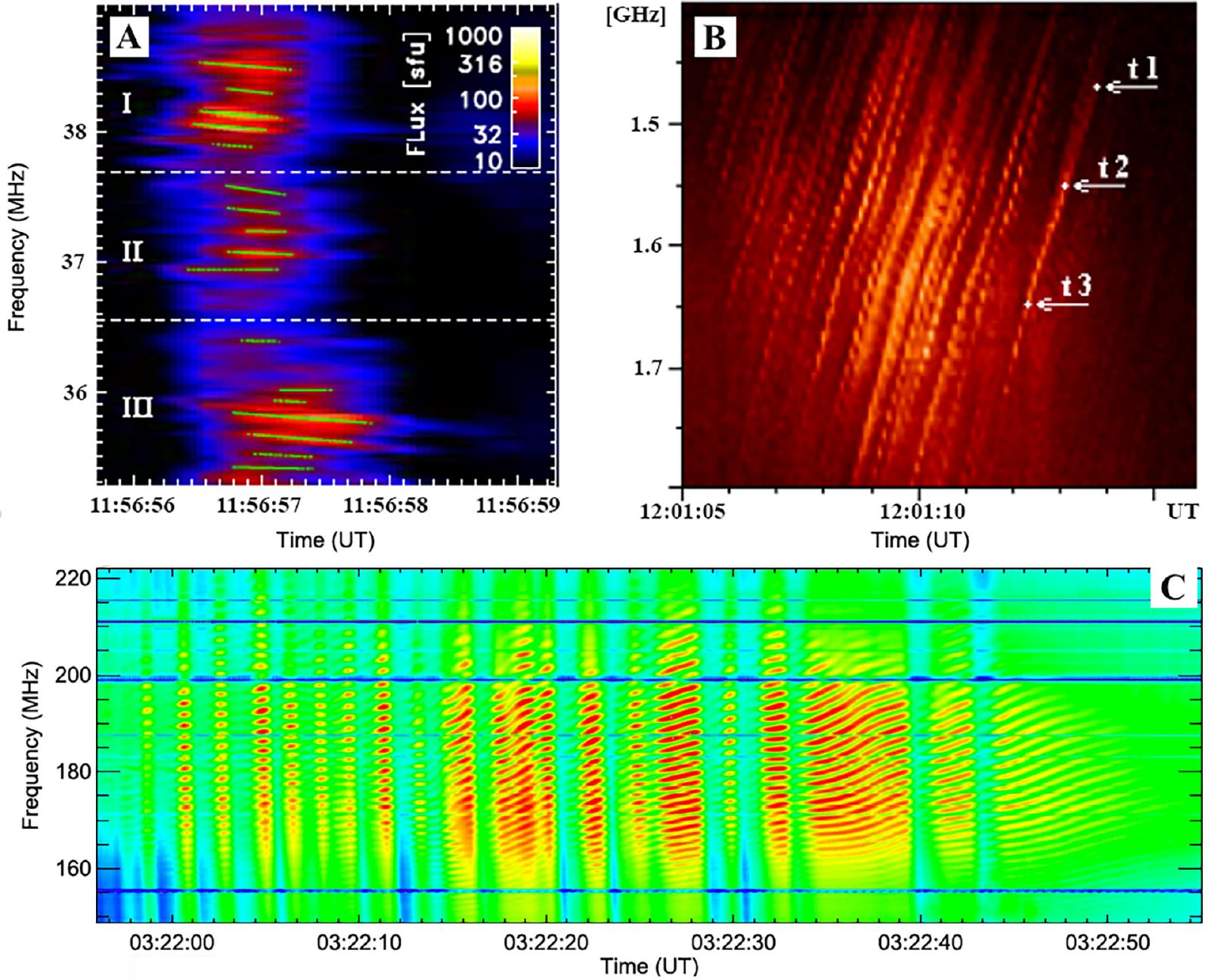}}
\caption{Candidate signatures in radio dynamic spectrums for coronal QFP wave trains. The upper left panel shows the dynamic spectrum of a type \uppercase\expandafter{\romannumeral 3} radio burst occurred on 2015 April 16 and observed by the LOFAR in the frequency band of 35--39 MHz, in which the fine horizontal striae that can be fitted by a linear function (green lines) are the so-called type \uppercase\expandafter{\romannumeral 3}b radio bursts. The regions of apparent clustering of the striae into three distinct groups are indicated by `` \uppercase\expandafter{\romannumeral 1},'' `` \uppercase\expandafter{\romannumeral 2},'' and `` \uppercase\expandafter{\romannumeral 3} and separated by the horizontal dashed lines \citep{2018ApJ...861...33K}. The upper right panel shows an example of radio fiber bursts on 1998 November 23 \citep{ 2013A&A...550A...1K}, which was  observed by the Ond\v{r}ejov radio spectrograph \citep{1993SoPh..147..203J}. The bottom panel shows an example of radio zebra patter structures in a type \uppercase\expandafter{\romannumeral 4} radio burst on 2011 June 21 \citep{2018ApJ...855L..29K}, which was observed by the Assembly of Metric-band Aperture Telescope and Real-time Analysis System \citep{2012SoPh..277..447I}.}
\label{fig8}
\end{figure}

In addition to direct imaging observations in EUV wavelength band, quasi-periodic patterns or fine structures in radio dynamic spectrum are generally thought to be the possible indirect signals of spatially-resolved QFP wave trains in EUV. In principle, quasi-periodic fine structures in radio dynamic spectrum can be produced by means of coherent modulating of the local coronal plasma density \citep{2010RAA....10..821C}, and this periodic modulation can be result from the propagation of QFP wave trains in the low corona \citep[][]{2013A&A...550A...1K,2013A&A...552A..90K,2018SoPh..293..115S,2018ApJ...861...33K}. \cite{1983Natur.305..688R} developed a theory to interpret the observed short period (a second or sub-second) pulsations in type \uppercase\expandafter{\romannumeral 4} radio bursts by means of studying the development and propagation of an impulsively generated QFP wave train within a dense coronal loop, and the authors provided that an impulsive disturbance (such as a flare) can naturally gives rise to quasi-periodic pulsations owning to the dispersive evolution of the disturbance \citep{1984ApJ...279..857R}. From then on, this theory has been applied to explain various quasi-periodic features in radio observations \citep[see][and references therein]{2020SSRv..216..136L}, such as type \uppercase\expandafter{\romannumeral 3}b bursts  \citep[see the upper left panel in \fig{fig8},][]{2018ApJ...861...33K}, fiber bursts \citep[see the upper right panel in \fig{fig8},][]{2011SoPh..273..393M,2013A&A...550A...1K}, and wiggly zebra patterns \citep[see the bottom panel in \fig{fig8},][]{2018ApJ...855L..29K}. Both fiber bursts and zebra patterns are particular quasi-periodic fine structures in solar type \uppercase\expandafter{\romannumeral 4} radio bursts, while type \uppercase\expandafter{\romannumeral 3}b bursts are fine spectral structuring in type \uppercase\expandafter{\romannumeral 3} bursts characterized by multiple narrowband bursts with slow frequency drift \citep[][]{1972A&A....20...55D, 2018SoPh..293..115S}. These fine structures in radio spectrum are believed to be important sources of information for probing coronal plasma parameters and diagnosing flare processes \citep[see][and references therein]{2006SSRv..127..195C}. 

\begin{figure}[!t]
\centerline{\includegraphics[width=0.9\textwidth]{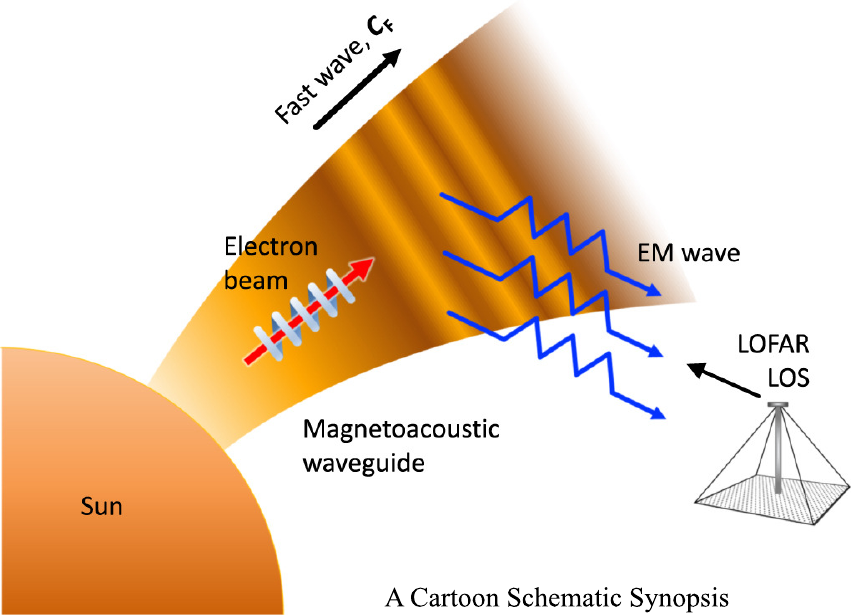}}
\caption{A schematic synopsis illustrates a scenario for the generation of quasi-periodic striations (type \uppercase\expandafter{\romannumeral 3}b bursts) in the dynamic spectrum of type \uppercase\expandafter{\romannumeral 3} bursts by a QFP wave train \citep{2018ApJ...861...33K}.}
\label{fig9}
\end{figure}

Solar radio observations typically have high temporal resolution but without spatial resolution. Even for compound interferometer observations, the spatial resolution is still very low. Therefore, the physical connections between various quasi-periodic fine structures in radio and spatially-resolved QFP wave trains in EUV are still unclear. Generally, one often connects quasi-periodic radio structures with QFP wave trains in EUV by comparing their physical parameters such as periods, speeds, and temporal correlation. In the series work published by M\'{e}sz\'{a}rosov\'{a} et al., the periods of the radio pulsations are in the range of 60--80 and 0.5--1.9 seconds. The longer periods are similar to those measured in spatially-resolved EUV observations of QFP wave trains, while the short ones are unclear because current AIA EUV observations can not detect periods lower than 24 seconds \citep{2014SoPh..289.3233L}. Similar physical parameters are also derived from the observations of type \uppercase\expandafter{\romannumeral 3}b radio bursts \citep{2018SoPh..293..115S}. For example, \cite{2018ApJ...861...33K} studied the type \uppercase\expandafter{\romannumeral 3}b radio bursts observed in a dynamic spectrum of a type \uppercase\expandafter{\romannumeral 3} radio burst \citep[see also][]{2013A&A...550A...1K, 2018SoPh..293..115S}. The authors proposed that the formation of the observed type \uppercase\expandafter{\romannumeral 3}b radio bursts were probably caused by the modulation of the field-aligned propagating electron beam by a QFP wave train along the same bundle of funnel-like coronal loops. Therefore, the observed radio emissions in the type \uppercase\expandafter{\romannumeral 3} radio burst also carry the same periodic information of the QFP wave train (see \fig{fig9}). Based on this scenario, the authors further derived the physical parameters including speed, period, and amplitude of the possible QFP wave train, and their corresponding values are respectively about \speed{657}, 3 seconds, and a few per cent, in agreement with those detected in spatially-resolved QFP wave trains in EUV observations.

Theoretically, the time signature of an impulsively generated QFP wave train propagating along coronal loops with different density contrast ratios is expected to produce a characteristic tadpole wavelet spectrum, i.e., a narrow-spectrum tail precedes a broad-band head, which indicates that the instantaneous period of the oscillations in the wave train decreases gradually with time \citep{2004MNRAS.349..705N}. In observation, the possible QFP wave train detected in the solar eclipse on 1999 August 11 does show such a special signature \citep{2003A&A...406..709K}. In some studies, if a tadpole wavelet spectrum can be observed in radio observations, one often speculates the appearance of a possible QFP wave train in the low corona, even though the wave signature does not observed in EUV imaging observations. For example, M\'{e}sza\'{a}rosov\'{a} et al. detected similar tadpole wavelet spectrum in solar decimetric type \uppercase\expandafter{\romannumeral 4} radio bursts, and they therefore interpreted the detected radio pulsations as the results of possible QFP wave trains traveling along loops through the radio source and modulating the gyrosynchrotron emissions \citep[e.g.,][]{2009AdSpR..43.1479M,2009ApJ...697L.108M,2011SoPh..273..393M}. In combination with imaging observations and radio interferometric maps, \cite{2013SoPh..283..473M} showed that a radio source that exhibits the wavelet tadpole feature was located at the null point of a fan-spine structure in the low corona, and the author suggested that this might imply the passage of a QFP wave train through there. 

In the above mentioned studies, although in radio observations the authors detected similar physical parameters (e.g., period and speed) as those observed in spatially-resolved QFP wave trains in EUV, and similar characteristic tadpole wavelet spectrum as predicted by the theory, it is also unclear whether various types of quasi-periodic radio features truly result from the modulation of the local coronal plasma by QFP wave trains. Firstly, in all the above studies, the authors did not observe the simultaneous appearance of spatially-resolved QFP wave trains. Vice versa, most QFP wave trains in EUV do not be accompany by quasi-periodic radio fine structures. Secondly, in practical observations, the wavelet spectrums of spatially-resolved QFP wave trains in EUV do not exhibit the tadpole feature.

\begin{figure} [!t]
\centerline{\includegraphics[width=0.9\textwidth]{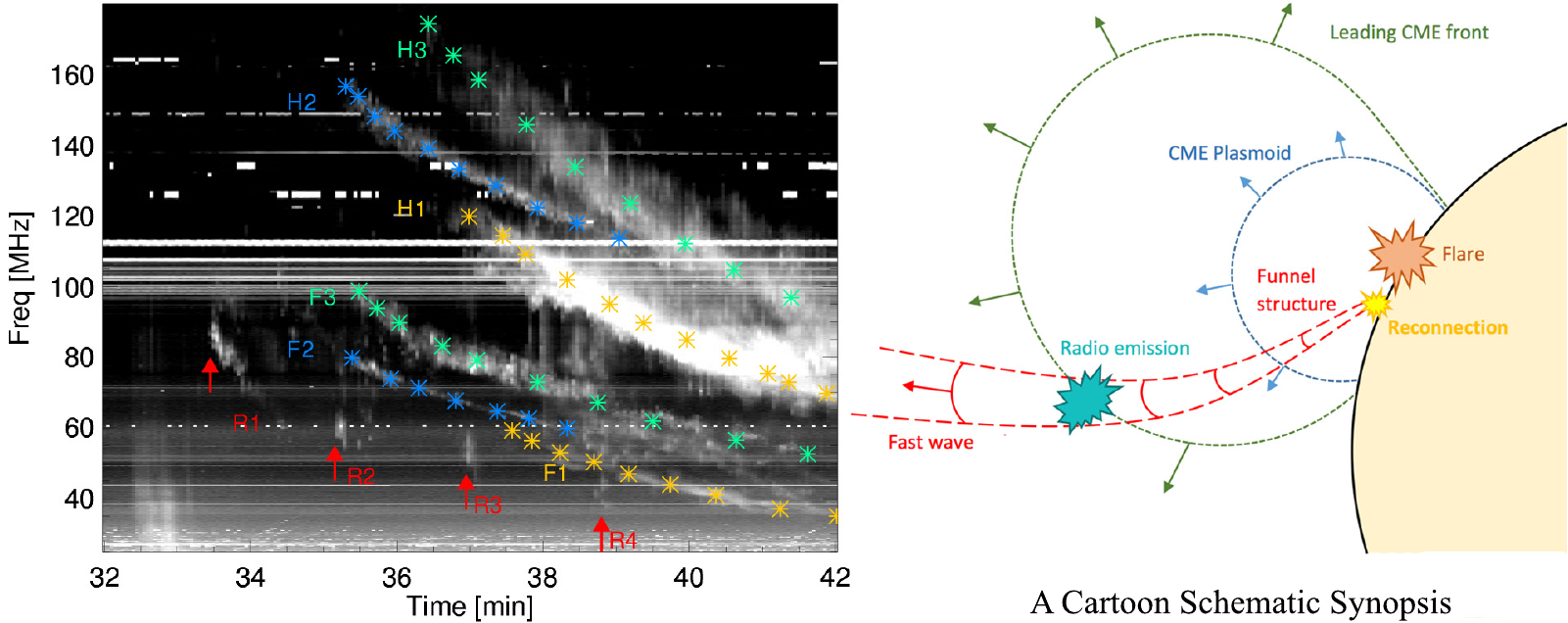}}
\caption{The left panel is the Learmonth radio spectra on 2014 November 03, which shows the discrete regions of enhanced emissions (radio sparks) in association with a type \uppercase\expandafter{\romannumeral 2} radio burst. These radio sparks are proposed to be caused by the interaction of a QFP wave train to the leading edge of the accompanying CME \citep{2016A&A...594A..96G}. The three lanes of fundamental type \uppercase\expandafter{\romannumeral 2} radio bursts are indicated by F1, F2 and F3, while the corresponding harmonic emission are indicated by H1, H2 and H3, respectively. The small radio sparks are indicated by the red arrows and symbols R1, R2, R3 and R4. The time axis refers to the time elapsed since 22:00 UT. The right panel is a schematic synopsis for illustrating the generation of the radio sparks in the radio spectrum.}
\label{fig10}
\end{figure}

Recently, \cite{2016A&A...594A..96G} observed a chain of discrete, quasi-periodic radio sparks preceding a type \uppercase\expandafter{\romannumeral 2} radio burst, which were evidenced to be associated with a CME and an ambiguous QFP wave train in the low corona. The authors found that the moving speeds and heights of the radio sparks are comparable to the CME leading edge in time, and the period of the radio sparks is similar to that of the QFP wave train. Therefore, they interpreted the observed radio sparks as the result of the interaction between the QFP wave train and the CME leading edge (see \fig{fig10}). In some spatially-resolved QFP wave trains in EUV, the generation of QFP wave trains are found to be highly correlated in start time with radio bursts \citep{2013AA...554A.144Y,2018ApJ...853....1S}, or their periods are similar to the associated quasi-periodic type \uppercase\expandafter{\romannumeral 3} radio bursts \citep{2017ApJ...844..149K}. Type \uppercase\expandafter{\romannumeral 3} radio bursts are typically associated with electron beams accelerated to small fractions of light speed by magnetic reconnection, and their appearance often suggests the bursty energy releases in the low corona. Therefore, in some studies the generation of QFP wave trains are suggested to be caused by the dispersive evolution of impulsively generated broadband disturbances \citep[e.g.,][]{2013AA...554A.144Y, 2017ApJ...844..149K}. 

\section{Theory and Modeling} \label{theo}
As a booming research field in solar physics, corresponding theory and numerical simulation have made significant achievements since the discovery of QFP wave trains. Although there are various aspects that have not yet been fully addressed, the current numerical and analytical results have been in reasonably good agreement with observations, including the morphology, periodicity, velocity, as well as other properties \citep[e.g.,][]{2011ApJ...740L..33O, 2018ApJ...860...54O, 2013A&A...560A..97P, 2017ApJ...847L..21P}. In terms of the generation mechanism, studies are mainly focussed on two interconnected scenarios similar to the generation of flare QPPs (see also Section \ref{sp}). The first scenario is that a QFP wave train can be formed due to the dispersive evolution of an impulsively generated broadband perturbation, and the wave periodicity is determined by the physical properties of the waveguide and its surrounding \citep[e.g.,][]{1983Natur.305..688R,1984ApJ...279..857R,1994SoPh..151..305M,2004MNRAS.349..705N}. The second scenario is that a QFP wave train can be attributed to pulsed energy release involving in the magnetic reconnection process, and the wave periodicity is basically determined by the wave source \citep[e.g.,][]{2015ApJ...800..111Y,2016ApJ...823..150T}.

\begin{figure}[!b]
\centerline{\includegraphics[width=0.9\textwidth]{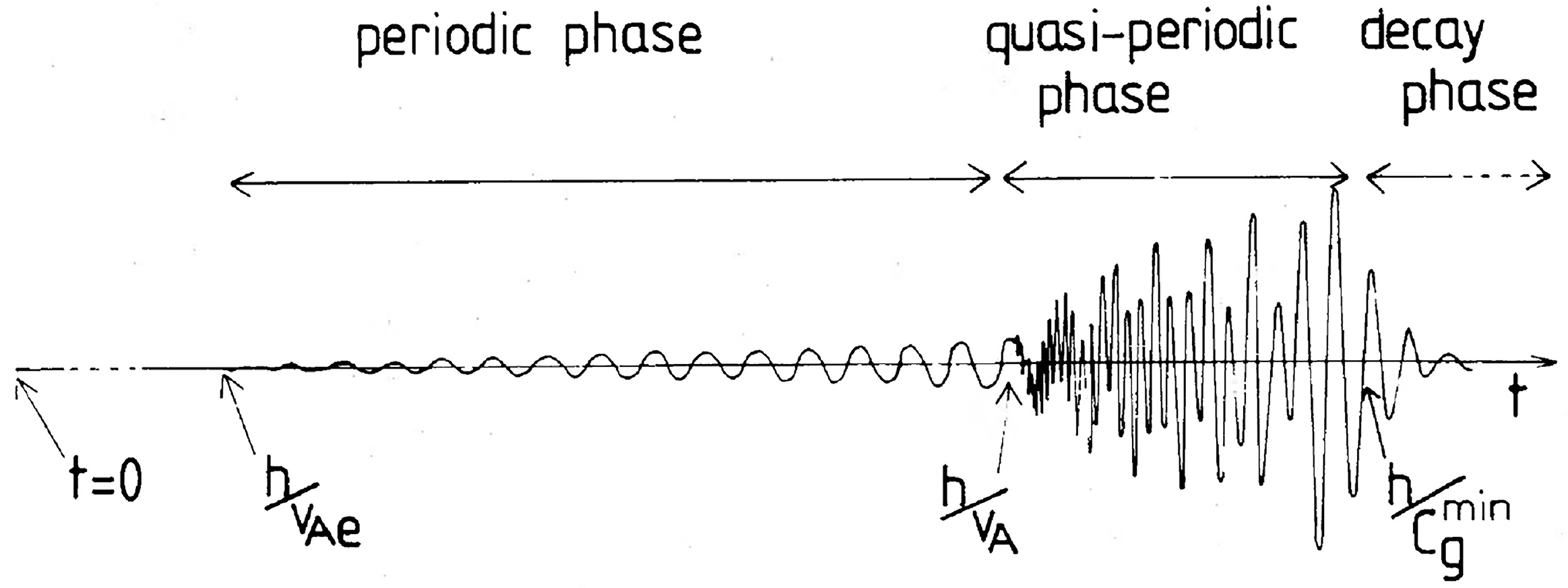}}
\caption{A sketch of the evolution of a fast sausage wave evolved from an impulsively generated perturbation in the low-$\beta$ extreme, which exhibits three distinct phases including periodic, quasi-periodic, and decay phases \citep{1984ApJ...279..857R}. $h$ is the distance away from the initial perturbation, $v_{\rm A}$ and $v_{\rm Ae}$ are respectively the internal and external Alfv\'{e}n speeds of the slab, and $c^{\rm min}_{\rm g}$ is the minimum in the group velocity.}
\label{fig11}
\end{figure}

\subsection{Dispersion Evolution Mechanism} \label{dem}
The corona hosts many filamentary structures of enhanced plasma density (low Alfv\'{e}n speed) with respect to the background, such as coronal loops, fibrils, and plumes. These coronal structures act as waveguides for fast propagating magnetosonic waves that are highly dispersive when their wavelengths  are comparable or longer than the widths of the waveguides, and the wave dispersion properties are seriously affected by the parameters of the waveguide and the surrounding \citep[e.g.,][]{2015ApJ...810...87L,2017ApJ...840...26L,2019MNRAS.488..660L}. Since a fast-mode propagating magnetosonic wave with different frequencies travel at different group speeds in an inhomogeneous structure, an impulsively generated broadband perturbation, i.e., a Fourier integral over all frequencies and wave numbers (wave packets; such as a flare), can naturally give rise to the generation of QFP wave trains in a waveguide at certain distance from the initial site \citep{1983Natur.305..688R}. In the coronal context, the speeds of fast propagating magnetosonic waves along coronal loops are on the order of Alfv\'{e}n speed, which can vary from the minimum Alfv\'{e}n speed inside of a loop to the maximum Alfv\'{e}n speed outside of the loop \citep{2005psci.book.....A}. \cite{1983Natur.305..688R,1984ApJ...279..857R} analytically analyzed the development of QFP wave trains in coronal loops that were modeled as straight slabs with sharp boundaries. The authors found that the group speeds of QFP wave trains with longer wavelength spectral components propagate faster than those with shorter ones, and they qualitatively predicted that a QFP wave train will experience three distinct phases including periodic, quasi-periodic, and decay phases (see \fig{fig11}).

The periodic phase starts at some distance $h$ from the perturbation source with low amplitude and constant frequency, whose start and end times are respectively at $h/v_{\rm Ae}$ and $h/v_{\rm A}$, where $v_{\rm Ae}$ and $v_{\rm A}$ are the external and internal Alfv\'{e}n speeds of the waveguide, respectively. During the periodic phase, the oscillation amplitude steadily grows, and the start (end) time represents the arriving time of the fastest (slowest) signal component of the perturbation. The quasi-periodic phase after the periodic phase but before the decay phase, which starts at the time $h/v_{\rm A}$ and ends at a time $h/c^{\rm min}_{\rm g}$, where $c^{\rm min}_{\rm g}$ is the minimum group speed. It can be seen that the end time of the quasi-periodic phase is determined by the minimum group speed of the perturbation. The quasi-periodic phase has a larger amplitude and high frequency than the earlier periodic phase, which makes itself most detectable in observations. After the quasi-periodic phase is the decay phase, during which the amplitude of the perturbation declines quickly (see \fig{fig11}). Initial numerical studies have been performed successfully to study these distinct phases of QFP wave trains \citep{1993SoPh..145...65M,1993SoPh..143...89M,1993SoPh..144..101M,1994SoPh..151..305M,1998SoPh..179..313M}, and the average periods are found to be of the order of the wave travel time across the waveguides, in agreement with previous analytical results \citep{1983Natur.305..688R, 1984ApJ...279..857R}.

\begin{figure}[!b]
\centerline{\includegraphics[width=0.9\textwidth]{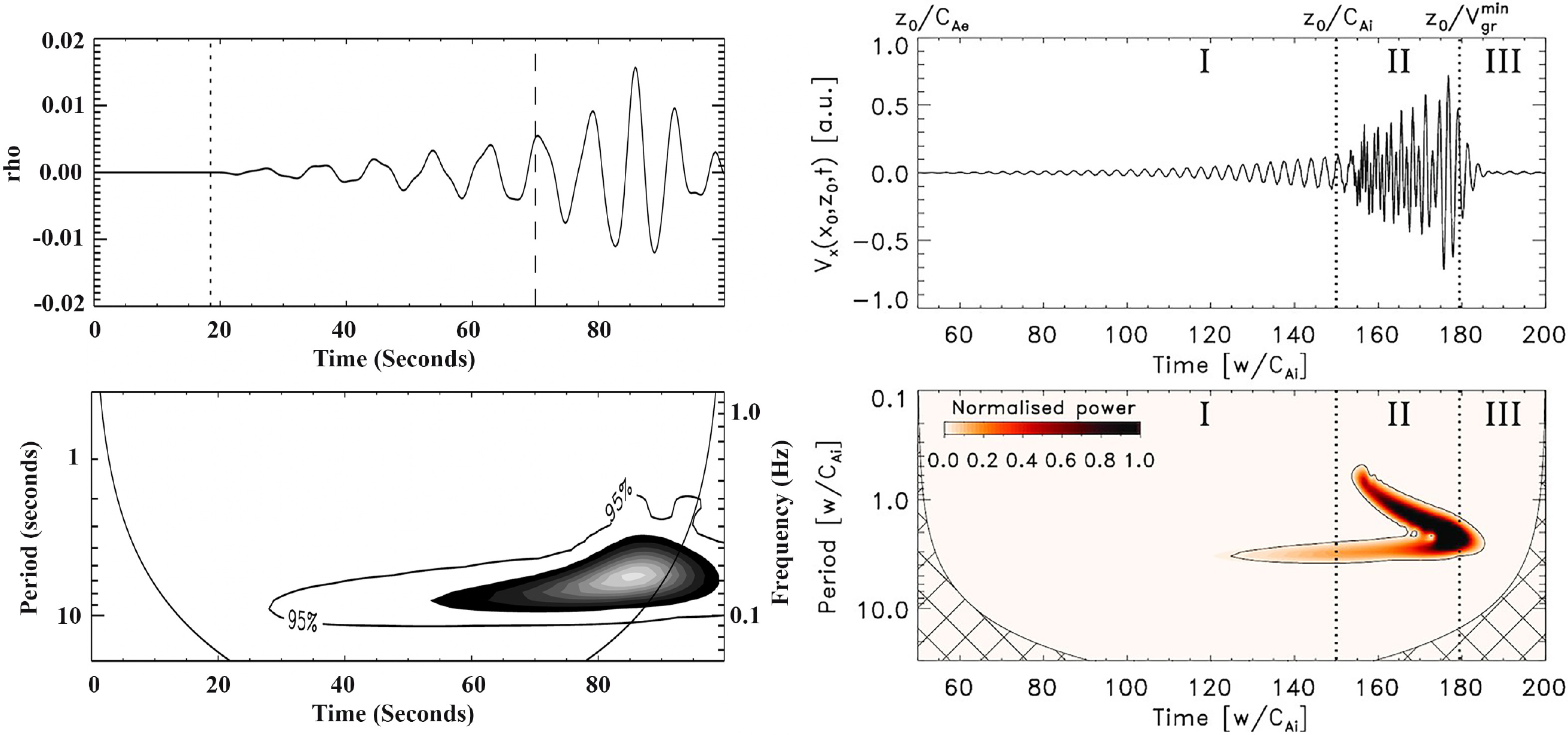}}
\caption{Wavelet power spectrums of dispersively formed QFP wave trains in waveguides. The left column is a numerical simulation of an impulsively generated QFP wave train along a coronal loop with a smooth boundary, in which the top panel shows the density variation profile of the wave train, while the bottom panel is the wavelet transform analysis of the signal, demonstrating the characteristic tadpole wavelet signature. The vertical lines in the top panel show the pulse arrival time if the density was uniform; the dotted line using the external density; and the dashed line the density at the center of the structure \citep{2004MNRAS.349..705N}. The right column shows the time profile (top) and wavelet power spectrum (bottom) of a fully developed fast sausage wave train in a steep plasma waveguide. The three distinct developing phases of the wave train are indicated in the figure, and the wavelet spectrum shows a boomerang shape \citep{2021MNRAS.505.3505K}.}
\label{fig12}
\end{figure}

\cite{2004MNRAS.349..705N} numerically modeled the developed stage of a QFP wave train in a smooth slab of a low $\beta$ plasma. They found that the quasi-periodicity is owning to the geometrical dispersion of the wave train and is determined by the transverse profile of the loop, and the period and the spectral amplitude are determined by the steepness of the transverse density profile and the density contrast ratio in the loop. In addition, the authors further analyzed the time-dependent power spectrum using the wavelet transform technique, which yields that the QFP wave train has a special tadpole shape in the Morlet wavelet spectrum, i.e., a narrow-spectrum tail precedes a broad-band head (see the left column of \fig{fig12}). Comparing with \cite{1984ApJ...279..857R}, the periodic and quasi-periodic phases correspond respectively to the tadpole tail and head, while the decay phase corresponds to the tadpole head maximum. The typical feature of tadpole wavelet spectrum has been used as a characteristic signature for identifying the presence of possible QFP wave trains in both observational and numerical studies, when direct imaging of QFP wave trains in EUV were unavailable  \citep[e.g.,][]{2009ApJ...697L.108M, 2013SoPh..283..473M, 2011A&A...529A..96K, 2013A&A...550A...1K, 2012A&A...546A..49J,2014ApJ...788...44M}. Recently, \cite{2021MNRAS.505.3505K} modeled the linear dispersively evolving of QFP wave trains in plasma slabs with varying steepness of the transverse density profile, in which they showed that the development of a QFP wave train evolved from an initial impulsive perturbation undergoes three distinct phases fully consistent with that qualitatively predicted by \cite{1983Natur.305..688R,1984ApJ...279..857R}. In contrast to wave trains in smooth waveguides that produce the tadpole structures \citep{2004MNRAS.349..705N}, it is interesting that the wavelet power spectrum develops into a boomerang structure that has two pronounced arms in the longer- and shorter-period parts of the spectrum (see the right column of \fig{fig12}). The authors further pointed out that the duration of different phases and how prominent they are in the whole time profile of the wave train depend on the parameters of the waveguide and the wave perturbation symmetry, and this characteristic signature can be used as a seismological indicator of the transverse structuring of a hosting plasma waveguide. It should be pointed out here that in practice most direct imaging of QFP wave trains in EUV do not show such a tadpole or boomerang structure in the wavelet spectrum. It seems that such a special tadpole wavelet spectrum is more preferable to appear in QFP wave trains with shorter periods of a few seconds \citep{2003A&A...406..709K}.

In a series of recent theoretical works, attentions are mainly payed to the geometric effects \citep[e.g.,][]{2012A&A...546A..49J, 2013A&A...560A..97P, 2014ApJ...788...44M, 2015ApJ...814..135S} and transverse plasma density structuring \citep[e.g.,][]{2015ApJ...814...60Y, 2016ApJ...833...51Y, 2017ApJ...836....1Y, 2018ApJ...855...53L} of the waveguide on the formation and evolution of QFP wave trains. In particular, \cite{2014ApJ...789...48O,2015ApJ...806...56O} analytically demonstrated that QFP wave trains experience stronger attenuation for longer axisymmetric (or shorter transverse) perturbations, while the internal to external density ratio has a smaller effect on the attenuation. For typical coronal loops, axisymmetric (transverse) wave trains travel at a speed of 0.75--1 (1.2) times of the Alfv\'{e}n speed of the waveguide and with periods of the order of seconds. To efficiently excite a QFP wave train, a larger spatial extent (compared to the waveguide width) and a longer temporal duration of the initial impulsive driver are probably necessary conditions \citep[e.g.,][]{2005SSRv..121..115N,2017ApJ...836....1Y,2019A&A...624L...4G}. \cite{2015ApJ...814..135S} concluded that the characteristics of QFP wave trains are depended on the fast-mode magnetosonic speed in both the internal and external mediums, the smoothness of the transverse profile of the equilibrium quantities, and also the spatial size of the initial impulsive perturbation.

\begin{figure}[!t]
\centerline{\includegraphics[width=0.9\textwidth]{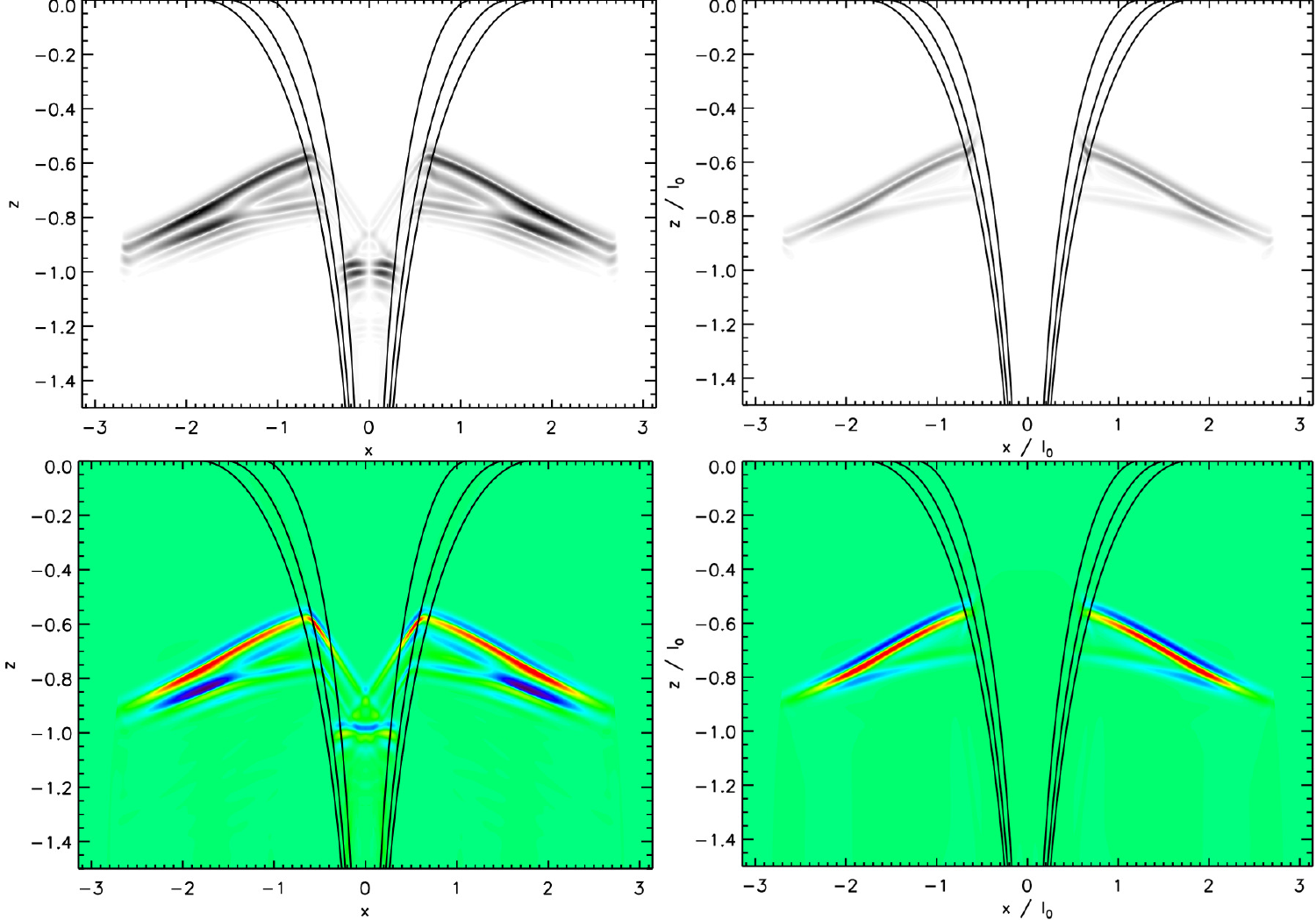}}
\caption{Numerical simulation results of developing QFP wave trains in funnel geometry overdense waveguide \citep[left column, from the paper of ][]{2013A&A...560A..97P} and underdense anti-waveguide \citep[right column, from the paper of ][]{2014A&A...568A..20P}. For each column, the top (bottom) panels shows the velocity (density) perturbations, while the line contours outline the equilibrium density profile.}
\label{fig13}
\end{figure}

\begin{figure}[!t]
\centerline{\includegraphics[width=0.9\textwidth]{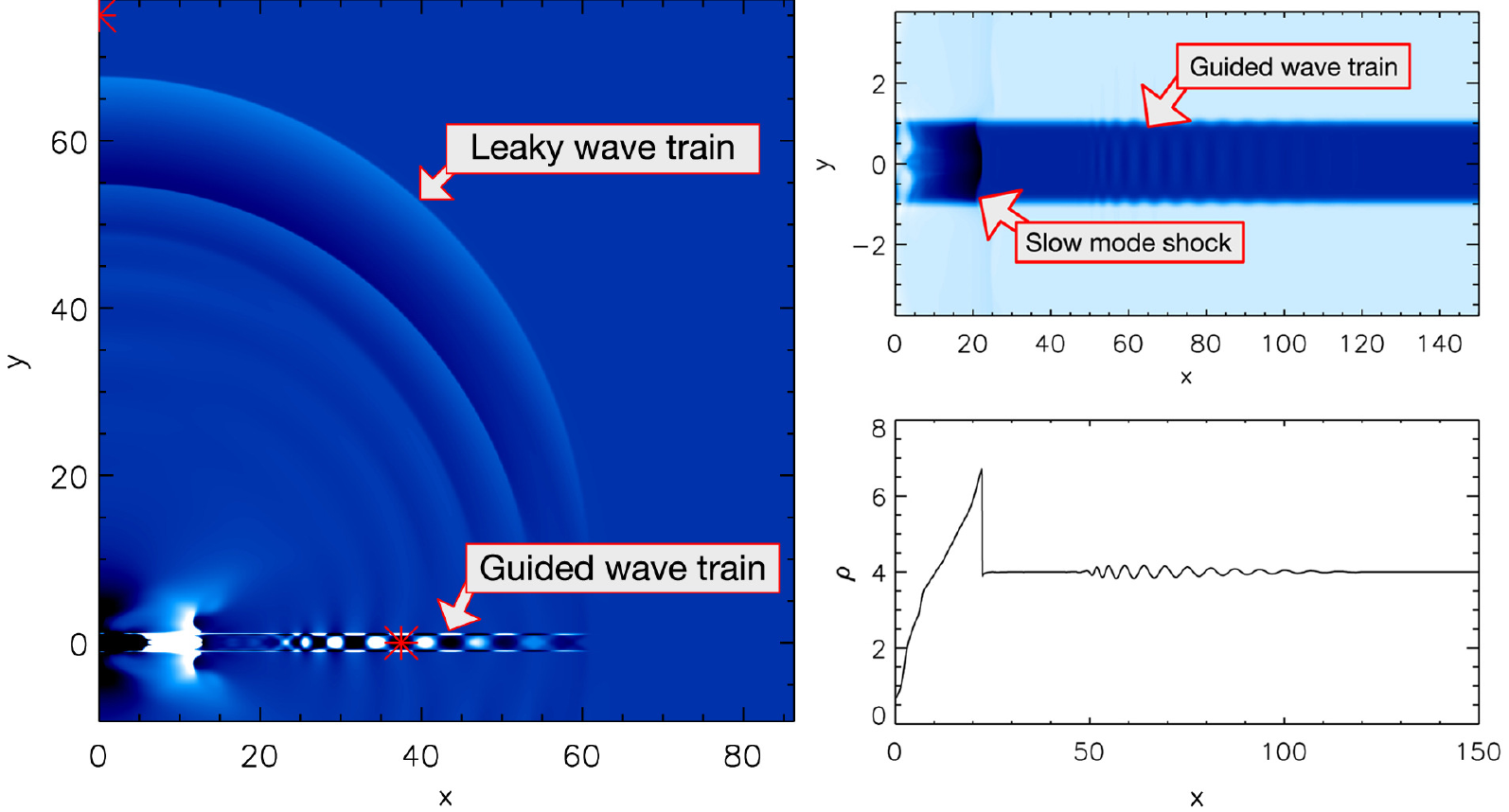}}
\caption{Numerical study on the nonlinear steepening of the trapped and leaky wave trains inside and outside a straight slab \citep{2017ApJ...847L..21P}. The left panel shows the density perturbations, in which the trapped and leaky wave trains are indicated by the two arrows. The upper right panel shows the density of the slab, in which guided a slow-mode shock and a fast-mode wave trains can be identified in opposite directions along the slab. The low right panel shows the intensity profile at the center of the slab (y=0) as shown in the upper right panel.}
\label{fig14}
\end{figure}

The propagation of QFP wave trains can be both trapped and leaky in nature, especially for axisymmetric sausage waves of long wavelengths in smooth slabs  \citep{1993SoPh..143...89M}. An initial impulsive perturbation can result in the propagation of both trapped and leaky waves inside and outside of a coronal loop, respectively. The trapped and leaky waves occur as a result of a total reflection and refraction around the boundary of a waveguide \citep{1994SoPh..151..305M,2014A&A...568A..20P}. In contrast to previous studies in which coronal loops are considered as straight slabs or cylinders, \cite{2013A&A...560A..97P,2014A&A...568A..20P} performed two-dimensional numerical simulations to study the evolution of impulsively generated QFP wave trains in a funnel geometry resembling active region coronal loops and coronal holes \citep[e.g.,][]{2011ApJ...736L..13L,2012ApJ...753...53S,2018MNRAS.477L...6S}, where the funnel expands with height and with a field-aligned enhanced or reduced plasma density in comparison to the surrounding. For both an overdense waveguide and an underdense anti-waveguide, trapped and leaky QFP wave trains appear respectively inside and outside of the waveguides, and the leaky QFP wave trains experience refraction that turns the local wave vector in the vertical direction due to the refraction effect caused by the variation in the magnetic field strength with height (see \fig{fig13}). In comparison, both the trapped and leaky wave trains propagate in perpendicular directions in the case of straight waveguides. In contrast to the case of an overdense waveguide, the leaky wave train in the case of an underdense anti-waveguide is much more pronounced than the corresponding trapped component. In addition, the trapped wave train in the case of an underdense anti-waveguide exhibits less dispersive evolution than that in the case of an overdense waveguide.

It has been evidenced in numerical simulation that the propagation properties of the trapped and leaky QFP wave trains are completely different. \cite{2017ApJ...847L..21P} showed that the nonlinear steepening of the trapped wave train is suppressed by the geometrical dispersion associated with the waveguide, while the leaky wave train does not undergo dispersion once it leaves the waveguide and therefore it can steepen into shock waves (see \fig{fig14}). The formation of shock waves from the leaky wave train could possibly account for the direct observation of broad QFP waves trains in the low corona \citep[e.g.,][]{2012ApJ...753...52L,2019ApJ...873...22S,2021SoPh..296..169Z,zhou2021b}, or quasi-periodic type \uppercase\expandafter{\romannumeral 2} radio bursts in association with one flare. \cite{2014AA...569A..12N} reported an interesting event in which both narrow and broad QFP wave trains are possibly simultaneously detected in one event, and their observations are thought to be consistent with the trapped and leaky wave trains as what had been identified in their numerical simulations \citep{2013A&A...560A..97P,2017ApJ...847L..21P}.

\subsection{Pulsed Energy Excitation Mechanism}\label{pm}
Pulsed energy excitation mechanisms of QFP wave trains relate to the magnetic reconnection process that converts magnetic field energy to plasma kinetic, thermal, and non-thermal high energy particle energies \citep[e.g.,][]{2011SSRv..159...19F,2011LRSP....8....6S,2015SSRv..194..237L}. Magnetic reconnection is a complex and highly nonlinear process referring to the breaking and reconnecting of oppositely directed magnetic field lines in a highly conducting plasma due to finite resistivity \citep{2002A&ARv..10..313P}, which is intrinsic to launch intermittent energy release pulses and therefore cause QPPs in light curves from radio to gamma-ray and QFP wave trains. In observations, some periods of QFP waves are found to be consistent with those of QPPs, which might suggest their common origins. In addition, this also implies the existence of an intimate physical relationship between QFP wave trains and nonlinear physical processes in magnetic reconnection (see Section \ref{sp}).

\begin{figure}[!t]
\centerline{\includegraphics[width=0.9\textwidth]{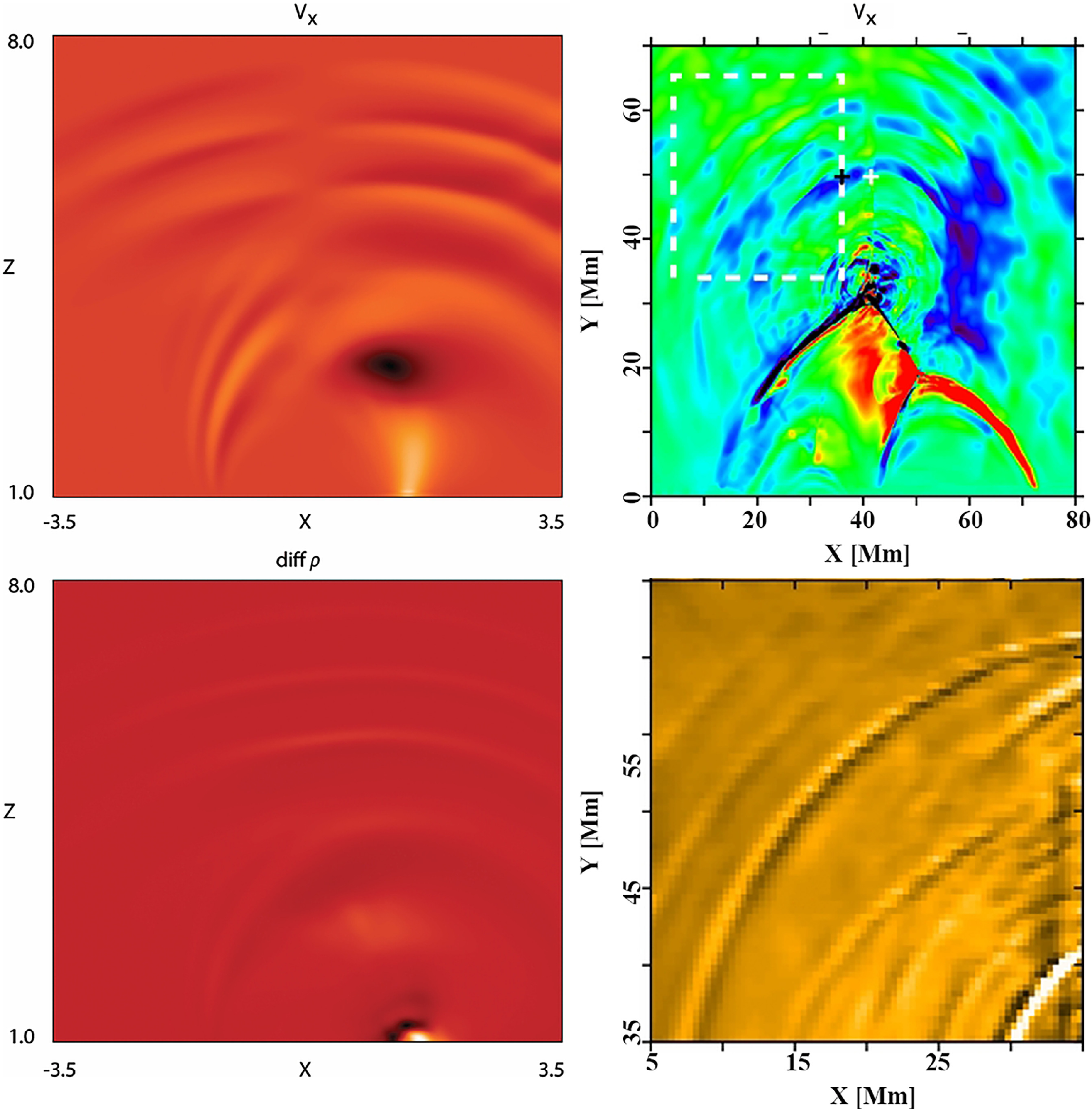}}
\caption{The left column show in simulation results presented by \cite{2011ApJ...740L..33O}, in which the top and the bottom panels display the velocity component $V_{x}$ and the density difference in the xz-plane at the center of the model, respectively. The right column shows the simulation results present by \cite{2015ApJ...800..111Y}, in which the top and the bottom panels show the horizontal velocity $V_{x}$ and the running difference of the synthesized real emission at 171 \AA\ wavelength, respectively. The white dashed box in the top panel indicates the field-of-view of the bottom panel.}
\label{fig15}
\end{figure}

Generally, theoretical and numerical studies have revealed that the launch of a fast magnetic reconnection requires the development of turbulence and the fragmentation of a thin current sheet into many small-scale plasmoids \citep[magnetic islands or flux ropes in three-dimension,][]{1963PhFl....6..459F,2001EP&S...53..473S,1999ApJ...517..700L}. The formation of plasmoids is owning to the tearing-mode or plasmoid instability of the current sheet when its Lundquist number and aspect ratio are large enough \citep[e.g.,][]{2012ApJ...758...20N,2015ApJ...799...79N}. Plasmoids in a current sheet are typically generated repetitively and exhibit characteristics such as coalescence and bi-directional outward ejections at about the Alfv\'{e}n speed. These motions reduce the magnetic flux in the current sheet, which in turn enables new magnetic flux to continuously enter the current sheet to achieve a fast reconnection speed \citep{2011LRSP....8....6S}. So far, many numerical simulations have successfully produced such a physical process; and the presence and dynamic characteristics of plasmoids are also observed indirectly in various solar eruptions from radio to gamma-rays \citep[see][and reference therein]{2016ASSL..427..373S,2020RSPSA.47690867N}. In some studies, flare QPPs have been related to the repetitive generation, coalescence and ejections of plasmoids in current sheets, in which plasmoids are considered as a trap for accelerated particles that can result in drifting pulsating structures in the radio spectrum \citep[e.g,][]{2000A&A...360..715K, 2004A&A...417..325K, 2007A&A...464..735K, 2008SoPh..253..173B, 2016ApJ...823..150T, 2020ApJ...905..165R}. Especially, \cite{2017ApJ...847...98J} numerically evidenced the merging of two plasmoids, and the resulting larger plasmoid oscillated with a period of about 25 seconds; in the meantime, the downward plasmoids interact with the underlying flare arcade and causes the oscillation of the latter with a period of about 35 seconds. These periods are consistent with those observed in flare QPPs and QFP wave trains. In addition, plasmoid contraction or squashing are suggested as a promising mechanism for particle acceleration \citep[e.g.,][]{2006Natur.443..553D,2016ApJ...820...60G}, and particles are shown to gain more energy in multiple X-points between plasmoids \citep{2012SoPh..279...91L, 2017A&A...605A.120L, 2018A&A...620A.121X}.

Recent numerical simulations have studied the physical relationship between the nonlinear processes in magnetic reconnection and the generation of QFP wave trains. \cite{2011ApJ...740L..33O} firstly performed a three-dimensional MHD model in which they identified that the observed QFP wave trains are fast magnetosonic waves driven by quasi-periodical drivers at the base of the flaring region. The simulated QFP wave trains driven by periodic velocity pulsations at lower coronal boundary propagate outward in a magnetic funnel and are evident through density fluctuations due to compressibility. The authors confirmed that the simulated QFP wave trains have similar physical properties as those obtained in real observations, including their amplitude, wavelength, and speeds (see the left column of \fig{fig15}). Using real observations as a guideline, \cite{2018ApJ...860...54O} investigated the excitation, propagation, nonlinearity, and interaction of counter-propagating QFP wave trains in a large-scale, trans-equatorial coronal loop system using time-dependent periodic boundary conditions at the two ends of the loop system. Besides QFP wave trains, trapped fast-(kink) and slow-mode waves are also identified in the closed loop system. These results suggest that the counter-propagating QFP wave trains in closed coronal loops can potentially lead to turbulent cascade that carries significant energy for coronal heating in low-corona magnetic structures. \cite{2015ApJ...800..111Y} performed a 2.5 dimensional numerical MHD simulation to study the generation of QFP wave trains using the interchange reconnection scenario, they found that QFP wave trains can be launched by the impingement of plasmoids ejected outwardly from the current sheet upon the ambient magnetic field in the outflow region, and an one-to-one correlation between the energy release and the wave generation can be identified. The wave properties are also found to be similar to the observed QFP wave trains (see the right column of \fig{fig15}). However, as pointed out by the authors, the simulated QFP wave train propagates isotropically from the wave source other than along funnel-like loop structures as narrow QFP wave trains. Therefore, QFP wave trains excited by the impingement of plasmoids upon the ambient magnetic field in the outflow region could possibly be used to explain the generation of broad QFP wave trains.

\begin{figure}[!t]
\centerline{\includegraphics[width=0.9\textwidth]{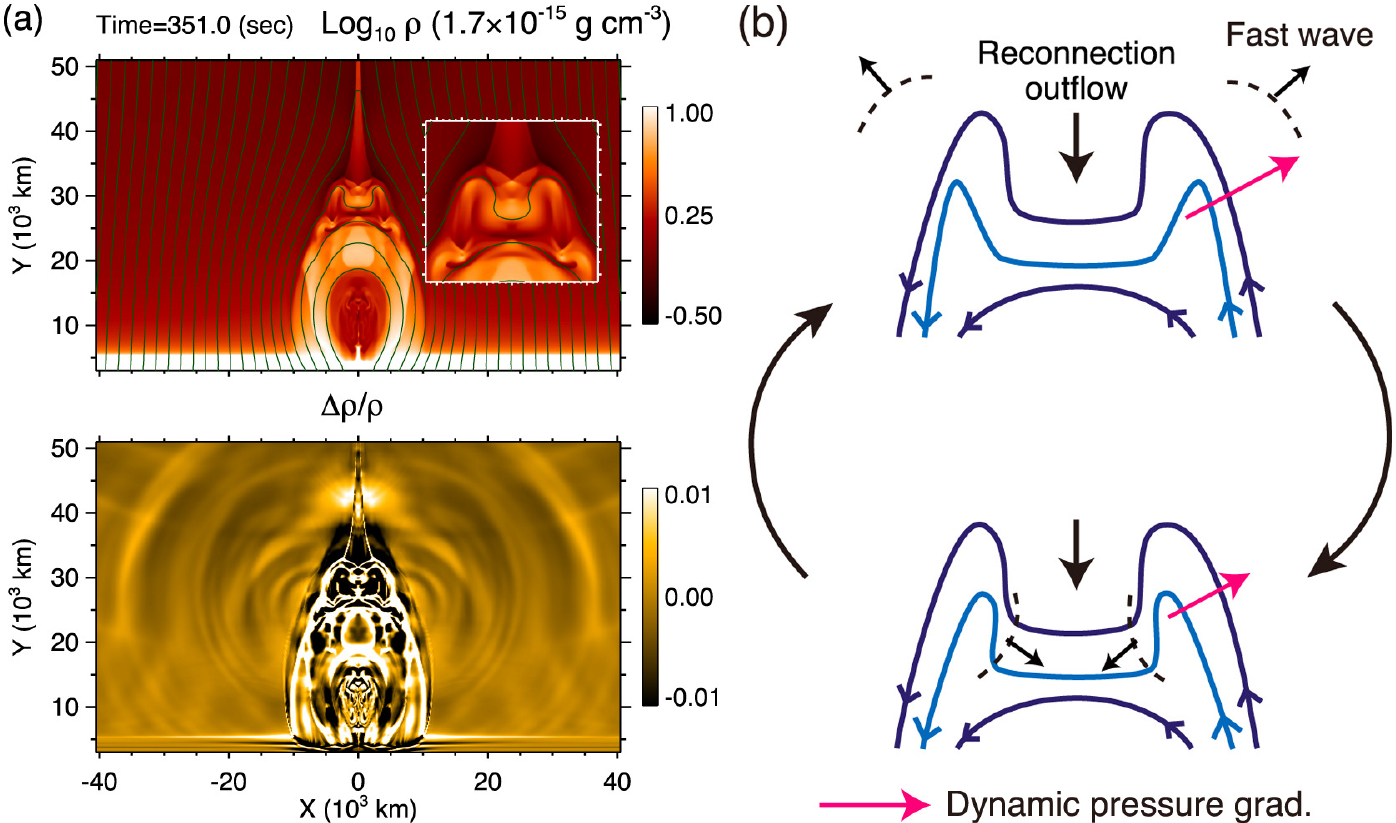}}
\caption{The numerical simulation results presented by \cite{2016ApJ...823..150T}. The upper left is the density map overlaid with magnetic field lines, and the above-the-loop-top region is plotted as an inset. The bottom left panel shows the running difference image of the density perturbation, in which multiple wavefronts can be clearly identified. The right panel is a schematic for illustrating the generation of QFP wave trains due to the above-the-loop-top oscillation, in which the pink arrows indicate the dynamic pressure gradient, the black vertical arrow indicates the downward reconnection outflow, and the short black arrows indicate the generated QFP wave trains.}
\label{fig16}
\end{figure}

\cite{2016ApJ...823..150T} described an alternative physical picture for the generation of QFP wave trains through a two-dimensional MHD simulation on the flare process, which includes essential physics such as magnetic reconnection, heat conduction, and chromospheric evaporation. It was found that QFP wave trains are spontaneously excited by the oscillating region filled with evaporated plasma above the flaring loop, and the oscillation of this region is controlled by the backflow of the reconnection outflow. Therefore, the authors claimed that the backflow of the reconnection outflow can act as an exciter of QFP wave trains (see \fig{fig16}). The oscillation region has an U-shaped structure due to the continuous impingement of the reconnection outflow, and therefore the generation process of QFP wave trains is similar to the sound wave generated by an externally driven tuning fork. \cite{2021ApJ...908L..37M} observed simultaneous bi-directional narrow QFP wave trains originating from the same flaring region, and the authors suggested that their observation might be a good example for supporting such a magnetic tuning fork model. Here, it should be noted that the propagation of the simulated QFP wave train in \cite{2016ApJ...823..150T} is also isotropically as that in \cite{2015ApJ...800..111Y}. It is hard to understand why the observed bi-directional narrow QFP wave trains in \cite{2021ApJ...908L..37M} can be interpreted by the magnetic tuning fork model. We think that this model should be more suitable for broad QFP wave trains, but it could also be used to interpret narrow QFP wave trains when the isotropic propagating wave train is captured by and therefore trapped in some inhomogeneous coronal structures such as coronal loops \citep[e.g.,][]{2019ApJ...873...22S}.

\begin{figure}[!t]
\centerline{\includegraphics[width=0.9\textwidth]{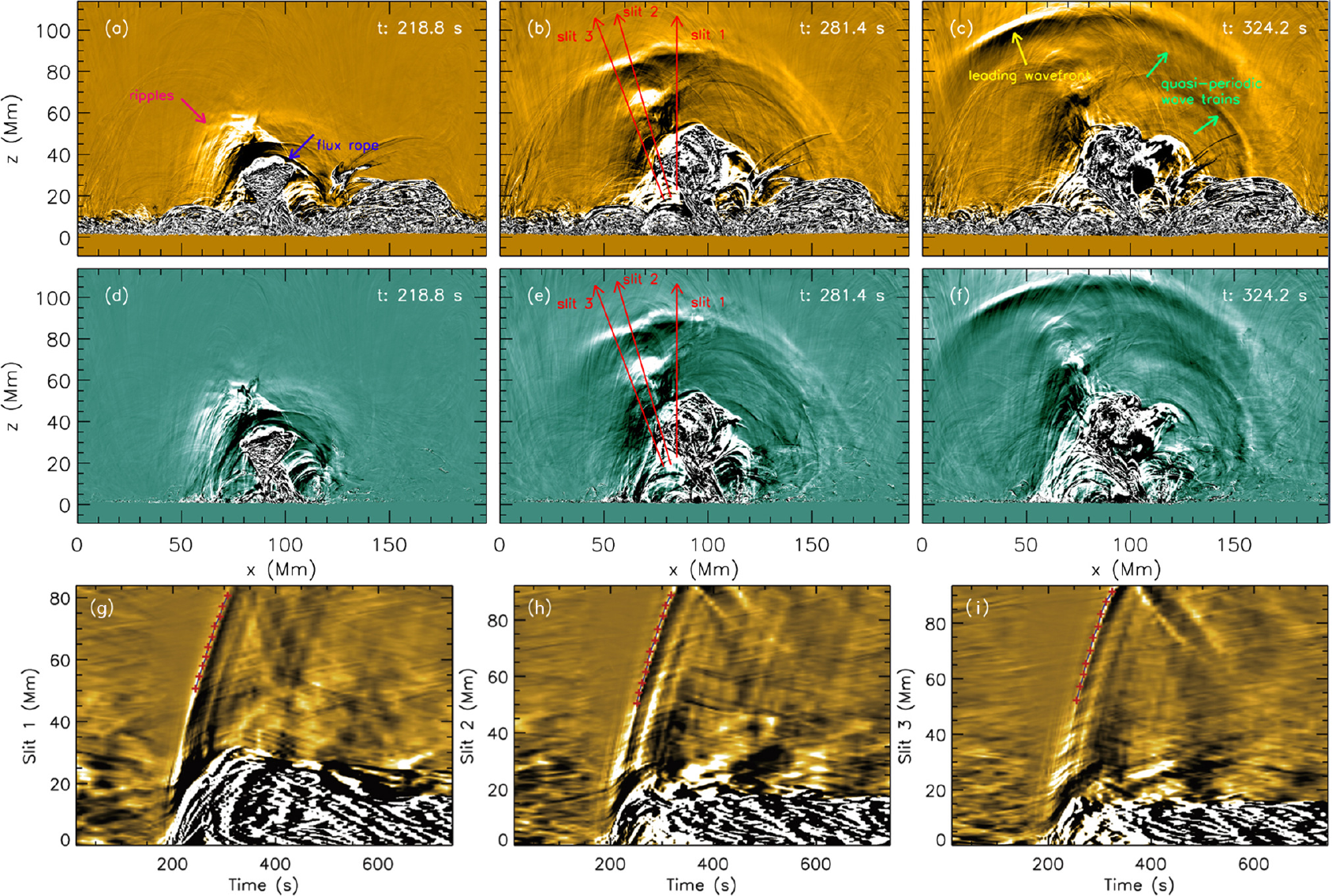}}
\caption{The simulation results presented by \cite{2021ApJ...911L...8W}. The top and middle rows display the synthetic 171 \AA\ and 94 \AA\ running difference images, respectively. The leading EUV  wavefront and the following QFP wave train are indicated respectively by the yellow and green arrows in  panel (c). The three red arrows in the panels (b) and (e) are used to generate the time-distance diagrams plotted in the bottom row (based on the synthetic 171 \AA\ running difference images).}
\label{fig17}
\end{figure}

\cite{2021ApJ...911L...8W} performed a three-dimensional radiative MHD simulation to model the formation of active regions through magnetic flux emergence from the convection zone to the corona, in which the eruption of a magnetic flux rope produced a C-class flare and a QFP wave train with a period of about 30 seconds in-between the erupting flux rope and a preceding global EUV wave that was driven by the erupting flux rope (see \fig{fig17}). Obviously, the propagation of the generated QFP wave train is a broad QFP wave train perpendicular to magnetic field lines \citep{2012ApJ...753...52L,2019ApJ...873...22S,2021SoPh..296..169Z,zhou2021b} rather than along magnetic field lines as narrow QFP wave trains \citep{2011ApJ...736L..13L,2012ApJ...753...53S,2013SoPh..288..585S,2018ApJ...853....1S}. Therefore, this simulation provided an additional numerical model for the generation of broad QFP wave trains, as well as the simultaneous preceding global EUV wave. The generation of the QFP wave train in \cite{2021ApJ...911L...8W} occurs spontaneously without any artifactual exciters as used in previous simulations \citep[e.g.,][]{2011ApJ...740L..33O,2013A&A...560A..97P}. The authors proposed that excitation of the QFP wave train was possibly due to pulsed energy release in the accompanying flare, as what had been proposed in \cite{2012ApJ...753...52L}. However, the authors also pointed out that the essential physical cause of the excitation mechanism still needs further investigation using higher spatiotemporal resolution three-dimensional simulations. This is true because there may be other excitation mechanisms for broad QFP wave trains. For example, \cite{2019ApJ...873...22S} proposed that the generation of broad QFP wave trains behind the CME-driven global EUV wave can possibly be driven by the pulsed energy release caused by the periodic unwinding and expanding twisted thin  threads in the erupting filament, because the period of the observed QFP wave train is similar to the unwinding filament threads instead of the QPPs in the accompanying flare. In addition, the generation of broad QFP wave trains is possibly in association with the fine structure of CMEs. We note the appearance of  large-scale quasi-periodic EUV wavefronts ahead of the CME in numerical simulations with a period of about 84--168 seconds (see Figure 1 of \cite{2002ApJ...572L..99C} and Figure 2 of \cite{2005ApJ...622.1202C} for details). Although the authors did not analyze these interesting wavefronts, their period is well consistent with those of broad QFP wave trains \citep{2019ApJ...873...22S}. Besides, \cite{2010A&A...522A.100P} also observed the appearance of broad QFP wave trains ahead of a CME, where the authors proposed that the wave train was excited by the fine expanding pulse-like lateral structures in the CME. Recently, \cite{shen2022} found the generation of a broad QFP wave train can be driven by the sequential stretching of expansion of the newly formed reconnected magnetic field lines, it is also a good observation supporting the scenario of pulsed energy release in magnetic reconnection.

Other nonlinear physical processes in association to pulsed energy release in magnetic reconnection include the mechanism of oscillatory reconnection which couples resistive diffusion at X-type null points to global advection of the outer fields \citep[e.g.,][]{1991ApJ...371L..41C, 2009A&A...493..227M, 2012A&A...548A..98M, 2012ApJ...749...30M,2019ApJ...874..146H,2019ApJ...874L..27X,2017ApJ...844....2T,2019A&A...621A.106T}, 3) patchy magnetic reconnection shows as supra-arcade downflows \citep[e.g.,][]{2006ApJ...642.1177L, 2009ApJ...697.1569M, 2012ApJ...747L..40S, 2020ApJ...898...88X, 2019MNRAS.489.3183C, 2020ApJ...905..165R}, and 4) the fluctuation of current sheets result from super-Alfv\'{e}nic beams and Kelvin-Helmholtz instability nonlinear oscillations \citep[e.g.,][]{2006ApJ...644L.149O, 2016ApJ...829L..33L}. In addition, periodicities in magnetic reconnection can also be launched by external quasi-periodic disturbances from lateral or lower layers of the solar atmosphere through interaction and therefore modulating the reconnection process \citep[e.g.,][]{2006A&A...452..343N, 2006SoPh..238..313C, 2009A&A...505..791S, 2012ApJ...753...53S, 2012ApJ...757..160J, 2019A&A...625A...3J}. Basically, all these possible physical processes are potentially to produce both flare QPPs and QFP wave trains. However, although a number of numerical and theoretical studies have been performed to investigate the excitation mechanism of flare QPPs based on these scenarios \citep{2009SSRv..149..119N,2016SoPh..291.3143V,2018SSRv..214...45M,2020STP.....6a...3K,2021SSRv..217...66Z}, the physical relationship between these  processes and the generation of QFP wave trains have not yet been established. Therefore, in the future more attentions should be paid to these candidate mechanisms for the generation of QFP waves.

\subsection{Discussion of the Current Models} \label{dpm}
The current models showed that the two kinds of possible generation mechanisms of QFP wave trains are both supported by some observational evidence \citep[e.g.,][]{2011ApJ...736L..13L, 2012ApJ...753...53S, 2018MNRAS.477L...6S, 2019ApJ...873...22S, 2013AA...554A.144Y, 2014AA...569A..12N}. However, for a particular event, the generation of the QFP wave train is owning to a specific mechanism or the combination of both is still unclear. Here, we would like to make some preliminary discussions about this problem based on previous observational and theoretical studies.

The dispersion evolution mechanism was firstly developed to interpret short period pulsations of a few seconds observed in radio emission, and these pulsations were thought to be the manifestation of plasma emission modulated by QFP wave trains propagating in inhomogeneous coronal structures such as coronal loops that act as overdense plasma tubes \citep{1983Natur.305..688R,1984ApJ...279..857R}. Obviously, QFP wave trains formed by the dispersion evolution mechanism require the presence of overdense waveguides. In actual observations, many spatially-resolved narrow QFP wave trains in the EUV can satisfy such a requirement, since they propagate along coronal loops. In addition, it has been widely accepted that such a dispersively formed QFP wave train should lead to a characteristic tadpole structure in the time-dependent wavelet spectrum \citep{2004MNRAS.349..705N}, if the driver is a broadband perturbation \citep{2005SSRv..121..115N}. Up to now, the characteristic tadpole structure has not been detected in all published narrow QFP wave trains observed by the AIA. This might be attributed to the relatively longer period (25--550 seconds, see \tbl{tbl1}) of the narrow QFP wave trains observed by the AIA, since we note that a tadpole wavelet spectrum did detect in the QFP wave train observed by SECIS during the total solar eclipse on 1999 August 11, where the period was about 6 seconds \cite{2003A&A...406..709K}. Besides, as what had been pointed out in \cite{2005SSRv..121..115N}, the absence of the tadpole wavelet spectrum of narrow QFP wave trains could also be due to the more monochromatic and narrowband driver.  

For broad QFP wave trains propagating in the homogeneous quiet-Sun region where the magnetic field has a strong vertical component, they are non-dispersive in nature and their propagation can be viewed as perpendicular to the magnetic field. As evidenced in simulations \citep[e.g.,][]{1993SoPh..143...89M, 2013A&A...560A..97P}, an impulsively perturbation can dispersively evolved into both trapped and leaky wave trains inside and outside of the waveguide. \cite{2017ApJ...847L..21P} showed that broad QFP wave trains can be formed through steepening the leaky component of a dispersively formed wave train in an overdense waveguide, while the trapped component does not experience nonlinear steepening; the trapped and leaky wave trains have the same periods of about 6 seconds, and their velocity amplitudes are estimated to be respectively about 30\% and 10\% with respect to the local Alfv\'{e}n speeds. This simulation might imply that broad QFP wave trains formed by the leaky component of dispersively formed QFP wave trains should also have relatively short periods as what we have discussed in the above paragraph. The pulsed energy excitation mechanism include various forms as what has been stated in Section \ref{pm}, in which the dynamic evolution of plasmoids and their interaction with magnetic structures in the reconnection outflow region often excite broad QFP wave trains with periods of dozens of seconds \citep[e.g.,][]{2015ApJ...800..111Y, 2016ApJ...823..150T, 2017ApJ...847...98J, 2021ApJ...911L...8W}, in quantitative agreement with the lower limit of the period of the published broad QFP wave trains observed by the AIA (36--240 seconds, see \tbl{tbl1}). In addition, broad QFP wave trains with longer periods of about several minutes are probably associated with the other kinds of pulsed energy release processes in association to magnetic reconnection. For example, the periodic untwisting motion of twisted erupting filament threads \citep{2019ApJ...873...22S}, the intermittent generation and stretching (or expansion) of reconnected magnetic field lines \citep{shen2022}, and the sequentially eruption of coronal loops \citep{2010A&A...522A.100P}. 

Based on the above discussions, it is noted that both narrow and broad QFP wave trains can be produced by the two different generation mechanisms. In general, it appears that the dispersion evolution mechanism seems more preferable for the generation of  QFP wave trains with short periods of about a few seconds, while the pulsed energy release excitation mechanism seems more preferable for the generation of QFP wave trains with relatively long periods typically of about dozens of seconds to a few minutes. In principle, the two different generation mechanisms do not contradict to each other. For the dispersion evolution mechanism, it requires that the initial perturbation should be broadband. For pulsed energy release excitation mechanism, the initial perturbation is more monochromatic. Here, we would like to point out that these preliminary thoughts are premature,  and they need to be verified with high spatiotemporal resolution observations and theoretical works in the future.

\section{Seismological Application}
Seismology is the study of earthquakes and seismic waves that move through and around the Earth. This technique has been extended to other areas of science such as helioseismology, stellar seismology, as well as MHD spectroscopy of laboratory plasma. Coronal seismology uses MHD waves and oscillations to probe unknown physical parameters of the solar corona \citep{2005LRSP....2....3N}, which was originally proposed by \cite{1970PASJ...22..341U} for global and \cite{1984ApJ...279..857R} for local seismology. In principle, coronal seismology requires the combined application of theoretical modeling knowledge and observational parameters of MHD waves and oscillations, which yields the mean parameters of the corona that are currently not accessible in the absence of in situ instruments, such as the magnetic field strength and Alfv\'{e}n velocity and coronal dissipative coefficients \citep{2005RSPTA.363.2743D,2012RSPTA.370.3193D}. So far, different types of MHD waves have been detected in the corona, and the technique of coronal seismology has also been successfully applied to estimate various coronal parameters \citep{2020ARA&A..58..441N}. In previous studies, particular attentions have been paid to derive the elusive coronal magnetic field, and the results are often comparable with those obtained by using other direct or indirect methods including polarimetric measurements using Zeeman and Hanle effects \citep{2000ApJ...541L..83L,2004ApJ...613L.177L}, extrapolations using photospheric magnetograms \citep{1994SoPh..151...91Z,2008ApJ...680.1496L}, and radio observations of gyrosynchrotron emission with a coronal density model \citep{1994ApJ...420..903G,1997SoPh..174...31W, 2010ApJ...711.1029R, 2010ASSP...19..482S}. Here, we only briefly review the applications of coronal seismology by using QFP wave trains, those for other types of waves can refer to several recent reviews \citep[e.g.,][]{2005RSPTA.363.2743D,2012RSPTA.370.3193D,2015SSRv..190..103J,2014SoPh..289.3233L,2020SSRv..216..136L,2005LRSP....2....3N,2020ARA&A..58..441N}.

For a linear fast-mode magnetosonic wave in a homogeneous medium, its propagation is weakly depends on the direction of the wave vector with respect to the magnetic field, which means that it propagates in any direction. The restoring force is the resultant force of the magnetic and the gas pressure gradient forces, and the speeds is combinedly determined by the Alfv\'{e}n speed and the sound speed of the local plasma medium. Theoretically, the speed of a fast-mode magnetosonic wave $v_{\rm f}$ in a uniform medium is written as 
\begin{equation}
\label{vf}
v_{\rm f}=[\frac{1}{2}(v^{2}_{\rm A}+c^{2}_{\rm s}+\sqrt{(v^{2}_{\rm A}+c^{2}_{\rm s})^{2}-4v^{2}_{\rm A}c^{2}_{\rm s}{\rm cos}^{2}\theta_{\rm B}})]^{1/2},
\end{equation}
where $c_{\rm s}$, $v_{\rm A}$, and $\theta_{\rm B}$ are the sound speed, Alfv\'{e}n speed, and the angle between the wave vector and the magnetic field, respectively. Specifically, the mathematical expressions of $c_{\rm s}$ and $v_{\rm A}$ are 
\begin{equation}
c_{\rm s} = \sqrt{\frac{\gamma \kappa T}{\bar{\mu}m_{\rm p}}}
\end{equation}
\begin{equation}
\label{va}
{\rm and}~v_{\rm A}=\frac{B}{\sqrt{4\pi\rho}}=\frac{B}{\sqrt{4\pi\bar{\mu}m_{\rm p}n}},
\end{equation}
respectively, where $\gamma=5/3$ is the adiabatic exponent for fully ionized plasmas, $\kappa$ the Boltzmann constant, $T$ the temperature, $\bar{\mu}$ the mean molecular weight, $m_{\rm p}$ the proton mass, $B$ the magnetic field strength, $\rho$ the mass density, and $n$ the total particle number density. According to \cite{1982soma.book.....P}, $\bar{\mu}$ and $n$ are often respectively taken as 0.6 and $1.92n_{e}$, with $n_{e}$ as the electron density.

Obviously, for a fast-mode magnetosonic wave traveling in a particular direction, its speed depends on coronal parameters including the temperature, plasma density, and magnetic field strength. Particularly, if a wave propagates perpendicular to the magnetic field (i.e., $\theta_{\rm B}=90^{\circ}$), Equation (\ref{vf}) reduces to a simple form of 
\begin{equation}
\label{vfh}
v_{\rm f} = \sqrt{v^{2}_{\rm A}+c^{2}_{\rm s}},
\end{equation}
and the magnetic field strength of the medium in which the wave propagates can be estimate through measuring the wave speed and coronal parameters including plasma density and temperature. In the case of $\theta_{\rm B}=0^{\circ}$, i.e., the wave propagates along the magnetic field, Equation (\ref{vf}) becomes as 
\begin{equation}
\label{vfp}
v_{\rm f} = v_{\rm A}=\frac{B}{\sqrt{4\pi\rho}},
\end{equation}
namely, the fast-mode magnetosonic wave speed is equal to the Aflv\'{e}n speed. Therefore, one can simply measure the wave speed and the plasma density to estimate the magnetic field strength of the waveguide.

In the corona, magnetic field lines are believed to be highlighted by coronal loops due to the coupling of hot plasma and magnetic field. Therefore, coronal loops  commonly manifest the orientation and distribution of the coronal magnetic field. In practice, since narrow QFP wave trains travel along coronal loops, their propagations are along magnetic fields. Therefore, one often uses Equation (\ref{vfp}) to estimate the magnetic field strength of the guiding magnetic field. \cite{2002MNRAS.336..747W} estimated that the magnetic field strength of an active region loop is about 25 Gauss. \cite{2011ApJ...736L..13L} obtained that the magnetic field strength of an active region funnel-like loop is greater than 8 Gauss. \cite{2019ApJ...873...22S} derived that the magnetic field strength of a closed transequatorial loop is about 6 Gauss. \cite{2021SoPh..296..169Z} estimated that the magnetic field strength of an interconnecting loop is about 5.6 Gauss, in agreement with the result (about 5.2 Gauss) derived from the simultaneous global EUV wave. \cite{2021ApJ...908L..37M} reported a bi-directional QFP wave event, in which simultaneous QFP wave trains are observed in two opposite funnel-like loops rooted in the same active region. The magnetic field strengths of the two funnel-like loops are estimated to be respectively about 12.8 and 11.3 Gauss, consistent with the results alternatively obtained by using magnetic field extrapolation. Radio observations of possible QFP wave trains were also used to estimate the magnetic field strengths of coronal loops, which are found to be in the range of 1.1--47.8 Gauss \citep{2011SoPh..273..393M,2018ApJ...861...33K}. It should be pointed out that these results are all obtained by using average parameters (plasma density and wave speed) along the entire loop structure. In practice, since narrow QFP wave trains decelerate fast as they propagate outward, it should be better to estimate the magnetic field strengths of the different sections of the waveguiding loop. According to this line of thought, \cite{2013SoPh..288..585S} obtained that the magnetic field strengths of the footpoint, middle, and outer sections of an active region loop are about 5.4, 4.5, and 2.2 Gauss, respectively. This indicates the fast decreasing of the magnetic field strength with the increasing of height of active region coronal loops. Here, it should be pointed out that the above magnetic field strength estimations based on Equation (\ref{vfp}) are only approximations but with a certain accuracy. As what has been introduced in Section \ref{dem} (for more details, one can refer to many books or reviews \citep[e.g.,][]{2005psci.book.....A, roberts2003, 2005LRSP....2....3N}), the speeds of QFP wave trains along coronal loops are on the order of Alfv\'{e}n speed, which is greater than the Alfv\'{e}n speed inside of a loop but less than the Alfv\'{e}n speed outside of the loop. Therefore, the derived values based on Equation (\ref{vfp}) should be approximately consistent with the lower limit of those estimated based on the theory of QFP wave trains along inhomogeneous waveguides.

In the simulation work performed by \cite{2004MNRAS.349..705N}, the authors found that the mean wavelength of the QFP wave train is comparable to the width of the guiding loop. Since the fast-mode wave speed is equal to the Alfv\'{e}n speed of the waveguide, the relationship among wavelength ($\lambda$), period ($P$) and wave speed ($v_{\rm f}$) can be written as 
\begin{equation}
P=\frac{\lambda}{v_{\rm f}} \simeq \frac{w}{v_{\rm A}},
\end{equation}
where $w$ and $v_{\rm A}$ are the width and Alfv\'{e}n speed of the guiding loop. Therefore, with the measurable physical parameters of period and wave speed, one can estimate the width of the guiding loop. For example, in the absence of imaging observations, \cite{2011SoPh..273..393M} and \cite{2013SoPh..283..473M} estimated the loop widths with the results derived from radio observations, which are in the range of about 1--30 Mm.

If simultaneous slow- and fast-mode waves are observed in the same waveguide, one can further estimate the plasma $\beta$ of the medium which is defined as the ratio of gas pressure to magnetic pressure \citep{2011ApJ...740...90V}. \cite{2015AA...581A..78Z} reported the first imaging observation of simultaneous slow- and fast-mode wave trains propagating along the same coronal loop at speeds were about \speed{80 and 900}, respectively. By assuming that the speeds of the observed slow- and fast-mode wave trains are respectively equal to the sound speed and Alfv\'{e}n speed of the waveguide, the plasma $\beta$ can be expressed with the slow- ($v_{\rm s}$) and fast-mode ($v_{\rm f†}$) characteristic speeds, i.e., 
\begin{equation}
\beta=\frac{2\mu p}{B^2} \approx \frac{2}{\gamma}(\frac{v_{\rm s}}{v_{\rm f}})^{2},
\end{equation}
where $p$, $\mu$, $\gamma$ and $B$ are the gas pressure, the permeability, the adiabatic index and the magnetic field magnitude, respectively. In the corona, the value of $\gamma$ often ranges from 1 to 5/3 for isothermal and adiabatic sases, respectively. For the case analyzed by \cite{2015AA...581A..78Z}, the authors derived that the value of the plasma $\beta$ ranges from 0.009 to 0.015, confirming the low $\beta$ nature of the low corona.

Broad QFP wave trains commonly travel parallel to the solar surface which has a strong vertical magnetic field component. Therefore, the propagation of broad QFP wave trains are assumed to be perpendicular to the magnetic field, and one often uses Equation (\ref{vfh}) to derive the magnetic field strength of the  supporting medium. This method is the same with the magneto-seismology by using global EUV waves \citep[see][and references therein]{2014SoPh..289.3233L,2015LRSP...12....3W}, and one need to firstly determine the sound speed and plasma density of the supporting medium. For example, \citep{zhou2021b} used the observation of a broad QFP wave train to estimate the magnetic field strength in the quest-Sun, which yields a result of about 4.7 Gauss. 

The large extent propagation of broad QFP wave trains is potential to trigger oscillations of remote coronal structures such as coronal loops and filaments. The transverse oscillation of these coronal structures can be interpreted as kink global standing mode of the loops, and one can use the measured oscillation parameters for coronal seismology \citep{1999ApJ...520..880A,1999Sci...285..862N,2005LRSP....2....3N}. According to \cite{2001A&A...372L..53N}, the observed wavelength of a global standing kink mode is double the length of the loop, one can estimate the phase speed $C_{\rm k}$ based on the observable period $P$ and loop length $L$ with the formula 
\begin{equation}
\label{pl}
P=\frac{2L} {C_{\rm k}}.
\end{equation}
Assuming that in the low $\beta$ coronal plasma the magnetic field is almost equal inside and outside the waveguide, the equation of the kink speed can be rewritten as 
\begin{equation}
\label{vkink}
C_{\rm k}=\sqrt{\frac{\rho_{i}v^{2}_{\rm Ai}+\rho_{\rm e}v^{2}_{\rm Ae}}{\rho_{i}+\rho_{e}}} \approx v_{\rm Ai}\sqrt{\frac{2}{1+\rho_{\rm e}/\rho_{\rm i}}},
\end{equation}
where $\rho_{\rm i}$ ($\rho_{\rm e}$) is the internal (external) density, $v_{\rm Ai}$ ($v_{\rm Ae}$) is the internal (external) Alfv\'en speed. As the density contrast $\rho_{\rm e}/\rho_{\rm i}$, the density inside the waveguide $\rho_{i}$ and the kink speed $C_{\rm k}$ can be measured from observations, one can estimate the magnetic field strength $B$ of the waveguide using equation (\ref{pl}) that can be rewritten as 
\begin{equation}
\label{magk}
B = v_{\rm Ai}\sqrt{4\pi \rho_{\rm i}}=\frac{L}{P}\sqrt{8\pi \rho_{\rm i}(1+\rho_{\rm e}/\rho_{\rm i})}.
\end{equation}
This formula can be written as a more convenient practical formula with measurable parameters of the distance between the footpoints of the loop $d$, the number density inside the loop $n_{\rm i}$, the number density contrast $n_{\rm e}/n_{\rm i}$, and the period of the loop oscillation $P$, i.e., 
\begin{equation}
B \approx 7.9 \times 10^{-13} \frac{d}{P}\sqrt{n_{\rm i} + n_{\rm e}},
\end{equation}
where the magnetic field $B$ is in Gauss, the distance $d$ is in meter, the number densities $n_{\rm i}$ and $n_{\rm e}$ are in m$^{-3}$, and the period $P$ is in seconds \citep{roberts2003}. 

\cite{2018ApJ...860...54O} studied the transverse oscillation of a coronal loop caused by the counter-propagating of two quasi-simultaneous narrow QFP wave trains in it. The authors firstly measured the width and length of the loop, and then they derived the background and loop density with the technique of differential emission measure \citep[DEM; see][for instance]{2015ApJ...807..143C}. With the knowledge of loop length, oscillation period, and the background and loop densities, the magnetic field strength of the loop is estimated to be about 5.3 Gauss with equation (\ref{magk}). Such a value is consistent with their numerical model that can produce similar observational characteristics to those obtained from the real observations. \cite{2019ApJ...873...22S} reported an interesting broad QFP wave train that propagated simultaneously along a transequatorial loop and on the solar surface, and the trapped part of the wave train result in the transverse oscillation of the loop system. Using the same methods as in \cite{2018ApJ...860...54O}, the authors estimated that the magnetic field strength of the transequatorial loop is about 6 Gauss. In addition, the authors also estimated the magnetic field strength of the loop with equation (\ref{vfp}) by using the physical property of the wave train, which yields a value of about 8.3 Gauss. This result is obviously inconsistent with that obtained by using the oscillation property of the loop. The different magnetic field strengths for the same loop derived from different methods are mainly because of that the broad QFP wave train was actually a shock rather than linear fast-mode magnetosonic wave. Therefore, the authors further derived the Alfv\'en Mach number, and then estimated the magnetic field strength of the loop by using the characteristic fast-mode speed obtained through dividing the measured wave speed by the Alfv\'en Mach number. Finally, the authors obtained the same result as that derived from the loop oscillation, which also confirmed the reliability of the two seismology methods.

Filaments (or prominence) oscillations include transverse and longitudinal oscillations, and their oscillation parameters have also been applied into prominence seismology with various inversion techniques \citep[see][and reference therein]{2018LRSP...15....3A}. In previous studies, filament oscillations are commonly observed to be caused by the interaction of global EUV waves \citep[e.g.,][]{2012ApJ...754....7S,2014ApJ...786..151S,2014ApJ...795..130S,2017ApJ...851..101S,2018ApJ...860..113Z}. \cite{2012ApJ...753...52L} observed the transverse oscillation of the limb cavity, as well as the hosting prominence caused the passing of a broad QFP wave train. Taking the oscillation as a global standing transverse oscillation as those observed in coronal loops, the authors derived the cavity's magnetic field strength is about 6 Gauss with a pitch angle of about $70^{\circ}$, suggesting that the observed cavity is a highly twisted flux rope. \cite{2019ApJ...873...22S} studied the transverse oscillation of a remote filament caused by the interaction of an on-disk propagating broad QFP wave train. The authors estimated the radial component of the magnetic field of the filament by using the method proposed by \cite{1966ZA.....63...78H} with the measured parameters of oscillation period and damping time, and the derived value is about 12.4 Gauss. These results are in agreement with those obtained by the inversion of full-Stokes observations \citep[e.g.,][]{2003ApJ...598L..67C}.

\section{Conclusion and Prospect}
As one of the new discoveries of {\em SDO}/AIA, spatially-resolved QFP wave trains in EUV wavelength band have attracted a lot of attentions in the past decade. In this paper, we have reviewed the observational properties, the possible formation mechanisms, and the associated coronal seismology applications of coronal QFP wave trains. Generally, a QFP wave train consists of multiple coherent and concentric wavefronts emanating successively near the epicenter of the accompanying flare and propagating outwardly either along or across coronal loops at fast-mode magnetosonic speed from several hundred to more than \speed{2000}. Based on the statistical study of the published QFP wave trains observed by the AIA, we propose that QFP wave trains could possibly be divided into two distinct categories including narrow and broad QFP wave trains. Although both narrow and broad QFP wave trains are fast-mode magnetosonic waves in the physical nature and with similar speeds, periods and wavelengths, they also show distinct differences including physical properties of observation wavelength, propagation direction, angular width, intensity amplitude and energy flux. The energy flux carried by QFP wave trains is  found to be enough for heating the local low corona plasma, and the measured parameters such as period, amplitude and speed can be used to seismological diagnosing of the currently undetectable coronal parameters such as magnetic field strength.

Observations suggest that the generation of QFP wave trains are intimately associated with flare QPPs owning to their similar period and close temporal relation, and the two different phenomena might manifest the different aspects of the same physical process. Detailed theoretical and numerical studies revealed that the periodicity origins of QFP wave trains should be diversified, but they can be summarized as two interconnected groups dubbed as dispersion evolution mechanism and pulsed energy excitation mechanism. The dispersion evolution mechanism refers to a QFP wave train that develops from the dispersive evolution of an impulsive generated broadband  perturbation in an inhomogeneous overdense waveguide, because for a wave packet which represents a Fourier integral over all frequencies and wave number, different frequencies propagates at different group speeds. In this regime, the periodicity of the wave train is not necessarily connected with the wave source, but can be created by the dispersive evolution of the initial perturbation based on the physical conditions inside and outside of the waveguide. For the pulsed energy excitation mechanism, it means that the generation of a QFP wave train is periodically driven by pulsed energy releases owning to some nonlinear physical processes in association to magnetic reconnection, such as the repetitive generation, coalescence and ejection of plasmoids, oscillatory reconnection, and the modulation of the magnetic reconnection by external disturbances. Quasi-periodic motions in solar eruptions such as the unwinding motion of erupting twisted filament threads and expansion of coronal loops can also launch broad QFP wave trains in the corona. In addition, it is also noted that some periods in QFP wave trains are possibly connected to leakage of photospheric and chromospheric three and five minute oscillations into the corona. The generation mechanism of QFP wave trains should be diversified and more complicated than we thought; therefore, it should be pointed out that for a specific QFP wave train, it might be generated by a single physical process or by the combination of different ones.

Despite significant progress achieved in both theoretical and observational aspects on the study of coronal QFP wave trains in the past decade, thanks to the high spatiotemporal resolution and full-disk, wide-temperature coverage observations taken by the {\em SDO} and the tremendous improvement in computing and calculation techniques, there are still many important open questions that deserve further in-depth investigations. The following is a list of some outstanding issues.

\begin{enumerate}
\item Statistical surveys by considering large samples should be performed to explore the common properties of QFP wave trains. So far, only \cite{2016AIPC.1720d0010L} performed a preliminary survey based on the database of global EUV waves, where the authors found the high occurrence rate of QFP wave trains. In addition, the present review, as well as \cite{2014SoPh..289.3233L}, also provides a simple statistical study of QFP wave trains observed by the AIA using the published events. Since the intensity variations caused by narrow QFP wave trains are too small to be observed in the direct EUV images, one should alternatively use the running-difference or running-ratio images to search narrow QFP wave trains. Coupled with the difficulties caused by the AIA's massive data base, one need to develop sophisticated automatic detection software to perform a complete survey and to obtain more reliable physical parameters and other properties of QFP wave trains.

\item The excitation mechanisms of QFP wave trains are still unclear, although various possible mechanisms have been proposed in previous studies. The high spatiotemporal resolution, multi-angle observations, three-dimensional radiate MHD simulations using more realistic initial conditions, and data-driven simulations that use multi-wavelength observations in tandem with MHD simulation are all required to clarify the real excitation mechanism of QFP wave trains, as well as the waveguide properties. In addition, more attentions should be paid to the possible excitation of QFP wave trains by the leakage of photospheric and chromospheric three and five minute oscillations into the corona. As what has been discussed in Section \ref{dpm}, one needs to consider which mechanism is more suitable for which kind of QFP wave trains, or are there any new generation mechanisms?

\item QFP wave trains are typically associated with flares, but not all flares cause QFP wave trains. In addition, QFP wave trains do not show any obvious dependence tendency on the flare energy class. It is worthy to investigate that what type of flares are more in favor of the occurrence of QFP wave trains. Our survey based on the published events suggests that broad QFP wave trains are associated with more energetic flares than the narrow ones. \cite{2016AIPC.1720d0010L} found an interesting trend of preferential association of QFP wave trains with successful solar eruptions accompanied by CMEs based on the survey of two flare productive active regions. \cite{2007ApJ...665.1428W} found that failed and successful solar eruptions tend to occur closer to the magnetic center and to the edge of active regions, respectively. Does QFP wave trains also have such a location preference to occur more frequently in association with flares close to the edge of active regions? These special trends of preferential association with flares need further statistical investigations by using large statistical samples of QFP wave trains.

\item The relationship between QFP wave trains and QPPs in solar and stellar flares deserves further in-depth investigations. These investigations can help us to diagnose the flaring process and physical properties of the waveguides, coronal and stellar crown conditions, as well as the generation mechanism of QFP wave trains.

\item Studies based on high temporal resolution radio observations combined with EUV imaging observations are important to investigate the fine physical process in the generation of QFP wave trains. The relationship between narrow and broad QFP wave trains is worthy to study to answer why they appear together in some events but separately in other individual ones. Does this mean different generation mechanisms or different propagation conditions? For broad QFP wave trains propagating in large-scale areas, they will inevitably interact with remote coronal structures such as coronal holes, active regions, filaments and coronal loops. The phenomena occurred during these interactions can be applied to coronal seismology to diagnose the physical properties of the structures and the local coronal conditions.

\item Since QFP wave trains carry energy away from the eruption source regions and propagate along or across magnetic field lines, it is important to investigate their possible roles in energy transport, coronal heating, and the acceleration of solar wind.
\end{enumerate}

Future studies of QFP wave trains will continuously benefit from the joint observations taken by ground-based and space-borne solar telescopes. Especially, the massive database of the {\em SDO} remains to be fully exploited with sophisticated automatic detection techniques. The {\em Solar Orbiter} launched in 2020 operates both in and out of the ecliptic plane and images the polar regions of the Sun \citep{2020A&A...642A...1M}; the EUV imager and spectrometer onboard it can make further contributions to the investigation of QFP wave trains. In addition, the combination of the {\em SDO} and the {\em Solar Orbiter} can make a stereoscopic diagnosing to QFP wave trains. Other solar telescopes including the 4-meter Daniel K. Inouye Solar Telescope \citep[DKIST;][]{2021SoPh..296...70R}, the {\em Advanced Space-based Solar Observatory} \citep[ASO-S;][]{2019RAA....19..156G}, the Goode Solar Telescope \citep[GST;][]{2010AN....331..636C,2010AN....331..620G}, the New Vacuum Solar Telescope \citep[NVST;][]{2014RAA....14..705L}, the {\em Interface Region Imaging Spectrograph} \citep[IRIS;][]{2014SoPh..289.2733D}, and the {\em Parker Solar Probe} \citep[PSP;][]{2016SSRv..204....7F} are all important for the diagnosis the eruption source region and the associated magnetic reconnection process. A combination of the measurements of magnetic field, spectroscopy, imaging and {\em in situ} observations provided by these solar telescopes and high temporal resolution radio telescopes will undoubtedly lead to a significant breakthrough in the comprehensive understanding of coronal QFP wave trains in the future.

\begin{acks}
This work is supported by the Natural Science Foundation of China (12173083, 11922307, 11773068, 11633008), the Yunnan Science Foundation for Distinguished Young Scholars (202101AV070004), the Yunnan Science Foundation (2017FB006), the National Key R\&D Program of China (2019YFA0405000), the Specialized Research Fund for State Key Laboratories, and the West Light Foundation of Chinese Academy of Sciences. The writing of the present review is based upon the invited talk presented by Y. Shen in the international workshop ``MHD Coronal Seismology 2020: Twenty Years of Probing the Sun's Corona with MHD Waves'' organized by Dr. D. Kolotkov, B. Li, S. Anfinogentov, K. Murawski, G. Nistico, D. Tsiklauri, and T. Van Doorsselaere in 2020, and it will be published in the Solar Physics's Topical Collection ``Magnetohydrodynamic (MHD) Waves and Oscillations in the Sun's Corona and MHD Coronal Seismology''. The authors would like to thank the hard works done by the organizers and the Editors (D. Kolotkov and B. Li) for the Topical Collection.
\end{acks}


\begin{thebibliography}{260}
\ifx\bisbn     \undefined \def\bisbn  #1{ISBN #1}\fi
\ifx\binits    \undefined \def\binits#1{#1}\fi
\ifx\bauthor   \undefined \def\bauthor#1{#1}\fi
\ifx\batitle   \undefined \def\batitle#1{#1}\fi
\ifx\bjtitle   \undefined \def\bjtitle#1{\textit{#1}}\fi
\ifx\bvolume   \undefined \def\bvolume#1{\textbf{#1}}\fi
\ifx\byear     \undefined \def\byear#1{#1}\fi
\ifx\bissue    \undefined \def\bissue#1{#1}\fi
\ifx\bfpage    \undefined \def\bfpage#1{#1}\fi
\ifx\blpage    \undefined \def\blpage #1{#1}\fi
\ifx\burl      \undefined \def\burl#1{#1}\fi
\ifx\href      \undefined \def\href#1#2{#2}\fi
\ifx\betal     \undefined \def\betal{et al.}\fi
\ifx\bctitle   \undefined \def\bctitle#1{#1}\fi
\ifx\beditor   \undefined \def\beditor#1{#1}\fi
\ifx\bbtitle   \undefined \def\bbtitle#1{\textit{#1}}\fi
\ifx\bedition  \undefined \def\bedition#1{#1}\fi
\ifx\bseriesno \undefined \def\bseriesno#1{\textbf{#1}}\fi
\ifx\blocation \undefined \def\blocation#1{#1}\fi
\ifx\bsertitle \undefined \def\bsertitle#1{\textit{#1}}\fi
\ifx\bsnm      \undefined \def\bsnm#1{#1}\fi
\ifx\bsuffix   \undefined \def\bsuffix#1{#1}\fi
\ifx\bparticle \undefined \def\bparticle#1{#1}\fi
\ifx\barticle  \undefined \def\barticle#1{}\fi
\ifx\binstitute  \undefined \def\binstitute#1{#1}\fi
\ifx\bpublisher  \undefined \def\bpublisher#1{#1}\fi
\ifx\doiurl    \undefined \def\doiurl#1{\href{#1}{DOI}}\fi
\makeatletter
\def\safeHref#1#2#3{\in@{http}{#2}\ifin@\href{#2}{#3}\else\href{#1#2}{#3}\fi}
\makeatother
\ifx\adsurl    \undefined
  \def\adsurl#1{\safeHref{https://ui.adsabs.harvard.edu/abs/}{#1}{ADS}}\fi
\ifx\arxivurl  \undefined
  \def\arxivurl#1{\safeHref{http://arxiv.org/abs/}{#1}{arXiv}}\fi
\ifx\botherref \undefined \def\botherref#1{}\fi
\ifx\url       \undefined \def\url#1{#1}\fi
\ifx\bchapter  \undefined \def\bchapter#1{}\fi
\ifx\bbook     \undefined \def\bbook#1{}\fi
\ifx\bcomment  \undefined \def\bcomment#1{#1}\fi
\ifx\oauthor   \undefined \def\oauthor#1{#1}\fi
\ifx\citeauthoryear \undefined\def \citeauthoryear#1{#1}\fi
\def\endbibitem {}
\ifx\bconflocation  \undefined \def\bconflocation#1{#1} \fi

\bibitem[\protect\citeauthoryear{{Arregui}, {Oliver}, and
  {Ballester}}{2018}]{2018LRSP...15....3A}
\begin{barticle}
\bauthor{\bsnm{{Arregui}}, \binits{I.}},
\bauthor{\bsnm{{Oliver}}, \binits{R.}},
\bauthor{\bsnm{{Ballester}}, \binits{J.L.}}:
\byear{2018},
\batitle{{Prominence oscillations}}.
\bjtitle{Living Reviews in Solar Physics}
\bvolume{15},
\bfpage{3}.
\doiurl{https://doi.org/10.1007/s41116-018-0012-6}.
\adsurl{2018LRSP...15....3A}.
\end{barticle}
\endbibitem

\bibitem[\protect\citeauthoryear{{Aschwanden}}{2004}]{2004ESASP.575...97A}
\begin{bchapter}
\bauthor{\bsnm{{Aschwanden}}, \binits{M.J.}}:
\byear{2004},
\bctitle{{The Role of Observed MHD Oscillations and Waves for Coronal
  Heating}}.
In: \beditor{\bsnm{{Walsh}}, \binits{R.W.}},
\beditor{\bsnm{{Ireland}}, \binits{J.}},
\beditor{\bsnm{{Danesy}}, \binits{D.}},
\beditor{\bsnm{{Fleck}}, \binits{B.}} (eds.)
\bbtitle{SOHO 15 Coronal Heating},
\bsertitle{ESA Special Publication}
\bseriesno{575},
\bfpage{97}.
\adsurl{2004ESASP.575...97A}.
\end{bchapter}
\endbibitem

\bibitem[\protect\citeauthoryear{{Aschwanden}}{2005}]{2005psci.book.....A}
\begin{bbook}
\bauthor{\bsnm{{Aschwanden}}, \binits{M.J.}}:
\byear{2005},
\bbtitle{{Physics of the Solar Corona. An Introduction with Problems and
  Solutions (2nd edition)}}.
\adsurl{2005psci.book.....A}.
\end{bbook}
\endbibitem

\bibitem[\protect\citeauthoryear{{Aschwanden}
  et~al.}{1999}]{1999ApJ...520..880A}
\begin{barticle}
\bauthor{\bsnm{{Aschwanden}}, \binits{M.J.}},
\bauthor{\bsnm{{Fletcher}}, \binits{L.}},
\bauthor{\bsnm{{Schrijver}}, \binits{C.J.}},
\bauthor{\bsnm{{Alexander}}, \binits{D.}}:
\byear{1999},
\batitle{{Coronal Loop Oscillations Observed with the Transition Region and
  Coronal Explorer}}.
\bjtitle{\apj}
\bvolume{520},
\bfpage{880}.
\doiurl{https://doi.org/10.1086/307502}.
\adsurl{1999ApJ...520..880A}.
\end{barticle}
\endbibitem

\bibitem[\protect\citeauthoryear{{B{\'a}rta}, {Karlick{\'y}}, and
  {{\v{Z}}emli{\v{c}}ka}}{2008}]{2008SoPh..253..173B}
\begin{barticle}
\bauthor{\bsnm{{B{\'a}rta}}, \binits{M.}},
\bauthor{\bsnm{{Karlick{\'y}}}, \binits{M.}},
\bauthor{\bsnm{{{\v{Z}}emli{\v{c}}ka}}, \binits{R.}}:
\byear{2008},
\batitle{{Plasmoid Dynamics in Flare Reconnection and the Frequency Drift of
  the Drifting Pulsating Structure}}.
\bjtitle{\solphys}
\bvolume{253},
\bfpage{173}.
\doiurl{https://doi.org/10.1007/s11207-008-9217-5}.
\adsurl{2008SoPh..253..173B}.
\end{barticle}
\endbibitem

\bibitem[\protect\citeauthoryear{{Beckers} and
  {Tallant}}{1969}]{1969SoPh....7..351B}
\begin{barticle}
\bauthor{\bsnm{{Beckers}}, \binits{J.M.}},
\bauthor{\bsnm{{Tallant}}, \binits{P.E.}}:
\byear{1969},
\batitle{{Chromospheric Inhomogeneities in Sunspot Umbrae}}.
\bjtitle{\solphys}
\bvolume{7},
\bfpage{351}.
\doiurl{https://doi.org/10.1007/BF00146140}.
\adsurl{1969SoPh....7..351B}.
\end{barticle}
\endbibitem

\bibitem[\protect\citeauthoryear{{Bogdan} et~al.}{2003}]{2003ApJ...599..626B}
\begin{barticle}
\bauthor{\bsnm{{Bogdan}}, \binits{T.J.}},
\bauthor{\bsnm{{Carlsson}}, \binits{M.}},
\bauthor{\bsnm{{Hansteen}}, \binits{V.H.}},
\bauthor{\bsnm{{McMurry}}, \binits{A.}},
\bauthor{\bsnm{{Rosenthal}}, \binits{C.S.}},
\bauthor{\bsnm{{Johnson}}, \binits{M.}},
\bauthor{\bsnm{{Petty-Powell}}, \binits{S.}},
\bauthor{\bsnm{{Zita}}, \binits{E.J.}},
\bauthor{\bsnm{{Stein}}, \binits{R.F.}},
\bauthor{\bsnm{{McIntosh}}, \binits{S.W.}},
\bauthor{\bsnm{{Nordlund}}, \binits{{\r{A}}.}}:
\byear{2003},
\batitle{{Waves in the Magnetized Solar Atmosphere. II. Waves from Localized
  Sources in Magnetic Flux Concentrations}}.
\bjtitle{\apj}
\bvolume{599},
\bfpage{626}.
\doiurl{https://doi.org/10.1086/378512}.
\adsurl{2003ApJ...599..626B}.
\end{barticle}
\endbibitem

\bibitem[\protect\citeauthoryear{{Cai} et~al.}{2019}]{2019MNRAS.489.3183C}
\begin{barticle}
\bauthor{\bsnm{{Cai}}, \binits{Q.}},
\bauthor{\bsnm{{Shen}}, \binits{C.}},
\bauthor{\bsnm{{Raymond}}, \binits{J.C.}},
\bauthor{\bsnm{{Mei}}, \binits{Z.}},
\bauthor{\bsnm{{Warmuth}}, \binits{A.}},
\bauthor{\bsnm{{Roussev}}, \binits{I.I.}},
\bauthor{\bsnm{{Lin}}, \binits{J.}}:
\byear{2019},
\batitle{{Investigations of a supra-arcade fan and termination shock above the
  top of the flare-loop system of the 2017 September 10 event}}.
\bjtitle{\mnras}
\bvolume{489},
\bfpage{3183}.
\doiurl{https://doi.org/10.1093/mnras/stz2167}.
\adsurl{2019MNRAS.489.3183C}.
\end{barticle}
\endbibitem

\bibitem[\protect\citeauthoryear{{Cao} et~al.}{2010}]{2010AN....331..636C}
\begin{barticle}
\bauthor{\bsnm{{Cao}}, \binits{W.}},
\bauthor{\bsnm{{Gorceix}}, \binits{N.}},
\bauthor{\bsnm{{Coulter}}, \binits{R.}},
\bauthor{\bsnm{{Ahn}}, \binits{K.}},
\bauthor{\bsnm{{Rimmele}}, \binits{T.R.}},
\bauthor{\bsnm{{Goode}}, \binits{P.R.}}:
\byear{2010},
\batitle{{Scientific instrumentation for the 1.6 m New Solar Telescope in Big
  Bear}}.
\bjtitle{Astronomische Nachrichten}
\bvolume{331},
\bfpage{636}.
\doiurl{https://doi.org/10.1002/asna.201011390}.
\adsurl{2010AN....331..636C}.
\end{barticle}
\endbibitem

\bibitem[\protect\citeauthoryear{{Casini} et~al.}{2003}]{2003ApJ...598L..67C}
\begin{barticle}
\bauthor{\bsnm{{Casini}}, \binits{R.}},
\bauthor{\bsnm{{L{\'o}pez Ariste}}, \binits{A.}},
\bauthor{\bsnm{{Tomczyk}}, \binits{S.}},
\bauthor{\bsnm{{Lites}}, \binits{B.W.}}:
\byear{2003},
\batitle{{Magnetic Maps of Prominences from Full Stokes Analysis of the He I D3
  Line}}.
\bjtitle{\apjl}
\bvolume{598},
\bfpage{L67}.
\doiurl{https://doi.org/10.1086/380496}.
\adsurl{2003ApJ...598L..67C}.
\end{barticle}
\endbibitem

\bibitem[\protect\citeauthoryear{{Chandra} et~al.}{2016}]{2016ApJ...822..106C}
\begin{barticle}
\bauthor{\bsnm{{Chandra}}, \binits{R.}},
\bauthor{\bsnm{{Chen}}, \binits{P.F.}},
\bauthor{\bsnm{{Fulara}}, \binits{A.}},
\bauthor{\bsnm{{Srivastava}}, \binits{A.K.}},
\bauthor{\bsnm{{Uddin}}, \binits{W.}}:
\byear{2016},
\batitle{{Peculiar Stationary EUV Wave Fronts in the Eruption on 2011 May 11}}.
\bjtitle{\apj}
\bvolume{822},
\bfpage{106}.
\doiurl{https://doi.org/10.3847/0004-637X/822/2/106}.
\adsurl{2016ApJ...822..106C}.
\end{barticle}
\endbibitem

\bibitem[\protect\citeauthoryear{{Chandra} et~al.}{2018}]{2018ApJ...863..101C}
\begin{barticle}
\bauthor{\bsnm{{Chandra}}, \binits{R.}},
\bauthor{\bsnm{{Chen}}, \binits{P.F.}},
\bauthor{\bsnm{{Joshi}}, \binits{R.}},
\bauthor{\bsnm{{Joshi}}, \binits{B.}},
\bauthor{\bsnm{{Schmieder}}, \binits{B.}}:
\byear{2018},
\batitle{{Observations of Two Successive EUV Waves and Their Mode Conversion}}.
\bjtitle{\apj}
\bvolume{863},
\bfpage{101}.
\doiurl{https://doi.org/10.3847/1538-4357/aad097}.
\adsurl{2018ApJ...863..101C}.
\end{barticle}
\endbibitem

\bibitem[\protect\citeauthoryear{{Chen}}{2011}]{2011LRSP....8....1C}
\begin{barticle}
\bauthor{\bsnm{{Chen}}, \binits{P.F.}}:
\byear{2011},
\batitle{{Coronal Mass Ejections: Models and Their Observational Basis}}.
\bjtitle{Living Reviews in Solar Physics}
\bvolume{8},
\bfpage{1}.
\doiurl{https://doi.org/10.12942/lrsp-2011-1}.
\adsurl{2011LRSP....8....1C}.
\end{barticle}
\endbibitem

\bibitem[\protect\citeauthoryear{{Chen}}{2016}]{2016GMS...216..381C}
\begin{barticle}
\bauthor{\bsnm{{Chen}}, \binits{P.F.}}:
\byear{2016},
\batitle{{Global Coronal Waves}}.
\bjtitle{Washington DC American Geophysical Union Geophysical Monograph Series}
\bvolume{216},
\bfpage{381}.
\doiurl{https://doi.org/10.1002/9781119055006.ch22}.
\adsurl{2016GMS...216..381C}.
\end{barticle}
\endbibitem

\bibitem[\protect\citeauthoryear{{Chen} and
  {Priest}}{2006}]{2006SoPh..238..313C}
\begin{barticle}
\bauthor{\bsnm{{Chen}}, \binits{P.F.}},
\bauthor{\bsnm{{Priest}}, \binits{E.R.}}:
\byear{2006},
\batitle{{Transition-Region Explosive Events: Reconnection Modulated by p-Mode
  Waves}}.
\bjtitle{\solphys}
\bvolume{238},
\bfpage{313}.
\doiurl{https://doi.org/10.1007/s11207-006-0215-1}.
\adsurl{2006SoPh..238..313C}.
\end{barticle}
\endbibitem

\bibitem[\protect\citeauthoryear{{Chen}, {Fang}, and
  {Shibata}}{2005}]{2005ApJ...622.1202C}
\begin{barticle}
\bauthor{\bsnm{{Chen}}, \binits{P.F.}},
\bauthor{\bsnm{{Fang}}, \binits{C.}},
\bauthor{\bsnm{{Shibata}}, \binits{K.}}:
\byear{2005},
\batitle{{A Full View of EIT Waves}}.
\bjtitle{\apj}
\bvolume{622},
\bfpage{1202}.
\doiurl{https://doi.org/10.1086/428084}.
\adsurl{2005ApJ...622.1202C}.
\end{barticle}
\endbibitem

\bibitem[\protect\citeauthoryear{{Chen} et~al.}{2002}]{2002ApJ...572L..99C}
\begin{barticle}
\bauthor{\bsnm{{Chen}}, \binits{P.F.}},
\bauthor{\bsnm{{Wu}}, \binits{S.T.}},
\bauthor{\bsnm{{Shibata}}, \binits{K.}},
\bauthor{\bsnm{{Fang}}, \binits{C.}}:
\byear{2002},
\batitle{{Evidence of EIT and Moreton Waves in Numerical Simulations}}.
\bjtitle{\apjl}
\bvolume{572},
\bfpage{L99}.
\doiurl{https://doi.org/10.1086/341486}.
\adsurl{2002ApJ...572L..99C}.
\end{barticle}
\endbibitem

\bibitem[\protect\citeauthoryear{{Chen} et~al.}{2016}]{2016SoPh..291.3195C}
\begin{barticle}
\bauthor{\bsnm{{Chen}}, \binits{P.F.}},
\bauthor{\bsnm{{Fang}}, \binits{C.}},
\bauthor{\bsnm{{Chandra}}, \binits{R.}},
\bauthor{\bsnm{{Srivastava}}, \binits{A.K.}}:
\byear{2016},
\batitle{{Can a Fast-Mode EUV Wave Generate a Stationary Front?}}
\bjtitle{\solphys}
\bvolume{291},
\bfpage{3195}.
\doiurl{https://doi.org/10.1007/s11207-016-0920-3}.
\adsurl{2016SoPh..291.3195C}.
\end{barticle}
\endbibitem

\bibitem[\protect\citeauthoryear{{Chen} et~al.}{2019}]{2019ApJ...878...78C}
\begin{barticle}
\bauthor{\bsnm{{Chen}}, \binits{X.}},
\bauthor{\bsnm{{Yan}}, \binits{Y.}},
\bauthor{\bsnm{{Tan}}, \binits{B.}},
\bauthor{\bsnm{{Huang}}, \binits{J.}},
\bauthor{\bsnm{{Wang}}, \binits{W.}},
\bauthor{\bsnm{{Chen}}, \binits{L.}},
\bauthor{\bsnm{{Zhang}}, \binits{Y.}},
\bauthor{\bsnm{{Tan}}, \binits{C.}},
\bauthor{\bsnm{{Liu}}, \binits{D.}},
\bauthor{\bsnm{{Masuda}}, \binits{S.}}:
\byear{2019},
\batitle{{Quasi-periodic Pulsations before and during a Solar Flare in AR
  12242}}.
\bjtitle{\apj}
\bvolume{878},
\bfpage{78}.
\doiurl{https://doi.org/10.3847/1538-4357/ab1d64}.
\adsurl{2019ApJ...878...78C}.
\end{barticle}
\endbibitem

\bibitem[\protect\citeauthoryear{{Cheng} et~al.}{2012}]{2012ApJ...745L...5C}
\begin{barticle}
\bauthor{\bsnm{{Cheng}}, \binits{X.}},
\bauthor{\bsnm{{Zhang}}, \binits{J.}},
\bauthor{\bsnm{{Olmedo}}, \binits{O.}},
\bauthor{\bsnm{{Vourlidas}}, \binits{A.}},
\bauthor{\bsnm{{Ding}}, \binits{M.D.}},
\bauthor{\bsnm{{Liu}}, \binits{Y.}}:
\byear{2012},
\batitle{{Investigation of the Formation and Separation of an
  Extreme-ultraviolet Wave from the Expansion of a Coronal Mass Ejection}}.
\bjtitle{\apjl}
\bvolume{745},
\bfpage{L5}.
\doiurl{https://doi.org/10.1088/2041-8205/745/1/L5}.
\adsurl{2012ApJ...745L...5C}.
\end{barticle}
\endbibitem

\bibitem[\protect\citeauthoryear{{Cheng} et~al.}{2018}]{2018ApJ...866...64C}
\begin{barticle}
\bauthor{\bsnm{{Cheng}}, \binits{X.}},
\bauthor{\bsnm{{Li}}, \binits{Y.}},
\bauthor{\bsnm{{Wan}}, \binits{L.F.}},
\bauthor{\bsnm{{Ding}}, \binits{M.D.}},
\bauthor{\bsnm{{Chen}}, \binits{P.F.}},
\bauthor{\bsnm{{Zhang}}, \binits{J.}},
\bauthor{\bsnm{{Liu}}, \binits{J.J.}}:
\byear{2018},
\batitle{{Observations of Turbulent Magnetic Reconnection within a Solar
  Current Sheet}}.
\bjtitle{\apj}
\bvolume{866},
\bfpage{64}.
\doiurl{https://doi.org/10.3847/1538-4357/aadd16}.
\adsurl{2018ApJ...866...64C}.
\end{barticle}
\endbibitem

\bibitem[\protect\citeauthoryear{{Chernov}}{2006}]{2006SSRv..127..195C}
\begin{barticle}
\bauthor{\bsnm{{Chernov}}, \binits{G.P.}}:
\byear{2006},
\batitle{{Solar Radio Bursts with Drifting Stripes in Emission and
  Absorption}}.
\bjtitle{\ssr}
\bvolume{127},
\bfpage{195}.
\doiurl{https://doi.org/10.1007/s11214-006-9141-7}.
\adsurl{2006SSRv..127..195C}.
\end{barticle}
\endbibitem

\bibitem[\protect\citeauthoryear{{Chernov}}{2010}]{2010RAA....10..821C}
\begin{barticle}
\bauthor{\bsnm{{Chernov}}, \binits{G.P.}}:
\byear{2010},
\batitle{{Recent results of zebra patterns in solar radio bursts}}.
\bjtitle{Research in Astronomy and Astrophysics}
\bvolume{10},
\bfpage{821}.
\doiurl{https://doi.org/10.1088/1674-4527/10/9/002}.
\adsurl{2010RAA....10..821C}.
\end{barticle}
\endbibitem

\bibitem[\protect\citeauthoryear{{Cheung} et~al.}{2015}]{2015ApJ...807..143C}
\begin{barticle}
\bauthor{\bsnm{{Cheung}}, \binits{M.C.M.}},
\bauthor{\bsnm{{Boerner}}, \binits{P.}},
\bauthor{\bsnm{{Schrijver}}, \binits{C.J.}},
\bauthor{\bsnm{{Testa}}, \binits{P.}},
\bauthor{\bsnm{{Chen}}, \binits{F.}},
\bauthor{\bsnm{{Peter}}, \binits{H.}},
\bauthor{\bsnm{{Malanushenko}}, \binits{A.}}:
\byear{2015},
\batitle{{Thermal Diagnostics with the Atmospheric Imaging Assembly on board
  the Solar Dynamics Observatory: A Validated Method for Differential Emission
  Measure Inversions}}.
\bjtitle{\apj}
\bvolume{807},
\bfpage{143}.
\doiurl{https://doi.org/10.1088/0004-637X/807/2/143}.
\adsurl{2015ApJ...807..143C}.
\end{barticle}
\endbibitem

\bibitem[\protect\citeauthoryear{{Clarke} et~al.}{2021}]{2021ApJ...910..123C}
\begin{barticle}
\bauthor{\bsnm{{Clarke}}, \binits{B.P.}},
\bauthor{\bsnm{{Hayes}}, \binits{L.A.}},
\bauthor{\bsnm{{Gallagher}}, \binits{P.T.}},
\bauthor{\bsnm{{Maloney}}, \binits{S.A.}},
\bauthor{\bsnm{{Carley}}, \binits{E.P.}}:
\byear{2021},
\batitle{{Quasi-periodic Particle Acceleration in a Solar Flare}}.
\bjtitle{\apj}
\bvolume{910},
\bfpage{123}.
\doiurl{https://doi.org/10.3847/1538-4357/abe463}.
\adsurl{2021ApJ...910..123C}.
\end{barticle}
\endbibitem

\bibitem[\protect\citeauthoryear{{Cooper}, {Nakariakov}, and
  {Williams}}{2003}]{2003A&A...409..325C}
\begin{barticle}
\bauthor{\bsnm{{Cooper}}, \binits{F.C.}},
\bauthor{\bsnm{{Nakariakov}}, \binits{V.M.}},
\bauthor{\bsnm{{Williams}}, \binits{D.R.}}:
\byear{2003},
\batitle{{Short period fast waves in solar coronal loops}}.
\bjtitle{\aap}
\bvolume{409},
\bfpage{325}.
\doiurl{https://doi.org/10.1051/0004-6361:20031071}.
\adsurl{2003A&A...409..325C}.
\end{barticle}
\endbibitem

\bibitem[\protect\citeauthoryear{{Cowsik} et~al.}{1999}]{1999SoPh..188...89C}
\begin{barticle}
\bauthor{\bsnm{{Cowsik}}, \binits{R.}},
\bauthor{\bsnm{{Singh}}, \binits{J.}},
\bauthor{\bsnm{{Saxena}}, \binits{A.K.}},
\bauthor{\bsnm{{Srinivasan}}, \binits{R.}},
\bauthor{\bsnm{{Raveendran}}, \binits{A.V.}}:
\byear{1999},
\batitle{{Short-period intensity oscillations in the solar corona observed
  during the total solar eclipse of 26 February 1998}}.
\bjtitle{\solphys}
\bvolume{188},
\bfpage{89}.
\doiurl{https://doi.org/10.1023/A:1005149303094}.
\adsurl{1999SoPh..188...89C}.
\end{barticle}
\endbibitem

\bibitem[\protect\citeauthoryear{{Craig} and
  {McClymont}}{1991}]{1991ApJ...371L..41C}
\begin{barticle}
\bauthor{\bsnm{{Craig}}, \binits{I.J.D.}},
\bauthor{\bsnm{{McClymont}}, \binits{A.N.}}:
\byear{1991},
\batitle{{Dynamic Magnetic Reconnection at an X-Type Neutral Point}}.
\bjtitle{\apjl}
\bvolume{371},
\bfpage{L41}.
\doiurl{https://doi.org/10.1086/185997}.
\adsurl{1991ApJ...371L..41C}.
\end{barticle}
\endbibitem

\bibitem[\protect\citeauthoryear{{de La Noe} and
  {Boischot}}{1972}]{1972A&A....20...55D}
\begin{barticle}
\bauthor{\bsnm{{de La Noe}}, \binits{J.}},
\bauthor{\bsnm{{Boischot}}, \binits{A.}}:
\byear{1972},
\batitle{{The Type III B Burst}}.
\bjtitle{\aap}
\bvolume{20},
\bfpage{55}.
\adsurl{1972A&A....20...55D}.
\end{barticle}
\endbibitem

\bibitem[\protect\citeauthoryear{{De Moortel}}{2005}]{2005RSPTA.363.2743D}
\begin{barticle}
\bauthor{\bsnm{{De Moortel}}, \binits{I.}}:
\byear{2005},
\batitle{{An overview of coronal seismology}}.
\bjtitle{Philosophical Transactions of the Royal Society of London Series A}
\bvolume{363},
\bfpage{2743}.
\doiurl{https://doi.org/10.1098/rsta.2005.1665}.
\adsurl{2005RSPTA.363.2743D}.
\end{barticle}
\endbibitem

\bibitem[\protect\citeauthoryear{{De Moortel} and
  {Nakariakov}}{2012}]{2012RSPTA.370.3193D}
\begin{barticle}
\bauthor{\bsnm{{De Moortel}}, \binits{I.}},
\bauthor{\bsnm{{Nakariakov}}, \binits{V.M.}}:
\byear{2012},
\batitle{{Magnetohydrodynamic waves and coronal seismology: an overview of
  recent results}}.
\bjtitle{Philosophical Transactions of the Royal Society of London Series A}
\bvolume{370},
\bfpage{3193}.
\doiurl{https://doi.org/10.1098/rsta.2011.0640}.
\adsurl{2012RSPTA.370.3193D}.
\end{barticle}
\endbibitem

\bibitem[\protect\citeauthoryear{{De Moortel}
  et~al.}{2002}]{2002A&A...387L..13D}
\begin{barticle}
\bauthor{\bsnm{{De Moortel}}, \binits{I.}},
\bauthor{\bsnm{{Ireland}}, \binits{J.}},
\bauthor{\bsnm{{Hood}}, \binits{A.W.}},
\bauthor{\bsnm{{Walsh}}, \binits{R.W.}}:
\byear{2002},
\batitle{{The detection of 3 \& 5 min period oscillations in coronal loops}}.
\bjtitle{\aap}
\bvolume{387},
\bfpage{L13}.
\doiurl{https://doi.org/10.1051/0004-6361:20020436}.
\adsurl{2002A&A...387L..13D}.
\end{barticle}
\endbibitem

\bibitem[\protect\citeauthoryear{{De Pontieu}
  et~al.}{2014}]{2014SoPh..289.2733D}
\begin{barticle}
\bauthor{\bsnm{{De Pontieu}}, \binits{B.}},
\bauthor{\bsnm{{Title}}, \binits{A.M.}},
\bauthor{\bsnm{{Lemen}}, \binits{J.R.}},
\bauthor{\bsnm{{Kushner}}, \binits{G.D.}},
\bauthor{\bsnm{{Akin}}, \binits{D.J.}},
\bauthor{\bsnm{{Allard}}, \binits{B.}},
\bauthor{\bsnm{{Berger}}, \binits{T.}},
\bauthor{\bsnm{{Boerner}}, \binits{P.}},
\bauthor{\bsnm{{Cheung}}, \binits{M.}},
\bauthor{\bsnm{{Chou}}, \binits{C.}},
\bauthor{\bsnm{{Drake}}, \binits{J.F.}},
\bauthor{\bsnm{{Duncan}}, \binits{D.W.}},
\bauthor{\bsnm{{Freeland}}, \binits{S.}},
\bauthor{\bsnm{{Heyman}}, \binits{G.F.}},
\bauthor{\bsnm{{Hoffman}}, \binits{C.}},
\bauthor{\bsnm{{Hurlburt}}, \binits{N.E.}},
\bauthor{\bsnm{{Lindgren}}, \binits{R.W.}},
\bauthor{\bsnm{{Mathur}}, \binits{D.}},
\bauthor{\bsnm{{Rehse}}, \binits{R.}},
\bauthor{\bsnm{{Sabolish}}, \binits{D.}},
\bauthor{\bsnm{{Seguin}}, \binits{R.}},
\bauthor{\bsnm{{Schrijver}}, \binits{C.J.}},
\bauthor{\bsnm{{Tarbell}}, \binits{T.D.}},
\bauthor{\bsnm{{W{\"u}lser}}, \binits{J.-P.}},
\bauthor{\bsnm{{Wolfson}}, \binits{C.J.}},
\bauthor{\bsnm{{Yanari}}, \binits{C.}},
\bauthor{\bsnm{{Mudge}}, \binits{J.}},
\bauthor{\bsnm{{Nguyen-Phuc}}, \binits{N.}},
\bauthor{\bsnm{{Timmons}}, \binits{R.}},
\bauthor{\bsnm{{van Bezooijen}}, \binits{R.}},
\bauthor{\bsnm{{Weingrod}}, \binits{I.}},
\bauthor{\bsnm{{Brookner}}, \binits{R.}},
\bauthor{\bsnm{{Butcher}}, \binits{G.}},
\bauthor{\bsnm{{Dougherty}}, \binits{B.}},
\bauthor{\bsnm{{Eder}}, \binits{J.}},
\bauthor{\bsnm{{Knagenhjelm}}, \binits{V.}},
\bauthor{\bsnm{{Larsen}}, \binits{S.}},
\bauthor{\bsnm{{Mansir}}, \binits{D.}},
\bauthor{\bsnm{{Phan}}, \binits{L.}},
\bauthor{\bsnm{{Boyle}}, \binits{P.}},
\bauthor{\bsnm{{Cheimets}}, \binits{P.N.}},
\bauthor{\bsnm{{DeLuca}}, \binits{E.E.}},
\bauthor{\bsnm{{Golub}}, \binits{L.}},
\bauthor{\bsnm{{Gates}}, \binits{R.}},
\bauthor{\bsnm{{Hertz}}, \binits{E.}},
\bauthor{\bsnm{{McKillop}}, \binits{S.}},
\bauthor{\bsnm{{Park}}, \binits{S.}},
\bauthor{\bsnm{{Perry}}, \binits{T.}},
\bauthor{\bsnm{{Podgorski}}, \binits{W.A.}},
\bauthor{\bsnm{{Reeves}}, \binits{K.}},
\bauthor{\bsnm{{Saar}}, \binits{S.}},
\bauthor{\bsnm{{Testa}}, \binits{P.}},
\bauthor{\bsnm{{Tian}}, \binits{H.}},
\bauthor{\bsnm{{Weber}}, \binits{M.}},
\bauthor{\bsnm{{Dunn}}, \binits{C.}},
\bauthor{\bsnm{{Eccles}}, \binits{S.}},
\bauthor{\bsnm{{Jaeggli}}, \binits{S.A.}},
\bauthor{\bsnm{{Kankelborg}}, \binits{C.C.}},
\bauthor{\bsnm{{Mashburn}}, \binits{K.}},
\bauthor{\bsnm{{Pust}}, \binits{N.}},
\bauthor{\bsnm{{Springer}}, \binits{L.}},
\bauthor{\bsnm{{Carvalho}}, \binits{R.}},
\bauthor{\bsnm{{Kleint}}, \binits{L.}},
\bauthor{\bsnm{{Marmie}}, \binits{J.}},
\bauthor{\bsnm{{Mazmanian}}, \binits{E.}},
\bauthor{\bsnm{{Pereira}}, \binits{T.M.D.}},
\bauthor{\bsnm{{Sawyer}}, \binits{S.}},
\bauthor{\bsnm{{Strong}}, \binits{J.}},
\bauthor{\bsnm{{Worden}}, \binits{S.P.}},
\bauthor{\bsnm{{Carlsson}}, \binits{M.}},
\bauthor{\bsnm{{Hansteen}}, \binits{V.H.}},
\bauthor{\bsnm{{Leenaarts}}, \binits{J.}},
\bauthor{\bsnm{{Wiesmann}}, \binits{M.}},
\bauthor{\bsnm{{Aloise}}, \binits{J.}},
\bauthor{\bsnm{{Chu}}, \binits{K.-C.}},
\bauthor{\bsnm{{Bush}}, \binits{R.I.}},
\bauthor{\bsnm{{Scherrer}}, \binits{P.H.}},
\bauthor{\bsnm{{Brekke}}, \binits{P.}},
\bauthor{\bsnm{{Martinez-Sykora}}, \binits{J.}},
\bauthor{\bsnm{{Lites}}, \binits{B.W.}},
\bauthor{\bsnm{{McIntosh}}, \binits{S.W.}},
\bauthor{\bsnm{{Uitenbroek}}, \binits{H.}},
\bauthor{\bsnm{{Okamoto}}, \binits{T.J.}},
\bauthor{\bsnm{{Gummin}}, \binits{M.A.}},
\bauthor{\bsnm{{Auker}}, \binits{G.}},
\bauthor{\bsnm{{Jerram}}, \binits{P.}},
\bauthor{\bsnm{{Pool}}, \binits{P.}},
\bauthor{\bsnm{{Waltham}}, \binits{N.}}:
\byear{2014},
\batitle{{The Interface Region Imaging Spectrograph (IRIS)}}.
\bjtitle{\solphys}
\bvolume{289},
\bfpage{2733}.
\doiurl{https://doi.org/10.1007/s11207-014-0485-y}.
\adsurl{2014SoPh..289.2733D}.
\end{barticle}
\endbibitem

\bibitem[\protect\citeauthoryear{{DeForest}}{2004}]{2004ApJ...617L..89D}
\begin{barticle}
\bauthor{\bsnm{{DeForest}}, \binits{C.E.}}:
\byear{2004},
\batitle{{High-Frequency Waves Detected in the Solar Atmosphere}}.
\bjtitle{\apjl}
\bvolume{617},
\bfpage{L89}.
\doiurl{https://doi.org/10.1086/427181}.
\adsurl{2004ApJ...617L..89D}.
\end{barticle}
\endbibitem

\bibitem[\protect\citeauthoryear{{Delaboudini{\`e}re}
  et~al.}{1995}]{1995SoPh..162..291D}
\begin{barticle}
\bauthor{\bsnm{{Delaboudini{\`e}re}}, \binits{J.-P.}},
\bauthor{\bsnm{{Artzner}}, \binits{G.E.}},
\bauthor{\bsnm{{Brunaud}}, \binits{J.}},
\bauthor{\bsnm{{Gabriel}}, \binits{A.H.}},
\bauthor{\bsnm{{Hochedez}}, \binits{J.F.}},
\bauthor{\bsnm{{Millier}}, \binits{F.}},
\bauthor{\bsnm{{Song}}, \binits{X.Y.}},
\bauthor{\bsnm{{Au}}, \binits{B.}},
\bauthor{\bsnm{{Dere}}, \binits{K.P.}},
\bauthor{\bsnm{{Howard}}, \binits{R.A.}},
\bauthor{\bsnm{{Kreplin}}, \binits{R.}},
\bauthor{\bsnm{{Michels}}, \binits{D.J.}},
\bauthor{\bsnm{{Moses}}, \binits{J.D.}},
\bauthor{\bsnm{{Defise}}, \binits{J.M.}},
\bauthor{\bsnm{{Jamar}}, \binits{C.}},
\bauthor{\bsnm{{Rochus}}, \binits{P.}},
\bauthor{\bsnm{{Chauvineau}}, \binits{J.P.}},
\bauthor{\bsnm{{Marioge}}, \binits{J.P.}},
\bauthor{\bsnm{{Catura}}, \binits{R.C.}},
\bauthor{\bsnm{{Lemen}}, \binits{J.R.}},
\bauthor{\bsnm{{Shing}}, \binits{L.}},
\bauthor{\bsnm{{Stern}}, \binits{R.A.}},
\bauthor{\bsnm{{Gurman}}, \binits{J.B.}},
\bauthor{\bsnm{{Neupert}}, \binits{W.M.}},
\bauthor{\bsnm{{Maucherat}}, \binits{A.}},
\bauthor{\bsnm{{Clette}}, \binits{F.}},
\bauthor{\bsnm{{Cugnon}}, \binits{P.}},
\bauthor{\bsnm{{van Dessel}}, \binits{E.L.}}:
\byear{1995},
\batitle{{EIT: Extreme-Ultraviolet Imaging Telescope for the SOHO Mission}}.
\bjtitle{\solphys}
\bvolume{162},
\bfpage{291}.
\doiurl{https://doi.org/10.1007/BF00733432}.
\adsurl{1995SoPh..162..291D}.
\end{barticle}
\endbibitem

\bibitem[\protect\citeauthoryear{{Delann{\'e}e} and
  {Aulanier}}{1999}]{1999SoPh..190..107D}
\begin{barticle}
\bauthor{\bsnm{{Delann{\'e}e}}, \binits{C.}},
\bauthor{\bsnm{{Aulanier}}, \binits{G.}}:
\byear{1999},
\batitle{{Cme Associated with Transequatorial Loops and a Bald Patch Flare}}.
\bjtitle{\solphys}
\bvolume{190},
\bfpage{107}.
\doiurl{https://doi.org/10.1023/A:1005249416605}.
\adsurl{1999SoPh..190..107D}.
\end{barticle}
\endbibitem

\bibitem[\protect\citeauthoryear{{Drake} et~al.}{2006}]{2006Natur.443..553D}
\begin{barticle}
\bauthor{\bsnm{{Drake}}, \binits{J.F.}},
\bauthor{\bsnm{{Swisdak}}, \binits{M.}},
\bauthor{\bsnm{{Che}}, \binits{H.}},
\bauthor{\bsnm{{Shay}}, \binits{M.A.}}:
\byear{2006},
\batitle{{Electron acceleration from contracting magnetic islands during
  reconnection}}.
\bjtitle{\nat}
\bvolume{443},
\bfpage{553}.
\doiurl{https://doi.org/10.1038/nature05116}.
\adsurl{2006Natur.443..553D}.
\end{barticle}
\endbibitem

\bibitem[\protect\citeauthoryear{{Duan} et~al.}{2021}]{duan2021}
\begin{botherref}
\oauthor{\bsnm{{Duan}}, \binits{Y.}},
\oauthor{\bsnm{{Shen}}, \binits{Y.}},
\oauthor{\bsnm{{Zhou}}, \binits{X.}},
\oauthor{\bsnm{{Tang}}, \binits{Z.}},
\oauthor{\bsnm{{Zhou}}, \binits{C.}},
\oauthor{\bsnm{{Tan}}, \binits{S.}}:
2021,
Homologous accelerated electron beams and an euv wave train associated with a
  fan-spine jet.
\textit{in prepare}.
\end{botherref}
\endbibitem

\bibitem[\protect\citeauthoryear{{Eto} et~al.}{2002}]{2002PASJ...54..481E}
\begin{barticle}
\bauthor{\bsnm{{Eto}}, \binits{S.}},
\bauthor{\bsnm{{Isobe}}, \binits{H.}},
\bauthor{\bsnm{{Narukage}}, \binits{N.}},
\bauthor{\bsnm{{Asai}}, \binits{A.}},
\bauthor{\bsnm{{Morimoto}}, \binits{T.}},
\bauthor{\bsnm{{Thompson}}, \binits{B.}},
\bauthor{\bsnm{{Yashiro}}, \binits{S.}},
\bauthor{\bsnm{{Wang}}, \binits{T.}},
\bauthor{\bsnm{{Kitai}}, \binits{R.}},
\bauthor{\bsnm{{Kurokawa}}, \binits{H.}},
\bauthor{\bsnm{{Shibata}}, \binits{K.}}:
\byear{2002},
\batitle{{Relation between a Moreton Wave and an EIT Wave Observed on 1997
  November 4}}.
\bjtitle{\pasj}
\bvolume{54},
\bfpage{481}.
\doiurl{https://doi.org/10.1093/pasj/54.3.481}.
\adsurl{2002PASJ...54..481E}.
\end{barticle}
\endbibitem

\bibitem[\protect\citeauthoryear{{Fletcher} et~al.}{2011}]{2011SSRv..159...19F}
\begin{barticle}
\bauthor{\bsnm{{Fletcher}}, \binits{L.}},
\bauthor{\bsnm{{Dennis}}, \binits{B.R.}},
\bauthor{\bsnm{{Hudson}}, \binits{H.S.}},
\bauthor{\bsnm{{Krucker}}, \binits{S.}},
\bauthor{\bsnm{{Phillips}}, \binits{K.}},
\bauthor{\bsnm{{Veronig}}, \binits{A.}},
\bauthor{\bsnm{{Battaglia}}, \binits{M.}},
\bauthor{\bsnm{{Bone}}, \binits{L.}},
\bauthor{\bsnm{{Caspi}}, \binits{A.}},
\bauthor{\bsnm{{Chen}}, \binits{Q.}},
\bauthor{\bsnm{{Gallagher}}, \binits{P.}},
\bauthor{\bsnm{{Grigis}}, \binits{P.T.}},
\bauthor{\bsnm{{Ji}}, \binits{H.}},
\bauthor{\bsnm{{Liu}}, \binits{W.}},
\bauthor{\bsnm{{Milligan}}, \binits{R.O.}},
\bauthor{\bsnm{{Temmer}}, \binits{M.}}:
\byear{2011},
\batitle{{An Observational Overview of Solar Flares}}.
\bjtitle{\ssr}
\bvolume{159},
\bfpage{19}.
\doiurl{https://doi.org/10.1007/s11214-010-9701-8}.
\adsurl{2011SSRv..159...19F}.
\end{barticle}
\endbibitem

\bibitem[\protect\citeauthoryear{{Foullon} et~al.}{2005}]{2005A&A...440L..59F}
\begin{barticle}
\bauthor{\bsnm{{Foullon}}, \binits{C.}},
\bauthor{\bsnm{{Verwichte}}, \binits{E.}},
\bauthor{\bsnm{{Nakariakov}}, \binits{V.M.}},
\bauthor{\bsnm{{Fletcher}}, \binits{L.}}:
\byear{2005},
\batitle{{X-ray quasi-periodic pulsations in solar flares as
  magnetohydrodynamic oscillations}}.
\bjtitle{\aap}
\bvolume{440},
\bfpage{L59}.
\doiurl{https://doi.org/10.1051/0004-6361:200500169}.
\adsurl{2005A&A...440L..59F}.
\end{barticle}
\endbibitem

\bibitem[\protect\citeauthoryear{{Fox} et~al.}{2016}]{2016SSRv..204....7F}
\begin{barticle}
\bauthor{\bsnm{{Fox}}, \binits{N.J.}},
\bauthor{\bsnm{{Velli}}, \binits{M.C.}},
\bauthor{\bsnm{{Bale}}, \binits{S.D.}},
\bauthor{\bsnm{{Decker}}, \binits{R.}},
\bauthor{\bsnm{{Driesman}}, \binits{A.}},
\bauthor{\bsnm{{Howard}}, \binits{R.A.}},
\bauthor{\bsnm{{Kasper}}, \binits{J.C.}},
\bauthor{\bsnm{{Kinnison}}, \binits{J.}},
\bauthor{\bsnm{{Kusterer}}, \binits{M.}},
\bauthor{\bsnm{{Lario}}, \binits{D.}},
\bauthor{\bsnm{{Lockwood}}, \binits{M.K.}},
\bauthor{\bsnm{{McComas}}, \binits{D.J.}},
\bauthor{\bsnm{{Raouafi}}, \binits{N.E.}},
\bauthor{\bsnm{{Szabo}}, \binits{A.}}:
\byear{2016},
\batitle{{The Solar Probe Plus Mission: Humanity's First Visit to Our Star}}.
\bjtitle{\ssr}
\bvolume{204},
\bfpage{7}.
\doiurl{https://doi.org/10.1007/s11214-015-0211-6}.
\adsurl{2016SSRv..204....7F}.
\end{barticle}
\endbibitem

\bibitem[\protect\citeauthoryear{{Furth}, {Killeen}, and
  {Rosenbluth}}{1963}]{1963PhFl....6..459F}
\begin{barticle}
\bauthor{\bsnm{{Furth}}, \binits{H.P.}},
\bauthor{\bsnm{{Killeen}}, \binits{J.}},
\bauthor{\bsnm{{Rosenbluth}}, \binits{M.N.}}:
\byear{1963},
\batitle{{Finite-Resistivity Instabilities of a Sheet Pinch}}.
\bjtitle{Physics of Fluids}
\bvolume{6},
\bfpage{459}.
\doiurl{https://doi.org/10.1063/1.1706761}.
\adsurl{1963PhFl....6..459F}.
\end{barticle}
\endbibitem

\bibitem[\protect\citeauthoryear{{Gan} et~al.}{2019}]{2019RAA....19..156G}
\begin{barticle}
\bauthor{\bsnm{{Gan}}, \binits{W.-Q.}},
\bauthor{\bsnm{{Zhu}}, \binits{C.}},
\bauthor{\bsnm{{Deng}}, \binits{Y.-Y.}},
\bauthor{\bsnm{{Li}}, \binits{H.}},
\bauthor{\bsnm{{Su}}, \binits{Y.}},
\bauthor{\bsnm{{Zhang}}, \binits{H.-Y.}},
\bauthor{\bsnm{{Chen}}, \binits{B.}},
\bauthor{\bsnm{{Zhang}}, \binits{Z.}},
\bauthor{\bsnm{{Wu}}, \binits{J.}},
\bauthor{\bsnm{{Deng}}, \binits{L.}},
\bauthor{\bsnm{{Huang}}, \binits{Y.}},
\bauthor{\bsnm{{Yang}}, \binits{J.-F.}},
\bauthor{\bsnm{{Cui}}, \binits{J.-J.}},
\bauthor{\bsnm{{Chang}}, \binits{J.}},
\bauthor{\bsnm{{Wang}}, \binits{C.}},
\bauthor{\bsnm{{Wu}}, \binits{J.}},
\bauthor{\bsnm{{Yin}}, \binits{Z.-S.}},
\bauthor{\bsnm{{Chen}}, \binits{W.}},
\bauthor{\bsnm{{Fang}}, \binits{C.}},
\bauthor{\bsnm{{Yan}}, \binits{Y.-H.}},
\bauthor{\bsnm{{Lin}}, \binits{J.}},
\bauthor{\bsnm{{Xiong}}, \binits{W.-M.}},
\bauthor{\bsnm{{Chen}}, \binits{B.}},
\bauthor{\bsnm{{Bao}}, \binits{H.-C.}},
\bauthor{\bsnm{{Cao}}, \binits{C.-X.}},
\bauthor{\bsnm{{Bai}}, \binits{Y.-P.}},
\bauthor{\bsnm{{Wang}}, \binits{T.}},
\bauthor{\bsnm{{Chen}}, \binits{B.-L.}},
\bauthor{\bsnm{{Li}}, \binits{X.-Y.}},
\bauthor{\bsnm{{Zhang}}, \binits{Y.}},
\bauthor{\bsnm{{Feng}}, \binits{L.}},
\bauthor{\bsnm{{Su}}, \binits{J.-T.}},
\bauthor{\bsnm{{Li}}, \binits{Y.}},
\bauthor{\bsnm{{Chen}}, \binits{W.}},
\bauthor{\bsnm{{Li}}, \binits{Y.-P.}},
\bauthor{\bsnm{{Su}}, \binits{Y.-N.}},
\bauthor{\bsnm{{Wu}}, \binits{H.-Y.}},
\bauthor{\bsnm{{Gu}}, \binits{M.}},
\bauthor{\bsnm{{Huang}}, \binits{L.}},
\bauthor{\bsnm{{Tang}}, \binits{X.-J.}}:
\byear{2019},
\batitle{{Advanced Space-based Solar Observatory (ASO-S): an overview}}.
\bjtitle{Research in Astronomy and Astrophysics}
\bvolume{19},
\bfpage{156}.
\doiurl{https://doi.org/10.1088/1674-4527/19/11/156}.
\adsurl{2019RAA....19..156G}.
\end{barticle}
\endbibitem

\bibitem[\protect\citeauthoryear{{Gary} and
  {Hurford}}{1994}]{1994ApJ...420..903G}
\begin{barticle}
\bauthor{\bsnm{{Gary}}, \binits{D.E.}},
\bauthor{\bsnm{{Hurford}}, \binits{G.J.}}:
\byear{1994},
\batitle{{Coronal Temperature, Density, and Magnetic Field Maps of a Solar
  Active Region Using the Owens Valley Solar Array}}.
\bjtitle{\apj}
\bvolume{420},
\bfpage{903}.
\doiurl{https://doi.org/10.1086/173614}.
\adsurl{1994ApJ...420..903G}.
\end{barticle}
\endbibitem

\bibitem[\protect\citeauthoryear{{Goddard}, {Nakariakov}, and
  {Pascoe}}{2019}]{2019A&A...624L...4G}
\begin{barticle}
\bauthor{\bsnm{{Goddard}}, \binits{C.R.}},
\bauthor{\bsnm{{Nakariakov}}, \binits{V.M.}},
\bauthor{\bsnm{{Pascoe}}, \binits{D.J.}}:
\byear{2019},
\batitle{{Fast magnetoacoustic wave trains with time-dependent drivers}}.
\bjtitle{\aap}
\bvolume{624},
\bfpage{L4}.
\doiurl{https://doi.org/10.1051/0004-6361/201935401}.
\adsurl{2019A&A...624L...4G}.
\end{barticle}
\endbibitem

\bibitem[\protect\citeauthoryear{{Goddard} et~al.}{2016}]{2016A&A...594A..96G}
\begin{barticle}
\bauthor{\bsnm{{Goddard}}, \binits{C.R.}},
\bauthor{\bsnm{{Nistic{\`o}}}, \binits{G.}},
\bauthor{\bsnm{{Nakariakov}}, \binits{V.M.}},
\bauthor{\bsnm{{Zimovets}}, \binits{I.V.}},
\bauthor{\bsnm{{White}}, \binits{S.M.}}:
\byear{2016},
\batitle{{Observation of quasi-periodic solar radio bursts associated with
  propagating fast-mode waves}}.
\bjtitle{\aap}
\bvolume{594},
\bfpage{A96}.
\doiurl{https://doi.org/10.1051/0004-6361/201628478}.
\adsurl{2016A&A...594A..96G}.
\end{barticle}
\endbibitem

\bibitem[\protect\citeauthoryear{{Goode} et~al.}{2010}]{2010AN....331..620G}
\begin{barticle}
\bauthor{\bsnm{{Goode}}, \binits{P.R.}},
\bauthor{\bsnm{{Coulter}}, \binits{R.}},
\bauthor{\bsnm{{Gorceix}}, \binits{N.}},
\bauthor{\bsnm{{Yurchyshyn}}, \binits{V.}},
\bauthor{\bsnm{{Cao}}, \binits{W.}}:
\byear{2010},
\batitle{{The NST: First results and some lessons for ATST and EST}}.
\bjtitle{Astronomische Nachrichten}
\bvolume{331},
\bfpage{620}.
\doiurl{https://doi.org/10.1002/asna.201011387}.
\adsurl{2010AN....331..620G}.
\end{barticle}
\endbibitem

\bibitem[\protect\citeauthoryear{{Gruszecki}, {Nakariakov}, and {Van
  Doorsselaere}}{2012}]{2012A&A...543A..12G}
\begin{barticle}
\bauthor{\bsnm{{Gruszecki}}, \binits{M.}},
\bauthor{\bsnm{{Nakariakov}}, \binits{V.M.}},
\bauthor{\bsnm{{Van Doorsselaere}}, \binits{T.}}:
\byear{2012},
\batitle{{Intensity variations associated with fast sausage modes}}.
\bjtitle{\aap}
\bvolume{543},
\bfpage{A12}.
\doiurl{https://doi.org/10.1051/0004-6361/201118168}.
\adsurl{2012A&A...543A..12G}.
\end{barticle}
\endbibitem

\bibitem[\protect\citeauthoryear{{Guidoni} et~al.}{2016}]{2016ApJ...820...60G}
\begin{barticle}
\bauthor{\bsnm{{Guidoni}}, \binits{S.E.}},
\bauthor{\bsnm{{DeVore}}, \binits{C.R.}},
\bauthor{\bsnm{{Karpen}}, \binits{J.T.}},
\bauthor{\bsnm{{Lynch}}, \binits{B.J.}}:
\byear{2016},
\batitle{{Magnetic-island Contraction and Particle Acceleration in Simulated
  Eruptive Solar Flares}}.
\bjtitle{\apj}
\bvolume{820},
\bfpage{60}.
\doiurl{https://doi.org/10.3847/0004-637X/820/1/60}.
\adsurl{2016ApJ...820...60G}.
\end{barticle}
\endbibitem

\bibitem[\protect\citeauthoryear{{Handy} et~al.}{1999}]{1999SoPh..187..229H}
\begin{barticle}
\bauthor{\bsnm{{Handy}}, \binits{B.N.}},
\bauthor{\bsnm{{Acton}}, \binits{L.W.}},
\bauthor{\bsnm{{Kankelborg}}, \binits{C.C.}},
\bauthor{\bsnm{{Wolfson}}, \binits{C.J.}},
\bauthor{\bsnm{{Akin}}, \binits{D.J.}},
\bauthor{\bsnm{{Bruner}}, \binits{M.E.}},
\bauthor{\bsnm{{Caravalho}}, \binits{R.}},
\bauthor{\bsnm{{Catura}}, \binits{R.C.}},
\bauthor{\bsnm{{Chevalier}}, \binits{R.}},
\bauthor{\bsnm{{Duncan}}, \binits{D.W.}},
\bauthor{\bsnm{{Edwards}}, \binits{C.G.}},
\bauthor{\bsnm{{Feinstein}}, \binits{C.N.}},
\bauthor{\bsnm{{Freeland}}, \binits{S.L.}},
\bauthor{\bsnm{{Friedlaender}}, \binits{F.M.}},
\bauthor{\bsnm{{Hoffmann}}, \binits{C.H.}},
\bauthor{\bsnm{{Hurlburt}}, \binits{N.E.}},
\bauthor{\bsnm{{Jurcevich}}, \binits{B.K.}},
\bauthor{\bsnm{{Katz}}, \binits{N.L.}},
\bauthor{\bsnm{{Kelly}}, \binits{G.A.}},
\bauthor{\bsnm{{Lemen}}, \binits{J.R.}},
\bauthor{\bsnm{{Levay}}, \binits{M.}},
\bauthor{\bsnm{{Lindgren}}, \binits{R.W.}},
\bauthor{\bsnm{{Mathur}}, \binits{D.P.}},
\bauthor{\bsnm{{Meyer}}, \binits{S.B.}},
\bauthor{\bsnm{{Morrison}}, \binits{S.J.}},
\bauthor{\bsnm{{Morrison}}, \binits{M.D.}},
\bauthor{\bsnm{{Nightingale}}, \binits{R.W.}},
\bauthor{\bsnm{{Pope}}, \binits{T.P.}},
\bauthor{\bsnm{{Rehse}}, \binits{R.A.}},
\bauthor{\bsnm{{Schrijver}}, \binits{C.J.}},
\bauthor{\bsnm{{Shine}}, \binits{R.A.}},
\bauthor{\bsnm{{Shing}}, \binits{L.}},
\bauthor{\bsnm{{Strong}}, \binits{K.T.}},
\bauthor{\bsnm{{Tarbell}}, \binits{T.D.}},
\bauthor{\bsnm{{Title}}, \binits{A.M.}},
\bauthor{\bsnm{{Torgerson}}, \binits{D.D.}},
\bauthor{\bsnm{{Golub}}, \binits{L.}},
\bauthor{\bsnm{{Bookbinder}}, \binits{J.A.}},
\bauthor{\bsnm{{Caldwell}}, \binits{D.}},
\bauthor{\bsnm{{Cheimets}}, \binits{P.N.}},
\bauthor{\bsnm{{Davis}}, \binits{W.N.}},
\bauthor{\bsnm{{Deluca}}, \binits{E.E.}},
\bauthor{\bsnm{{McMullen}}, \binits{R.A.}},
\bauthor{\bsnm{{Warren}}, \binits{H.P.}},
\bauthor{\bsnm{{Amato}}, \binits{D.}},
\bauthor{\bsnm{{Fisher}}, \binits{R.}},
\bauthor{\bsnm{{Maldonado}}, \binits{H.}},
\bauthor{\bsnm{{Parkinson}}, \binits{C.}}:
\byear{1999},
\batitle{{The transition region and coronal explorer}}.
\bjtitle{\solphys}
\bvolume{187},
\bfpage{229}.
\doiurl{https://doi.org/10.1023/A:1005166902804}.
\adsurl{1999SoPh..187..229H}.
\end{barticle}
\endbibitem

\bibitem[\protect\citeauthoryear{{Hayes} et~al.}{2020}]{2020ApJ...895...50H}
\begin{barticle}
\bauthor{\bsnm{{Hayes}}, \binits{L.A.}},
\bauthor{\bsnm{{Inglis}}, \binits{A.R.}},
\bauthor{\bsnm{{Christe}}, \binits{S.}},
\bauthor{\bsnm{{Dennis}}, \binits{B.}},
\bauthor{\bsnm{{Gallagher}}, \binits{P.T.}}:
\byear{2020},
\batitle{{Statistical Study of GOES X-Ray Quasi-periodic Pulsations in Solar
  Flares}}.
\bjtitle{\apj}
\bvolume{895},
\bfpage{50}.
\doiurl{https://doi.org/10.3847/1538-4357/ab8d40}.
\adsurl{2020ApJ...895...50H}.
\end{barticle}
\endbibitem

\bibitem[\protect\citeauthoryear{{Hong} et~al.}{2019}]{2019ApJ...874..146H}
\begin{barticle}
\bauthor{\bsnm{{Hong}}, \binits{J.}},
\bauthor{\bsnm{{Yang}}, \binits{J.}},
\bauthor{\bsnm{{Chen}}, \binits{H.}},
\bauthor{\bsnm{{Bi}}, \binits{Y.}},
\bauthor{\bsnm{{Yang}}, \binits{B.}},
\bauthor{\bsnm{{Chen}}, \binits{H.}}:
\byear{2019},
\batitle{{Observation of a Reversal of Breakout Reconnection Preceding a Jet:
  Evidence of Oscillatory Magnetic Reconnection?}}
\bjtitle{\apj}
\bvolume{874},
\bfpage{146}.
\doiurl{https://doi.org/10.3847/1538-4357/ab0c9d}.
\adsurl{2019ApJ...874..146H}.
\end{barticle}
\endbibitem

\bibitem[\protect\citeauthoryear{{Hyder}}{1966}]{1966ZA.....63...78H}
\begin{barticle}
\bauthor{\bsnm{{Hyder}}, \binits{C.L.}}:
\byear{1966},
\batitle{{Winking Filaments and Prominence and Coronal Magnetic Fields}}.
\bjtitle{\zap}
\bvolume{63},
\bfpage{78}.
\adsurl{1966ZA.....63...78H}.
\end{barticle}
\endbibitem

\bibitem[\protect\citeauthoryear{{Iwai} et~al.}{2012}]{2012SoPh..277..447I}
\begin{barticle}
\bauthor{\bsnm{{Iwai}}, \binits{K.}},
\bauthor{\bsnm{{Tsuchiya}}, \binits{F.}},
\bauthor{\bsnm{{Morioka}}, \binits{A.}},
\bauthor{\bsnm{{Misawa}}, \binits{H.}}:
\byear{2012},
\batitle{{IPRT/AMATERAS: A New Metric Spectrum Observation System for Solar
  Radio Bursts}}.
\bjtitle{\solphys}
\bvolume{277},
\bfpage{447}.
\doiurl{https://doi.org/10.1007/s11207-011-9919-y}.
\adsurl{2012SoPh..277..447I}.
\end{barticle}
\endbibitem

\bibitem[\protect\citeauthoryear{{Jel{\'\i}nek} and
  {Karlick{\'y}}}{2019}]{2019A&A...625A...3J}
\begin{barticle}
\bauthor{\bsnm{{Jel{\'\i}nek}}, \binits{P.}},
\bauthor{\bsnm{{Karlick{\'y}}}, \binits{M.}}:
\byear{2019},
\batitle{{Pulse-beam heating of deep atmospheric layers, their oscillations and
  shocks modulating the flare reconnection}}.
\bjtitle{\aap}
\bvolume{625},
\bfpage{A3}.
\doiurl{https://doi.org/10.1051/0004-6361/201935188}.
\adsurl{2019A&A...625A...3J}.
\end{barticle}
\endbibitem

\bibitem[\protect\citeauthoryear{{Jel{\'\i}nek}, {Karlick{\'y}}, and
  {Murawski}}{2012}]{2012A&A...546A..49J}
\begin{barticle}
\bauthor{\bsnm{{Jel{\'\i}nek}}, \binits{P.}},
\bauthor{\bsnm{{Karlick{\'y}}}, \binits{M.}},
\bauthor{\bsnm{{Murawski}}, \binits{K.}}:
\byear{2012},
\batitle{{Magnetoacoustic waves in a vertical flare current-sheet in a
  gravitationally stratified solar atmosphere}}.
\bjtitle{\aap}
\bvolume{546},
\bfpage{A49}.
\doiurl{https://doi.org/10.1051/0004-6361/201219891}.
\adsurl{2012A&A...546A..49J}.
\end{barticle}
\endbibitem

\bibitem[\protect\citeauthoryear{{Jel{\'\i}nek}
  et~al.}{2017}]{2017ApJ...847...98J}
\begin{barticle}
\bauthor{\bsnm{{Jel{\'\i}nek}}, \binits{P.}},
\bauthor{\bsnm{{Karlick{\'y}}}, \binits{M.}},
\bauthor{\bsnm{{Van Doorsselaere}}, \binits{T.}},
\bauthor{\bsnm{{B{\'a}rta}}, \binits{M.}}:
\byear{2017},
\batitle{{Oscillations Excited by Plasmoids Formed During Magnetic Reconnection
  in a Vertical Gravitationally Stratified Current Sheet}}.
\bjtitle{\apj}
\bvolume{847},
\bfpage{98}.
\doiurl{https://doi.org/10.3847/1538-4357/aa88a6}.
\adsurl{2017ApJ...847...98J}.
\end{barticle}
\endbibitem

\bibitem[\protect\citeauthoryear{{Jess} et~al.}{2012}]{2012ApJ...757..160J}
\begin{barticle}
\bauthor{\bsnm{{Jess}}, \binits{D.B.}},
\bauthor{\bsnm{{De Moortel}}, \binits{I.}},
\bauthor{\bsnm{{Mathioudakis}}, \binits{M.}},
\bauthor{\bsnm{{Christian}}, \binits{D.J.}},
\bauthor{\bsnm{{Reardon}}, \binits{K.P.}},
\bauthor{\bsnm{{Keys}}, \binits{P.H.}},
\bauthor{\bsnm{{Keenan}}, \binits{F.P.}}:
\byear{2012},
\batitle{{The Source of 3 Minute Magnetoacoustic Oscillations in Coronal
  Fans}}.
\bjtitle{\apj}
\bvolume{757},
\bfpage{160}.
\doiurl{https://doi.org/10.1088/0004-637X/757/2/160}.
\adsurl{2012ApJ...757..160J}.
\end{barticle}
\endbibitem

\bibitem[\protect\citeauthoryear{{Jess} et~al.}{2015}]{2015SSRv..190..103J}
\begin{barticle}
\bauthor{\bsnm{{Jess}}, \binits{D.B.}},
\bauthor{\bsnm{{Morton}}, \binits{R.J.}},
\bauthor{\bsnm{{Verth}}, \binits{G.}},
\bauthor{\bsnm{{Fedun}}, \binits{V.}},
\bauthor{\bsnm{{Grant}}, \binits{S.D.T.}},
\bauthor{\bsnm{{Giagkiozis}}, \binits{I.}}:
\byear{2015},
\batitle{{Multiwavelength Studies of MHD Waves in the Solar Chromosphere. An
  Overview of Recent Results}}.
\bjtitle{\ssr}
\bvolume{190},
\bfpage{103}.
\doiurl{https://doi.org/10.1007/s11214-015-0141-3}.
\adsurl{2015SSRv..190..103J}.
\end{barticle}
\endbibitem

\bibitem[\protect\citeauthoryear{{Jiricka} et~al.}{1993}]{1993SoPh..147..203J}
\begin{barticle}
\bauthor{\bsnm{{Jiricka}}, \binits{K.}},
\bauthor{\bsnm{{Karlicky}}, \binits{M.}},
\bauthor{\bsnm{{Kepka}}, \binits{O.}},
\bauthor{\bsnm{{Tlamicha}}, \binits{A.}}:
\byear{1993},
\batitle{{Fast drift burst observations with the new Ond{\v{r}}ejov
  radiospectrograph}}.
\bjtitle{\solphys}
\bvolume{147},
\bfpage{203}.
\doiurl{https://doi.org/10.1007/BF00675495}.
\adsurl{1993SoPh..147..203J}.
\end{barticle}
\endbibitem

\bibitem[\protect\citeauthoryear{{Kaiser} et~al.}{2008}]{2008SSRv..136....5K}
\begin{barticle}
\bauthor{\bsnm{{Kaiser}}, \binits{M.L.}},
\bauthor{\bsnm{{Kucera}}, \binits{T.A.}},
\bauthor{\bsnm{{Davila}}, \binits{J.M.}},
\bauthor{\bsnm{{St. Cyr}}, \binits{O.C.}},
\bauthor{\bsnm{{Guhathakurta}}, \binits{M.}},
\bauthor{\bsnm{{Christian}}, \binits{E.}}:
\byear{2008},
\batitle{{The STEREO Mission: An Introduction}}.
\bjtitle{\ssr}
\bvolume{136},
\bfpage{5}.
\doiurl{https://doi.org/10.1007/s11214-007-9277-0}.
\adsurl{2008SSRv..136....5K}.
\end{barticle}
\endbibitem

\bibitem[\protect\citeauthoryear{{Kane} et~al.}{1983}]{1983ApJ...271..376K}
\begin{barticle}
\bauthor{\bsnm{{Kane}}, \binits{S.R.}},
\bauthor{\bsnm{{Kai}}, \binits{K.}},
\bauthor{\bsnm{{Kosugi}}, \binits{T.}},
\bauthor{\bsnm{{Enome}}, \binits{S.}},
\bauthor{\bsnm{{Landecker}}, \binits{P.B.}},
\bauthor{\bsnm{{McKenzie}}, \binits{D.L.}}:
\byear{1983},
\batitle{{Acceleration and confinement of energetic particles in the 1980 June
  7 solar flare}}.
\bjtitle{\apj}
\bvolume{271},
\bfpage{376}.
\doiurl{https://doi.org/10.1086/161203}.
\adsurl{1983ApJ...271..376K}.
\end{barticle}
\endbibitem

\bibitem[\protect\citeauthoryear{{Kaneda} et~al.}{2018}]{2018ApJ...855L..29K}
\begin{barticle}
\bauthor{\bsnm{{Kaneda}}, \binits{K.}},
\bauthor{\bsnm{{Misawa}}, \binits{H.}},
\bauthor{\bsnm{{Iwai}}, \binits{K.}},
\bauthor{\bsnm{{Masuda}}, \binits{S.}},
\bauthor{\bsnm{{Tsuchiya}}, \binits{F.}},
\bauthor{\bsnm{{Katoh}}, \binits{Y.}},
\bauthor{\bsnm{{Obara}}, \binits{T.}}:
\byear{2018},
\batitle{{Detection of Propagating Fast Sausage Waves through Detailed Analysis
  of a Zebra-pattern Fine Structure in a Solar Radio Burst}}.
\bjtitle{\apjl}
\bvolume{855},
\bfpage{L29}.
\doiurl{https://doi.org/10.3847/2041-8213/aab2a5}.
\adsurl{2018ApJ...855L..29K}.
\end{barticle}
\endbibitem

\bibitem[\protect\citeauthoryear{{Karlick{\'y}}}{2004}]{2004A&A...417..325K}
\begin{barticle}
\bauthor{\bsnm{{Karlick{\'y}}}, \binits{M.}}:
\byear{2004},
\batitle{{Series of high-frequency slowly drifting structures mapping the flare
  magnetic field reconnection}}.
\bjtitle{\aap}
\bvolume{417},
\bfpage{325}.
\doiurl{https://doi.org/10.1051/0004-6361:20034249}.
\adsurl{2004A&A...417..325K}.
\end{barticle}
\endbibitem

\bibitem[\protect\citeauthoryear{{Karlick{\'y}}}{2013}]{2013A&A...552A..90K}
\begin{barticle}
\bauthor{\bsnm{{Karlick{\'y}}}, \binits{M.}}:
\byear{2013},
\batitle{{Radio continua modulated by waves: Zebra patterns in solar and pulsar
  radio spectra?}}
\bjtitle{\aap}
\bvolume{552},
\bfpage{A90}.
\doiurl{https://doi.org/10.1051/0004-6361/201321356}.
\adsurl{2013A&A...552A..90K}.
\end{barticle}
\endbibitem

\bibitem[\protect\citeauthoryear{{Karlick{\'y}} and
  {B{\'a}rta}}{2007}]{2007A&A...464..735K}
\begin{barticle}
\bauthor{\bsnm{{Karlick{\'y}}}, \binits{M.}},
\bauthor{\bsnm{{B{\'a}rta}}, \binits{M.}}:
\byear{2007},
\batitle{{Drifting pulsating structures generated during tearing and
  coalescence processes in a flare current sheet}}.
\bjtitle{\aap}
\bvolume{464},
\bfpage{735}.
\doiurl{https://doi.org/10.1051/0004-6361:20065983}.
\adsurl{2007A&A...464..735K}.
\end{barticle}
\endbibitem

\bibitem[\protect\citeauthoryear{{Karlick{\'y}}, {Jel{\'\i}nek}, and
  {M{\'e}sz{\'a}rosov{\'a}}}{2011}]{2011A&A...529A..96K}
\begin{barticle}
\bauthor{\bsnm{{Karlick{\'y}}}, \binits{M.}},
\bauthor{\bsnm{{Jel{\'\i}nek}}, \binits{P.}},
\bauthor{\bsnm{{M{\'e}sz{\'a}rosov{\'a}}}, \binits{H.}}:
\byear{2011},
\batitle{{Magnetoacoustic waves in the narrowband dm-spikes sources}}.
\bjtitle{\aap}
\bvolume{529},
\bfpage{A96}.
\doiurl{https://doi.org/10.1051/0004-6361/201016171}.
\adsurl{2011A&A...529A..96K}.
\end{barticle}
\endbibitem

\bibitem[\protect\citeauthoryear{{Karlick{\'y}}, {M{\'e}sz{\'a}rosov{\'a}}, and
  {Jel{\'\i}nek}}{2013}]{2013A&A...550A...1K}
\begin{barticle}
\bauthor{\bsnm{{Karlick{\'y}}}, \binits{M.}},
\bauthor{\bsnm{{M{\'e}sz{\'a}rosov{\'a}}}, \binits{H.}},
\bauthor{\bsnm{{Jel{\'\i}nek}}, \binits{P.}}:
\byear{2013},
\batitle{{Radio fiber bursts and fast magnetoacoustic wave trains}}.
\bjtitle{\aap}
\bvolume{550},
\bfpage{A1}.
\doiurl{https://doi.org/10.1051/0004-6361/201220296}.
\adsurl{2013A&A...550A...1K}.
\end{barticle}
\endbibitem

\bibitem[\protect\citeauthoryear{{Kashapova}
  et~al.}{2020}]{2020A&A...642A.195K}
\begin{barticle}
\bauthor{\bsnm{{Kashapova}}, \binits{L.K.}},
\bauthor{\bsnm{{Kupriyanova}}, \binits{E.G.}},
\bauthor{\bsnm{{Xu}}, \binits{Z.}},
\bauthor{\bsnm{{Reid}}, \binits{H.A.S.}},
\bauthor{\bsnm{{Kolotkov}}, \binits{D.Y.}}:
\byear{2020},
\batitle{{The origin of quasi-periodicities during circular ribbon flares}}.
\bjtitle{\aap}
\bvolume{642},
\bfpage{A195}.
\doiurl{https://doi.org/10.1051/0004-6361/201833947}.
\adsurl{2020A&A...642A.195K}.
\end{barticle}
\endbibitem

\bibitem[\protect\citeauthoryear{{Katsiyannis}
  et~al.}{2003}]{2003A&A...406..709K}
\begin{barticle}
\bauthor{\bsnm{{Katsiyannis}}, \binits{A.C.}},
\bauthor{\bsnm{{Williams}}, \binits{D.R.}},
\bauthor{\bsnm{{McAteer}}, \binits{R.T.J.}},
\bauthor{\bsnm{{Gallagher}}, \binits{P.T.}},
\bauthor{\bsnm{{Keenan}}, \binits{F.P.}},
\bauthor{\bsnm{{Murtagh}}, \binits{F.}}:
\byear{2003},
\batitle{{Eclipse observations of high-frequency oscillations in active region
  coronal loops}}.
\bjtitle{\aap}
\bvolume{406},
\bfpage{709}.
\doiurl{https://doi.org/10.1051/0004-6361:20030458}.
\adsurl{2003A&A...406..709K}.
\end{barticle}
\endbibitem

\bibitem[\protect\citeauthoryear{{Klassen} et~al.}{2000}]{2000A&AS..141..357K}
\begin{barticle}
\bauthor{\bsnm{{Klassen}}, \binits{A.}},
\bauthor{\bsnm{{Aurass}}, \binits{H.}},
\bauthor{\bsnm{{Mann}}, \binits{G.}},
\bauthor{\bsnm{{Thompson}}, \binits{B.J.}}:
\byear{2000},
\batitle{{Catalogue of the 1997 SOHO-EIT coronal transient waves and associated
  type II radio burst spectra}}.
\bjtitle{\aaps}
\bvolume{141},
\bfpage{357}.
\doiurl{https://doi.org/10.1051/aas:2000125}.
\adsurl{2000A&AS..141..357K}.
\end{barticle}
\endbibitem

\bibitem[\protect\citeauthoryear{{Kliem}, {Karlick{\'y}}, and
  {Benz}}{2000}]{2000A&A...360..715K}
\begin{barticle}
\bauthor{\bsnm{{Kliem}}, \binits{B.}},
\bauthor{\bsnm{{Karlick{\'y}}}, \binits{M.}},
\bauthor{\bsnm{{Benz}}, \binits{A.O.}}:
\byear{2000},
\batitle{{Solar flare radio pulsations as a signature of dynamic magnetic
  reconnection}}.
\bjtitle{\aap}
\bvolume{360},
\bfpage{715}.
\adsurl{2000A&A...360..715K}.
\end{barticle}
\endbibitem

\bibitem[\protect\citeauthoryear{{Kolotkov}, {Nakariakov}, and
  {Kontar}}{2018}]{2018ApJ...861...33K}
\begin{barticle}
\bauthor{\bsnm{{Kolotkov}}, \binits{D.Y.}},
\bauthor{\bsnm{{Nakariakov}}, \binits{V.M.}},
\bauthor{\bsnm{{Kontar}}, \binits{E.P.}}:
\byear{2018},
\batitle{{Origin of the Modulation of the Radio Emission from the Solar Corona
  by a Fast Magnetoacoustic Wave}}.
\bjtitle{\apj}
\bvolume{861},
\bfpage{33}.
\doiurl{https://doi.org/10.3847/1538-4357/aac77e}.
\adsurl{2018ApJ...861...33K}.
\end{barticle}
\endbibitem

\bibitem[\protect\citeauthoryear{{Kolotkov} et~al.}{2021}]{2021MNRAS.505.3505K}
\begin{barticle}
\bauthor{\bsnm{{Kolotkov}}, \binits{D.Y.}},
\bauthor{\bsnm{{Nakariakov}}, \binits{V.M.}},
\bauthor{\bsnm{{Moss}}, \binits{G.}},
\bauthor{\bsnm{{Shellard}}, \binits{P.}}:
\byear{2021},
\batitle{{Fast magnetoacoustic wave trains: from tadpoles to boomerangs}}.
\bjtitle{\mnras}
\bvolume{505},
\bfpage{3505}.
\doiurl{https://doi.org/10.1093/mnras/stab1587}.
\adsurl{2021MNRAS.505.3505K}.
\end{barticle}
\endbibitem

\bibitem[\protect\citeauthoryear{{Koutchmy}, {Zhugzhda}, and
  {Locans}}{1983}]{1983A&A...120..185K}
\begin{barticle}
\bauthor{\bsnm{{Koutchmy}}, \binits{S.}},
\bauthor{\bsnm{{Zhugzhda}}, \binits{I.D.}},
\bauthor{\bsnm{{Locans}}, \binits{V.}}:
\byear{1983},
\batitle{{Short period coronal oscillations - Observation and interpretation}}.
\bjtitle{\aap}
\bvolume{120},
\bfpage{185}.
\adsurl{1983A&A...120..185K}.
\end{barticle}
\endbibitem

\bibitem[\protect\citeauthoryear{{Kumar} and
  {Innes}}{2015}]{2015ApJ...803L..23K}
\begin{barticle}
\bauthor{\bsnm{{Kumar}}, \binits{P.}},
\bauthor{\bsnm{{Innes}}, \binits{D.E.}}:
\byear{2015},
\batitle{{Partial Reflection and Trapping of a Fast-mode Wave in Solar Coronal
  Arcade Loops}}.
\bjtitle{\apjl}
\bvolume{803},
\bfpage{L23}.
\doiurl{https://doi.org/10.1088/2041-8205/803/2/L23}.
\adsurl{2015ApJ...803L..23K}.
\end{barticle}
\endbibitem

\bibitem[\protect\citeauthoryear{{Kumar} and
  {Manoharan}}{2013}]{2013AA...553A.109K}
\begin{barticle}
\bauthor{\bsnm{{Kumar}}, \binits{P.}},
\bauthor{\bsnm{{Manoharan}}, \binits{P.K.}}:
\byear{2013},
\batitle{{Eruption of a plasma blob, associated M-class flare, and large-scale
  extreme-ultraviolet wave observed by SDO}}.
\bjtitle{\aap}
\bvolume{553},
\bfpage{A109}.
\doiurl{https://doi.org/10.1051/0004-6361/201220283}.
\adsurl{2013A&A...553A.109K}.
\end{barticle}
\endbibitem

\bibitem[\protect\citeauthoryear{{Kumar}, {Nakariakov}, and
  {Cho}}{2016}]{2016ApJ...822....7K}
\begin{barticle}
\bauthor{\bsnm{{Kumar}}, \binits{P.}},
\bauthor{\bsnm{{Nakariakov}}, \binits{V.M.}},
\bauthor{\bsnm{{Cho}}, \binits{K.-S.}}:
\byear{2016},
\batitle{{Observation of a Quasiperiodic Pulsation in Hard X-Ray, Radio, and
  Extreme-ultraviolet Wavelengths}}.
\bjtitle{\apj}
\bvolume{822},
\bfpage{7}.
\doiurl{https://doi.org/10.3847/0004-637X/822/1/7}.
\adsurl{2016ApJ...822....7K}.
\end{barticle}
\endbibitem

\bibitem[\protect\citeauthoryear{{Kumar}, {Nakariakov}, and
  {Cho}}{2017}]{2017ApJ...844..149K}
\begin{barticle}
\bauthor{\bsnm{{Kumar}}, \binits{P.}},
\bauthor{\bsnm{{Nakariakov}}, \binits{V.M.}},
\bauthor{\bsnm{{Cho}}, \binits{K.-S.}}:
\byear{2017},
\batitle{{Quasi-periodic Radio Bursts Associated with Fast-mode Waves near a
  Magnetic Null Point}}.
\bjtitle{\apj}
\bvolume{844},
\bfpage{149}.
\doiurl{https://doi.org/10.3847/1538-4357/aa7d53}.
\adsurl{2017ApJ...844..149K}.
\end{barticle}
\endbibitem

\bibitem[\protect\citeauthoryear{{Kupriyanova}
  et~al.}{2010}]{2010SoPh..267..329K}
\begin{barticle}
\bauthor{\bsnm{{Kupriyanova}}, \binits{E.G.}},
\bauthor{\bsnm{{Melnikov}}, \binits{V.F.}},
\bauthor{\bsnm{{Nakariakov}}, \binits{V.M.}},
\bauthor{\bsnm{{Shibasaki}}, \binits{K.}}:
\byear{2010},
\batitle{{Types of Microwave Quasi-Periodic Pulsations in Single Flaring
  Loops}}.
\bjtitle{\solphys}
\bvolume{267},
\bfpage{329}.
\doiurl{https://doi.org/10.1007/s11207-010-9642-0}.
\adsurl{2010SoPh..267..329K}.
\end{barticle}
\endbibitem

\bibitem[\protect\citeauthoryear{{Kupriyanova}
  et~al.}{2020}]{2020STP.....6a...3K}
\begin{barticle}
\bauthor{\bsnm{{Kupriyanova}}, \binits{E.}},
\bauthor{\bsnm{{Kolotkov}}, \binits{D.}},
\bauthor{\bsnm{{Nakariakov}}, \binits{V.}},
\bauthor{\bsnm{{Kaufman}}, \binits{A.}}:
\byear{2020},
\batitle{{Quasi-Periodic Pulsations in Solar and Stellar Flares. Review}}.
\bjtitle{Solar-Terrestrial Physics}
\bvolume{6},
\bfpage{3}.
\doiurl{https://doi.org/10.12737/stp-61202001}.
\adsurl{2020STP.....6a...3K}.
\end{barticle}
\endbibitem

\bibitem[\protect\citeauthoryear{{Lazarian} and
  {Vishniac}}{1999}]{1999ApJ...517..700L}
\begin{barticle}
\bauthor{\bsnm{{Lazarian}}, \binits{A.}},
\bauthor{\bsnm{{Vishniac}}, \binits{E.T.}}:
\byear{1999},
\batitle{{Reconnection in a Weakly Stochastic Field}}.
\bjtitle{\apj}
\bvolume{517},
\bfpage{700}.
\doiurl{https://doi.org/10.1086/307233}.
\adsurl{1999ApJ...517..700L}.
\end{barticle}
\endbibitem

\bibitem[\protect\citeauthoryear{{Lemen} et~al.}{2012}]{2012SoPh..275...17L}
\begin{barticle}
\bauthor{\bsnm{{Lemen}}, \binits{J.R.}},
\bauthor{\bsnm{{Title}}, \binits{A.M.}},
\bauthor{\bsnm{{Akin}}, \binits{D.J.}},
\bauthor{\bsnm{{Boerner}}, \binits{P.F.}},
\bauthor{\bsnm{{Chou}}, \binits{C.}},
\bauthor{\bsnm{{Drake}}, \binits{J.F.}},
\bauthor{\bsnm{{Duncan}}, \binits{D.W.}},
\bauthor{\bsnm{{Edwards}}, \binits{C.G.}},
\bauthor{\bsnm{{Friedlaender}}, \binits{F.M.}},
\bauthor{\bsnm{{Heyman}}, \binits{G.F.}},
\bauthor{\bsnm{{Hurlburt}}, \binits{N.E.}},
\bauthor{\bsnm{{Katz}}, \binits{N.L.}},
\bauthor{\bsnm{{Kushner}}, \binits{G.D.}},
\bauthor{\bsnm{{Levay}}, \binits{M.}},
\bauthor{\bsnm{{Lindgren}}, \binits{R.W.}},
\bauthor{\bsnm{{Mathur}}, \binits{D.P.}},
\bauthor{\bsnm{{McFeaters}}, \binits{E.L.}},
\bauthor{\bsnm{{Mitchell}}, \binits{S.}},
\bauthor{\bsnm{{Rehse}}, \binits{R.A.}},
\bauthor{\bsnm{{Schrijver}}, \binits{C.J.}},
\bauthor{\bsnm{{Springer}}, \binits{L.A.}},
\bauthor{\bsnm{{Stern}}, \binits{R.A.}},
\bauthor{\bsnm{{Tarbell}}, \binits{T.D.}},
\bauthor{\bsnm{{Wuelser}}, \binits{J.-P.}},
\bauthor{\bsnm{{Wolfson}}, \binits{C.J.}},
\bauthor{\bsnm{{Yanari}}, \binits{C.}},
\bauthor{\bsnm{{Bookbinder}}, \binits{J.A.}},
\bauthor{\bsnm{{Cheimets}}, \binits{P.N.}},
\bauthor{\bsnm{{Caldwell}}, \binits{D.}},
\bauthor{\bsnm{{Deluca}}, \binits{E.E.}},
\bauthor{\bsnm{{Gates}}, \binits{R.}},
\bauthor{\bsnm{{Golub}}, \binits{L.}},
\bauthor{\bsnm{{Park}}, \binits{S.}},
\bauthor{\bsnm{{Podgorski}}, \binits{W.A.}},
\bauthor{\bsnm{{Bush}}, \binits{R.I.}},
\bauthor{\bsnm{{Scherrer}}, \binits{P.H.}},
\bauthor{\bsnm{{Gummin}}, \binits{M.A.}},
\bauthor{\bsnm{{Smith}}, \binits{P.}},
\bauthor{\bsnm{{Auker}}, \binits{G.}},
\bauthor{\bsnm{{Jerram}}, \binits{P.}},
\bauthor{\bsnm{{Pool}}, \binits{P.}},
\bauthor{\bsnm{{Soufli}}, \binits{R.}},
\bauthor{\bsnm{{Windt}}, \binits{D.L.}},
\bauthor{\bsnm{{Beardsley}}, \binits{S.}},
\bauthor{\bsnm{{Clapp}}, \binits{M.}},
\bauthor{\bsnm{{Lang}}, \binits{J.}},
\bauthor{\bsnm{{Waltham}}, \binits{N.}}:
\byear{2012},
\batitle{{The Atmospheric Imaging Assembly (AIA) on the Solar Dynamics
  Observatory (SDO)}}.
\bjtitle{\solphys}
\bvolume{275},
\bfpage{17}.
\doiurl{https://doi.org/10.1007/s11207-011-9776-8}.
\adsurl{2012SoPh..275...17L}.
\end{barticle}
\endbibitem

\bibitem[\protect\citeauthoryear{{Li} et~al.}{2018a}]{2018ApJ...855...53L}
\begin{barticle}
\bauthor{\bsnm{{Li}}, \binits{B.}},
\bauthor{\bsnm{{Guo}}, \binits{M.-Z.}},
\bauthor{\bsnm{{Yu}}, \binits{H.}},
\bauthor{\bsnm{{Chen}}, \binits{S.-X.}}:
\byear{2018}a,
\batitle{{Impulsively Generated Wave Trains in Coronal Structures. II. Effects
  of Transverse Structuring on Sausage Waves in Pressurelesss Slabs}}.
\bjtitle{\apj}
\bvolume{855},
\bfpage{53}.
\doiurl{https://doi.org/10.3847/1538-4357/aaaf19}.
\adsurl{2018ApJ...855...53L}.
\end{barticle}
\endbibitem

\bibitem[\protect\citeauthoryear{{Li} et~al.}{2020a}]{2020SSRv..216..136L}
\begin{barticle}
\bauthor{\bsnm{{Li}}, \binits{B.}},
\bauthor{\bsnm{{Antolin}}, \binits{P.}},
\bauthor{\bsnm{{Guo}}, \binits{M.-Z.}},
\bauthor{\bsnm{{Kuznetsov}}, \binits{A.A.}},
\bauthor{\bsnm{{Pascoe}}, \binits{D.J.}},
\bauthor{\bsnm{{Van Doorsselaere}}, \binits{T.}},
\bauthor{\bsnm{{Vasheghani Farahani}}, \binits{S.}}:
\byear{2020}a,
\batitle{{Magnetohydrodynamic Fast Sausage Waves in the Solar Corona}}.
\bjtitle{\ssr}
\bvolume{216},
\bfpage{136}.
\doiurl{https://doi.org/10.1007/s11214-020-00761-z}.
\adsurl{2020SSRv..216..136L}.
\end{barticle}
\endbibitem

\bibitem[\protect\citeauthoryear{Li}{2021}]{Li:2021vc}
\begin{botherref}
\oauthor{\bsnm{Li}, \binits{D.}}:
2021,
Quasi-periodic pulsations with double periods observed in
  ly{\ensuremath{\alpha}} emission during solar flares.
\textit{Science China Technological Sciences}.
1869-1900.
\doiurl{https://doi.org/10.1007/s11431-020-1771-7}.
\url{https://doi.org/10.1007/s11431-020-1771-7}.
\end{botherref}
\endbibitem

\bibitem[\protect\citeauthoryear{{Li} et~al.}{2020b}]{2020ApJ...893L..17L}
\begin{barticle}
\bauthor{\bsnm{{Li}}, \binits{D.}},
\bauthor{\bsnm{{Li}}, \binits{Y.}},
\bauthor{\bsnm{{Lu}}, \binits{L.}},
\bauthor{\bsnm{{Zhang}}, \binits{Q.}},
\bauthor{\bsnm{{Ning}}, \binits{Z.}},
\bauthor{\bsnm{{Anfinogentov}}, \binits{S.}}:
\byear{2020}b,
\batitle{{Observations of a Quasi-periodic Pulsation in the Coronal Loop and
  Microwave Flux during a Solar Preflare Phase}}.
\bjtitle{\apjl}
\bvolume{893},
\bfpage{L17}.
\doiurl{https://doi.org/10.3847/2041-8213/ab830c}.
\adsurl{2020ApJ...893L..17L}.
\end{barticle}
\endbibitem

\bibitem[\protect\citeauthoryear{{Li} et~al.}{2020c}]{2020A&A...639L...5L}
\begin{barticle}
\bauthor{\bsnm{{Li}}, \binits{D.}},
\bauthor{\bsnm{{Feng}}, \binits{S.}},
\bauthor{\bsnm{{Su}}, \binits{W.}},
\bauthor{\bsnm{{Huang}}, \binits{Y.}}:
\byear{2020}c,
\batitle{{Preflare very long-periodic pulsations observed in
  H{\ensuremath{\alpha}} emission before the onset of a solar flare}}.
\bjtitle{\aap}
\bvolume{639},
\bfpage{L5}.
\doiurl{https://doi.org/10.1051/0004-6361/202038398}.
\adsurl{2020A&A...639L...5L}.
\end{barticle}
\endbibitem

\bibitem[\protect\citeauthoryear{{Li} et~al.}{2020d}]{2020ApJ...893....7L}
\begin{barticle}
\bauthor{\bsnm{{Li}}, \binits{D.}},
\bauthor{\bsnm{{Lu}}, \binits{L.}},
\bauthor{\bsnm{{Ning}}, \binits{Z.}},
\bauthor{\bsnm{{Feng}}, \binits{L.}},
\bauthor{\bsnm{{Gan}}, \binits{W.}},
\bauthor{\bsnm{{Li}}, \binits{H.}}:
\byear{2020}d,
\batitle{{Quasi-periodic Pulsation Detected in Ly{\ensuremath{\alpha}} Emission
  During Solar Flares}}.
\bjtitle{\apj}
\bvolume{893},
\bfpage{7}.
\doiurl{https://doi.org/10.3847/1538-4357/ab7cd1}.
\adsurl{2020ApJ...893....7L}.
\end{barticle}
\endbibitem

\bibitem[\protect\citeauthoryear{{Li} et~al.}{2020e}]{2020ApJ...888...53L}
\begin{barticle}
\bauthor{\bsnm{{Li}}, \binits{D.}},
\bauthor{\bsnm{{Kolotkov}}, \binits{D.Y.}},
\bauthor{\bsnm{{Nakariakov}}, \binits{V.M.}},
\bauthor{\bsnm{{Lu}}, \binits{L.}},
\bauthor{\bsnm{{Ning}}, \binits{Z.J.}}:
\byear{2020}e,
\batitle{{Quasi-periodic Pulsations of Gamma-Ray Emissions from a Solar Flare
  on 2017 September 6}}.
\bjtitle{\apj}
\bvolume{888},
\bfpage{53}.
\doiurl{https://doi.org/10.3847/1538-4357/ab5e86}.
\adsurl{2020ApJ...888...53L}.
\end{barticle}
\endbibitem

\bibitem[\protect\citeauthoryear{{Li} et~al.}{2021a}]{2021RAA....21...66L}
\begin{barticle}
\bauthor{\bsnm{{Li}}, \binits{D.}},
\bauthor{\bsnm{{Warmuth}}, \binits{A.}},
\bauthor{\bsnm{{Lu}}, \binits{L.}},
\bauthor{\bsnm{{Ning}}, \binits{Z.}}:
\byear{2021}a,
\batitle{{An investigation of flare emissions at multiple wavelengths}}.
\bjtitle{Research in Astronomy and Astrophysics}
\bvolume{21},
\bfpage{066}.
\doiurl{https://doi.org/10.1088/1674-4527/21/3/66}.
\adsurl{2021RAA....21...66L}.
\end{barticle}
\endbibitem

\bibitem[\protect\citeauthoryear{{Li} et~al.}{2021b}]{2021ApJ...921..179L}
\begin{barticle}
\bauthor{\bsnm{{Li}}, \binits{D.}},
\bauthor{\bsnm{{Ge}}, \binits{M.}},
\bauthor{\bsnm{{Dominique}}, \binits{M.}},
\bauthor{\bsnm{{Zhao}}, \binits{H.}},
\bauthor{\bsnm{{Li}}, \binits{G.}},
\bauthor{\bsnm{{Li}}, \binits{X.}},
\bauthor{\bsnm{{Zhang}}, \binits{S.}},
\bauthor{\bsnm{{Lu}}, \binits{F.}},
\bauthor{\bsnm{{Gan}}, \binits{W.}},
\bauthor{\bsnm{{Ning}}, \binits{Z.}}:
\byear{2021}b,
\batitle{{Detection of Flare Multiperiodic Pulsations in Mid-ultraviolet Balmer
  Continuum, Ly{\ensuremath{\alpha}}, Hard X-Ray, and Radio Emissions
  Simultaneously}}.
\bjtitle{\apj}
\bvolume{921},
\bfpage{179}.
\doiurl{https://doi.org/10.3847/1538-4357/ac1c05}.
\adsurl{2021ApJ...921..179L}.
\end{barticle}
\endbibitem

\bibitem[\protect\citeauthoryear{{Li} et~al.}{2016}]{2016ApJ...829L..33L}
\begin{barticle}
\bauthor{\bsnm{{Li}}, \binits{L.P.}},
\bauthor{\bsnm{{Zhang}}, \binits{J.}},
\bauthor{\bsnm{{Su}}, \binits{J.T.}},
\bauthor{\bsnm{{Liu}}, \binits{Y.}}:
\byear{2016},
\batitle{{Oscillation of Current Sheets in the Wake of a Flux Rope Eruption
  Observed by the Solar Dynamics Observatory}}.
\bjtitle{\apjl}
\bvolume{829},
\bfpage{L33}.
\doiurl{https://doi.org/10.3847/2041-8205/829/2/L33}.
\adsurl{2016ApJ...829L..33L}.
\end{barticle}
\endbibitem

\bibitem[\protect\citeauthoryear{{Li} et~al.}{2018b}]{2018ApJ...868L..33L}
\begin{barticle}
\bauthor{\bsnm{{Li}}, \binits{L.}},
\bauthor{\bsnm{{Zhang}}, \binits{J.}},
\bauthor{\bsnm{{Peter}}, \binits{H.}},
\bauthor{\bsnm{{Chitta}}, \binits{L.P.}},
\bauthor{\bsnm{{Su}}, \binits{J.}},
\bauthor{\bsnm{{Song}}, \binits{H.}},
\bauthor{\bsnm{{Xia}}, \binits{C.}},
\bauthor{\bsnm{{Hou}}, \binits{Y.}}:
\byear{2018}b,
\batitle{{Quasi-periodic Fast Propagating Magnetoacoustic Waves during the
  Magnetic Reconnection Between Solar Coronal Loops}}.
\bjtitle{\apjl}
\bvolume{868},
\bfpage{L33}.
\doiurl{https://doi.org/10.3847/2041-8213/aaf167}.
\adsurl{2018ApJ...868L..33L}.
\end{barticle}
\endbibitem

\bibitem[\protect\citeauthoryear{{Li} et~al.}{2012}]{2012ApJ...746...13L}
\begin{barticle}
\bauthor{\bsnm{{Li}}, \binits{T.}},
\bauthor{\bsnm{{Zhang}}, \binits{J.}},
\bauthor{\bsnm{{Yang}}, \binits{S.}},
\bauthor{\bsnm{{Liu}}, \binits{W.}}:
\byear{2012},
\batitle{{SDO/AIA Observations of Secondary Waves Generated by Interaction of
  the 2011 June 7 Global EUV Wave with Solar Coronal Structures}}.
\bjtitle{\apj}
\bvolume{746},
\bfpage{13}.
\doiurl{https://doi.org/10.1088/0004-637X/746/1/13}.
\adsurl{2012ApJ...746...13L}.
\end{barticle}
\endbibitem

\bibitem[\protect\citeauthoryear{{Li} and {Lin}}{2012}]{2012SoPh..279...91L}
\begin{barticle}
\bauthor{\bsnm{{Li}}, \binits{Y.}},
\bauthor{\bsnm{{Lin}}, \binits{J.}}:
\byear{2012},
\batitle{{Acceleration of Electrons and Protons in Reconnecting Current Sheets
  Including Single or Multiple X-points}}.
\bjtitle{\solphys}
\bvolume{279},
\bfpage{91}.
\doiurl{https://doi.org/10.1007/s11207-012-9956-1}.
\adsurl{2012SoPh..279...91L}.
\end{barticle}
\endbibitem

\bibitem[\protect\citeauthoryear{{Li}, {Wu}, and
  {Lin}}{2017}]{2017A&A...605A.120L}
\begin{barticle}
\bauthor{\bsnm{{Li}}, \binits{Y.}},
\bauthor{\bsnm{{Wu}}, \binits{N.}},
\bauthor{\bsnm{{Lin}}, \binits{J.}}:
\byear{2017},
\batitle{{Charged-particle acceleration in a reconnecting current sheet
  including multiple magnetic islands and a nonuniform background magnetic
  field}}.
\bjtitle{\aap}
\bvolume{605},
\bfpage{A120}.
\doiurl{https://doi.org/10.1051/0004-6361/201630026}.
\adsurl{2017A&A...605A.120L}.
\end{barticle}
\endbibitem

\bibitem[\protect\citeauthoryear{{Lin}, {Kuhn}, and
  {Coulter}}{2004}]{2004ApJ...613L.177L}
\begin{barticle}
\bauthor{\bsnm{{Lin}}, \binits{H.}},
\bauthor{\bsnm{{Kuhn}}, \binits{J.R.}},
\bauthor{\bsnm{{Coulter}}, \binits{R.}}:
\byear{2004},
\batitle{{Coronal Magnetic Field Measurements}}.
\bjtitle{\apjl}
\bvolume{613},
\bfpage{L177}.
\doiurl{https://doi.org/10.1086/425217}.
\adsurl{2004ApJ...613L.177L}.
\end{barticle}
\endbibitem

\bibitem[\protect\citeauthoryear{{Lin}, {Penn}, and
  {Tomczyk}}{2000}]{2000ApJ...541L..83L}
\begin{barticle}
\bauthor{\bsnm{{Lin}}, \binits{H.}},
\bauthor{\bsnm{{Penn}}, \binits{M.J.}},
\bauthor{\bsnm{{Tomczyk}}, \binits{S.}}:
\byear{2000},
\batitle{{A New Precise Measurement of the Coronal Magnetic Field Strength}}.
\bjtitle{\apjl}
\bvolume{541},
\bfpage{L83}.
\doiurl{https://doi.org/10.1086/312900}.
\adsurl{2000ApJ...541L..83L}.
\end{barticle}
\endbibitem

\bibitem[\protect\citeauthoryear{{Lin} et~al.}{2015}]{2015SSRv..194..237L}
\begin{barticle}
\bauthor{\bsnm{{Lin}}, \binits{J.}},
\bauthor{\bsnm{{Murphy}}, \binits{N.A.}},
\bauthor{\bsnm{{Shen}}, \binits{C.}},
\bauthor{\bsnm{{Raymond}}, \binits{J.C.}},
\bauthor{\bsnm{{Reeves}}, \binits{K.K.}},
\bauthor{\bsnm{{Zhong}}, \binits{J.}},
\bauthor{\bsnm{{Wu}}, \binits{N.}},
\bauthor{\bsnm{{Li}}, \binits{Y.}}:
\byear{2015},
\batitle{{Review on Current Sheets in CME Development: Theories and
  Observations}}.
\bjtitle{\ssr}
\bvolume{194},
\bfpage{237}.
\doiurl{https://doi.org/10.1007/s11214-015-0209-0}.
\adsurl{2015SSRv..194..237L}.
\end{barticle}
\endbibitem

\bibitem[\protect\citeauthoryear{{Linton} and
  {Longcope}}{2006}]{2006ApJ...642.1177L}
\begin{barticle}
\bauthor{\bsnm{{Linton}}, \binits{M.G.}},
\bauthor{\bsnm{{Longcope}}, \binits{D.W.}}:
\byear{2006},
\batitle{{A Model for Patchy Reconnection in Three Dimensions}}.
\bjtitle{\apj}
\bvolume{642},
\bfpage{1177}.
\doiurl{https://doi.org/10.1086/500965}.
\adsurl{2006ApJ...642.1177L}.
\end{barticle}
\endbibitem

\bibitem[\protect\citeauthoryear{{Liu} and {Ofman}}{2014}]{2014SoPh..289.3233L}
\begin{barticle}
\bauthor{\bsnm{{Liu}}, \binits{W.}},
\bauthor{\bsnm{{Ofman}}, \binits{L.}}:
\byear{2014},
\batitle{{Advances in Observing Various Coronal EUV Waves in the SDO Era and
  Their Seismological Applications (Invited Review)}}.
\bjtitle{\solphys}
\bvolume{289},
\bfpage{3233}.
\doiurl{https://doi.org/10.1007/s11207-014-0528-4}.
\adsurl{2014SoPh..289.3233L}.
\end{barticle}
\endbibitem

\bibitem[\protect\citeauthoryear{{Liu}, {Chen}, and
  {Petrosian}}{2013}]{2013ApJ...767..168L}
\begin{barticle}
\bauthor{\bsnm{{Liu}}, \binits{W.}},
\bauthor{\bsnm{{Chen}}, \binits{Q.}},
\bauthor{\bsnm{{Petrosian}}, \binits{V.}}:
\byear{2013},
\batitle{{Plasmoid Ejections and Loop Contractions in an Eruptive M7.7 Solar
  Flare: Evidence of Particle Acceleration and Heating in Magnetic Reconnection
  Outflows}}.
\bjtitle{\apj}
\bvolume{767},
\bfpage{168}.
\doiurl{https://doi.org/10.1088/0004-637X/767/2/168}.
\adsurl{2013ApJ...767..168L}.
\end{barticle}
\endbibitem

\bibitem[\protect\citeauthoryear{{Liu} et~al.}{2010}]{2010ApJ...723L..53L}
\begin{barticle}
\bauthor{\bsnm{{Liu}}, \binits{W.}},
\bauthor{\bsnm{{Nitta}}, \binits{N.V.}},
\bauthor{\bsnm{{Schrijver}}, \binits{C.J.}},
\bauthor{\bsnm{{Title}}, \binits{A.M.}},
\bauthor{\bsnm{{Tarbell}}, \binits{T.D.}}:
\byear{2010},
\batitle{{First SDO AIA Observations of a Global Coronal EUV ``Wave'': Multiple
  Components and ``Ripples''}}.
\bjtitle{\apjl}
\bvolume{723},
\bfpage{L53}.
\doiurl{https://doi.org/10.1088/2041-8205/723/1/L53}.
\adsurl{2010ApJ...723L..53L}.
\end{barticle}
\endbibitem

\bibitem[\protect\citeauthoryear{{Liu} et~al.}{2011}]{2011ApJ...736L..13L}
\begin{barticle}
\bauthor{\bsnm{{Liu}}, \binits{W.}},
\bauthor{\bsnm{{Title}}, \binits{A.M.}},
\bauthor{\bsnm{{Zhao}}, \binits{J.}},
\bauthor{\bsnm{{Ofman}}, \binits{L.}},
\bauthor{\bsnm{{Schrijver}}, \binits{C.J.}},
\bauthor{\bsnm{{Aschwanden}}, \binits{M.J.}},
\bauthor{\bsnm{{De Pontieu}}, \binits{B.}},
\bauthor{\bsnm{{Tarbell}}, \binits{T.D.}}:
\byear{2011},
\batitle{{Direct Imaging of Quasi-periodic Fast Propagating Waves of
  \raisebox{-0.5ex}\textasciitilde2000 km s$^{-1}$ in the Low Solar Corona by
  the Solar Dynamics Observatory Atmospheric Imaging Assembly}}.
\bjtitle{\apjl}
\bvolume{736},
\bfpage{L13}.
\doiurl{https://doi.org/10.1088/2041-8205/736/1/L13}.
\adsurl{2011ApJ...736L..13L}.
\end{barticle}
\endbibitem

\bibitem[\protect\citeauthoryear{{Liu} et~al.}{2012}]{2012ApJ...753...52L}
\begin{barticle}
\bauthor{\bsnm{{Liu}}, \binits{W.}},
\bauthor{\bsnm{{Ofman}}, \binits{L.}},
\bauthor{\bsnm{{Nitta}}, \binits{N.V.}},
\bauthor{\bsnm{{Aschwanden}}, \binits{M.J.}},
\bauthor{\bsnm{{Schrijver}}, \binits{C.J.}},
\bauthor{\bsnm{{Title}}, \binits{A.M.}},
\bauthor{\bsnm{{Tarbell}}, \binits{T.D.}}:
\byear{2012},
\batitle{{Quasi-periodic Fast-mode Wave Trains within a Global EUV Wave and
  Sequential Transverse Oscillations Detected by SDO/AIA}}.
\bjtitle{\apj}
\bvolume{753},
\bfpage{52}.
\doiurl{https://doi.org/10.1088/0004-637X/753/1/52}.
\adsurl{2012ApJ...753...52L}.
\end{barticle}
\endbibitem

\bibitem[\protect\citeauthoryear{{Liu} et~al.}{2016}]{2016AIPC.1720d0010L}
\begin{bchapter}
\bauthor{\bsnm{{Liu}}, \binits{W.}},
\bauthor{\bsnm{{Ofman}}, \binits{L.}},
\bauthor{\bsnm{{Broder}}, \binits{B.}},
\bauthor{\bsnm{{Karlick{\'y}}}, \binits{M.}},
\bauthor{\bsnm{{Downs}}, \binits{C.}}:
\byear{2016},
\bctitle{{Quasi-periodic fast-mode magnetosonic wave trains within coronal
  waveguides associated with flares and CMEs}}.
In: \bbtitle{Solar Wind 14},
\bsertitle{American Institute of Physics Conference Series}
\bseriesno{1720},
\bfpage{040010}.
\doiurl{https://doi.org/10.1063/1.4943821}.
\adsurl{2016AIPC.1720d0010L}.
\end{bchapter}
\endbibitem

\bibitem[\protect\citeauthoryear{{Liu} and {Lin}}{2008}]{2008ApJ...680.1496L}
\begin{barticle}
\bauthor{\bsnm{{Liu}}, \binits{Y.}},
\bauthor{\bsnm{{Lin}}, \binits{H.}}:
\byear{2008},
\batitle{{Observational Test of Coronal Magnetic Field Models. I. Comparison
  with Potential Field Model}}.
\bjtitle{\apj}
\bvolume{680},
\bfpage{1496}.
\doiurl{https://doi.org/10.1086/588645}.
\adsurl{2008ApJ...680.1496L}.
\end{barticle}
\endbibitem

\bibitem[\protect\citeauthoryear{{Liu} et~al.}{2014}]{2014RAA....14..705L}
\begin{barticle}
\bauthor{\bsnm{{Liu}}, \binits{Z.}},
\bauthor{\bsnm{{Xu}}, \binits{J.}},
\bauthor{\bsnm{{Gu}}, \binits{B.-Z.}},
\bauthor{\bsnm{{Wang}}, \binits{S.}},
\bauthor{\bsnm{{You}}, \binits{J.-Q.}},
\bauthor{\bsnm{{Shen}}, \binits{L.-X.}},
\bauthor{\bsnm{{Lu}}, \binits{R.-W.}},
\bauthor{\bsnm{{Jin}}, \binits{Z.-Y.}},
\bauthor{\bsnm{{Chen}}, \binits{L.-F.}},
\bauthor{\bsnm{{Lou}}, \binits{K.}},
\bauthor{\bsnm{{Li}}, \binits{Z.}},
\bauthor{\bsnm{{Liu}}, \binits{G.-Q.}},
\bauthor{\bsnm{{Xu}}, \binits{Z.}},
\bauthor{\bsnm{{Rao}}, \binits{C.-H.}},
\bauthor{\bsnm{{Hu}}, \binits{Q.-Q.}},
\bauthor{\bsnm{{Li}}, \binits{R.-F.}},
\bauthor{\bsnm{{Fu}}, \binits{H.-W.}},
\bauthor{\bsnm{{Wang}}, \binits{F.}},
\bauthor{\bsnm{{Bao}}, \binits{M.-X.}},
\bauthor{\bsnm{{Wu}}, \binits{M.-C.}},
\bauthor{\bsnm{{Zhang}}, \binits{B.-R.}}:
\byear{2014},
\batitle{{New vacuum solar telescope and observations with high resolution}}.
\bjtitle{Research in Astronomy and Astrophysics}
\bvolume{14},
\bfpage{705}.
\doiurl{https://doi.org/10.1088/1674-4527/14/6/009}.
\adsurl{2014RAA....14..705L}.
\end{barticle}
\endbibitem

\bibitem[\protect\citeauthoryear{{Long} et~al.}{2017a}]{2017SoPh..292..185L}
\begin{barticle}
\bauthor{\bsnm{{Long}}, \binits{D.M.}},
\bauthor{\bsnm{{Murphy}}, \binits{P.}},
\bauthor{\bsnm{{Graham}}, \binits{G.}},
\bauthor{\bsnm{{Carley}}, \binits{E.P.}},
\bauthor{\bsnm{{P{\'e}rez-Su{\'a}rez}}, \binits{D.}}:
\byear{2017}a,
\batitle{{A Statistical Analysis of the Solar Phenomena Associated with Global
  EUV Waves}}.
\bjtitle{\solphys}
\bvolume{292},
\bfpage{185}.
\doiurl{https://doi.org/10.1007/s11207-017-1206-0}.
\adsurl{2017SoPh..292..185L}.
\end{barticle}
\endbibitem

\bibitem[\protect\citeauthoryear{{Long} et~al.}{2017b}]{2017SoPh..292....7L}
\begin{barticle}
\bauthor{\bsnm{{Long}}, \binits{D.M.}},
\bauthor{\bsnm{{Bloomfield}}, \binits{D.S.}},
\bauthor{\bsnm{{Chen}}, \binits{P.F.}},
\bauthor{\bsnm{{Downs}}, \binits{C.}},
\bauthor{\bsnm{{Gallagher}}, \binits{P.T.}},
\bauthor{\bsnm{{Kwon}}, \binits{R.-Y.}},
\bauthor{\bsnm{{Vanninathan}}, \binits{K.}},
\bauthor{\bsnm{{Veronig}}, \binits{A.M.}},
\bauthor{\bsnm{{Vourlidas}}, \binits{A.}},
\bauthor{\bsnm{{Vr{\v{s}}nak}}, \binits{B.}},
\bauthor{\bsnm{{Warmuth}}, \binits{A.}},
\bauthor{\bsnm{{{\v{Z}}ic}}, \binits{T.}}:
\byear{2017}b,
\batitle{{Understanding the Physical Nature of Coronal ``EIT Waves''}}.
\bjtitle{\solphys}
\bvolume{292},
\bfpage{7}.
\doiurl{https://doi.org/10.1007/s11207-016-1030-y}.
\adsurl{2017SoPh..292....7L}.
\end{barticle}
\endbibitem

\bibitem[\protect\citeauthoryear{{Lopin} and
  {Nagorny}}{2015}]{2015ApJ...810...87L}
\begin{barticle}
\bauthor{\bsnm{{Lopin}}, \binits{I.}},
\bauthor{\bsnm{{Nagorny}}, \binits{I.}}:
\byear{2015},
\batitle{{Sausage Waves in Transversely Nonuniform Monolithic Coronal Tubes}}.
\bjtitle{\apj}
\bvolume{810},
\bfpage{87}.
\doiurl{https://doi.org/10.1088/0004-637X/810/2/87}.
\adsurl{2015ApJ...810...87L}.
\end{barticle}
\endbibitem

\bibitem[\protect\citeauthoryear{{Lopin} and
  {Nagorny}}{2017}]{2017ApJ...840...26L}
\begin{barticle}
\bauthor{\bsnm{{Lopin}}, \binits{I.}},
\bauthor{\bsnm{{Nagorny}}, \binits{I.}}:
\byear{2017},
\batitle{{Kink Waves in Thin Stratified Magnetically Twisted Flux Tubes}}.
\bjtitle{\apj}
\bvolume{840},
\bfpage{26}.
\doiurl{https://doi.org/10.3847/1538-4357/aa6c5a}.
\adsurl{2017ApJ...840...26L}.
\end{barticle}
\endbibitem

\bibitem[\protect\citeauthoryear{{Lopin} and
  {Nagorny}}{2019}]{2019MNRAS.488..660L}
\begin{barticle}
\bauthor{\bsnm{{Lopin}}, \binits{I.}},
\bauthor{\bsnm{{Nagorny}}, \binits{I.}}:
\byear{2019},
\batitle{{Dispersion of sausage waves in coronal waveguides with transverse
  density structuring}}.
\bjtitle{\mnras}
\bvolume{488},
\bfpage{660}.
\doiurl{https://doi.org/10.1093/mnras/stz1737}.
\adsurl{2019MNRAS.488..660L}.
\end{barticle}
\endbibitem

\bibitem[\protect\citeauthoryear{{Lu} et~al.}{2021}]{2021SoPh..296..130L}
\begin{barticle}
\bauthor{\bsnm{{Lu}}, \binits{L.}},
\bauthor{\bsnm{{Li}}, \binits{D.}},
\bauthor{\bsnm{{Ning}}, \binits{Z.}},
\bauthor{\bsnm{{Feng}}, \binits{L.}},
\bauthor{\bsnm{{Gan}}, \binits{W.}}:
\byear{2021},
\batitle{{Quasi-Periodic Pulsations Detected in Ly {\ensuremath{\alpha}} and
  Nonthermal Emissions During Solar Flares}}.
\bjtitle{\solphys}
\bvolume{296},
\bfpage{130}.
\doiurl{https://doi.org/10.1007/s11207-021-01876-4}.
\adsurl{2021SoPh..296..130L}.
\end{barticle}
\endbibitem

\bibitem[\protect\citeauthoryear{{Ma} et~al.}{2011}]{2011ApJ...738..160M}
\begin{barticle}
\bauthor{\bsnm{{Ma}}, \binits{S.}},
\bauthor{\bsnm{{Raymond}}, \binits{J.C.}},
\bauthor{\bsnm{{Golub}}, \binits{L.}},
\bauthor{\bsnm{{Lin}}, \binits{J.}},
\bauthor{\bsnm{{Chen}}, \binits{H.}},
\bauthor{\bsnm{{Grigis}}, \binits{P.}},
\bauthor{\bsnm{{Testa}}, \binits{P.}},
\bauthor{\bsnm{{Long}}, \binits{D.}}:
\byear{2011},
\batitle{{Observations and Interpretation of a Low Coronal Shock Wave Observed
  in the EUV by the SDO/AIA}}.
\bjtitle{\apj}
\bvolume{738},
\bfpage{160}.
\doiurl{https://doi.org/10.1088/0004-637X/738/2/160}.
\adsurl{2011ApJ...738..160M}.
\end{barticle}
\endbibitem

\bibitem[\protect\citeauthoryear{{Mackay} et~al.}{2010}]{2010SSRv..151..333M}
\begin{barticle}
\bauthor{\bsnm{{Mackay}}, \binits{D.H.}},
\bauthor{\bsnm{{Karpen}}, \binits{J.T.}},
\bauthor{\bsnm{{Ballester}}, \binits{J.L.}},
\bauthor{\bsnm{{Schmieder}}, \binits{B.}},
\bauthor{\bsnm{{Aulanier}}, \binits{G.}}:
\byear{2010},
\batitle{{Physics of Solar Prominences: II{\textemdash}Magnetic Structure and
  Dynamics}}.
\bjtitle{\ssr}
\bvolume{151},
\bfpage{333}.
\doiurl{https://doi.org/10.1007/s11214-010-9628-0}.
\adsurl{2010SSRv..151..333M}.
\end{barticle}
\endbibitem

\bibitem[\protect\citeauthoryear{{Mann} et~al.}{1999}]{1999A&A...348..614M}
\begin{barticle}
\bauthor{\bsnm{{Mann}}, \binits{G.}},
\bauthor{\bsnm{{Jansen}}, \binits{F.}},
\bauthor{\bsnm{{MacDowall}}, \binits{R.J.}},
\bauthor{\bsnm{{Kaiser}}, \binits{M.L.}},
\bauthor{\bsnm{{Stone}}, \binits{R.G.}}:
\byear{1999},
\batitle{{A heliospheric density model and type III radio bursts}}.
\bjtitle{\aap}
\bvolume{348},
\bfpage{614}.
\adsurl{1999A&A...348..614M}.
\end{barticle}
\endbibitem

\bibitem[\protect\citeauthoryear{{McKenzie} and
  {Savage}}{2009}]{2009ApJ...697.1569M}
\begin{barticle}
\bauthor{\bsnm{{McKenzie}}, \binits{D.E.}},
\bauthor{\bsnm{{Savage}}, \binits{S.L.}}:
\byear{2009},
\batitle{{Quantitative Examination of Supra-arcade Downflows in Eruptive Solar
  Flares}}.
\bjtitle{\apj}
\bvolume{697},
\bfpage{1569}.
\doiurl{https://doi.org/10.1088/0004-637X/697/2/1569}.
\adsurl{2009ApJ...697.1569M}.
\end{barticle}
\endbibitem

\bibitem[\protect\citeauthoryear{{McLaughlin}, {Thurgood}, and
  {MacTaggart}}{2012}]{2012A&A...548A..98M}
\begin{barticle}
\bauthor{\bsnm{{McLaughlin}}, \binits{J.A.}},
\bauthor{\bsnm{{Thurgood}}, \binits{J.O.}},
\bauthor{\bsnm{{MacTaggart}}, \binits{D.}}:
\byear{2012},
\batitle{{On the periodicity of oscillatory reconnection}}.
\bjtitle{\aap}
\bvolume{548},
\bfpage{A98}.
\doiurl{https://doi.org/10.1051/0004-6361/201220234}.
\adsurl{2012A&A...548A..98M}.
\end{barticle}
\endbibitem

\bibitem[\protect\citeauthoryear{{McLaughlin}
  et~al.}{2009}]{2009A&A...493..227M}
\begin{barticle}
\bauthor{\bsnm{{McLaughlin}}, \binits{J.A.}},
\bauthor{\bsnm{{De Moortel}}, \binits{I.}},
\bauthor{\bsnm{{Hood}}, \binits{A.W.}},
\bauthor{\bsnm{{Brady}}, \binits{C.S.}}:
\byear{2009},
\batitle{{Nonlinear fast magnetoacoustic wave propagation in the neighbourhood
  of a 2D magnetic X-point: oscillatory reconnection}}.
\bjtitle{\aap}
\bvolume{493},
\bfpage{227}.
\doiurl{https://doi.org/10.1051/0004-6361:200810465}.
\adsurl{2009A&A...493..227M}.
\end{barticle}
\endbibitem

\bibitem[\protect\citeauthoryear{{McLaughlin}
  et~al.}{2012}]{2012ApJ...749...30M}
\begin{barticle}
\bauthor{\bsnm{{McLaughlin}}, \binits{J.A.}},
\bauthor{\bsnm{{Verth}}, \binits{G.}},
\bauthor{\bsnm{{Fedun}}, \binits{V.}},
\bauthor{\bsnm{{Erd{\'e}lyi}}, \binits{R.}}:
\byear{2012},
\batitle{{Generation of Quasi-periodic Waves and Flows in the Solar Atmosphere
  by Oscillatory Reconnection}}.
\bjtitle{\apj}
\bvolume{749},
\bfpage{30}.
\doiurl{https://doi.org/10.1088/0004-637X/749/1/30}.
\adsurl{2012ApJ...749...30M}.
\end{barticle}
\endbibitem

\bibitem[\protect\citeauthoryear{{McLaughlin}
  et~al.}{2018}]{2018SSRv..214...45M}
\begin{barticle}
\bauthor{\bsnm{{McLaughlin}}, \binits{J.A.}},
\bauthor{\bsnm{{Nakariakov}}, \binits{V.M.}},
\bauthor{\bsnm{{Dominique}}, \binits{M.}},
\bauthor{\bsnm{{Jel{\'\i}nek}}, \binits{P.}},
\bauthor{\bsnm{{Takasao}}, \binits{S.}}:
\byear{2018},
\batitle{{Modelling Quasi-Periodic Pulsations in Solar and Stellar Flares}}.
\bjtitle{\ssr}
\bvolume{214},
\bfpage{45}.
\doiurl{https://doi.org/10.1007/s11214-018-0478-5}.
\adsurl{2018SSRv..214...45M}.
\end{barticle}
\endbibitem

\bibitem[\protect\citeauthoryear{{M{\'e}sz{\'a}rosov{\'a}}, {Karlick{\'y}}, and
  {Ryb{\'a}k}}{2011}]{2011SoPh..273..393M}
\begin{barticle}
\bauthor{\bsnm{{M{\'e}sz{\'a}rosov{\'a}}}, \binits{H.}},
\bauthor{\bsnm{{Karlick{\'y}}}, \binits{M.}},
\bauthor{\bsnm{{Ryb{\'a}k}}, \binits{J.}}:
\byear{2011},
\batitle{{Magnetoacoustic Wave Trains in the 11 July 2005 Radio Event with
  Fiber Bursts}}.
\bjtitle{\solphys}
\bvolume{273},
\bfpage{393}.
\doiurl{https://doi.org/10.1007/s11207-011-9794-6}.
\adsurl{2011SoPh..273..393M}.
\end{barticle}
\endbibitem

\bibitem[\protect\citeauthoryear{{M{\'e}sz{\'a}rosov{\'a}}
  et~al.}{2009a}]{2009AdSpR..43.1479M}
\begin{barticle}
\bauthor{\bsnm{{M{\'e}sz{\'a}rosov{\'a}}}, \binits{H.}},
\bauthor{\bsnm{{Sawant}}, \binits{H.S.}},
\bauthor{\bsnm{{Cecatto}}, \binits{J.R.}},
\bauthor{\bsnm{{Ryb{\'a}k}}, \binits{J.}},
\bauthor{\bsnm{{Karlick{\'y}}}, \binits{M.}},
\bauthor{\bsnm{{Fernandes}}, \binits{F.C.R.}},
\bauthor{\bsnm{{de Andrade}}, \binits{M.C.}},
\bauthor{\bsnm{{Ji{\v{r}}i{\v{c}}ka}}, \binits{K.}}:
\byear{2009}a,
\batitle{{Coronal fast wave trains of the decimetric type IV radio event
  observed during the decay phase of the June 6, 2000 flare}}.
\bjtitle{Advances in Space Research}
\bvolume{43},
\bfpage{1479}.
\doiurl{https://doi.org/10.1016/j.asr.2009.01.032}.
\adsurl{2009AdSpR..43.1479M}.
\end{barticle}
\endbibitem

\bibitem[\protect\citeauthoryear{{M{\'e}sz{\'a}rosov{\'a}}
  et~al.}{2009b}]{2009ApJ...697L.108M}
\begin{barticle}
\bauthor{\bsnm{{M{\'e}sz{\'a}rosov{\'a}}}, \binits{H.}},
\bauthor{\bsnm{{Karlick{\'y}}}, \binits{M.}},
\bauthor{\bsnm{{Ryb{\'a}k}}, \binits{J.}},
\bauthor{\bsnm{{Ji{\v{r}}i{\v{c}}ka}}, \binits{K.}}:
\byear{2009}b,
\batitle{{Tadpoles in Wavelet Spectra of a Solar Decimetric Radio Burst}}.
\bjtitle{\apjl}
\bvolume{697},
\bfpage{L108}.
\doiurl{https://doi.org/10.1088/0004-637X/697/2/L108}.
\adsurl{2009ApJ...697L.108M}.
\end{barticle}
\endbibitem

\bibitem[\protect\citeauthoryear{{M{\'e}sz{\'a}rosov{\'a}}
  et~al.}{2013}]{2013SoPh..283..473M}
\begin{barticle}
\bauthor{\bsnm{{M{\'e}sz{\'a}rosov{\'a}}}, \binits{H.}},
\bauthor{\bsnm{{Dud{\'\i}k}}, \binits{J.}},
\bauthor{\bsnm{{Karlick{\'y}}}, \binits{M.}},
\bauthor{\bsnm{{Madsen}}, \binits{F.R.H.}},
\bauthor{\bsnm{{Sawant}}, \binits{H.S.}}:
\byear{2013},
\batitle{{Fast Magnetoacoustic Waves in a Fan Structure Above a Coronal
  Magnetic Null Point}}.
\bjtitle{\solphys}
\bvolume{283},
\bfpage{473}.
\doiurl{https://doi.org/10.1007/s11207-013-0243-6}.
\adsurl{2013SoPh..283..473M}.
\end{barticle}
\endbibitem

\bibitem[\protect\citeauthoryear{{M{\'e}sz{\'a}rosov{\'a}}
  et~al.}{2014}]{2014ApJ...788...44M}
\begin{barticle}
\bauthor{\bsnm{{M{\'e}sz{\'a}rosov{\'a}}}, \binits{H.}},
\bauthor{\bsnm{{Karlick{\'y}}}, \binits{M.}},
\bauthor{\bsnm{{Jel{\'\i}nek}}, \binits{P.}},
\bauthor{\bsnm{{Ryb{\'a}k}}, \binits{J.}}:
\byear{2014},
\batitle{{Magnetoacoustic Waves Propagating along a Dense Slab and Harris
  Current Sheet and their Wavelet Spectra}}.
\bjtitle{\apj}
\bvolume{788},
\bfpage{44}.
\doiurl{https://doi.org/10.1088/0004-637X/788/1/44}.
\adsurl{2014ApJ...788...44M}.
\end{barticle}
\endbibitem

\bibitem[\protect\citeauthoryear{{Miao} et~al.}{2019}]{2019ApJ...871L...2M}
\begin{barticle}
\bauthor{\bsnm{{Miao}}, \binits{Y.H.}},
\bauthor{\bsnm{{Liu}}, \binits{Y.}},
\bauthor{\bsnm{{Shen}}, \binits{Y.D.}},
\bauthor{\bsnm{{Li}}, \binits{H.B.}},
\bauthor{\bsnm{{Abidin}}, \binits{Z.Z.}},
\bauthor{\bsnm{{Elmhamdi}}, \binits{A.}},
\bauthor{\bsnm{{Kordi}}, \binits{A.S.}}:
\byear{2019},
\batitle{{A Quasi-periodic Propagating Wave and Extreme-ultraviolet Waves
  Excited Simultaneously in a Solar Eruption Event}}.
\bjtitle{\apjl}
\bvolume{871},
\bfpage{L2}.
\doiurl{https://doi.org/10.3847/2041-8213/aafaf9}.
\adsurl{2019ApJ...871L...2M}.
\end{barticle}
\endbibitem

\bibitem[\protect\citeauthoryear{{Miao} et~al.}{2020}]{2020ApJ...889..139M}
\begin{barticle}
\bauthor{\bsnm{{Miao}}, \binits{Y.}},
\bauthor{\bsnm{{Liu}}, \binits{Y.}},
\bauthor{\bsnm{{Elmhamdi}}, \binits{A.}},
\bauthor{\bsnm{{Kordi}}, \binits{A.S.}},
\bauthor{\bsnm{{Shen}}, \binits{Y.D.}},
\bauthor{\bsnm{{Al-Shammari}}, \binits{R.}},
\bauthor{\bsnm{{Al-Mosabeh}}, \binits{K.}},
\bauthor{\bsnm{{Jiang}}, \binits{C.}},
\bauthor{\bsnm{{Yuan}}, \binits{D.}}:
\byear{2020},
\batitle{{Two Quasi-periodic Fast-propagating Magnetosonic Wave Events Observed
  in Active Region NOAA 11167}}.
\bjtitle{\apj}
\bvolume{889},
\bfpage{139}.
\doiurl{https://doi.org/10.3847/1538-4357/ab655f}.
\adsurl{2020ApJ...889..139M}.
\end{barticle}
\endbibitem

\bibitem[\protect\citeauthoryear{{Miao} et~al.}{2021}]{2021ApJ...908L..37M}
\begin{barticle}
\bauthor{\bsnm{{Miao}}, \binits{Y.}},
\bauthor{\bsnm{{Li}}, \binits{D.}},
\bauthor{\bsnm{{Yuan}}, \binits{D.}},
\bauthor{\bsnm{{Jiang}}, \binits{C.}},
\bauthor{\bsnm{{Elmhamdi}}, \binits{A.}},
\bauthor{\bsnm{{Zhao}}, \binits{M.}},
\bauthor{\bsnm{{Anfinogentov}}, \binits{S.}}:
\byear{2021},
\batitle{{Diagnosing a Solar Flaring Core with Bidirectional Quasi-periodic
  Fast Propagating Magnetoacoustic Waves}}.
\bjtitle{\apjl}
\bvolume{908},
\bfpage{L37}.
\doiurl{https://doi.org/10.3847/2041-8213/abdfce}.
\adsurl{2021ApJ...908L..37M}.
\end{barticle}
\endbibitem

\bibitem[\protect\citeauthoryear{{Milligan} et~al.}{2017}]{2017ApJ...848L...8M}
\begin{barticle}
\bauthor{\bsnm{{Milligan}}, \binits{R.O.}},
\bauthor{\bsnm{{Fleck}}, \binits{B.}},
\bauthor{\bsnm{{Ireland}}, \binits{J.}},
\bauthor{\bsnm{{Fletcher}}, \binits{L.}},
\bauthor{\bsnm{{Dennis}}, \binits{B.R.}}:
\byear{2017},
\batitle{{Detection of Three-minute Oscillations in Full-disk
  Ly{\ensuremath{\alpha}} Emission during a Solar Flare}}.
\bjtitle{\apjl}
\bvolume{848},
\bfpage{L8}.
\doiurl{https://doi.org/10.3847/2041-8213/aa8f3a}.
\adsurl{2017ApJ...848L...8M}.
\end{barticle}
\endbibitem

\bibitem[\protect\citeauthoryear{{Moreton}}{1960}]{1960AJ.....65U.494M}
\begin{barticle}
\bauthor{\bsnm{{Moreton}}, \binits{G.E.}}:
\byear{1960},
\batitle{{H{\ensuremath{\alpha}} Observations of Flare-Initiated Disturbances
  with Velocities \raisebox{-0.5ex}\textasciitilde1000 km/sec.}}
\bjtitle{\aj}
\bvolume{65},
\bfpage{494}.
\doiurl{https://doi.org/10.1086/108346}.
\adsurl{1960AJ.....65U.494M}.
\end{barticle}
\endbibitem

\bibitem[\protect\citeauthoryear{{Moreton} and
  {Ramsey}}{1960}]{1960PASP...72..357M}
\begin{barticle}
\bauthor{\bsnm{{Moreton}}, \binits{G.E.}},
\bauthor{\bsnm{{Ramsey}}, \binits{H.E.}}:
\byear{1960},
\batitle{{Recent Observations of Dynamical Phenomena Associated with Solar
  Flares}}.
\bjtitle{\pasp}
\bvolume{72},
\bfpage{357}.
\doiurl{https://doi.org/10.1086/127549}.
\adsurl{1960PASP...72..357M}.
\end{barticle}
\endbibitem

\bibitem[\protect\citeauthoryear{{Moses} et~al.}{1997}]{1997SoPh..175..571M}
\begin{barticle}
\bauthor{\bsnm{{Moses}}, \binits{D.}},
\bauthor{\bsnm{{Clette}}, \binits{F.}},
\bauthor{\bsnm{{Delaboudini{\`e}re}}, \binits{J.-P.}},
\bauthor{\bsnm{{Artzner}}, \binits{G.E.}},
\bauthor{\bsnm{{Bougnet}}, \binits{M.}},
\bauthor{\bsnm{{Brunaud}}, \binits{J.}},
\bauthor{\bsnm{{Carabetian}}, \binits{C.}},
\bauthor{\bsnm{{Gabriel}}, \binits{A.H.}},
\bauthor{\bsnm{{Hochedez}}, \binits{J.F.}},
\bauthor{\bsnm{{Millier}}, \binits{F.}},
\bauthor{\bsnm{{Song}}, \binits{X.Y.}},
\bauthor{\bsnm{{Au}}, \binits{B.}},
\bauthor{\bsnm{{Dere}}, \binits{K.P.}},
\bauthor{\bsnm{{Howard}}, \binits{R.A.}},
\bauthor{\bsnm{{Kreplin}}, \binits{R.}},
\bauthor{\bsnm{{Michels}}, \binits{D.J.}},
\bauthor{\bsnm{{Defise}}, \binits{J.M.}},
\bauthor{\bsnm{{Jamar}}, \binits{C.}},
\bauthor{\bsnm{{Rochus}}, \binits{P.}},
\bauthor{\bsnm{{Chauvineau}}, \binits{J.P.}},
\bauthor{\bsnm{{Marioge}}, \binits{J.P.}},
\bauthor{\bsnm{{Catura}}, \binits{R.C.}},
\bauthor{\bsnm{{Lemen}}, \binits{J.R.}},
\bauthor{\bsnm{{Shing}}, \binits{L.}},
\bauthor{\bsnm{{Stern}}, \binits{R.A.}},
\bauthor{\bsnm{{Gurman}}, \binits{J.B.}},
\bauthor{\bsnm{{Neupert}}, \binits{W.M.}},
\bauthor{\bsnm{{Newmark}}, \binits{J.}},
\bauthor{\bsnm{{Thompson}}, \binits{B.}},
\bauthor{\bsnm{{Maucherat}}, \binits{A.}},
\bauthor{\bsnm{{Portier-Fozzani}}, \binits{F.}},
\bauthor{\bsnm{{Berghmans}}, \binits{D.}},
\bauthor{\bsnm{{Cugnon}}, \binits{P.}},
\bauthor{\bsnm{{van Dessel}}, \binits{E.L.}},
\bauthor{\bsnm{{Gabryl}}, \binits{J.R.}}:
\byear{1997},
\batitle{{EIT Observations of the Extreme Ultraviolet Sun}}.
\bjtitle{\solphys}
\bvolume{175},
\bfpage{571}.
\doiurl{https://doi.org/10.1023/A:1004902913117}.
\adsurl{1997SoPh..175..571M}.
\end{barticle}
\endbibitem

\bibitem[\protect\citeauthoryear{{M{\"u}ller}
  et~al.}{2020}]{2020A&A...642A...1M}
\begin{barticle}
\bauthor{\bsnm{{M{\"u}ller}}, \binits{D.}},
\bauthor{\bsnm{{St. Cyr}}, \binits{O.C.}},
\bauthor{\bsnm{{Zouganelis}}, \binits{I.}},
\bauthor{\bsnm{{Gilbert}}, \binits{H.R.}},
\bauthor{\bsnm{{Marsden}}, \binits{R.}},
\bauthor{\bsnm{{Nieves-Chinchilla}}, \binits{T.}},
\bauthor{\bsnm{{Antonucci}}, \binits{E.}},
\bauthor{\bsnm{{Auch{\`e}re}}, \binits{F.}},
\bauthor{\bsnm{{Berghmans}}, \binits{D.}},
\bauthor{\bsnm{{Horbury}}, \binits{T.S.}},
\bauthor{\bsnm{{Howard}}, \binits{R.A.}},
\bauthor{\bsnm{{Krucker}}, \binits{S.}},
\bauthor{\bsnm{{Maksimovic}}, \binits{M.}},
\bauthor{\bsnm{{Owen}}, \binits{C.J.}},
\bauthor{\bsnm{{Rochus}}, \binits{P.}},
\bauthor{\bsnm{{Rodriguez-Pacheco}}, \binits{J.}},
\bauthor{\bsnm{{Romoli}}, \binits{M.}},
\bauthor{\bsnm{{Solanki}}, \binits{S.K.}},
\bauthor{\bsnm{{Bruno}}, \binits{R.}},
\bauthor{\bsnm{{Carlsson}}, \binits{M.}},
\bauthor{\bsnm{{Fludra}}, \binits{A.}},
\bauthor{\bsnm{{Harra}}, \binits{L.}},
\bauthor{\bsnm{{Hassler}}, \binits{D.M.}},
\bauthor{\bsnm{{Livi}}, \binits{S.}},
\bauthor{\bsnm{{Louarn}}, \binits{P.}},
\bauthor{\bsnm{{Peter}}, \binits{H.}},
\bauthor{\bsnm{{Sch{\"u}hle}}, \binits{U.}},
\bauthor{\bsnm{{Teriaca}}, \binits{L.}},
\bauthor{\bsnm{{del Toro Iniesta}}, \binits{J.C.}},
\bauthor{\bsnm{{Wimmer-Schweingruber}}, \binits{R.F.}},
\bauthor{\bsnm{{Marsch}}, \binits{E.}},
\bauthor{\bsnm{{Velli}}, \binits{M.}},
\bauthor{\bsnm{{De Groof}}, \binits{A.}},
\bauthor{\bsnm{{Walsh}}, \binits{A.}},
\bauthor{\bsnm{{Williams}}, \binits{D.}}:
\byear{2020},
\batitle{{The Solar Orbiter mission. Science overview}}.
\bjtitle{\aap}
\bvolume{642},
\bfpage{A1}.
\doiurl{https://doi.org/10.1051/0004-6361/202038467}.
\adsurl{2020A&A...642A...1M}.
\end{barticle}
\endbibitem

\bibitem[\protect\citeauthoryear{{Murawski} and
  {Roberts}}{1993a}]{1993SoPh..145...65M}
\begin{barticle}
\bauthor{\bsnm{{Murawski}}, \binits{K.}},
\bauthor{\bsnm{{Roberts}}, \binits{B.}}:
\byear{1993}a,
\batitle{{Numerical Simulations of Fast Magnetohydrodynamic Waves in a Coronal
  Plasma - Part Four}}.
\bjtitle{\solphys}
\bvolume{145},
\bfpage{65}.
\doiurl{https://doi.org/10.1007/BF00627983}.
\adsurl{1993SoPh..145...65M}.
\end{barticle}
\endbibitem

\bibitem[\protect\citeauthoryear{{Murawski} and
  {Roberts}}{1993b}]{1993SoPh..143...89M}
\begin{barticle}
\bauthor{\bsnm{{Murawski}}, \binits{K.}},
\bauthor{\bsnm{{Roberts}}, \binits{B.}}:
\byear{1993}b,
\batitle{{Numerical Simulations of Fast Magnetohydrodynamic Waves in a Coronal
  Plasma - Part One}}.
\bjtitle{\solphys}
\bvolume{143},
\bfpage{89}.
\doiurl{https://doi.org/10.1007/BF00619098}.
\adsurl{1993SoPh..143...89M}.
\end{barticle}
\endbibitem

\bibitem[\protect\citeauthoryear{{Murawski} and
  {Roberts}}{1993c}]{1993SoPh..144..101M}
\begin{barticle}
\bauthor{\bsnm{{Murawski}}, \binits{K.}},
\bauthor{\bsnm{{Roberts}}, \binits{B.}}:
\byear{1993}c,
\batitle{{Numerical Simulations of Fast Magnetohydrodynamic Waves in a Coronal
  Plasma - Part Two}}.
\bjtitle{\solphys}
\bvolume{144},
\bfpage{101}.
\doiurl{https://doi.org/10.1007/BF00667986}.
\adsurl{1993SoPh..144..101M}.
\end{barticle}
\endbibitem

\bibitem[\protect\citeauthoryear{{Murawski} and
  {Roberts}}{1994}]{1994SoPh..151..305M}
\begin{barticle}
\bauthor{\bsnm{{Murawski}}, \binits{K.}},
\bauthor{\bsnm{{Roberts}}, \binits{B.}}:
\byear{1994},
\batitle{{Time Signatures of Impulsively Generated Waves in a Coronal Plasma}}.
\bjtitle{\solphys}
\bvolume{151},
\bfpage{305}.
\doiurl{https://doi.org/10.1007/BF00679077}.
\adsurl{1994SoPh..151..305M}.
\end{barticle}
\endbibitem

\bibitem[\protect\citeauthoryear{{Murawski}, {Aschwanden}, and
  {Smith}}{1998}]{1998SoPh..179..313M}
\begin{barticle}
\bauthor{\bsnm{{Murawski}}, \binits{K.}},
\bauthor{\bsnm{{Aschwanden}}, \binits{M.J.}},
\bauthor{\bsnm{{Smith}}, \binits{J.M.}}:
\byear{1998},
\batitle{{Impulsively generated MHD waves and their detectability in solar
  coronal loops}}.
\bjtitle{\solphys}
\bvolume{179},
\bfpage{313}.
\doiurl{https://doi.org/10.1023/A:1005012309703}.
\adsurl{1998SoPh..179..313M}.
\end{barticle}
\endbibitem

\bibitem[\protect\citeauthoryear{{Nakariakov} and
  {Kolotkov}}{2020}]{2020ARA&A..58..441N}
\begin{barticle}
\bauthor{\bsnm{{Nakariakov}}, \binits{V.M.}},
\bauthor{\bsnm{{Kolotkov}}, \binits{D.Y.}}:
\byear{2020},
\batitle{{Magnetohydrodynamic Waves in the Solar Corona}}.
\bjtitle{\araa}
\bvolume{58},
\bfpage{441}.
\doiurl{https://doi.org/10.1146/annurev-astro-032320-042940}.
\adsurl{2020ARA&A..58..441N}.
\end{barticle}
\endbibitem

\bibitem[\protect\citeauthoryear{{Nakariakov} and
  {Melnikov}}{2009}]{2009SSRv..149..119N}
\begin{barticle}
\bauthor{\bsnm{{Nakariakov}}, \binits{V.M.}},
\bauthor{\bsnm{{Melnikov}}, \binits{V.F.}}:
\byear{2009},
\batitle{{Quasi-Periodic Pulsations in Solar Flares}}.
\bjtitle{\ssr}
\bvolume{149},
\bfpage{119}.
\doiurl{https://doi.org/10.1007/s11214-009-9536-3}.
\adsurl{2009SSRv..149..119N}.
\end{barticle}
\endbibitem

\bibitem[\protect\citeauthoryear{{Nakariakov} and
  {Ofman}}{2001}]{2001A&A...372L..53N}
\begin{barticle}
\bauthor{\bsnm{{Nakariakov}}, \binits{V.M.}},
\bauthor{\bsnm{{Ofman}}, \binits{L.}}:
\byear{2001},
\batitle{{Determination of the coronal magnetic field by coronal loop
  oscillations}}.
\bjtitle{\aap}
\bvolume{372},
\bfpage{L53}.
\doiurl{https://doi.org/10.1051/0004-6361:20010607}.
\adsurl{2001A&A...372L..53N}.
\end{barticle}
\endbibitem

\bibitem[\protect\citeauthoryear{{Nakariakov} and
  {Verwichte}}{2005}]{2005LRSP....2....3N}
\begin{barticle}
\bauthor{\bsnm{{Nakariakov}}, \binits{V.M.}},
\bauthor{\bsnm{{Verwichte}}, \binits{E.}}:
\byear{2005},
\batitle{{Coronal Waves and Oscillations}}.
\bjtitle{Living Reviews in Solar Physics}
\bvolume{2},
\bfpage{3}.
\doiurl{https://doi.org/10.12942/lrsp-2005-3}.
\adsurl{2005LRSP....2....3N}.
\end{barticle}
\endbibitem

\bibitem[\protect\citeauthoryear{{Nakariakov}, {Pascoe}, and
  {Arber}}{2005}]{2005SSRv..121..115N}
\begin{barticle}
\bauthor{\bsnm{{Nakariakov}}, \binits{V.M.}},
\bauthor{\bsnm{{Pascoe}}, \binits{D.J.}},
\bauthor{\bsnm{{Arber}}, \binits{T.D.}}:
\byear{2005},
\batitle{{Short Quasi-Periodic MHD Waves in Coronal Structures}}.
\bjtitle{\ssr}
\bvolume{121},
\bfpage{115}.
\doiurl{https://doi.org/10.1007/s11214-006-4718-8}.
\adsurl{2005SSRv..121..115N}.
\end{barticle}
\endbibitem

\bibitem[\protect\citeauthoryear{{Nakariakov}
  et~al.}{1999}]{1999Sci...285..862N}
\begin{barticle}
\bauthor{\bsnm{{Nakariakov}}, \binits{V.M.}},
\bauthor{\bsnm{{Ofman}}, \binits{L.}},
\bauthor{\bsnm{{Deluca}}, \binits{E.E.}},
\bauthor{\bsnm{{Roberts}}, \binits{B.}},
\bauthor{\bsnm{{Davila}}, \binits{J.M.}}:
\byear{1999},
\batitle{{TRACE observation of damped coronal loop oscillations: Implications
  for coronal heating}}.
\bjtitle{Science}
\bvolume{285},
\bfpage{862}.
\doiurl{https://doi.org/10.1126/science.285.5429.862}.
\adsurl{1999Sci...285..862N}.
\end{barticle}
\endbibitem

\bibitem[\protect\citeauthoryear{{Nakariakov}
  et~al.}{2004}]{2004MNRAS.349..705N}
\begin{barticle}
\bauthor{\bsnm{{Nakariakov}}, \binits{V.M.}},
\bauthor{\bsnm{{Arber}}, \binits{T.D.}},
\bauthor{\bsnm{{Ault}}, \binits{C.E.}},
\bauthor{\bsnm{{Katsiyannis}}, \binits{A.C.}},
\bauthor{\bsnm{{Williams}}, \binits{D.R.}},
\bauthor{\bsnm{{Keenan}}, \binits{F.P.}}:
\byear{2004},
\batitle{{Time signatures of impulsively generated coronal fast wave trains}}.
\bjtitle{\mnras}
\bvolume{349},
\bfpage{705}.
\doiurl{https://doi.org/10.1111/j.1365-2966.2004.07537.x}.
\adsurl{2004MNRAS.349..705N}.
\end{barticle}
\endbibitem

\bibitem[\protect\citeauthoryear{{Nakariakov}
  et~al.}{2006}]{2006A&A...452..343N}
\begin{barticle}
\bauthor{\bsnm{{Nakariakov}}, \binits{V.M.}},
\bauthor{\bsnm{{Foullon}}, \binits{C.}},
\bauthor{\bsnm{{Verwichte}}, \binits{E.}},
\bauthor{\bsnm{{Young}}, \binits{N.P.}}:
\byear{2006},
\batitle{{Quasi-periodic modulation of solar and stellar flaring emission by
  magnetohydrodynamic oscillations in a nearby loop}}.
\bjtitle{\aap}
\bvolume{452},
\bfpage{343}.
\doiurl{https://doi.org/10.1051/0004-6361:20054608}.
\adsurl{2006A&A...452..343N}.
\end{barticle}
\endbibitem

\bibitem[\protect\citeauthoryear{{Nakariakov}
  et~al.}{2010}]{2010ApJ...708L..47N}
\begin{barticle}
\bauthor{\bsnm{{Nakariakov}}, \binits{V.M.}},
\bauthor{\bsnm{{Foullon}}, \binits{C.}},
\bauthor{\bsnm{{Myagkova}}, \binits{I.N.}},
\bauthor{\bsnm{{Inglis}}, \binits{A.R.}}:
\byear{2010},
\batitle{{Quasi-Periodic Pulsations in the Gamma-Ray Emission of a Solar
  Flare}}.
\bjtitle{\apjl}
\bvolume{708},
\bfpage{L47}.
\doiurl{https://doi.org/10.1088/2041-8205/708/1/L47}.
\adsurl{2010ApJ...708L..47N}.
\end{barticle}
\endbibitem

\bibitem[\protect\citeauthoryear{{Nakariakov}
  et~al.}{2016}]{2016SSRv..200...75N}
\begin{barticle}
\bauthor{\bsnm{{Nakariakov}}, \binits{V.M.}},
\bauthor{\bsnm{{Pilipenko}}, \binits{V.}},
\bauthor{\bsnm{{Heilig}}, \binits{B.}},
\bauthor{\bsnm{{Jel{\'\i}nek}}, \binits{P.}},
\bauthor{\bsnm{{Karlick{\'y}}}, \binits{M.}},
\bauthor{\bsnm{{Klimushkin}}, \binits{D.Y.}},
\bauthor{\bsnm{{Kolotkov}}, \binits{D.Y.}},
\bauthor{\bsnm{{Lee}}, \binits{D.-H.}},
\bauthor{\bsnm{{Nistic{\`o}}}, \binits{G.}},
\bauthor{\bsnm{{Van Doorsselaere}}, \binits{T.}},
\bauthor{\bsnm{{Verth}}, \binits{G.}},
\bauthor{\bsnm{{Zimovets}}, \binits{I.V.}}:
\byear{2016},
\batitle{{Magnetohydrodynamic Oscillations in the Solar Corona and Earth's
  Magnetosphere: Towards Consolidated Understanding}}.
\bjtitle{\ssr}
\bvolume{200},
\bfpage{75}.
\doiurl{https://doi.org/10.1007/s11214-015-0233-0}.
\adsurl{2016SSRv..200...75N}.
\end{barticle}
\endbibitem

\bibitem[\protect\citeauthoryear{{Nakariakov}
  et~al.}{2019}]{2019PPCF...61a4024N}
\begin{barticle}
\bauthor{\bsnm{{Nakariakov}}, \binits{V.M.}},
\bauthor{\bsnm{{Kolotkov}}, \binits{D.Y.}},
\bauthor{\bsnm{{Kupriyanova}}, \binits{E.G.}},
\bauthor{\bsnm{{Mehta}}, \binits{T.}},
\bauthor{\bsnm{{Pugh}}, \binits{C.E.}},
\bauthor{\bsnm{{Lee}}, \binits{D.-H.}},
\bauthor{\bsnm{{Broomhall}}, \binits{A.-M.}}:
\byear{2019},
\batitle{{Non-stationary quasi-periodic pulsations in solar and stellar
  flares}}.
\bjtitle{Plasma Physics and Controlled Fusion}
\bvolume{61},
\bfpage{014024}.
\doiurl{https://doi.org/10.1088/1361-6587/aad97c}.
\adsurl{2019PPCF...61a4024N}.
\end{barticle}
\endbibitem

\bibitem[\protect\citeauthoryear{{Nakariakov}
  et~al.}{2021}]{2021SSRv..217...73N}
\begin{barticle}
\bauthor{\bsnm{{Nakariakov}}, \binits{V.M.}},
\bauthor{\bsnm{{Anfinogentov}}, \binits{S.A.}},
\bauthor{\bsnm{{Antolin}}, \binits{P.}},
\bauthor{\bsnm{{Jain}}, \binits{R.}},
\bauthor{\bsnm{{Kolotkov}}, \binits{D.Y.}},
\bauthor{\bsnm{{Kupriyanova}}, \binits{E.G.}},
\bauthor{\bsnm{{Li}}, \binits{D.}},
\bauthor{\bsnm{{Magyar}}, \binits{N.}},
\bauthor{\bsnm{{Nistic{\`o}}}, \binits{G.}},
\bauthor{\bsnm{{Pascoe}}, \binits{D.J.}},
\bauthor{\bsnm{{Srivastava}}, \binits{A.K.}},
\bauthor{\bsnm{{Terradas}}, \binits{J.}},
\bauthor{\bsnm{{Vasheghani Farahani}}, \binits{S.}},
\bauthor{\bsnm{{Verth}}, \binits{G.}},
\bauthor{\bsnm{{Yuan}}, \binits{D.}},
\bauthor{\bsnm{{Zimovets}}, \binits{I.V.}}:
\byear{2021},
\batitle{{Kink Oscillations of Coronal Loops}}.
\bjtitle{\ssr}
\bvolume{217},
\bfpage{73}.
\doiurl{https://doi.org/10.1007/s11214-021-00847-2}.
\adsurl{2021SSRv..217...73N}.
\end{barticle}
\endbibitem

\bibitem[\protect\citeauthoryear{{Ni} et~al.}{2012}]{2012ApJ...758...20N}
\begin{barticle}
\bauthor{\bsnm{{Ni}}, \binits{L.}},
\bauthor{\bsnm{{Roussev}}, \binits{I.I.}},
\bauthor{\bsnm{{Lin}}, \binits{J.}},
\bauthor{\bsnm{{Ziegler}}, \binits{U.}}:
\byear{2012},
\batitle{{Impact of Temperature-dependent Resistivity and Thermal Conduction on
  Plasmoid Instabilities in Current Sheets in the Solar Corona}}.
\bjtitle{\apj}
\bvolume{758},
\bfpage{20}.
\doiurl{https://doi.org/10.1088/0004-637X/758/1/20}.
\adsurl{2012ApJ...758...20N}.
\end{barticle}
\endbibitem

\bibitem[\protect\citeauthoryear{{Ni} et~al.}{2015}]{2015ApJ...799...79N}
\begin{barticle}
\bauthor{\bsnm{{Ni}}, \binits{L.}},
\bauthor{\bsnm{{Kliem}}, \binits{B.}},
\bauthor{\bsnm{{Lin}}, \binits{J.}},
\bauthor{\bsnm{{Wu}}, \binits{N.}}:
\byear{2015},
\batitle{{Fast Magnetic Reconnection in the Solar Chromosphere Mediated by the
  Plasmoid Instability}}.
\bjtitle{\apj}
\bvolume{799},
\bfpage{79}.
\doiurl{https://doi.org/10.1088/0004-637X/799/1/79}.
\adsurl{2015ApJ...799...79N}.
\end{barticle}
\endbibitem

\bibitem[\protect\citeauthoryear{{Ni} et~al.}{2020}]{2020RSPSA.47690867N}
\begin{barticle}
\bauthor{\bsnm{{Ni}}, \binits{L.}},
\bauthor{\bsnm{{Ji}}, \binits{H.}},
\bauthor{\bsnm{{Murphy}}, \binits{N.A.}},
\bauthor{\bsnm{{Jara-Almonte}}, \binits{J.}}:
\byear{2020},
\batitle{{Magnetic reconnection in partially ionized plasmas}}.
\bjtitle{Proceedings of the Royal Society of London Series A}
\bvolume{476},
\bfpage{20190867}.
\doiurl{https://doi.org/10.1098/rspa.2019.0867}.
\adsurl{2020RSPSA.47690867N}.
\end{barticle}
\endbibitem

\bibitem[\protect\citeauthoryear{{Ning}}{2014}]{2014SoPh..289.1239N}
\begin{barticle}
\bauthor{\bsnm{{Ning}}, \binits{Z.}}:
\byear{2014},
\batitle{{Imaging Observations of X-Ray Quasi-periodic Oscillations at 3 - 6
  keV in the 26 December 2002 Solar Flare}}.
\bjtitle{\solphys}
\bvolume{289},
\bfpage{1239}.
\doiurl{https://doi.org/10.1007/s11207-013-0405-6}.
\adsurl{2014SoPh..289.1239N}.
\end{barticle}
\endbibitem

\bibitem[\protect\citeauthoryear{{Nistic{\`o}}, {Pascoe}, and
  {Nakariakov}}{2014}]{2014AA...569A..12N}
\begin{barticle}
\bauthor{\bsnm{{Nistic{\`o}}}, \binits{G.}},
\bauthor{\bsnm{{Pascoe}}, \binits{D.J.}},
\bauthor{\bsnm{{Nakariakov}}, \binits{V.M.}}:
\byear{2014},
\batitle{{Observation of a high-quality quasi-periodic rapidly propagating wave
  train using SDO/AIA}}.
\bjtitle{\aap}
\bvolume{569},
\bfpage{A12}.
\doiurl{https://doi.org/10.1051/0004-6361/201423763}.
\adsurl{2014A&A...569A..12N}.
\end{barticle}
\endbibitem

\bibitem[\protect\citeauthoryear{{Nitta} et~al.}{2013}]{2013ApJ...776...58N}
\begin{barticle}
\bauthor{\bsnm{{Nitta}}, \binits{N.V.}},
\bauthor{\bsnm{{Schrijver}}, \binits{C.J.}},
\bauthor{\bsnm{{Title}}, \binits{A.M.}},
\bauthor{\bsnm{{Liu}}, \binits{W.}}:
\byear{2013},
\batitle{{Large-scale Coronal Propagating Fronts in Solar Eruptions as Observed
  by the Atmospheric Imaging Assembly on Board the Solar Dynamics
  Observatory{\textemdash}an Ensemble Study}}.
\bjtitle{\apj}
\bvolume{776},
\bfpage{58}.
\doiurl{https://doi.org/10.1088/0004-637X/776/1/58}.
\adsurl{2013ApJ...776...58N}.
\end{barticle}
\endbibitem

\bibitem[\protect\citeauthoryear{{Ofman} and {Liu}}{2018}]{2018ApJ...860...54O}
\begin{barticle}
\bauthor{\bsnm{{Ofman}}, \binits{L.}},
\bauthor{\bsnm{{Liu}}, \binits{W.}}:
\byear{2018},
\batitle{{Quasi-periodic Counter-propagating Fast Magnetosonic Wave Trains from
  Neighboring Flares: SDO/AIA Observations and 3D MHD Modeling}}.
\bjtitle{\apj}
\bvolume{860},
\bfpage{54}.
\doiurl{https://doi.org/10.3847/1538-4357/aac2e8}.
\adsurl{2018ApJ...860...54O}.
\end{barticle}
\endbibitem

\bibitem[\protect\citeauthoryear{{Ofman} and {Sui}}{2006}]{2006ApJ...644L.149O}
\begin{barticle}
\bauthor{\bsnm{{Ofman}}, \binits{L.}},
\bauthor{\bsnm{{Sui}}, \binits{L.}}:
\byear{2006},
\batitle{{Oscillations of Hard X-Ray Flare Emission Observed by RHESSI: Effects
  of Super-Alfv{\'e}nic Beams?}}
\bjtitle{\apjl}
\bvolume{644},
\bfpage{L149}.
\doiurl{https://doi.org/10.1086/505622}.
\adsurl{2006ApJ...644L.149O}.
\end{barticle}
\endbibitem

\bibitem[\protect\citeauthoryear{{Ofman} et~al.}{1997}]{1997ApJ...491L.111O}
\begin{barticle}
\bauthor{\bsnm{{Ofman}}, \binits{L.}},
\bauthor{\bsnm{{Romoli}}, \binits{M.}},
\bauthor{\bsnm{{Poletto}}, \binits{G.}},
\bauthor{\bsnm{{Noci}}, \binits{G.}},
\bauthor{\bsnm{{Kohl}}, \binits{J.L.}}:
\byear{1997},
\batitle{{Ultraviolet Coronagraph Spectrometer Observations of Density
  Fluctuations in the Solar Wind}}.
\bjtitle{\apjl}
\bvolume{491},
\bfpage{L111}.
\doiurl{https://doi.org/10.1086/311067}.
\adsurl{1997ApJ...491L.111O}.
\end{barticle}
\endbibitem

\bibitem[\protect\citeauthoryear{{Ofman} et~al.}{2011}]{2011ApJ...740L..33O}
\begin{barticle}
\bauthor{\bsnm{{Ofman}}, \binits{L.}},
\bauthor{\bsnm{{Liu}}, \binits{W.}},
\bauthor{\bsnm{{Title}}, \binits{A.}},
\bauthor{\bsnm{{Aschwanden}}, \binits{M.}}:
\byear{2011},
\batitle{{Modeling Super-fast Magnetosonic Waves Observed by SDO in Active
  Region Funnels}}.
\bjtitle{\apjl}
\bvolume{740},
\bfpage{L33}.
\doiurl{https://doi.org/10.1088/2041-8205/740/2/L33}.
\adsurl{2011ApJ...740L..33O}.
\end{barticle}
\endbibitem

\bibitem[\protect\citeauthoryear{{Oliver}, {Ruderman}, and
  {Terradas}}{2014}]{2014ApJ...789...48O}
\begin{barticle}
\bauthor{\bsnm{{Oliver}}, \binits{R.}},
\bauthor{\bsnm{{Ruderman}}, \binits{M.S.}},
\bauthor{\bsnm{{Terradas}}, \binits{J.}}:
\byear{2014},
\batitle{{Propagation and Dispersion of Transverse Wave Trains in Magnetic Flux
  Tubes}}.
\bjtitle{\apj}
\bvolume{789},
\bfpage{48}.
\doiurl{https://doi.org/10.1088/0004-637X/789/1/48}.
\adsurl{2014ApJ...789...48O}.
\end{barticle}
\endbibitem

\bibitem[\protect\citeauthoryear{{Oliver}, {Ruderman}, and
  {Terradas}}{2015}]{2015ApJ...806...56O}
\begin{barticle}
\bauthor{\bsnm{{Oliver}}, \binits{R.}},
\bauthor{\bsnm{{Ruderman}}, \binits{M.S.}},
\bauthor{\bsnm{{Terradas}}, \binits{J.}}:
\byear{2015},
\batitle{{Propagation and Dispersion of Sausage Wave Trains in Magnetic Flux
  Tubes}}.
\bjtitle{\apj}
\bvolume{806},
\bfpage{56}.
\doiurl{https://doi.org/10.1088/0004-637X/806/1/56}.
\adsurl{2015ApJ...806...56O}.
\end{barticle}
\endbibitem

\bibitem[\protect\citeauthoryear{{Pant} et~al.}{2016}]{2016SoPh..291.3303P}
\begin{barticle}
\bauthor{\bsnm{{Pant}}, \binits{V.}},
\bauthor{\bsnm{{Mazumder}}, \binits{R.}},
\bauthor{\bsnm{{Yuan}}, \binits{D.}},
\bauthor{\bsnm{{Banerjee}}, \binits{D.}},
\bauthor{\bsnm{{Srivastava}}, \binits{A.K.}},
\bauthor{\bsnm{{Shen}}, \binits{Y.}}:
\byear{2016},
\batitle{{Simultaneous Longitudinal and Transverse Oscillations in an
  Active-Region Filament}}.
\bjtitle{\solphys}
\bvolume{291},
\bfpage{3303}.
\doiurl{https://doi.org/10.1007/s11207-016-1018-7}.
\adsurl{2016SoPh..291.3303P}.
\end{barticle}
\endbibitem

\bibitem[\protect\citeauthoryear{{Parker}}{1988}]{1988ApJ...330..474P}
\begin{barticle}
\bauthor{\bsnm{{Parker}}, \binits{E.N.}}:
\byear{1988},
\batitle{{Nanoflares and the Solar X-Ray Corona}}.
\bjtitle{\apj}
\bvolume{330},
\bfpage{474}.
\doiurl{https://doi.org/10.1086/166485}.
\adsurl{1988ApJ...330..474P}.
\end{barticle}
\endbibitem

\bibitem[\protect\citeauthoryear{{Parks} and
  {Winckler}}{1969}]{1969ApJ...155L.117P}
\begin{barticle}
\bauthor{\bsnm{{Parks}}, \binits{G.K.}},
\bauthor{\bsnm{{Winckler}}, \binits{J.R.}}:
\byear{1969},
\batitle{{Sixteen-Second Periodic Pulsations Observed in the Correlated
  Microwave and Energetic X-Ray Emission from a Solar Flare}}.
\bjtitle{\apjl}
\bvolume{155},
\bfpage{L117}.
\doiurl{https://doi.org/10.1086/180315}.
\adsurl{1969ApJ...155L.117P}.
\end{barticle}
\endbibitem

\bibitem[\protect\citeauthoryear{{Pasachoff} and
  {Landman}}{1984}]{1984SoPh...90..325P}
\begin{barticle}
\bauthor{\bsnm{{Pasachoff}}, \binits{J.M.}},
\bauthor{\bsnm{{Landman}}, \binits{D.A.}}:
\byear{1984},
\batitle{{High Frequency Coronal Oscillations and Coronal Heating}}.
\bjtitle{\solphys}
\bvolume{90},
\bfpage{325}.
\doiurl{https://doi.org/10.1007/BF00173960}.
\adsurl{1984SoPh...90..325P}.
\end{barticle}
\endbibitem

\bibitem[\protect\citeauthoryear{{Pasachoff}
  et~al.}{2002}]{2002SoPh..207..241P}
\begin{barticle}
\bauthor{\bsnm{{Pasachoff}}, \binits{J.M.}},
\bauthor{\bsnm{{Babcock}}, \binits{B.A.}},
\bauthor{\bsnm{{Russell}}, \binits{K.D.}},
\bauthor{\bsnm{{Seaton}}, \binits{D.B.}}:
\byear{2002},
\batitle{{Short-Period Waves That Heat the Corona Detected at the 1999
  Eclipse}}.
\bjtitle{\solphys}
\bvolume{207},
\bfpage{241}.
\doiurl{https://doi.org/10.1023/A:1016297800478}.
\adsurl{2002SoPh..207..241P}.
\end{barticle}
\endbibitem

\bibitem[\protect\citeauthoryear{{Pascoe}, {Goddard}, and
  {Nakariakov}}{2017}]{2017ApJ...847L..21P}
\begin{barticle}
\bauthor{\bsnm{{Pascoe}}, \binits{D.J.}},
\bauthor{\bsnm{{Goddard}}, \binits{C.R.}},
\bauthor{\bsnm{{Nakariakov}}, \binits{V.M.}}:
\byear{2017},
\batitle{{Dispersive Evolution of Nonlinear Fast Magnetoacoustic Wave Trains}}.
\bjtitle{\apjl}
\bvolume{847},
\bfpage{L21}.
\doiurl{https://doi.org/10.3847/2041-8213/aa8db8}.
\adsurl{2017ApJ...847L..21P}.
\end{barticle}
\endbibitem

\bibitem[\protect\citeauthoryear{{Pascoe}, {Nakariakov}, and
  {Kupriyanova}}{2013}]{2013A&A...560A..97P}
\begin{barticle}
\bauthor{\bsnm{{Pascoe}}, \binits{D.J.}},
\bauthor{\bsnm{{Nakariakov}}, \binits{V.M.}},
\bauthor{\bsnm{{Kupriyanova}}, \binits{E.G.}}:
\byear{2013},
\batitle{{Fast magnetoacoustic wave trains in magnetic funnels of the solar
  corona}}.
\bjtitle{\aap}
\bvolume{560},
\bfpage{A97}.
\doiurl{https://doi.org/10.1051/0004-6361/201322678}.
\adsurl{2013A&A...560A..97P}.
\end{barticle}
\endbibitem

\bibitem[\protect\citeauthoryear{{Pascoe}, {Nakariakov}, and
  {Kupriyanova}}{2014}]{2014A&A...568A..20P}
\begin{barticle}
\bauthor{\bsnm{{Pascoe}}, \binits{D.J.}},
\bauthor{\bsnm{{Nakariakov}}, \binits{V.M.}},
\bauthor{\bsnm{{Kupriyanova}}, \binits{E.G.}}:
\byear{2014},
\batitle{{Fast magnetoacoustic wave trains in coronal holes}}.
\bjtitle{\aap}
\bvolume{568},
\bfpage{A20}.
\doiurl{https://doi.org/10.1051/0004-6361/201423931}.
\adsurl{2014A&A...568A..20P}.
\end{barticle}
\endbibitem

\bibitem[\protect\citeauthoryear{{Patsourakos}, {Vourlidas}, and
  {Kliem}}{2010}]{2010A&A...522A.100P}
\begin{barticle}
\bauthor{\bsnm{{Patsourakos}}, \binits{S.}},
\bauthor{\bsnm{{Vourlidas}}, \binits{A.}},
\bauthor{\bsnm{{Kliem}}, \binits{B.}}:
\byear{2010},
\batitle{{Toward understanding the early stages of an impulsively accelerated
  coronal mass ejection. SECCHI observations}}.
\bjtitle{\aap}
\bvolume{522},
\bfpage{A100}.
\doiurl{https://doi.org/10.1051/0004-6361/200913599}.
\adsurl{2010A&A...522A.100P}.
\end{barticle}
\endbibitem

\bibitem[\protect\citeauthoryear{{Porter}, {Klimchuk}, and
  {Sturrock}}{1994}]{1994ApJ...435..482P}
\begin{barticle}
\bauthor{\bsnm{{Porter}}, \binits{L.J.}},
\bauthor{\bsnm{{Klimchuk}}, \binits{J.A.}},
\bauthor{\bsnm{{Sturrock}}, \binits{P.A.}}:
\byear{1994},
\batitle{{The Possible Role of MHD Waves in Heating the Solar Corona}}.
\bjtitle{\apj}
\bvolume{435},
\bfpage{482}.
\doiurl{https://doi.org/10.1086/174830}.
\adsurl{1994ApJ...435..482P}.
\end{barticle}
\endbibitem

\bibitem[\protect\citeauthoryear{{Priest}}{1982}]{1982soma.book.....P}
\begin{bbook}
\bauthor{\bsnm{{Priest}}, \binits{E.R.}}:
\byear{1982},
\bbtitle{{Solar magneto-hydrodynamics}}.
\adsurl{1982soma.book.....P}.
\end{bbook}
\endbibitem

\bibitem[\protect\citeauthoryear{{Priest} and
  {Forbes}}{2002}]{2002A&ARv..10..313P}
\begin{barticle}
\bauthor{\bsnm{{Priest}}, \binits{E.R.}},
\bauthor{\bsnm{{Forbes}}, \binits{T.G.}}:
\byear{2002},
\batitle{{The magnetic nature of solar flares}}.
\bjtitle{\aapr}
\bvolume{10},
\bfpage{313}.
\doiurl{https://doi.org/10.1007/s001590100013}.
\adsurl{2002A&ARv..10..313P}.
\end{barticle}
\endbibitem

\bibitem[\protect\citeauthoryear{{Qu}, {Jiang}, and
  {Chen}}{2017}]{2017ApJ...851...41Q}
\begin{barticle}
\bauthor{\bsnm{{Qu}}, \binits{Z.N.}},
\bauthor{\bsnm{{Jiang}}, \binits{L.Q.}},
\bauthor{\bsnm{{Chen}}, \binits{S.L.}}:
\byear{2017},
\batitle{{Observations of a Fast-mode Magnetosonic Wave Propagating along a
  Curving Coronal Loop on 2011 November 11}}.
\bjtitle{\apj}
\bvolume{851},
\bfpage{41}.
\doiurl{https://doi.org/10.3847/1538-4357/aa9beb}.
\adsurl{2017ApJ...851...41Q}.
\end{barticle}
\endbibitem

\bibitem[\protect\citeauthoryear{{Ramesh}, {Kathiravan}, and
  {Sastry}}{2010}]{2010ApJ...711.1029R}
\begin{barticle}
\bauthor{\bsnm{{Ramesh}}, \binits{R.}},
\bauthor{\bsnm{{Kathiravan}}, \binits{C.}},
\bauthor{\bsnm{{Sastry}}, \binits{C.V.}}:
\byear{2010},
\batitle{{Estimation of Magnetic Field in the Solar Coronal Streamers Through
  Low Frequency Radio Observations}}.
\bjtitle{\apj}
\bvolume{711},
\bfpage{1029}.
\doiurl{https://doi.org/10.1088/0004-637X/711/2/1029}.
\adsurl{2010ApJ...711.1029R}.
\end{barticle}
\endbibitem

\bibitem[\protect\citeauthoryear{{Rast} et~al.}{2021}]{2021SoPh..296...70R}
\begin{barticle}
\bauthor{\bsnm{{Rast}}, \binits{M.P.}},
\bauthor{\bsnm{{Bello Gonz{\'a}lez}}, \binits{N.}},
\bauthor{\bsnm{{Bellot Rubio}}, \binits{L.}},
\bauthor{\bsnm{{Cao}}, \binits{W.}},
\bauthor{\bsnm{{Cauzzi}}, \binits{G.}},
\bauthor{\bsnm{{Deluca}}, \binits{E.}},
\bauthor{\bsnm{{de Pontieu}}, \binits{B.}},
\bauthor{\bsnm{{Fletcher}}, \binits{L.}},
\bauthor{\bsnm{{Gibson}}, \binits{S.E.}},
\bauthor{\bsnm{{Judge}}, \binits{P.G.}},
\bauthor{\bsnm{{Katsukawa}}, \binits{Y.}},
\bauthor{\bsnm{{Kazachenko}}, \binits{M.D.}},
\bauthor{\bsnm{{Khomenko}}, \binits{E.}},
\bauthor{\bsnm{{Landi}}, \binits{E.}},
\bauthor{\bsnm{{Mart{\'\i}nez Pillet}}, \binits{V.}},
\bauthor{\bsnm{{Petrie}}, \binits{G.J.D.}},
\bauthor{\bsnm{{Qiu}}, \binits{J.}},
\bauthor{\bsnm{{Rachmeler}}, \binits{L.A.}},
\bauthor{\bsnm{{Rempel}}, \binits{M.}},
\bauthor{\bsnm{{Schmidt}}, \binits{W.}},
\bauthor{\bsnm{{Scullion}}, \binits{E.}},
\bauthor{\bsnm{{Sun}}, \binits{X.}},
\bauthor{\bsnm{{Welsch}}, \binits{B.T.}},
\bauthor{\bsnm{{Andretta}}, \binits{V.}},
\bauthor{\bsnm{{Antolin}}, \binits{P.}},
\bauthor{\bsnm{{Ayres}}, \binits{T.R.}},
\bauthor{\bsnm{{Balasubramaniam}}, \binits{K.S.}},
\bauthor{\bsnm{{Ballai}}, \binits{I.}},
\bauthor{\bsnm{{Berger}}, \binits{T.E.}},
\bauthor{\bsnm{{Bradshaw}}, \binits{S.J.}},
\bauthor{\bsnm{{Campbell}}, \binits{R.J.}},
\bauthor{\bsnm{{Carlsson}}, \binits{M.}},
\bauthor{\bsnm{{Casini}}, \binits{R.}},
\bauthor{\bsnm{{Centeno}}, \binits{R.}},
\bauthor{\bsnm{{Cranmer}}, \binits{S.R.}},
\bauthor{\bsnm{{Criscuoli}}, \binits{S.}},
\bauthor{\bsnm{{Deforest}}, \binits{C.}},
\bauthor{\bsnm{{Deng}}, \binits{Y.}},
\bauthor{\bsnm{{Erd{\'e}lyi}}, \binits{R.}},
\bauthor{\bsnm{{Fedun}}, \binits{V.}},
\bauthor{\bsnm{{Fischer}}, \binits{C.E.}},
\bauthor{\bsnm{{Gonz{\'a}lez Manrique}}, \binits{S.J.}},
\bauthor{\bsnm{{Hahn}}, \binits{M.}},
\bauthor{\bsnm{{Harra}}, \binits{L.}},
\bauthor{\bsnm{{Henriques}}, \binits{V.M.J.}},
\bauthor{\bsnm{{Hurlburt}}, \binits{N.E.}},
\bauthor{\bsnm{{Jaeggli}}, \binits{S.}},
\bauthor{\bsnm{{Jafarzadeh}}, \binits{S.}},
\bauthor{\bsnm{{Jain}}, \binits{R.}},
\bauthor{\bsnm{{Jefferies}}, \binits{S.M.}},
\bauthor{\bsnm{{Keys}}, \binits{P.H.}},
\bauthor{\bsnm{{Kowalski}}, \binits{A.F.}},
\bauthor{\bsnm{{Kuckein}}, \binits{C.}},
\bauthor{\bsnm{{Kuhn}}, \binits{J.R.}},
\bauthor{\bsnm{{Kuridze}}, \binits{D.}},
\bauthor{\bsnm{{Liu}}, \binits{J.}},
\bauthor{\bsnm{{Liu}}, \binits{W.}},
\bauthor{\bsnm{{Longcope}}, \binits{D.}},
\bauthor{\bsnm{{Mathioudakis}}, \binits{M.}},
\bauthor{\bsnm{{McAteer}}, \binits{R.T.J.}},
\bauthor{\bsnm{{McIntosh}}, \binits{S.W.}},
\bauthor{\bsnm{{McKenzie}}, \binits{D.E.}},
\bauthor{\bsnm{{Miralles}}, \binits{M.P.}},
\bauthor{\bsnm{{Morton}}, \binits{R.J.}},
\bauthor{\bsnm{{Muglach}}, \binits{K.}},
\bauthor{\bsnm{{Nelson}}, \binits{C.J.}},
\bauthor{\bsnm{{Panesar}}, \binits{N.K.}},
\bauthor{\bsnm{{Parenti}}, \binits{S.}},
\bauthor{\bsnm{{Parnell}}, \binits{C.E.}},
\bauthor{\bsnm{{Poduval}}, \binits{B.}},
\bauthor{\bsnm{{Reardon}}, \binits{K.P.}},
\bauthor{\bsnm{{Reep}}, \binits{J.W.}},
\bauthor{\bsnm{{Schad}}, \binits{T.A.}},
\bauthor{\bsnm{{Schmit}}, \binits{D.}},
\bauthor{\bsnm{{Sharma}}, \binits{R.}},
\bauthor{\bsnm{{Socas-Navarro}}, \binits{H.}},
\bauthor{\bsnm{{Srivastava}}, \binits{A.K.}},
\bauthor{\bsnm{{Sterling}}, \binits{A.C.}},
\bauthor{\bsnm{{Suematsu}}, \binits{Y.}},
\bauthor{\bsnm{{Tarr}}, \binits{L.A.}},
\bauthor{\bsnm{{Tiwari}}, \binits{S.}},
\bauthor{\bsnm{{Tritschler}}, \binits{A.}},
\bauthor{\bsnm{{Verth}}, \binits{G.}},
\bauthor{\bsnm{{Vourlidas}}, \binits{A.}},
\bauthor{\bsnm{{Wang}}, \binits{H.}},
\bauthor{\bsnm{{Wang}}, \binits{Y.-M.}},
\bauthor{\bsnm{{NSO and DKIST Project}}},
\bauthor{\bsnm{{DKIST Instrument Scientists}}},
\bauthor{\bsnm{{DKIST Science Working Group}}},
\bauthor{\bsnm{{DKIST Critical Science Plan Community}}}:
\byear{2021},
\batitle{{Critical Science Plan for the Daniel K. Inouye Solar Telescope
  (DKIST)}}.
\bjtitle{\solphys}
\bvolume{296},
\bfpage{70}.
\doiurl{https://doi.org/10.1007/s11207-021-01789-2}.
\adsurl{2021SoPh..296...70R}.
\end{barticle}
\endbibitem

\bibitem[\protect\citeauthoryear{{Reeves} et~al.}{2020}]{2020ApJ...905..165R}
\begin{barticle}
\bauthor{\bsnm{{Reeves}}, \binits{K.K.}},
\bauthor{\bsnm{{Polito}}, \binits{V.}},
\bauthor{\bsnm{{Chen}}, \binits{B.}},
\bauthor{\bsnm{{Galan}}, \binits{G.}},
\bauthor{\bsnm{{Yu}}, \binits{S.}},
\bauthor{\bsnm{{Liu}}, \binits{W.}},
\bauthor{\bsnm{{Li}}, \binits{G.}}:
\byear{2020},
\batitle{{Hot Plasma Flows and Oscillations in the Loop-top Region During the
  2017 September 10 X8.2 Solar Flare}}.
\bjtitle{\apj}
\bvolume{905},
\bfpage{165}.
\doiurl{https://doi.org/10.3847/1538-4357/abc4e0}.
\adsurl{2020ApJ...905..165R}.
\end{barticle}
\endbibitem

\bibitem[\protect\citeauthoryear{{Roberts} and
  {Nakariakov}}{2003}]{roberts2003}
\begin{bchapter}
\bauthor{\bsnm{{Roberts}}, \binits{B.}},
\bauthor{\bsnm{{Nakariakov}}, \binits{V.M.}}:
\byear{2003},
\bctitle{Theory of mhd waves in the solar corona}.
In: \beditor{\bsnm{{Erd{\'e}lyi}}, \binits{K.} \bsuffix{R.~adn~{Petrovay}}},
\beditor{\bsnm{{Roberts}}, \binits{B.}},
\beditor{\bsnm{{Aschwanden}}, \binits{M.}} (eds.)
\bbtitle{Turbulence, Waves and Instabilities in the Solar Plasma},
\bsertitle{NATO Science Series II: Mathematics, Physics and Chemistry}
\bseriesno{124},
\bpublisher{Springer Dordrecht},
\bfpage{167}.
\end{bchapter}
\endbibitem

\bibitem[\protect\citeauthoryear{{Roberts}, {Edwin}, and
  {Benz}}{1983}]{1983Natur.305..688R}
\begin{barticle}
\bauthor{\bsnm{{Roberts}}, \binits{B.}},
\bauthor{\bsnm{{Edwin}}, \binits{P.M.}},
\bauthor{\bsnm{{Benz}}, \binits{A.O.}}:
\byear{1983},
\batitle{{Fast pulsations in the solar corona}}.
\bjtitle{\nat}
\bvolume{305},
\bfpage{688}.
\doiurl{https://doi.org/10.1038/305688a0}.
\adsurl{1983Natur.305..688R}.
\end{barticle}
\endbibitem

\bibitem[\protect\citeauthoryear{{Roberts}, {Edwin}, and
  {Benz}}{1984}]{1984ApJ...279..857R}
\begin{barticle}
\bauthor{\bsnm{{Roberts}}, \binits{B.}},
\bauthor{\bsnm{{Edwin}}, \binits{P.M.}},
\bauthor{\bsnm{{Benz}}, \binits{A.O.}}:
\byear{1984},
\batitle{{On coronal oscillations}}.
\bjtitle{\apj}
\bvolume{279},
\bfpage{857}.
\doiurl{https://doi.org/10.1086/161956}.
\adsurl{1984ApJ...279..857R}.
\end{barticle}
\endbibitem

\bibitem[\protect\citeauthoryear{{Sakurai} et~al.}{2002}]{2002SoPh..209..265S}
\begin{barticle}
\bauthor{\bsnm{{Sakurai}}, \binits{T.}},
\bauthor{\bsnm{{Ichimoto}}, \binits{K.}},
\bauthor{\bsnm{{Raju}}, \binits{K.P.}},
\bauthor{\bsnm{{Singh}}, \binits{J.}}:
\byear{2002},
\batitle{{Spectroscopic Observation of Coronal Waves}}.
\bjtitle{\solphys}
\bvolume{209},
\bfpage{265}.
\doiurl{https://doi.org/10.1023/A:1021297313448}.
\adsurl{2002SoPh..209..265S}.
\end{barticle}
\endbibitem

\bibitem[\protect\citeauthoryear{{Samanta} et~al.}{2016}]{2016SoPh..291..155S}
\begin{barticle}
\bauthor{\bsnm{{Samanta}}, \binits{T.}},
\bauthor{\bsnm{{Singh}}, \binits{J.}},
\bauthor{\bsnm{{Sindhuja}}, \binits{G.}},
\bauthor{\bsnm{{Banerjee}}, \binits{D.}}:
\byear{2016},
\batitle{{Detection of High-Frequency Oscillations and Damping from Multi-slit
  Spectroscopic Observations of the Corona}}.
\bjtitle{\solphys}
\bvolume{291},
\bfpage{155}.
\doiurl{https://doi.org/10.1007/s11207-015-0821-x}.
\adsurl{2016SoPh..291..155S}.
\end{barticle}
\endbibitem

\bibitem[\protect\citeauthoryear{{Savage}, {McKenzie}, and
  {Reeves}}{2012}]{2012ApJ...747L..40S}
\begin{barticle}
\bauthor{\bsnm{{Savage}}, \binits{S.L.}},
\bauthor{\bsnm{{McKenzie}}, \binits{D.E.}},
\bauthor{\bsnm{{Reeves}}, \binits{K.K.}}:
\byear{2012},
\batitle{{Re-interpretation of Supra-arcade Downflows in Solar Flares}}.
\bjtitle{\apjl}
\bvolume{747},
\bfpage{L40}.
\doiurl{https://doi.org/10.1088/2041-8205/747/2/L40}.
\adsurl{2012ApJ...747L..40S}.
\end{barticle}
\endbibitem

\bibitem[\protect\citeauthoryear{{Schrijver}
  et~al.}{2011}]{2011ApJ...738..167S}
\begin{barticle}
\bauthor{\bsnm{{Schrijver}}, \binits{C.J.}},
\bauthor{\bsnm{{Aulanier}}, \binits{G.}},
\bauthor{\bsnm{{Title}}, \binits{A.M.}},
\bauthor{\bsnm{{Pariat}}, \binits{E.}},
\bauthor{\bsnm{{Delann{\'e}e}}, \binits{C.}}:
\byear{2011},
\batitle{{The 2011 February 15 X2 Flare, Ribbons, Coronal Front, and Mass
  Ejection: Interpreting the Three-dimensional Views from the Solar Dynamics
  Observatory and STEREO Guided by Magnetohydrodynamic Flux-rope Modeling}}.
\bjtitle{\apj}
\bvolume{738},
\bfpage{167}.
\doiurl{https://doi.org/10.1088/0004-637X/738/2/167}.
\adsurl{2011ApJ...738..167S}.
\end{barticle}
\endbibitem

\bibitem[\protect\citeauthoryear{{Sharykin}, {Kontar}, and
  {Kuznetsov}}{2018}]{2018SoPh..293..115S}
\begin{barticle}
\bauthor{\bsnm{{Sharykin}}, \binits{I.N.}},
\bauthor{\bsnm{{Kontar}}, \binits{E.P.}},
\bauthor{\bsnm{{Kuznetsov}}, \binits{A.A.}}:
\byear{2018},
\batitle{{LOFAR Observations of Fine Spectral Structure Dynamics in Type IIIb
  Radio Bursts}}.
\bjtitle{\solphys}
\bvolume{293},
\bfpage{115}.
\doiurl{https://doi.org/10.1007/s11207-018-1333-2}.
\adsurl{2018SoPh..293..115S}.
\end{barticle}
\endbibitem

\bibitem[\protect\citeauthoryear{{Shen}}{2021}]{2021RSPSA.47700217S}
\begin{barticle}
\bauthor{\bsnm{{Shen}}, \binits{Y.}}:
\byear{2021},
\batitle{{Observation and modelling of solar jets}}.
\bjtitle{Proceedings of the Royal Society of London Series A}
\bvolume{477},
\bfpage{217}.
\doiurl{https://doi.org/10.1098/rspa.2020.0217}.
\adsurl{2021RSPSA.47700217S}.
\end{barticle}
\endbibitem

\bibitem[\protect\citeauthoryear{{Shen} and {Liu}}{2012a}]{2012ApJ...754....7S}
\begin{barticle}
\bauthor{\bsnm{{Shen}}, \binits{Y.}},
\bauthor{\bsnm{{Liu}}, \binits{Y.}}:
\byear{2012}a,
\batitle{{Evidence for the Wave Nature of an Extreme Ultraviolet Wave Observed
  by the Atmospheric Imaging Assembly on Board the Solar Dynamics
  Observatory}}.
\bjtitle{\apj}
\bvolume{754},
\bfpage{7}.
\doiurl{https://doi.org/10.1088/0004-637X/754/1/7}.
\adsurl{2012ApJ...754....7S}.
\end{barticle}
\endbibitem

\bibitem[\protect\citeauthoryear{{Shen} and {Liu}}{2012b}]{2012ApJ...753...53S}
\begin{barticle}
\bauthor{\bsnm{{Shen}}, \binits{Y.}},
\bauthor{\bsnm{{Liu}}, \binits{Y.}}:
\byear{2012}b,
\batitle{{Observational Study of the Quasi-periodic Fast-propagating
  Magnetosonic Waves and the Associated Flare on 2011 May 30}}.
\bjtitle{\apj}
\bvolume{753},
\bfpage{53}.
\doiurl{https://doi.org/10.1088/0004-637X/753/1/53}.
\adsurl{2012ApJ...753...53S}.
\end{barticle}
\endbibitem

\bibitem[\protect\citeauthoryear{{Shen} and {Liu}}{2012c}]{2012ApJ...752L..23S}
\begin{barticle}
\bauthor{\bsnm{{Shen}}, \binits{Y.}},
\bauthor{\bsnm{{Liu}}, \binits{Y.}}:
\byear{2012}c,
\batitle{{Simultaneous Observations of a Large-scale Wave Event in the Solar
  Atmosphere: From Photosphere to Corona}}.
\bjtitle{\apjl}
\bvolume{752},
\bfpage{L23}.
\doiurl{https://doi.org/10.1088/2041-8205/752/2/L23}.
\adsurl{2012ApJ...752L..23S}.
\end{barticle}
\endbibitem

\bibitem[\protect\citeauthoryear{{Shen}, {Song}, and
  {Liu}}{2018}]{2018MNRAS.477L...6S}
\begin{barticle}
\bauthor{\bsnm{{Shen}}, \binits{Y.}},
\bauthor{\bsnm{{Song}}, \binits{T.}},
\bauthor{\bsnm{{Liu}}, \binits{Y.}}:
\byear{2018},
\batitle{{Dispersively formed quasi-periodic fast magnetosonic wavefronts due
  to the eruption of a nearby mini-filament}}.
\bjtitle{\mnras}
\bvolume{477},
\bfpage{L6}.
\doiurl{https://doi.org/10.1093/mnrasl/sly044}.
\adsurl{2018MNRAS.477L...6S}.
\end{barticle}
\endbibitem

\bibitem[\protect\citeauthoryear{{Shen} et~al.}{2013a}]{2013SoPh..288..585S}
\begin{barticle}
\bauthor{\bsnm{{Shen}}, \binits{Y.-D.}},
\bauthor{\bsnm{{Liu}}, \binits{Y.}},
\bauthor{\bsnm{{Su}}, \binits{J.-T.}},
\bauthor{\bsnm{{Li}}, \binits{H.}},
\bauthor{\bsnm{{Zhang}}, \binits{X.-F.}},
\bauthor{\bsnm{{Tian}}, \binits{Z.-J.}},
\bauthor{\bsnm{{Zhao}}, \binits{R.-J.}},
\bauthor{\bsnm{{Elmhamdi}}, \binits{A.}}:
\byear{2013}a,
\batitle{{Observations of a Quasi-periodic, Fast-Propagating Magnetosonic Wave
  in Multiple Wavelengths and Its Interaction with Other Magnetic Structures}}.
\bjtitle{\solphys}
\bvolume{288},
\bfpage{585}.
\doiurl{https://doi.org/10.1007/s11207-013-0395-4}.
\adsurl{2013SoPh..288..585S}.
\end{barticle}
\endbibitem

\bibitem[\protect\citeauthoryear{{Shen} et~al.}{2020}]{Shen2020}
\begin{barticle}
\bauthor{\bsnm{{Shen}}, \binits{Y.D.}},
\bauthor{\bsnm{{Li}}, \binits{B.}},
\bauthor{\bsnm{{Chen}}, \binits{P.F.}},
\bauthor{\bsnm{{Zhou}}, \binits{X.P.}},
\bauthor{\bsnm{{Liu}}, \binits{Y.}}:
\byear{2020},
\batitle{{Research progress on coronal extreme ultraviolet waves (in
  Chinese)}}.
\bjtitle{Chin Sci Bull}
\bvolume{65},
\bfpage{3909}.
\doiurl{https://doi.org/10.1360/TB-2020-0748}.
\end{barticle}
\endbibitem

\bibitem[\protect\citeauthoryear{{Shen} et~al.}{2013b}]{2013ApJ...773L..33S}
\begin{barticle}
\bauthor{\bsnm{{Shen}}, \binits{Y.}},
\bauthor{\bsnm{{Liu}}, \binits{Y.}},
\bauthor{\bsnm{{Su}}, \binits{J.}},
\bauthor{\bsnm{{Li}}, \binits{H.}},
\bauthor{\bsnm{{Zhao}}, \binits{R.}},
\bauthor{\bsnm{{Tian}}, \binits{Z.}},
\bauthor{\bsnm{{Ichimoto}}, \binits{K.}},
\bauthor{\bsnm{{Shibata}}, \binits{K.}}:
\byear{2013}b,
\batitle{{Diffraction, Refraction, and Reflection of an Extreme-ultraviolet
  Wave Observed during Its Interactions with Remote Active Regions}}.
\bjtitle{\apjl}
\bvolume{773},
\bfpage{L33}.
\doiurl{https://doi.org/10.1088/2041-8205/773/2/L33}.
\adsurl{2013ApJ...773L..33S}.
\end{barticle}
\endbibitem

\bibitem[\protect\citeauthoryear{{Shen} et~al.}{2014a}]{2014ApJ...786..151S}
\begin{barticle}
\bauthor{\bsnm{{Shen}}, \binits{Y.}},
\bauthor{\bsnm{{Ichimoto}}, \binits{K.}},
\bauthor{\bsnm{{Ishii}}, \binits{T.T.}},
\bauthor{\bsnm{{Tian}}, \binits{Z.}},
\bauthor{\bsnm{{Zhao}}, \binits{R.}},
\bauthor{\bsnm{{Shibata}}, \binits{K.}}:
\byear{2014}a,
\batitle{{A Chain of Winking (Oscillating) Filaments Triggered by an Invisible
  Extreme-ultraviolet Wave}}.
\bjtitle{\apj}
\bvolume{786},
\bfpage{151}.
\doiurl{https://doi.org/10.1088/0004-637X/786/2/151}.
\adsurl{2014ApJ...786..151S}.
\end{barticle}
\endbibitem

\bibitem[\protect\citeauthoryear{{Shen} et~al.}{2014b}]{2014ApJ...795..130S}
\begin{barticle}
\bauthor{\bsnm{{Shen}}, \binits{Y.}},
\bauthor{\bsnm{{Liu}}, \binits{Y.D.}},
\bauthor{\bsnm{{Chen}}, \binits{P.F.}},
\bauthor{\bsnm{{Ichimoto}}, \binits{K.}}:
\byear{2014}b,
\batitle{{Simultaneous Transverse Oscillations of a Prominence and a Filament
  and Longitudinal Oscillation of Another Filament Induced by a Single Shock
  Wave}}.
\bjtitle{\apj}
\bvolume{795},
\bfpage{130}.
\doiurl{https://doi.org/10.1088/0004-637X/795/2/130}.
\adsurl{2014ApJ...795..130S}.
\end{barticle}
\endbibitem

\bibitem[\protect\citeauthoryear{{Shen} et~al.}{2017}]{2017ApJ...851..101S}
\begin{barticle}
\bauthor{\bsnm{{Shen}}, \binits{Y.}},
\bauthor{\bsnm{{Liu}}, \binits{Y.}},
\bauthor{\bsnm{{Tian}}, \binits{Z.}},
\bauthor{\bsnm{{Qu}}, \binits{Z.}}:
\byear{2017},
\batitle{{On a Small-scale EUV Wave: The Driving Mechanism and the Associated
  Oscillating Filament}}.
\bjtitle{\apj}
\bvolume{851},
\bfpage{101}.
\doiurl{https://doi.org/10.3847/1538-4357/aa9af0}.
\adsurl{2017ApJ...851..101S}.
\end{barticle}
\endbibitem

\bibitem[\protect\citeauthoryear{{Shen} et~al.}{2018a}]{2018ApJ...853....1S}
\begin{barticle}
\bauthor{\bsnm{{Shen}}, \binits{Y.}},
\bauthor{\bsnm{{Liu}}, \binits{Y.}},
\bauthor{\bsnm{{Song}}, \binits{T.}},
\bauthor{\bsnm{{Tian}}, \binits{Z.}}:
\byear{2018}a,
\batitle{{A Quasi-periodic Fast-propagating Magnetosonic Wave Associated with
  the Eruption of a Magnetic Flux Rope}}.
\bjtitle{\apj}
\bvolume{853},
\bfpage{1}.
\doiurl{https://doi.org/10.3847/1538-4357/aaa3ff}.
\adsurl{2018ApJ...853....1S}.
\end{barticle}
\endbibitem

\bibitem[\protect\citeauthoryear{{Shen} et~al.}{2018b}]{2018MNRAS.480L..63S}
\begin{barticle}
\bauthor{\bsnm{{Shen}}, \binits{Y.}},
\bauthor{\bsnm{{Tang}}, \binits{Z.}},
\bauthor{\bsnm{{Li}}, \binits{H.}},
\bauthor{\bsnm{{Liu}}, \binits{Y.}}:
\byear{2018}b,
\batitle{{Coronal EUV, QFP, and kink waves simultaneously launched during the
  course of jet-loop interaction}}.
\bjtitle{\mnras}
\bvolume{480},
\bfpage{L63}.
\doiurl{https://doi.org/10.1093/mnrasl/sly127}.
\adsurl{2018MNRAS.480L..63S}.
\end{barticle}
\endbibitem

\bibitem[\protect\citeauthoryear{{Shen} et~al.}{2018c}]{2018ApJ...860L...8S}
\begin{barticle}
\bauthor{\bsnm{{Shen}}, \binits{Y.}},
\bauthor{\bsnm{{Tang}}, \binits{Z.}},
\bauthor{\bsnm{{Miao}}, \binits{Y.}},
\bauthor{\bsnm{{Su}}, \binits{J.}},
\bauthor{\bsnm{{Liu}}, \binits{Y.}}:
\byear{2018}c,
\batitle{{EUV Waves Driven by the Sudden Expansion of Transequatorial Loops
  Caused by Coronal Jets}}.
\bjtitle{\apjl}
\bvolume{860},
\bfpage{L8}.
\doiurl{https://doi.org/10.3847/2041-8213/aac8dd}.
\adsurl{2018ApJ...860L...8S}.
\end{barticle}
\endbibitem

\bibitem[\protect\citeauthoryear{{Shen} et~al.}{2019}]{2019ApJ...873...22S}
\begin{barticle}
\bauthor{\bsnm{{Shen}}, \binits{Y.}},
\bauthor{\bsnm{{Chen}}, \binits{P.F.}},
\bauthor{\bsnm{{Liu}}, \binits{Y.D.}},
\bauthor{\bsnm{{Shibata}}, \binits{K.}},
\bauthor{\bsnm{{Tang}}, \binits{Z.}},
\bauthor{\bsnm{{Liu}}, \binits{Y.}}:
\byear{2019},
\batitle{{First Unambiguous Imaging of Large-scale Quasi-periodic
  Extreme-ultraviolet Wave or Shock}}.
\bjtitle{\apj}
\bvolume{873},
\bfpage{22}.
\doiurl{https://doi.org/10.3847/1538-4357/ab01dd}.
\adsurl{2019ApJ...873...22S}.
\end{barticle}
\endbibitem

\bibitem[\protect\citeauthoryear{{Shen} et~al.}{2022}]{shen2022}
\begin{botherref}
\oauthor{\bsnm{{Shen}}, \binits{Y.}},
\oauthor{\bsnm{{Zhou}}, \binits{X.}},
\oauthor{\bsnm{{Tang}}, \binits{Z.}},
\oauthor{\bsnm{{Duan}}, \binits{Y.}},
\oauthor{\bsnm{{Zhou}}, \binits{C.}},
\oauthor{\bsnm{{Tan}}, \binits{S.}}:
2022,
Coronagraph white-light observation of a broad qfp wave train associated with a
  failed breakout eruption.
\textit{in prepare}.
\end{botherref}
\endbibitem

\bibitem[\protect\citeauthoryear{{Shestov}, {Nakariakov}, and
  {Kuzin}}{2015}]{2015ApJ...814..135S}
\begin{barticle}
\bauthor{\bsnm{{Shestov}}, \binits{S.}},
\bauthor{\bsnm{{Nakariakov}}, \binits{V.M.}},
\bauthor{\bsnm{{Kuzin}}, \binits{S.}}:
\byear{2015},
\batitle{{Fast Magnetoacoustic Wave Trains of Sausage Symmetry in Cylindrical
  Waveguides of the Solar Corona}}.
\bjtitle{\apj}
\bvolume{814},
\bfpage{135}.
\doiurl{https://doi.org/10.1088/0004-637X/814/2/135}.
\adsurl{2015ApJ...814..135S}.
\end{barticle}
\endbibitem

\bibitem[\protect\citeauthoryear{{Shibata} and
  {Magara}}{2011}]{2011LRSP....8....6S}
\begin{barticle}
\bauthor{\bsnm{{Shibata}}, \binits{K.}},
\bauthor{\bsnm{{Magara}}, \binits{T.}}:
\byear{2011},
\batitle{{Solar Flares: Magnetohydrodynamic Processes}}.
\bjtitle{Living Reviews in Solar Physics}
\bvolume{8},
\bfpage{6}.
\doiurl{https://doi.org/10.12942/lrsp-2011-6}.
\adsurl{2011LRSP....8....6S}.
\end{barticle}
\endbibitem

\bibitem[\protect\citeauthoryear{{Shibata} and
  {Takasao}}{2016}]{2016ASSL..427..373S}
\begin{bbook}
\bauthor{\bsnm{{Shibata}}, \binits{K.}},
\bauthor{\bsnm{{Takasao}}, \binits{S.}}:
\byear{2016},
In: \beditor{\bsnm{{Gonzalez}}, \binits{W.}},
\beditor{\bsnm{{Parker}}, \binits{E.}} (eds.)
\bbtitle{{Fractal Reconnection in Solar and Stellar Environments}}
\bseriesno{427},
\bfpage{373}.
\doiurl{https://doi.org/10.1007/978-3-319-26432-5\_10}.
\adsurl{2016ASSL..427..373S}.
\end{bbook}
\endbibitem

\bibitem[\protect\citeauthoryear{{Shibata} and
  {Tanuma}}{2001}]{2001EP&S...53..473S}
\begin{barticle}
\bauthor{\bsnm{{Shibata}}, \binits{K.}},
\bauthor{\bsnm{{Tanuma}}, \binits{S.}}:
\byear{2001},
\batitle{{Plasmoid-induced-reconnection and fractal reconnection}}.
\bjtitle{Earth, Planets, and Space}
\bvolume{53},
\bfpage{473}.
\doiurl{https://doi.org/10.1186/BF03353258}.
\adsurl{2001EP&S...53..473S}.
\end{barticle}
\endbibitem

\bibitem[\protect\citeauthoryear{{Singh} et~al.}{1997}]{1997SoPh..170..235S}
\begin{barticle}
\bauthor{\bsnm{{Singh}}, \binits{J.}},
\bauthor{\bsnm{{Cowsik}}, \binits{R.}},
\bauthor{\bsnm{{Raveendran}}, \binits{A.V.}},
\bauthor{\bsnm{{Bagare}}, \binits{S.P.}},
\bauthor{\bsnm{{Saxena}}, \binits{A.K.}},
\bauthor{\bsnm{{Sundararaman}}, \binits{K.}},
\bauthor{\bsnm{{Krishan}}, \binits{V.}},
\bauthor{\bsnm{{Naidu}}, \binits{N.}},
\bauthor{\bsnm{{Samson}}, \binits{J.P.A.}},
\bauthor{\bsnm{{Gabriel}}, \binits{F.}}:
\byear{1997},
\batitle{{Detection of Short-Period Coronal Oscillations during the Total Solar
  Eclipse of 24 October, 1995}}.
\bjtitle{\solphys}
\bvolume{170},
\bfpage{235}.
\doiurl{https://doi.org/10.1023/A:1004943924584}.
\adsurl{1997SoPh..170..235S}.
\end{barticle}
\endbibitem

\bibitem[\protect\citeauthoryear{{Subramanian}, {Ebenezer}, and
  {Raveesha}}{2010}]{2010ASSP...19..482S}
\begin{barticle}
\bauthor{\bsnm{{Subramanian}}, \binits{K.R.}},
\bauthor{\bsnm{{Ebenezer}}, \binits{E.}},
\bauthor{\bsnm{{Raveesha}}, \binits{K.H.}}:
\byear{2010},
\batitle{{Coronal Magnetic Field Estimation Using Type-II Radio Bursts}}.
\bjtitle{Astrophysics and Space Science Proceedings}
\bvolume{19},
\bfpage{482}.
\doiurl{https://doi.org/10.1007/978-3-642-02859-5\_62}.
\adsurl{2010ASSP...19..482S}.
\end{barticle}
\endbibitem

\bibitem[\protect\citeauthoryear{{Sych} et~al.}{2009}]{2009A&A...505..791S}
\begin{barticle}
\bauthor{\bsnm{{Sych}}, \binits{R.}},
\bauthor{\bsnm{{Nakariakov}}, \binits{V.M.}},
\bauthor{\bsnm{{Karlicky}}, \binits{M.}},
\bauthor{\bsnm{{Anfinogentov}}, \binits{S.}}:
\byear{2009},
\batitle{{Relationship between wave processes in sunspots and quasi-periodic
  pulsations in active region flares}}.
\bjtitle{\aap}
\bvolume{505},
\bfpage{791}.
\doiurl{https://doi.org/10.1051/0004-6361/200912132}.
\adsurl{2009A&A...505..791S}.
\end{barticle}
\endbibitem

\bibitem[\protect\citeauthoryear{{Takasao} and
  {Shibata}}{2016}]{2016ApJ...823..150T}
\begin{barticle}
\bauthor{\bsnm{{Takasao}}, \binits{S.}},
\bauthor{\bsnm{{Shibata}}, \binits{K.}}:
\byear{2016},
\batitle{{Above-the-loop-top Oscillation and Quasi-periodic Coronal Wave
  Generation in Solar Flares}}.
\bjtitle{\apj}
\bvolume{823},
\bfpage{150}.
\doiurl{https://doi.org/10.3847/0004-637X/823/2/150}.
\adsurl{2016ApJ...823..150T}.
\end{barticle}
\endbibitem

\bibitem[\protect\citeauthoryear{{Thompson} et~al.}{1998}]{1998GeoRL..25.2465T}
\begin{barticle}
\bauthor{\bsnm{{Thompson}}, \binits{B.J.}},
\bauthor{\bsnm{{Plunkett}}, \binits{S.P.}},
\bauthor{\bsnm{{Gurman}}, \binits{J.B.}},
\bauthor{\bsnm{{Newmark}}, \binits{J.S.}},
\bauthor{\bsnm{{St. Cyr}}, \binits{O.C.}},
\bauthor{\bsnm{{Michels}}, \binits{D.J.}}:
\byear{1998},
\batitle{{SOHO/EIT observations of an Earth-directed coronal mass ejection on
  May 12, 1997}}.
\bjtitle{\grl}
\bvolume{25},
\bfpage{2465}.
\doiurl{https://doi.org/10.1029/98GL50429}.
\adsurl{1998GeoRL..25.2465T}.
\end{barticle}
\endbibitem

\bibitem[\protect\citeauthoryear{{Thompson} et~al.}{1999}]{1999ApJ...517L.151T}
\begin{barticle}
\bauthor{\bsnm{{Thompson}}, \binits{B.J.}},
\bauthor{\bsnm{{Gurman}}, \binits{J.B.}},
\bauthor{\bsnm{{Neupert}}, \binits{W.M.}},
\bauthor{\bsnm{{Newmark}}, \binits{J.S.}},
\bauthor{\bsnm{{Delaboudini{\`e}re}}, \binits{J.-P.}},
\bauthor{\bsnm{{Cyr}}, \binits{O.C.S.}},
\bauthor{\bsnm{{Stezelberger}}, \binits{S.}},
\bauthor{\bsnm{{Dere}}, \binits{K.P.}},
\bauthor{\bsnm{{Howard}}, \binits{R.A.}},
\bauthor{\bsnm{{Michels}}, \binits{D.J.}}:
\byear{1999},
\batitle{{SOHO/EIT Observations of the 1997 April 7 Coronal Transient: Possible
  Evidence of Coronal Moreton Waves}}.
\bjtitle{\apjl}
\bvolume{517},
\bfpage{L151}.
\doiurl{https://doi.org/10.1086/312030}.
\adsurl{1999ApJ...517L.151T}.
\end{barticle}
\endbibitem

\bibitem[\protect\citeauthoryear{{Thurgood}, {Pontin}, and
  {McLaughlin}}{2017}]{2017ApJ...844....2T}
\begin{barticle}
\bauthor{\bsnm{{Thurgood}}, \binits{J.O.}},
\bauthor{\bsnm{{Pontin}}, \binits{D.I.}},
\bauthor{\bsnm{{McLaughlin}}, \binits{J.A.}}:
\byear{2017},
\batitle{{Three-dimensional Oscillatory Magnetic Reconnection}}.
\bjtitle{\apj}
\bvolume{844},
\bfpage{2}.
\doiurl{https://doi.org/10.3847/1538-4357/aa79fa}.
\adsurl{2017ApJ...844....2T}.
\end{barticle}
\endbibitem

\bibitem[\protect\citeauthoryear{{Thurgood}, {Pontin}, and
  {McLaughlin}}{2019}]{2019A&A...621A.106T}
\begin{barticle}
\bauthor{\bsnm{{Thurgood}}, \binits{J.O.}},
\bauthor{\bsnm{{Pontin}}, \binits{D.I.}},
\bauthor{\bsnm{{McLaughlin}}, \binits{J.A.}}:
\byear{2019},
\batitle{{On the periodicity of linear and nonlinear oscillatory
  reconnection}}.
\bjtitle{\aap}
\bvolume{621},
\bfpage{A106}.
\doiurl{https://doi.org/10.1051/0004-6361/201834369}.
\adsurl{2019A&A...621A.106T}.
\end{barticle}
\endbibitem

\bibitem[\protect\citeauthoryear{{Tian} et~al.}{2021}]{2021SoPh..296...47T}
\begin{barticle}
\bauthor{\bsnm{{Tian}}, \binits{H.}},
\bauthor{\bsnm{{Harra}}, \binits{L.}},
\bauthor{\bsnm{{Baker}}, \binits{D.}},
\bauthor{\bsnm{{Brooks}}, \binits{D.H.}},
\bauthor{\bsnm{{Xia}}, \binits{L.}}:
\byear{2021},
\batitle{{Upflows in the Upper Solar Atmosphere}}.
\bjtitle{\solphys}
\bvolume{296},
\bfpage{47}.
\doiurl{https://doi.org/10.1007/s11207-021-01792-7}.
\adsurl{2021SoPh..296...47T}.
\end{barticle}
\endbibitem

\bibitem[\protect\citeauthoryear{{Torrence} and
  {Compo}}{1998}]{1998BAMS...79...61T}
\begin{barticle}
\bauthor{\bsnm{{Torrence}}, \binits{C.}},
\bauthor{\bsnm{{Compo}}, \binits{G.P.}}:
\byear{1998},
\batitle{{A Practical Guide to Wavelet Analysis.}}
\bjtitle{Bulletin of the American Meteorological Society}
\bvolume{79},
\bfpage{61}.
\adsurl{1998BAMS...79...61T}.
\end{barticle}
\endbibitem

\bibitem[\protect\citeauthoryear{{Uchida}}{1968}]{1968SoPh....4...30U}
\begin{barticle}
\bauthor{\bsnm{{Uchida}}, \binits{Y.}}:
\byear{1968},
\batitle{{Propagation of Hydromagnetic Disturbances in the Solar Corona and
  Moreton's Wave Phenomenon}}.
\bjtitle{\solphys}
\bvolume{4},
\bfpage{30}.
\doiurl{https://doi.org/10.1007/BF00146996}.
\adsurl{1968SoPh....4...30U}.
\end{barticle}
\endbibitem

\bibitem[\protect\citeauthoryear{{Uchida}}{1970}]{1970PASJ...22..341U}
\begin{barticle}
\bauthor{\bsnm{{Uchida}}, \binits{Y.}}:
\byear{1970},
\batitle{{Diagnosis of Coronal Magnetic Structure by Flare-Associated
  Hydromagnetic Disturbances}}.
\bjtitle{\pasj}
\bvolume{22},
\bfpage{341}.
\adsurl{1970PASJ...22..341U}.
\end{barticle}
\endbibitem

\bibitem[\protect\citeauthoryear{{Van Doorsselaere}, {Kupriyanova}, and
  {Yuan}}{2016}]{2016SoPh..291.3143V}
\begin{barticle}
\bauthor{\bsnm{{Van Doorsselaere}}, \binits{T.}},
\bauthor{\bsnm{{Kupriyanova}}, \binits{E.G.}},
\bauthor{\bsnm{{Yuan}}, \binits{D.}}:
\byear{2016},
\batitle{{Quasi-periodic Pulsations in Solar and Stellar Flares: An Overview of
  Recent Results (Invited Review)}}.
\bjtitle{\solphys}
\bvolume{291},
\bfpage{3143}.
\doiurl{https://doi.org/10.1007/s11207-016-0977-z}.
\adsurl{2016SoPh..291.3143V}.
\end{barticle}
\endbibitem

\bibitem[\protect\citeauthoryear{{Van Doorsselaere}
  et~al.}{2011}]{2011ApJ...740...90V}
\begin{barticle}
\bauthor{\bsnm{{Van Doorsselaere}}, \binits{T.}},
\bauthor{\bsnm{{De Groof}}, \binits{A.}},
\bauthor{\bsnm{{Zender}}, \binits{J.}},
\bauthor{\bsnm{{Berghmans}}, \binits{D.}},
\bauthor{\bsnm{{Goossens}}, \binits{M.}}:
\byear{2011},
\batitle{{LYRA Observations of Two Oscillation Modes in a Single Flare}}.
\bjtitle{\apj}
\bvolume{740},
\bfpage{90}.
\doiurl{https://doi.org/10.1088/0004-637X/740/2/90}.
\adsurl{2011ApJ...740...90V}.
\end{barticle}
\endbibitem

\bibitem[\protect\citeauthoryear{{Van Doorsselaere}
  et~al.}{2020}]{2020SSRv..216..140V}
\begin{barticle}
\bauthor{\bsnm{{Van Doorsselaere}}, \binits{T.}},
\bauthor{\bsnm{{Srivastava}}, \binits{A.K.}},
\bauthor{\bsnm{{Antolin}}, \binits{P.}},
\bauthor{\bsnm{{Magyar}}, \binits{N.}},
\bauthor{\bsnm{{Vasheghani Farahani}}, \binits{S.}},
\bauthor{\bsnm{{Tian}}, \binits{H.}},
\bauthor{\bsnm{{Kolotkov}}, \binits{D.}},
\bauthor{\bsnm{{Ofman}}, \binits{L.}},
\bauthor{\bsnm{{Guo}}, \binits{M.}},
\bauthor{\bsnm{{Arregui}}, \binits{I.}},
\bauthor{\bsnm{{De Moortel}}, \binits{I.}},
\bauthor{\bsnm{{Pascoe}}, \binits{D.}}:
\byear{2020},
\batitle{{Coronal Heating by MHD Waves}}.
\bjtitle{\ssr}
\bvolume{216},
\bfpage{140}.
\doiurl{https://doi.org/10.1007/s11214-020-00770-y}.
\adsurl{2020SSRv..216..140V}.
\end{barticle}
\endbibitem

\bibitem[\protect\citeauthoryear{{Verwichte}, {Nakariakov}, and
  {Cooper}}{2005}]{2005A&A...430L..65V}
\begin{barticle}
\bauthor{\bsnm{{Verwichte}}, \binits{E.}},
\bauthor{\bsnm{{Nakariakov}}, \binits{V.M.}},
\bauthor{\bsnm{{Cooper}}, \binits{F.C.}}:
\byear{2005},
\batitle{{Transverse waves in a post-flare supra-arcade}}.
\bjtitle{\aap}
\bvolume{430},
\bfpage{L65}.
\doiurl{https://doi.org/10.1051/0004-6361:200400133}.
\adsurl{2005A&A...430L..65V}.
\end{barticle}
\endbibitem

\bibitem[\protect\citeauthoryear{{Wang}, {Chen}, and
  {Ding}}{2021}]{2021ApJ...911L...8W}
\begin{barticle}
\bauthor{\bsnm{{Wang}}, \binits{C.}},
\bauthor{\bsnm{{Chen}}, \binits{F.}},
\bauthor{\bsnm{{Ding}}, \binits{M.}}:
\byear{2021},
\batitle{{Exploring the Nature of EUV Waves in a Radiative Magnetohydrodynamic
  Simulation}}.
\bjtitle{\apjl}
\bvolume{911},
\bfpage{L8}.
\doiurl{https://doi.org/10.3847/2041-8213/abefe6}.
\adsurl{2021ApJ...911L...8W}.
\end{barticle}
\endbibitem

\bibitem[\protect\citeauthoryear{{Wang} et~al.}{2021}]{2021SSRv..217...34W}
\begin{barticle}
\bauthor{\bsnm{{Wang}}, \binits{T.}},
\bauthor{\bsnm{{Ofman}}, \binits{L.}},
\bauthor{\bsnm{{Yuan}}, \binits{D.}},
\bauthor{\bsnm{{Reale}}, \binits{F.}},
\bauthor{\bsnm{{Kolotkov}}, \binits{D.Y.}},
\bauthor{\bsnm{{Srivastava}}, \binits{A.K.}}:
\byear{2021},
\batitle{{Slow-Mode Magnetoacoustic Waves in Coronal Loops}}.
\bjtitle{\ssr}
\bvolume{217},
\bfpage{34}.
\doiurl{https://doi.org/10.1007/s11214-021-00811-0}.
\adsurl{2021SSRv..217...34W}.
\end{barticle}
\endbibitem

\bibitem[\protect\citeauthoryear{{Wang} and
  {Zhang}}{2007}]{2007ApJ...665.1428W}
\begin{barticle}
\bauthor{\bsnm{{Wang}}, \binits{Y.}},
\bauthor{\bsnm{{Zhang}}, \binits{J.}}:
\byear{2007},
\batitle{{A Comparative Study between Eruptive X-Class Flares Associated with
  Coronal Mass Ejections and Confined X-Class Flares}}.
\bjtitle{\apj}
\bvolume{665},
\bfpage{1428}.
\doiurl{https://doi.org/10.1086/519765}.
\adsurl{2007ApJ...665.1428W}.
\end{barticle}
\endbibitem

\bibitem[\protect\citeauthoryear{{Wang}}{2000}]{2000ApJ...543L..89W}
\begin{barticle}
\bauthor{\bsnm{{Wang}}, \binits{Y.-M.}}:
\byear{2000},
\batitle{{EIT Waves and Fast-Mode Propagation in the Solar Corona}}.
\bjtitle{\apjl}
\bvolume{543},
\bfpage{L89}.
\doiurl{https://doi.org/10.1086/318178}.
\adsurl{2000ApJ...543L..89W}.
\end{barticle}
\endbibitem

\bibitem[\protect\citeauthoryear{{Warmuth}}{2015}]{2015LRSP...12....3W}
\begin{barticle}
\bauthor{\bsnm{{Warmuth}}, \binits{A.}}:
\byear{2015},
\batitle{{Large-scale Globally Propagating Coronal Waves}}.
\bjtitle{Living Reviews in Solar Physics}
\bvolume{12},
\bfpage{3}.
\doiurl{https://doi.org/10.1007/lrsp-2015-3}.
\adsurl{2015LRSP...12....3W}.
\end{barticle}
\endbibitem

\bibitem[\protect\citeauthoryear{{White} and
  {Kundu}}{1997}]{1997SoPh..174...31W}
\begin{barticle}
\bauthor{\bsnm{{White}}, \binits{S.M.}},
\bauthor{\bsnm{{Kundu}}, \binits{M.R.}}:
\byear{1997},
\batitle{{Radio Observations of Gyroresonance Emission from Coronal Magnetic
  Fields}}.
\bjtitle{\solphys}
\bvolume{174},
\bfpage{31}.
\doiurl{https://doi.org/10.1023/A:1004975528106}.
\adsurl{1997SoPh..174...31W}.
\end{barticle}
\endbibitem

\bibitem[\protect\citeauthoryear{{Williams} et~al.}{2001}]{2001MNRAS.326..428W}
\begin{barticle}
\bauthor{\bsnm{{Williams}}, \binits{D.R.}},
\bauthor{\bsnm{{Phillips}}, \binits{K.J.H.}},
\bauthor{\bsnm{{Rudawy}}, \binits{P.}},
\bauthor{\bsnm{{Mathioudakis}}, \binits{M.}},
\bauthor{\bsnm{{Gallagher}}, \binits{P.T.}},
\bauthor{\bsnm{{O'Shea}}, \binits{E.}},
\bauthor{\bsnm{{Keenan}}, \binits{F.P.}},
\bauthor{\bsnm{{Read}}, \binits{P.}},
\bauthor{\bsnm{{Rompolt}}, \binits{B.}}:
\byear{2001},
\batitle{{High-frequency oscillations in a solar active region coronal loop}}.
\bjtitle{\mnras}
\bvolume{326},
\bfpage{428}.
\doiurl{https://doi.org/10.1046/j.1365-8711.2001.04491.x}.
\adsurl{2001MNRAS.326..428W}.
\end{barticle}
\endbibitem

\bibitem[\protect\citeauthoryear{{Williams} et~al.}{2002}]{2002MNRAS.336..747W}
\begin{barticle}
\bauthor{\bsnm{{Williams}}, \binits{D.R.}},
\bauthor{\bsnm{{Mathioudakis}}, \binits{M.}},
\bauthor{\bsnm{{Gallagher}}, \binits{P.T.}},
\bauthor{\bsnm{{Phillips}}, \binits{K.J.H.}},
\bauthor{\bsnm{{McAteer}}, \binits{R.T.J.}},
\bauthor{\bsnm{{Keenan}}, \binits{F.P.}},
\bauthor{\bsnm{{Rudawy}}, \binits{P.}},
\bauthor{\bsnm{{Katsiyannis}}, \binits{A.C.}}:
\byear{2002},
\batitle{{An observational study of a magneto-acoustic wave in the solar
  corona}}.
\bjtitle{\mnras}
\bvolume{336},
\bfpage{747}.
\doiurl{https://doi.org/10.1046/j.1365-8711.2002.05764.x}.
\adsurl{2002MNRAS.336..747W}.
\end{barticle}
\endbibitem

\bibitem[\protect\citeauthoryear{{Withbroe} and
  {Noyes}}{1977}]{1977ARA&A..15..363W}
\begin{barticle}
\bauthor{\bsnm{{Withbroe}}, \binits{G.L.}},
\bauthor{\bsnm{{Noyes}}, \binits{R.W.}}:
\byear{1977},
\batitle{{Mass and energy flow in the solar chromosphere and corona.}}
\bjtitle{\araa}
\bvolume{15},
\bfpage{363}.
\doiurl{https://doi.org/10.1146/annurev.aa.15.090177.002051}.
\adsurl{1977ARA&A..15..363W}.
\end{barticle}
\endbibitem

\bibitem[\protect\citeauthoryear{{Wu} et~al.}{2001}]{2001JGR...10625089W}
\begin{barticle}
\bauthor{\bsnm{{Wu}}, \binits{S.T.}},
\bauthor{\bsnm{{Zheng}}, \binits{H.}},
\bauthor{\bsnm{{Wang}}, \binits{S.}},
\bauthor{\bsnm{{Thompson}}, \binits{B.J.}},
\bauthor{\bsnm{{Plunkett}}, \binits{S.P.}},
\bauthor{\bsnm{{Zhao}}, \binits{X.P.}},
\bauthor{\bsnm{{Dryer}}, \binits{M.}}:
\byear{2001},
\batitle{{Three-dimensional numerical simulation of MHD waves observed by the
  Extreme Ultraviolet Imaging Telescope}}.
\bjtitle{\jgr}
\bvolume{106},
\bfpage{25089}.
\doiurl{https://doi.org/10.1029/2000JA000447}.
\adsurl{2001JGR...10625089W}.
\end{barticle}
\endbibitem

\bibitem[\protect\citeauthoryear{{Wuelser} et~al.}{2004}]{2004SPIE.5171..111W}
\begin{bchapter}
\bauthor{\bsnm{{Wuelser}}, \binits{J.-P.}},
\bauthor{\bsnm{{Lemen}}, \binits{J.R.}},
\bauthor{\bsnm{{Tarbell}}, \binits{T.D.}},
\bauthor{\bsnm{{Wolfson}}, \binits{C.J.}},
\bauthor{\bsnm{{Cannon}}, \binits{J.C.}},
\bauthor{\bsnm{{Carpenter}}, \binits{B.A.}},
\bauthor{\bsnm{{Duncan}}, \binits{D.W.}},
\bauthor{\bsnm{{Gradwohl}}, \binits{G.S.}},
\bauthor{\bsnm{{Meyer}}, \binits{S.B.}},
\bauthor{\bsnm{{Moore}}, \binits{A.S.}},
\bauthor{\bsnm{{Navarro}}, \binits{R.L.}},
\bauthor{\bsnm{{Pearson}}, \binits{J.D.}},
\bauthor{\bsnm{{Rossi}}, \binits{G.R.}},
\bauthor{\bsnm{{Springer}}, \binits{L.A.}},
\bauthor{\bsnm{{Howard}}, \binits{R.A.}},
\bauthor{\bsnm{{Moses}}, \binits{J.D.}},
\bauthor{\bsnm{{Newmark}}, \binits{J.S.}},
\bauthor{\bsnm{{Delaboudiniere}}, \binits{J.-P.}},
\bauthor{\bsnm{{Artzner}}, \binits{G.E.}},
\bauthor{\bsnm{{Auchere}}, \binits{F.}},
\bauthor{\bsnm{{Bougnet}}, \binits{M.}},
\bauthor{\bsnm{{Bouyries}}, \binits{P.}},
\bauthor{\bsnm{{Bridou}}, \binits{F.}},
\bauthor{\bsnm{{Clotaire}}, \binits{J.-Y.}},
\bauthor{\bsnm{{Colas}}, \binits{G.}},
\bauthor{\bsnm{{Delmotte}}, \binits{F.}},
\bauthor{\bsnm{{Jerome}}, \binits{A.}},
\bauthor{\bsnm{{Lamare}}, \binits{M.}},
\bauthor{\bsnm{{Mercier}}, \binits{R.}},
\bauthor{\bsnm{{Mullot}}, \binits{M.}},
\bauthor{\bsnm{{Ravet}}, \binits{M.-F.}},
\bauthor{\bsnm{{Song}}, \binits{X.}},
\bauthor{\bsnm{{Bothmer}}, \binits{V.}},
\bauthor{\bsnm{{Deutsch}}, \binits{W.}}:
\byear{2004},
\bctitle{{EUVI: the STEREO-SECCHI extreme ultraviolet imager}}.
In: \beditor{\bsnm{{Fineschi}}, \binits{S.}},
\beditor{\bsnm{{Gummin}}, \binits{M.A.}} (eds.)
\bbtitle{Telescopes and Instrumentation for Solar Astrophysics},
\bsertitle{Society of Photo-Optical Instrumentation Engineers (SPIE) Conference
  Series}
\bseriesno{5171},
\bfpage{111}.
\doiurl{https://doi.org/10.1117/12.506877}.
\adsurl{2004SPIE.5171..111W}.
\end{bchapter}
\endbibitem

\bibitem[\protect\citeauthoryear{{Xia} and
  {Zharkova}}{2018}]{2018A&A...620A.121X}
\begin{barticle}
\bauthor{\bsnm{{Xia}}, \binits{Q.}},
\bauthor{\bsnm{{Zharkova}}, \binits{V.}}:
\byear{2018},
\batitle{{Particle acceleration in coalescent and squashed magnetic islands. I.
  Test particle approach}}.
\bjtitle{\aap}
\bvolume{620},
\bfpage{A121}.
\doiurl{https://doi.org/10.1051/0004-6361/201833599}.
\adsurl{2018A&A...620A.121X}.
\end{barticle}
\endbibitem

\bibitem[\protect\citeauthoryear{{Xue} et~al.}{2020}]{2020ApJ...898...88X}
\begin{barticle}
\bauthor{\bsnm{{Xue}}, \binits{J.}},
\bauthor{\bsnm{{Su}}, \binits{Y.}},
\bauthor{\bsnm{{Li}}, \binits{H.}},
\bauthor{\bsnm{{Zhao}}, \binits{X.}}:
\byear{2020},
\batitle{{Thermodynamical Evolution of Supra-arcade Downflows}}.
\bjtitle{\apj}
\bvolume{898},
\bfpage{88}.
\doiurl{https://doi.org/10.3847/1538-4357/ab9a3d}.
\adsurl{2020ApJ...898...88X}.
\end{barticle}
\endbibitem

\bibitem[\protect\citeauthoryear{{Xue} et~al.}{2019}]{2019ApJ...874L..27X}
\begin{barticle}
\bauthor{\bsnm{{Xue}}, \binits{Z.}},
\bauthor{\bsnm{{Yan}}, \binits{X.}},
\bauthor{\bsnm{{Jin}}, \binits{C.}},
\bauthor{\bsnm{{Yang}}, \binits{L.}},
\bauthor{\bsnm{{Wang}}, \binits{J.}},
\bauthor{\bsnm{{Li}}, \binits{Q.}},
\bauthor{\bsnm{{Zhao}}, \binits{L.}}:
\byear{2019},
\batitle{{A Small-scale Oscillatory Reconnection and the Associated Formation
  and Disappearance of a Solar Flux Rope}}.
\bjtitle{\apjl}
\bvolume{874},
\bfpage{L27}.
\doiurl{https://doi.org/10.3847/2041-8213/ab1135}.
\adsurl{2019ApJ...874L..27X}.
\end{barticle}
\endbibitem

\bibitem[\protect\citeauthoryear{{Yang} et~al.}{2013}]{2013ApJ...775...39Y}
\begin{barticle}
\bauthor{\bsnm{{Yang}}, \binits{L.}},
\bauthor{\bsnm{{Zhang}}, \binits{J.}},
\bauthor{\bsnm{{Liu}}, \binits{W.}},
\bauthor{\bsnm{{Li}}, \binits{T.}},
\bauthor{\bsnm{{Shen}}, \binits{Y.}}:
\byear{2013},
\batitle{{SDO/AIA and Hinode/EIS Observations of Interaction between an EUV
  Wave and Active Region Loops}}.
\bjtitle{\apj}
\bvolume{775},
\bfpage{39}.
\doiurl{https://doi.org/10.1088/0004-637X/775/1/39}.
\adsurl{2013ApJ...775...39Y}.
\end{barticle}
\endbibitem

\bibitem[\protect\citeauthoryear{{Yang} et~al.}{2015}]{2015ApJ...800..111Y}
\begin{barticle}
\bauthor{\bsnm{{Yang}}, \binits{L.}},
\bauthor{\bsnm{{Zhang}}, \binits{L.}},
\bauthor{\bsnm{{He}}, \binits{J.}},
\bauthor{\bsnm{{Peter}}, \binits{H.}},
\bauthor{\bsnm{{Tu}}, \binits{C.}},
\bauthor{\bsnm{{Wang}}, \binits{L.}},
\bauthor{\bsnm{{Zhang}}, \binits{S.}},
\bauthor{\bsnm{{Feng}}, \binits{X.}}:
\byear{2015},
\batitle{{Numerical Simulation of Fast-mode Magnetosonic Waves Excited by
  Plasmoid Ejections in the Solar Corona}}.
\bjtitle{\apj}
\bvolume{800},
\bfpage{111}.
\doiurl{https://doi.org/10.1088/0004-637X/800/2/111}.
\adsurl{2015ApJ...800..111Y}.
\end{barticle}
\endbibitem

\bibitem[\protect\citeauthoryear{{Young} et~al.}{1961}]{1961ApJ...133..243Y}
\begin{barticle}
\bauthor{\bsnm{{Young}}, \binits{C.W.}},
\bauthor{\bsnm{{Spencer}}, \binits{C.L.}},
\bauthor{\bsnm{{Moreton}}, \binits{G.E.}},
\bauthor{\bsnm{{Roberts}}, \binits{J.A.}}:
\byear{1961},
\batitle{{A Preliminary Study of the Dynamic Spectra of Solar Radio Bursts in
  the Frequency Range 500-950 Mc/s.}}
\bjtitle{\apj}
\bvolume{133},
\bfpage{243}.
\doiurl{https://doi.org/10.1086/147019}.
\adsurl{1961ApJ...133..243Y}.
\end{barticle}
\endbibitem

\bibitem[\protect\citeauthoryear{{Yu} et~al.}{2015}]{2015ApJ...814...60Y}
\begin{barticle}
\bauthor{\bsnm{{Yu}}, \binits{H.}},
\bauthor{\bsnm{{Li}}, \binits{B.}},
\bauthor{\bsnm{{Chen}}, \binits{S.-X.}},
\bauthor{\bsnm{{Guo}}, \binits{M.-Z.}}:
\byear{2015},
\batitle{{Kink and Sausage Modes in Nonuniform Magnetic Slabs with Continuous
  Transverse Density Distributions}}.
\bjtitle{\apj}
\bvolume{814},
\bfpage{60}.
\doiurl{https://doi.org/10.1088/0004-637X/814/1/60}.
\adsurl{2015ApJ...814...60Y}.
\end{barticle}
\endbibitem

\bibitem[\protect\citeauthoryear{{Yu} et~al.}{2016}]{2016ApJ...833...51Y}
\begin{barticle}
\bauthor{\bsnm{{Yu}}, \binits{H.}},
\bauthor{\bsnm{{Li}}, \binits{B.}},
\bauthor{\bsnm{{Chen}}, \binits{S.-X.}},
\bauthor{\bsnm{{Xiong}}, \binits{M.}},
\bauthor{\bsnm{{Guo}}, \binits{M.-Z.}}:
\byear{2016},
\batitle{{Impulsively Generated Sausage Waves in Coronal Tubes with
  Transversally Continuous Structuring}}.
\bjtitle{\apj}
\bvolume{833},
\bfpage{51}.
\doiurl{https://doi.org/10.3847/1538-4357/833/1/51}.
\adsurl{2016ApJ...833...51Y}.
\end{barticle}
\endbibitem

\bibitem[\protect\citeauthoryear{{Yu} et~al.}{2017}]{2017ApJ...836....1Y}
\begin{barticle}
\bauthor{\bsnm{{Yu}}, \binits{H.}},
\bauthor{\bsnm{{Li}}, \binits{B.}},
\bauthor{\bsnm{{Chen}}, \binits{S.-X.}},
\bauthor{\bsnm{{Xiong}}, \binits{M.}},
\bauthor{\bsnm{{Guo}}, \binits{M.-Z.}}:
\byear{2017},
\batitle{{Impulsively Generated Wave Trains in Coronal Structures. I. Effects
  of Transverse Structuring on Sausage Waves in Pressureless Tubes}}.
\bjtitle{\apj}
\bvolume{836},
\bfpage{1}.
\doiurl{https://doi.org/10.3847/1538-4357/836/1/1}.
\adsurl{2017ApJ...836....1Y}.
\end{barticle}
\endbibitem

\bibitem[\protect\citeauthoryear{{Yu} and {Chen}}{2019}]{2019ApJ...872...71Y}
\begin{barticle}
\bauthor{\bsnm{{Yu}}, \binits{S.}},
\bauthor{\bsnm{{Chen}}, \binits{B.}}:
\byear{2019},
\batitle{{Possible Detection of Subsecond-period Propagating
  Magnetohydrodynamics Waves in Post-reconnection Magnetic Loops during a
  Two-ribbon Solar Flare}}.
\bjtitle{\apj}
\bvolume{872},
\bfpage{71}.
\doiurl{https://doi.org/10.3847/1538-4357/aaff6d}.
\adsurl{2019ApJ...872...71Y}.
\end{barticle}
\endbibitem

\bibitem[\protect\citeauthoryear{{Yuan}, {Li}, and
  {Walsh}}{2016}]{2016ApJ...828...17Y}
\begin{barticle}
\bauthor{\bsnm{{Yuan}}, \binits{D.}},
\bauthor{\bsnm{{Li}}, \binits{B.}},
\bauthor{\bsnm{{Walsh}}, \binits{R.W.}}:
\byear{2016},
\batitle{{Secondary Fast Magnetoacoustic Waves Trapped in Randomly Structured
  Plasmas}}.
\bjtitle{\apj}
\bvolume{828},
\bfpage{17}.
\doiurl{https://doi.org/10.3847/0004-637X/828/1/17}.
\adsurl{2016ApJ...828...17Y}.
\end{barticle}
\endbibitem

\bibitem[\protect\citeauthoryear{{Yuan} et~al.}{2013}]{2013AA...554A.144Y}
\begin{barticle}
\bauthor{\bsnm{{Yuan}}, \binits{D.}},
\bauthor{\bsnm{{Shen}}, \binits{Y.}},
\bauthor{\bsnm{{Liu}}, \binits{Y.}},
\bauthor{\bsnm{{Nakariakov}}, \binits{V.M.}},
\bauthor{\bsnm{{Tan}}, \binits{B.}},
\bauthor{\bsnm{{Huang}}, \binits{J.}}:
\byear{2013},
\batitle{{Distinct propagating fast wave trains associated with flaring energy
  releases}}.
\bjtitle{\aap}
\bvolume{554},
\bfpage{A144}.
\doiurl{https://doi.org/10.1051/0004-6361/201321435}.
\adsurl{2013A&A...554A.144Y}.
\end{barticle}
\endbibitem

\bibitem[\protect\citeauthoryear{{Yuan} et~al.}{2015}]{2015ApJ...799..221Y}
\begin{barticle}
\bauthor{\bsnm{{Yuan}}, \binits{D.}},
\bauthor{\bsnm{{Pascoe}}, \binits{D.J.}},
\bauthor{\bsnm{{Nakariakov}}, \binits{V.M.}},
\bauthor{\bsnm{{Li}}, \binits{B.}},
\bauthor{\bsnm{{Keppens}}, \binits{R.}}:
\byear{2015},
\batitle{{Evolution of Fast Magnetoacoustic Pulses in Randomly Structured
  Coronal Plasmas}}.
\bjtitle{\apj}
\bvolume{799},
\bfpage{221}.
\doiurl{https://doi.org/10.1088/0004-637X/799/2/221}.
\adsurl{2015ApJ...799..221Y}.
\end{barticle}
\endbibitem

\bibitem[\protect\citeauthoryear{{Yuan} et~al.}{2019}]{2019ApJ...886L..25Y}
\begin{barticle}
\bauthor{\bsnm{{Yuan}}, \binits{D.}},
\bauthor{\bsnm{{Feng}}, \binits{S.}},
\bauthor{\bsnm{{Li}}, \binits{D.}},
\bauthor{\bsnm{{Ning}}, \binits{Z.}},
\bauthor{\bsnm{{Tan}}, \binits{B.}}:
\byear{2019},
\batitle{{A Compact Source for Quasi-periodic Pulsation in an M-class Solar
  Flare}}.
\bjtitle{\apjl}
\bvolume{886},
\bfpage{L25}.
\doiurl{https://doi.org/10.3847/2041-8213/ab5648}.
\adsurl{2019ApJ...886L..25Y}.
\end{barticle}
\endbibitem

\bibitem[\protect\citeauthoryear{{Zhang} and {Ji}}{2018}]{2018ApJ...860..113Z}
\begin{barticle}
\bauthor{\bsnm{{Zhang}}, \binits{Q.M.}},
\bauthor{\bsnm{{Ji}}, \binits{H.S.}}:
\byear{2018},
\batitle{{Vertical Oscillation of a Coronal Cavity Triggered by an EUV Wave}}.
\bjtitle{\apj}
\bvolume{860},
\bfpage{113}.
\doiurl{https://doi.org/10.3847/1538-4357/aac37e}.
\adsurl{2018ApJ...860..113Z}.
\end{barticle}
\endbibitem

\bibitem[\protect\citeauthoryear{{Zhang}, {Li}, and
  {Ning}}{2016}]{2016ApJ...832...65Z}
\begin{barticle}
\bauthor{\bsnm{{Zhang}}, \binits{Q.M.}},
\bauthor{\bsnm{{Li}}, \binits{D.}},
\bauthor{\bsnm{{Ning}}, \binits{Z.J.}}:
\byear{2016},
\batitle{{Chromospheric Condensation and Quasi-periodic Pulsations in a
  Circular-ribbon Flare}}.
\bjtitle{\apj}
\bvolume{832},
\bfpage{65}.
\doiurl{https://doi.org/10.3847/0004-637X/832/1/65}.
\adsurl{2016ApJ...832...65Z}.
\end{barticle}
\endbibitem

\bibitem[\protect\citeauthoryear{{Zhang} et~al.}{2015}]{2015AA...581A..78Z}
\begin{barticle}
\bauthor{\bsnm{{Zhang}}, \binits{Y.}},
\bauthor{\bsnm{{Zhang}}, \binits{J.}},
\bauthor{\bsnm{{Wang}}, \binits{J.}},
\bauthor{\bsnm{{Nakariakov}}, \binits{V.M.}}:
\byear{2015},
\batitle{{Coexisting fast and slow propagating waves of the extreme-UV
  intensity in solar coronal plasma structures}}.
\bjtitle{\aap}
\bvolume{581},
\bfpage{A78}.
\doiurl{https://doi.org/10.1051/0004-6361/201525621}.
\adsurl{2015A&A...581A..78Z}.
\end{barticle}
\endbibitem

\bibitem[\protect\citeauthoryear{{Zhao} and
  {Hoeksema}}{1994}]{1994SoPh..151...91Z}
\begin{barticle}
\bauthor{\bsnm{{Zhao}}, \binits{X.}},
\bauthor{\bsnm{{Hoeksema}}, \binits{J.T.}}:
\byear{1994},
\batitle{{A Coronal Magnetic Field Model with Horizontal Volume and Sheet
  Currents}}.
\bjtitle{\solphys}
\bvolume{151},
\bfpage{91}.
\doiurl{https://doi.org/10.1007/BF00654084}.
\adsurl{1994SoPh..151...91Z}.
\end{barticle}
\endbibitem

\bibitem[\protect\citeauthoryear{{Zhou} et~al.}{2021a}]{2021arXiv210909285Z}
\begin{botherref}
\oauthor{\bsnm{{Zhou}}, \binits{C.}},
\oauthor{\bsnm{{Shen}}, \binits{Y.}},
\oauthor{\bsnm{{Zhou}}, \binits{X.}},
\oauthor{\bsnm{{Tang}}, \binits{Z.}},
\oauthor{\bsnm{{Duan}}, \binits{Y.}},
\oauthor{\bsnm{{Tan}}, \binits{S.}}:
2021a,
{Sympathetic Filament Eruptions within a Fan-spine Magnetic System}.
\textit{arXiv e-prints},
arXiv:2109.09285.
\adsurl{2021arXiv210909285Z}.
\end{botherref}
\endbibitem

\bibitem[\protect\citeauthoryear{{Zhou} et~al.}{2021b}]{2021SoPh..296..169Z}
\begin{barticle}
\bauthor{\bsnm{{Zhou}}, \binits{X.}},
\bauthor{\bsnm{{Shen}}, \binits{Y.}},
\bauthor{\bsnm{{Su}}, \binits{J.}},
\bauthor{\bsnm{{Tang}}, \binits{Z.}},
\bauthor{\bsnm{{Zhou}}, \binits{C.}},
\bauthor{\bsnm{{Duan}}, \binits{Y.}},
\bauthor{\bsnm{{Tan}}, \binits{S.}}:
\byear{2021}b,
\batitle{{CME-Driven and Flare-Ignited Fast Magnetosonic Waves Detected in a
  Solar Eruption}}.
\bjtitle{\solphys}
\bvolume{296},
\bfpage{169}.
\doiurl{https://doi.org/10.1007/s11207-021-01913-2}.
\adsurl{2021SoPh..296..169Z}.
\end{barticle}
\endbibitem

\bibitem[\protect\citeauthoryear{{Zhou} et~al.}{2021c}]{zhou2021b}
\begin{botherref}
\oauthor{\bsnm{{Zhou}}, \binits{X.}},
\oauthor{\bsnm{{Shen}}, \binits{Y.}},
\oauthor{\bsnm{{Hu}}, \binits{H.}},
\oauthor{\bsnm{{Liu}}, \binits{Y.D.}},
\oauthor{\bsnm{{Su}}, \binits{J.}},
\oauthor{\bsnm{{Tang}}, \binits{Z.}},
\oauthor{\bsnm{{Zhou}}, \binits{C.}},
\oauthor{\bsnm{{Duan}}, \binits{Y.}}:
2021c,
{Observations of a Quasi-periodic Large-scale EUV Wave Driven by Flare Pressure
  Pulses}.
\textit{Astrophys. J. Lett. to be submitted}.
\end{botherref}
\endbibitem

\bibitem[\protect\citeauthoryear{{Zimovets} et~al.}{2021}]{2021SSRv..217...66Z}
\begin{barticle}
\bauthor{\bsnm{{Zimovets}}, \binits{I.V.}},
\bauthor{\bsnm{{McLaughlin}}, \binits{J.A.}},
\bauthor{\bsnm{{Srivastava}}, \binits{A.K.}},
\bauthor{\bsnm{{Kolotkov}}, \binits{D.Y.}},
\bauthor{\bsnm{{Kuznetsov}}, \binits{A.A.}},
\bauthor{\bsnm{{Kupriyanova}}, \binits{E.G.}},
\bauthor{\bsnm{{Cho}}, \binits{I.-H.}},
\bauthor{\bsnm{{Inglis}}, \binits{A.R.}},
\bauthor{\bsnm{{Reale}}, \binits{F.}},
\bauthor{\bsnm{{Pascoe}}, \binits{D.J.}},
\bauthor{\bsnm{{Tian}}, \binits{H.}},
\bauthor{\bsnm{{Yuan}}, \binits{D.}},
\bauthor{\bsnm{{Li}}, \binits{D.}},
\bauthor{\bsnm{{Zhang}}, \binits{Q.M.}}:
\byear{2021},
\batitle{{Quasi-Periodic Pulsations in Solar and Stellar Flares: A Review of
  Underpinning Physical Mechanisms and Their Predicted Observational
  Signatures}}.
\bjtitle{\ssr}
\bvolume{217},
\bfpage{66}.
\doiurl{https://doi.org/10.1007/s11214-021-00840-9}.
\adsurl{2021SSRv..217...66Z}.
\end{barticle}
\endbibitem

\bibitem[\protect\citeauthoryear{{Zong} and {Dai}}{2017}]{2017ApJ...834L..15Z}
\begin{barticle}
\bauthor{\bsnm{{Zong}}, \binits{W.}},
\bauthor{\bsnm{{Dai}}, \binits{Y.}}:
\byear{2017},
\batitle{{Mode Conversion of a Solar Extreme-ultraviolet Wave over a Coronal
  Cavity}}.
\bjtitle{\apjl}
\bvolume{834},
\bfpage{L15}.
\doiurl{https://doi.org/10.3847/2041-8213/834/2/L15}.
\adsurl{2017ApJ...834L..15Z}.
\end{barticle}
\endbibitem

\end{thebibliography}

\end{article} 
\end{document}